\pdfoutput=1
\PassOptionsToPackage{table}{xcolor}
\documentclass[11pt]{article}
\usepackage{amssymb}
\usepackage{acl}

\usepackage{times}
\usepackage{latexsym}
\usepackage[T1]{fontenc}
\usepackage[utf8]{inputenc}
\usepackage{microtype}
\usepackage{float}

\usepackage{xspace}
\usepackage[table]{xcolor}
\usepackage[normalem]{ulem} %
\usepackage{enumitem}
\usepackage{amsmath}

\usepackage{booktabs}
\usepackage{multirow}
\usepackage{makecell}
\usepackage{ragged2e}

\usepackage{listings}

\usepackage{tikz}
\usetikzlibrary{calc}
\usetikzlibrary{positioning}
\usetikzlibrary{patterns}
\usetikzlibrary{shapes}

\usepackage{datetime}

\newcommand{\XSpace}[1]{}
\newcommand{\XComment}[1]{}
\newcommand{\CRNew}[1]{#1}
\newcommand{\DefMacro}[2]{\expandafter\newcommand\csname rmk-#1\endcsname{#2}}
\newcommand{\UseMacro}[1]{\csname rmk-#1\endcsname}

\newcommand{\MyPara}[1]{\noindent\textbf{#1}.}

\newcommand{\InputWithSpace}[1]{\bgroup\def\arraystretch{1.1}\input{#1}\egroup}
\newcommand{\Code}[1]{{\ifmmode{\mathtt{#1}}\else$\mathtt{#1}$\fi}}
\newcommand{\CodeIn}[1]{{\ifmmode{\mathtt{#1}}\else$\mathtt{#1}$\fi}}

\newcolumntype{R}[1]{>{\RaggedLeft\arraybackslash}p{#1}}
\newcolumntype{L}[1]{>{\RaggedRight\arraybackslash}p{#1}}

\makeatletter
\newcommand*{\Rom}[1]{\expandafter\@slowromancap\romannumeral #1@}
\makeatother

\newcommand{\sunion}{\cup} %

\makeatletter
\newenvironment{btHighlight}[1][]
{\begingroup\tikzset{bt@Highlight@par/.style={#1}}\begin{lrbox}{\@tempboxa}}
{\end{lrbox}\bt@HL@box[bt@Highlight@par]{\@tempboxa}\endgroup}

\newcommand\btHL[1][]{%
  \begin{btHighlight}[#1]\bgroup\aftergroup\bt@HL@endenv%
}
\def\bt@HL@endenv{%
  \end{btHighlight}%
  \egroup
}
\newcommand{\bt@HL@box}[2][]{%
  \tikz[#1]{%
    \pgfpathrectangle{\pgfpoint{1pt}{0pt}}{\pgfpoint{\wd #2}{\ht #2}}%
    \pgfusepath{use as bounding box}%
    \node[anchor=base west, fill=orange!30,outer sep=0pt,inner xsep=1pt, inner ysep=0pt, rounded corners=1pt, minimum height=\ht\strutbox,#1]{\raisebox{1pt}{\strut}\strut\usebox{#2}};
  }%
}

\newenvironment{btHighlightLine}[1][]
{\begingroup\tikzset{bt@HighlightLine@par/.style={#1}}\begin{lrbox}{\@tempboxa}}
{\end{lrbox}\bt@HLLine@box[bt@HighlightLine@par]{\@tempboxa}\endgroup}

\newcommand\btHLLine[1][]{%
  \begin{btHighlightLine}[#1]\bgroup\aftergroup\bt@HLLine@endenv%
}
\def\bt@HLLine@endenv{%
  \end{btHighlightLine}%
  \egroup
}
\newcommand{\bt@HLLine@box}[2][]{%
  \tikz[#1]{%
    \pgfpathrectangle{\pgfpoint{0pt}{-1pt}}{\pgfpoint{\wd #2}{\ht #2}}%
    \pgfusepath{use as bounding box}%
    \node[anchor=base west, fill=orange!30,outer sep=0pt,inner xsep=0pt, inner ysep=0pt, minimum height=\ht\strutbox+3pt, minimum width=\linewidth,#1] (line-bg) {};
    \node[right = 0 of line-bg.west, outer sep=0pt, inner xsep=0pt, inner ysep=0pt]{\raisebox{0pt}{\strut}\strut\usebox{#2}};
  }%
}
\makeatother

\makeatletter
\global\let\tikz@ensure@dollar@catcode=\relax
\makeatother

\definecolor{gray}{RGB}{211,211,211}
\newcommand{\jbasicstyle}{\small\sffamily} %

\newcommand{\jnumberstyle}{\scriptsize}

\lstdefinelanguage{pseudo}
{
  morekeywords={},
  keywordstyle=\bfseries,
  lineskip=-0.1em,
  numbers=left, %
  numberstyle=\jnumberstyle,
  numbersep=4pt,
  basicstyle=\jbasicstyle,
  breaklines=true,
  breakautoindent=true,
  tabsize=2,
  columns=fullflexible,
  morecomment=*[l][\textsl]{//},
  mathescape=true,
  xleftmargin=10pt,
}

\lstdefinelanguage{todo-comment}
{
  morekeywords={},
  keywordstyle=\bfseries,
  lineskip=-0.1em,
  numbers=none,
  basicstyle=\jbasicstyle,
  breaklines=true,
  breakautoindent=true,
  tabsize=2,
  columns=fullflexible,
  morecomment=*[l][\textsl]{//},
  mathescape=true,
  xleftmargin=-10pt,
}

\lstdefinelanguage{java-pretty}
{
  language=java,
  numbers=left,
  basicstyle=\scriptsize\ttfamily,
  numberstyle=\scriptsize,
  breaklines=true,
  columns=fullflexible,
  xleftmargin=6pt,
  numbersep=2pt,
  showstringspaces=false,
  moredelim=**[il][{\btHLLine[fill=blue!10]}]{<HLL/>},
  moredelim=**[is][{\btHL[fill=blue!20]}]{<HL>}{<HL/>},
}

\newcommand{\Title}{Impact of Evaluation Methodologies on Code Summarization}

\newcommand{\RPURL}{\url{https://github.com/EngineeringSoftware/time-segmented-evaluation}\xspace}

\newcommand{\mixedproj}{mixed-project\xspace}
\newcommand{\Mixedproj}{Mixed-project\xspace}
\newcommand{\MixedProj}{Mixed-Project\xspace}
\newcommand{\Amp}{MP\xspace}
\newcommand{\crossproj}{cross-project\xspace}
\newcommand{\Crossproj}{Cross-project\xspace}
\newcommand{\CrossProj}{Cross-Project\xspace}
\newcommand{\Acp}{CP\xspace}
\newcommand{\temporal}{time-segmented\xspace}  %
\newcommand{\Temporal}{Time-segmented\xspace}
\newcommand{\TemporalX}{Time-Segmented\xspace}
\newcommand{\At}{T\xspace}
\newcommand{\mpxcp}{$\text{\Amp}\cap\text{\Acp}$\xspace}
\newcommand{\mpxt}{$\text{\Amp}\cap\text{\At}$\xspace}
\newcommand{\cpxt}{$\text{\Acp}\cap\text{\At}$\xspace}

\newcommand{\example}{sample\xspace}  %
\newcommand{\examples}{samples\xspace}
\newcommand{\Example}{Sample\xspace}

\newcommand{\codecom}{(code, comment)\xspace}

\newcommand{\train}{training\xspace}
\newcommand{\Train}{Training\xspace}
\newcommand{\ATrain}{Train\xspace}
\newcommand{\val}{validation\xspace}
\newcommand{\Val}{Validation\xspace}
\newcommand{\AVal}{Val\xspace}
\newcommand{\test}{test\xspace}
\newcommand{\Test}{Test\xspace}
\newcommand{\ATest}{Test\xspace}
\newcommand{\testc}{common test\xspace}

\newcommand{\ATestC}{TestC\xspace}
\newcommand{\tests}{standard test\xspace}  %

\newcommand{\ATestS}{TestS\xspace}

\newcommand{\realistic}{realistic\xspace}  %
\newcommand{\Csharp}{{\settoheight{\dimen0}{C}C\kern-.05em \resizebox{!}{\dimen0}{\raisebox{\depth}{\#}}}\xspace}

\newcommand{\comgen}{comment generation\xspace}
\newcommand{\Comgen}{Comment generation\xspace}
\newcommand{\ComGen}{Comment Generation\xspace}
\newcommand{\method}{method\xspace}
\newcommand{\Method}{Method\xspace}
\newcommand{\methods}{methods\xspace}

\newcommand{\metnam}{\method naming\xspace}
\newcommand{\Metnam}{\Method naming\xspace}
\newcommand{\MetNam}{\Method Naming\xspace}

\newcommand{\javadoc}{JavaDoc\xspace}

\newcommand{\subtok}{subtoken\xspace}

\newcommand{\subtoks}{subtokens\xspace}

\newcommand{\methodology}{methodology\xspace}
\newcommand{\emethodology}{evaluation methodology\xspace}
\newcommand{\Methodology}{Methodology\xspace}
\newcommand{\methodologies}{methodologies\xspace}
\newcommand{\emethodologies}{evaluation methodologies\xspace}

\newcommand{\Methodologies}{Methodologies\xspace}
\newcommand{\bleu}{BLEU\xspace}
\newcommand{\BLEU}{BLEU\xspace}
\newcommand{\meteor}{METEOR\xspace}
\newcommand{\METEOR}{METEOR\xspace}
\newcommand{\xmatch}{EM\xspace}
\newcommand{\Xmatch}{EM\xspace}
\newcommand{\XMatch}{EM\xspace}

\newcommand{\TokAcc}{StAcc\xspace}

\newcommand{\ROUGEf}{ROUGE-L\xspace}

\newcommand{\Precision}{Precision\xspace}

\newcommand{\Recall}{Recall\xspace}

\newcommand{\Fone}{F1\xspace}
\newcommand{\bmode}{batch-mode\xspace}
\newcommand{\ipmode}{in-project batch-mode\xspace}
\newcommand{\pmode}{cross-project batch-mode\xspace}
\newcommand{\cmode}{con\-tin\-u\-ous-mode\xspace}
\newcommand{\Cmode}{Continuous-mode\xspace}

\DefMacro{nd_same_code}{\CodeIn{same\_code}\xspace}
\DefMacro{nd_same_nl}{\CodeIn{same\_summary}\xspace}
\DefMacro{nd_sim_90}{\CodeIn{high\_similarity}\xspace}
\DefMacro{task_MN}{\metnam{}\xspace}
\DefMacro{task_CG}{\comgen{}\xspace}
\DefMacro{TaskC_MN}{\Metnam{}\xspace}
\DefMacro{TaskC_CG}{\Comgen{}\xspace}
\DefMacro{Task_MN}{\MetNam{}\xspace}
\DefMacro{Task_CG}{\ComGen{}\xspace}
\DefMacro{TaskM_MN}{\makecell[c]{\scriptsize Method\\\scriptsize Naming}}
\DefMacro{TaskM_CG}{\makecell[c]{\scriptsize Comment\\\scriptsize Generation}}

\newcommand{\na}{n/a\xspace}

\DefMacro{ds-train}{\Train\xspace}
\DefMacro{ds-val}{\Val\xspace}
\DefMacro{ds-test}{\Test\xspace}
\DefMacro{ds-all}{All\xspace}

\DefMacro{model-DeepComHybridESE19}{DeepComHybrid\xspace}
\DefMacro{model-RNNBaseline}{Seq2Seq\xspace}
\DefMacro{model-TransformerACL20}{Transformer\xspace}
\DefMacro{model-Code2SeqICLR19}{Code2Seq\xspace}
\DefMacro{model-Code2VecPOPL19}{Code2Vec\xspace}

\DefMacro{CG-DeepCom}{\Fix{INVALID}\xspace}
\DefMacro{CG-Seq2seq}{\Fix{INVALID}\xspace}
\DefMacro{CG-Seq2seqAtt}{\Fix{INVALID}\xspace}
\DefMacro{MN-Code2Seq}{\Fix{INVALID}\xspace}
\DefMacro{MN-Bi-LSTM}{\Fix{INVALID}\xspace}
\DefMacro{MN-no-split-Bi-LSTM}{\Fix{INVALID}\xspace}

\def\checkmark{\tikz\fill[scale=0.25,color=black](0,.35) -- (.25,0) -- (1,.7) -- (.25,.15) -- cycle;}

\newcommand{\mycheckmark}{{\normalsize \checkmark}\xspace}
\newcommand{\mycross}{$\mathbin{\tikz [x=1.4ex,y=1.4ex,line width=.2ex] \draw (0,0) -- (1,1) (0,1) -- (1,0);}$\xspace\xspace}

\newcommand{\aexamples}{\ensuremath{\mathcal{E}}\xspace}
\newcommand{\atime}{\ensuremath{\tau}\xspace}
\newcommand{\atimep}{\ensuremath{\tau^{-1}}\xspace}
\newcommand{\atimepp}{\ensuremath{\tau^{-2}}\xspace}
\newcommand{\atimeppp}{\ensuremath{\tau^{-3}}\xspace}

\newcommand{\aproject}{\ensuremath{p}\xspace}
\newcommand{\aprojecttrain}{\ensuremath{\mathcal{P}_{train}}\xspace}
\newcommand{\aprojectval}{\ensuremath{\mathcal{P}_{val}}\xspace}
\newcommand{\aprojecttest}{\ensuremath{\mathcal{P}_{test}}\xspace}
\newcommand{\aprojects}{\ensuremath{\mathcal{P}}\xspace}

\newcommand{\ashuffle}{\ensuremath{\mathsf{shuffle}}\xspace}
\newcommand{\asplit}{\ensuremath{\mathsf{split}}\xspace}
\newcommand{\asplitprj}{\ensuremath{\mathsf{splitprj}}\xspace}
\newcommand{\ajanone}{\ensuremath{\text{Jan }1^{st}}\xspace}
\newcommand{\aclean}{\ensuremath{\mathsf{clean}}\xspace}
\newcommand{\apcttrain}{\ensuremath{r_x}\xspace}
\newcommand{\apctval}{\ensuremath{r_y}\xspace}
\newcommand{\apcttest}{\ensuremath{r_z}\xspace}

\tikzset{dsTrain/.style={preaction={fill=blue!60}}}
\tikzset{dsVal/.style={pattern=vertical lines, preaction={fill=yellow}}}
\tikzset{dsTest/.style={pattern=horizontal lines, preaction={fill=red!70}}}
\tikzset{dsTestC/.style={pattern=horizontal lines, preaction={fill=orange!70}}}
\tikzset{sProject/.style={rounded box, minimum width=4em, minimum height=4em}}
\tikzset{sExample/.style={circle, minimum size=.8em, inner sep=0, draw=black}}
\newcommand{\wSepProject}{1em}

\newcommand{\TCPriorWorkCodeSum}{\Methodologies used in prior work on code summarization; we use the highlighted lines in our experiments. \label{table:prior-work-code-sum}\vspace{-0pt}}
\newcommand{\TVPriorWorkCodeSum}{-0pt}

\newcommand{\TCPrelim}{Definitions of symbols and functions. \label{table:prelim}}
\newcommand{\TVPrelim}{-0pt}

\newcommand{\TCEvalSettings}{The formulas (at steps~4~and~5) to get
  the \train (\ATrain), \val (\AVal), and \tests (\ATestS) sets for
  each \methodology, and the \testc (\ATestC) set for each pair of
  \methodologies. \label{table:eval-settings}}
\newcommand{\TVEvalSettings}{-0pt}

\DefMacro{TC-dataset-metrics-CG}{\Comgen dataset metrics.
\label{table:dataset-metrics-cg}}
\DefMacro{TV-dataset-metrics-CG}{-0pt}
\DefMacro{TC-dataset-metrics-MN}{\Metnam dataset metrics.
\label{table:dataset-metrics-mn}}
\DefMacro{TV-dataset-metrics-MN}{-0pt}
\DefMacro{TC-dataset-metrics-small}{Number of \examples in our
datasets. \label{table:dataset-metrics-small}\vspace{-5pt}}
\DefMacro{TV-dataset-metrics-small}{-10pt}

\DefMacro{TC-results-CG}{\Comgen models' results on \ATestC sets. 
\CRNew{The six results in each block are comparable because they use
the same set and metric, where results marked with the same
Greek letter are \emph{not} statistically significantly different.}
  Depending on the \methodology, we may or may not observe
  statistically significant differences results between
  models.\label{table:results-cg}\vspace{-8pt}}
\DefMacro{TV-results-CG}{-6pt}

\DefMacro{TC-results-MN}{\Metnam models' results on and \ATestC sets.
\CRNew{The four results in each block are comparable because they use
the same set and metric, where results marked with the same
Greek letter are \emph{not} statistically significantly different.}
  Surprisingly, \UseMacro{model-Code2SeqICLR19} outperforms
  \UseMacro{model-Code2VecPOPL19} in the \CrossProj \methodology but
  the opposite holds in the other two
  \methodologies.\label{table:results-mn}\vspace{-8pt}}
\DefMacro{TV-results-MN}{-6pt}

\DefMacro{TC-results-aux-CG}{\Comgen models' results on \AVal and
  \ATestS sets.\label{table:results-aux-cg}}
\DefMacro{TV-results-aux-CG}{-0pt}

\DefMacro{TC-results-aux-MN}{\Metnam models' results on \AVal and
\ATestS sets.\label{table:results-aux-mn}}
\DefMacro{TV-results-aux-MN}{-0pt}

\DefMacro{TH-test-on}{\textbf{Test}\xspace}
\DefMacro{TH-train-on}{\textbf{Train}\xspace}
\DefMacro{TH-MP-CP}{\textbf{\mpxcp}\xspace}
\DefMacro{TH-MP-T}{\textbf{\mpxt}\xspace}
\DefMacro{TH-CP-T}{\textbf{\cpxt}\xspace}
\DefMacro{TH-MP}{\textbf{\Amp}\xspace}
\DefMacro{TH-CP}{\textbf{\Acp}\xspace}
\DefMacro{TH-T}{\textbf{\At}\xspace}
\DefMacro{TH-val}{\textbf{\AVal}\xspace}
\DefMacro{TH-test_standard}{\textbf{\ATestS}\xspace}

\DefMacro{TH-model-DeepComHybridESE19}{\UseMacro{model-DeepComHybridESE19}\xspace} %
\DefMacro{TH-model-RNNBaseline}{\UseMacro{model-RNNBaseline}\xspace} %
\DefMacro{TH-model-TransformerACL20}{\UseMacro{model-TransformerACL20}\xspace} %
\DefMacro{TH-model-Code2SeqICLR19}{\UseMacro{model-Code2SeqICLR19}\xspace} %
\DefMacro{TH-model-Code2VecPOPL19}{\UseMacro{model-Code2VecPOPL19}\xspace} %

\newcommand{\THDSStat}{\textbf{Statistic}\xspace}

\DefMacro{TH-metric-bleu}{\textbf{\BLEU}\xspace}
\DefMacro{TH-metric-meteor}{\textbf{\METEOR}\xspace}
\DefMacro{TH-metric-exact_match}{\textbf{EM}\xspace}
\DefMacro{TH-metric-rouge_l_f}{\textbf{\ROUGEf}\xspace}
\DefMacro{TH-metric-token_acc}{\textbf{\makecell[r]{Subtoken-Level\\Accuracy}}\xspace}
\DefMacro{TH-metric-set_match_p}{\textbf{\Precision}\xspace}
\DefMacro{TH-metric-set_match_r}{\textbf{\Recall}\xspace}
\DefMacro{TH-metric-set_match_f}{\textbf{\Fone}\xspace}

\DefMacro{TH-metric-table-bleu}{$\blacktriangledown$ \textbf{\BLEU [\%]}\xspace}
\DefMacro{TH-metric-table-meteor}{$\blacktriangledown$ \textbf{\METEOR [\%]}\xspace}
\DefMacro{TH-metric-table-exact_match}{$\blacktriangledown$ \textbf{\XMatch [\%]}\xspace}
\DefMacro{TH-metric-table-rouge_l_f}{$\blacktriangledown$ \textbf{\ROUGEf [\%]}\xspace}
\DefMacro{TH-metric-table-token_acc}{$\blacktriangledown$ \textbf{\TokAcc [\%]}\xspace}
\DefMacro{TH-metric-table-set_match_p}{$\blacktriangledown$ \textbf{\Precision [\%]}\xspace}
\DefMacro{TH-metric-table-set_match_r}{$\blacktriangledown$ \textbf{\Recall [\%]}\xspace}
\DefMacro{TH-metric-table-set_match_f}{$\blacktriangledown$ \textbf{\Fone [\%]}\xspace}

\DefMacro{TH-exp-crossproj-2020}{\textbf{\Crossproj}\xspace}
\DefMacro{TH-exp-mixedproj-2020}{\textbf{\Mixedproj}\xspace}
\DefMacro{TH-exp-evo-2020}{\textbf{\Temporal}\xspace}

\DefMacro{TH-ds-all}{\textbf{All}\xspace}
\DefMacro{TH-ds-MP}{\textbf{\Amp}\xspace}
\DefMacro{TH-ds-CP}{\textbf{\Acp}\xspace}
\DefMacro{TH-ds-T}{\textbf{\At}\xspace}
\DefMacro{TH-ds-MP-CP}{\textbf{\mpxcp}\xspace}
\DefMacro{TH-ds-MP-T}{\textbf{\mpxt}\xspace}
\DefMacro{TH-ds-CP-T}{\textbf{\cpxt}\xspace}
\DefMacro{TH-ds-train}{\textbf{\ATrain}\xspace}
\DefMacro{TH-ds-val}{\textbf{\AVal}\xspace}
\DefMacro{TH-ds-test_standard}{\textbf{\ATestS}\xspace}
\DefMacro{TH-ds-test_common}{\textbf{\ATestC}\xspace}
\DefMacro{TH-ds-num-project}{\textbf{\#Proj}\xspace}
\DefMacro{TH-ds-num-data}{\textbf{\#\Example}\xspace}
\DefMacro{TH-ds-len-code}{\rotatebox{90}{\textbf{len(code)}}\xspace}
\DefMacro{TH-ds-len-code-avg}{avg\xspace}
\DefMacro{TH-ds-len-code-mode}{mode\xspace}
\DefMacro{TH-ds-len-code-median}{median\xspace}
\DefMacro{TH-ds-len-code-le100}{$\le$100[\%]\xspace}
\DefMacro{TH-ds-len-code-le150}{$\le$150[\%]\xspace}
\DefMacro{TH-ds-len-code-le200}{$\le$200[\%]\xspace}
\DefMacro{TH-ds-len-comment}{\rotatebox{90}{\textbf{len(com)}}\xspace}
\DefMacro{TH-ds-len-comment-avg}{avg\xspace}
\DefMacro{TH-ds-len-comment-mode}{mode\xspace}
\DefMacro{TH-ds-len-comment-median}{median\xspace}
\DefMacro{TH-ds-len-comment-le20}{$\le$20[\%]\xspace}
\DefMacro{TH-ds-len-comment-le30}{$\le$30[\%]\xspace}
\DefMacro{TH-ds-len-comment-le50}{$\le$50[\%]\xspace}
\DefMacro{TH-ds-len-name}{\rotatebox{90}{\textbf{len(name)}}\xspace}
\DefMacro{TH-ds-len-name-avg}{avg\xspace}
\DefMacro{TH-ds-len-name-mode}{mode\xspace}
\DefMacro{TH-ds-len-name-median}{median\xspace}
\DefMacro{TH-ds-len-name-le2}{$\le$2[\%]\xspace}
\DefMacro{TH-ds-len-name-le3}{$\le$3[\%]\xspace}
\DefMacro{TH-ds-len-name-le6}{$\le$6[\%]\xspace}

\DefMacro{ds-filter-code_non_english}{157}
\DefMacro{ds-filter-code_too_long}{60}
\DefMacro{ds-filter-comment_non_english}{5,139}
\DefMacro{ds-filter-empty_body}{77,769}
\DefMacro{ds-filter-empty_comment}{681,453}
\DefMacro{ds-filter-empty_comment_summary}{21,779}
\DefMacro{ds-num_prj_train}{111}
\DefMacro{ds-num_prj_split_train}{50,879}
\DefMacro{ds-num_prj_val}{17}
\DefMacro{ds-num_prj_split_val}{9,460}
\DefMacro{ds-num_prj_test}{32}
\DefMacro{ds-num_prj_split_test}{17,136}
\DefMacro{ds-num_inprj_split_train}{54,141}
\DefMacro{ds-num_inprj_split_val}{7,778}
\DefMacro{ds-num_inprj_split_test}{15,556}
\DefMacro{ds-num_year_split_train}{55,460}
\DefMacro{ds-num_year_split_val}{11,958}
\DefMacro{ds-num_year_split_test}{10,057}
\DefMacro{ds-num_MP_train}{50,879}
\DefMacro{ds-num_MP_val}{7,778}
\DefMacro{ds-num_MP_test_standard}{15,556}
\DefMacro{ds-num_MP-CP_test_common}{3,438}
\DefMacro{ds-num_MP-T_test_common}{2,025}
\DefMacro{ds-num_CP_train}{50,879}
\DefMacro{ds-num_CP_val}{9,460}
\DefMacro{ds-num_CP_test_standard}{17,136}
\DefMacro{ds-num_CP-T_test_common}{2,259}
\DefMacro{ds-num_T_train}{50,879}
\DefMacro{ds-num_T_val}{11,958}
\DefMacro{ds-num_T_test_standard}{10,057}
\DefMacro{ds-num_prj_all}{160}
\DefMacro{ds-all_num-data}{77,475}
\DefMacro{ds-all_len-code-le-100}{79.54}
\DefMacro{ds-all_len-code-le-150}{88.13}
\DefMacro{ds-all_len-code-le-200}{92.55}
\DefMacro{ds-all_len-comment-le-20}{90.26}
\DefMacro{ds-all_len-comment-le-30}{97.63}
\DefMacro{ds-all_len-comment-le-50}{99.69}
\DefMacro{ds-all_len-name-le-2}{59.08}
\DefMacro{ds-all_len-name-le-3}{82.71}
\DefMacro{ds-all_len-name-le-6}{98.90}
\DefMacro{ds-all_len-code-AVG}{74.39}
\DefMacro{ds-all_len-code-SUM}{5,763,511}
\DefMacro{ds-all_len-code-MAX}{2,225}
\DefMacro{ds-all_len-code-MIN}{9}
\DefMacro{ds-all_len-code-MEDIAN}{39.00}
\DefMacro{ds-all_len-code-STDEV}{107.12}
\DefMacro{ds-all_len-code-MODE}{10}
\DefMacro{ds-all_len-code-CNT}{77,475}
\DefMacro{ds-all_len-comment-AVG}{11.54}
\DefMacro{ds-all_len-comment-SUM}{894,200}
\DefMacro{ds-all_len-comment-MAX}{172}
\DefMacro{ds-all_len-comment-MIN}{1}
\DefMacro{ds-all_len-comment-MEDIAN}{10.00}
\DefMacro{ds-all_len-comment-STDEV}{7.62}
\DefMacro{ds-all_len-comment-MODE}{9}
\DefMacro{ds-all_len-comment-CNT}{77,475}
\DefMacro{ds-all_len-name-AVG}{2.46}
\DefMacro{ds-all_len-name-SUM}{190,536}
\DefMacro{ds-all_len-name-MAX}{15}
\DefMacro{ds-all_len-name-MIN}{1}
\DefMacro{ds-all_len-name-MEDIAN}{2.00}
\DefMacro{ds-all_len-name-STDEV}{1.32}
\DefMacro{ds-all_len-name-MODE}{2}
\DefMacro{ds-all_len-name-CNT}{77,475}
\DefMacro{ds-MP-train_num-data}{50,879}
\DefMacro{ds-MP-train_len-code-le-100}{79.69}
\DefMacro{ds-MP-train_len-code-le-150}{88.30}
\DefMacro{ds-MP-train_len-code-le-200}{92.68}
\DefMacro{ds-MP-train_len-comment-le-20}{90.12}
\DefMacro{ds-MP-train_len-comment-le-30}{97.56}
\DefMacro{ds-MP-train_len-comment-le-50}{99.67}
\DefMacro{ds-MP-train_len-name-le-2}{59.36}
\DefMacro{ds-MP-train_len-name-le-3}{82.73}
\DefMacro{ds-MP-train_len-name-le-6}{98.89}
\DefMacro{ds-MP-train_len-code-AVG}{73.76}
\DefMacro{ds-MP-train_len-code-SUM}{3,752,804}
\DefMacro{ds-MP-train_len-code-MAX}{1,817}
\DefMacro{ds-MP-train_len-code-MIN}{9}
\DefMacro{ds-MP-train_len-code-MEDIAN}{39.00}
\DefMacro{ds-MP-train_len-code-STDEV}{105.23}
\DefMacro{ds-MP-train_len-code-MODE}{10}
\DefMacro{ds-MP-train_len-code-CNT}{50,879}
\DefMacro{ds-MP-train_len-comment-AVG}{11.55}
\DefMacro{ds-MP-train_len-comment-SUM}{587,630}
\DefMacro{ds-MP-train_len-comment-MAX}{172}
\DefMacro{ds-MP-train_len-comment-MIN}{1}
\DefMacro{ds-MP-train_len-comment-MEDIAN}{10.00}
\DefMacro{ds-MP-train_len-comment-STDEV}{7.67}
\DefMacro{ds-MP-train_len-comment-MODE}{9}
\DefMacro{ds-MP-train_len-comment-CNT}{50,879}
\DefMacro{ds-MP-train_len-name-AVG}{2.46}
\DefMacro{ds-MP-train_len-name-SUM}{124,913}
\DefMacro{ds-MP-train_len-name-MAX}{15}
\DefMacro{ds-MP-train_len-name-MIN}{1}
\DefMacro{ds-MP-train_len-name-MEDIAN}{2.00}
\DefMacro{ds-MP-train_len-name-STDEV}{1.32}
\DefMacro{ds-MP-train_len-name-MODE}{2}
\DefMacro{ds-MP-train_len-name-CNT}{50,879}
\DefMacro{ds-CP-train_num-data}{50,879}
\DefMacro{ds-CP-train_len-code-le-100}{80.94}
\DefMacro{ds-CP-train_len-code-le-150}{88.85}
\DefMacro{ds-CP-train_len-code-le-200}{93.07}
\DefMacro{ds-CP-train_len-comment-le-20}{90.10}
\DefMacro{ds-CP-train_len-comment-le-30}{97.61}
\DefMacro{ds-CP-train_len-comment-le-50}{99.68}
\DefMacro{ds-CP-train_len-name-le-2}{59.40}
\DefMacro{ds-CP-train_len-name-le-3}{82.91}
\DefMacro{ds-CP-train_len-name-le-6}{98.78}
\DefMacro{ds-CP-train_len-code-AVG}{71.16}
\DefMacro{ds-CP-train_len-code-SUM}{3,620,357}
\DefMacro{ds-CP-train_len-code-MAX}{2,225}
\DefMacro{ds-CP-train_len-code-MIN}{9}
\DefMacro{ds-CP-train_len-code-MEDIAN}{36.00}
\DefMacro{ds-CP-train_len-code-STDEV}{104.94}
\DefMacro{ds-CP-train_len-code-MODE}{10}
\DefMacro{ds-CP-train_len-code-CNT}{50,879}
\DefMacro{ds-CP-train_len-comment-AVG}{11.50}
\DefMacro{ds-CP-train_len-comment-SUM}{584,992}
\DefMacro{ds-CP-train_len-comment-MAX}{172}
\DefMacro{ds-CP-train_len-comment-MIN}{1}
\DefMacro{ds-CP-train_len-comment-MEDIAN}{10.00}
\DefMacro{ds-CP-train_len-comment-STDEV}{7.76}
\DefMacro{ds-CP-train_len-comment-MODE}{5}
\DefMacro{ds-CP-train_len-comment-CNT}{50,879}
\DefMacro{ds-CP-train_len-name-AVG}{2.45}
\DefMacro{ds-CP-train_len-name-SUM}{124,612}
\DefMacro{ds-CP-train_len-name-MAX}{15}
\DefMacro{ds-CP-train_len-name-MIN}{1}
\DefMacro{ds-CP-train_len-name-MEDIAN}{2.00}
\DefMacro{ds-CP-train_len-name-STDEV}{1.33}
\DefMacro{ds-CP-train_len-name-MODE}{2}
\DefMacro{ds-CP-train_len-name-CNT}{50,879}
\DefMacro{ds-T-train_num-data}{50,879}
\DefMacro{ds-T-train_len-code-le-100}{81.03}
\DefMacro{ds-T-train_len-code-le-150}{89.17}
\DefMacro{ds-T-train_len-code-le-200}{93.27}
\DefMacro{ds-T-train_len-comment-le-20}{90.16}
\DefMacro{ds-T-train_len-comment-le-30}{97.73}
\DefMacro{ds-T-train_len-comment-le-50}{99.74}
\DefMacro{ds-T-train_len-name-le-2}{59.70}
\DefMacro{ds-T-train_len-name-le-3}{83.50}
\DefMacro{ds-T-train_len-name-le-6}{99.14}
\DefMacro{ds-T-train_len-code-AVG}{70.78}
\DefMacro{ds-T-train_len-code-SUM}{3,601,218}
\DefMacro{ds-T-train_len-code-MAX}{2,225}
\DefMacro{ds-T-train_len-code-MIN}{9}
\DefMacro{ds-T-train_len-code-MEDIAN}{37.00}
\DefMacro{ds-T-train_len-code-STDEV}{103.28}
\DefMacro{ds-T-train_len-code-MODE}{10}
\DefMacro{ds-T-train_len-code-CNT}{50,879}
\DefMacro{ds-T-train_len-comment-AVG}{11.58}
\DefMacro{ds-T-train_len-comment-SUM}{589,207}
\DefMacro{ds-T-train_len-comment-MAX}{172}
\DefMacro{ds-T-train_len-comment-MIN}{1}
\DefMacro{ds-T-train_len-comment-MEDIAN}{10.00}
\DefMacro{ds-T-train_len-comment-STDEV}{7.47}
\DefMacro{ds-T-train_len-comment-MODE}{8}
\DefMacro{ds-T-train_len-comment-CNT}{50,879}
\DefMacro{ds-T-train_len-name-AVG}{2.43}
\DefMacro{ds-T-train_len-name-SUM}{123,611}
\DefMacro{ds-T-train_len-name-MAX}{15}
\DefMacro{ds-T-train_len-name-MIN}{1}
\DefMacro{ds-T-train_len-name-MEDIAN}{2.00}
\DefMacro{ds-T-train_len-name-STDEV}{1.28}
\DefMacro{ds-T-train_len-name-MODE}{2}
\DefMacro{ds-T-train_len-name-CNT}{50,879}
\DefMacro{ds-MP-val_num-data}{7,778}
\DefMacro{ds-MP-val_len-code-le-100}{79.09}
\DefMacro{ds-MP-val_len-code-le-150}{87.86}
\DefMacro{ds-MP-val_len-code-le-200}{92.36}
\DefMacro{ds-MP-val_len-comment-le-20}{90.32}
\DefMacro{ds-MP-val_len-comment-le-30}{97.74}
\DefMacro{ds-MP-val_len-comment-le-50}{99.74}
\DefMacro{ds-MP-val_len-name-le-2}{58.95}
\DefMacro{ds-MP-val_len-name-le-3}{83.11}
\DefMacro{ds-MP-val_len-name-le-6}{98.96}
\DefMacro{ds-MP-val_len-code-AVG}{75.69}
\DefMacro{ds-MP-val_len-code-SUM}{588,708}
\DefMacro{ds-MP-val_len-code-MAX}{1,643}
\DefMacro{ds-MP-val_len-code-MIN}{9}
\DefMacro{ds-MP-val_len-code-MEDIAN}{39.00}
\DefMacro{ds-MP-val_len-code-STDEV}{109.73}
\DefMacro{ds-MP-val_len-code-MODE}{10}
\DefMacro{ds-MP-val_len-code-CNT}{7,778}
\DefMacro{ds-MP-val_len-comment-AVG}{11.56}
\DefMacro{ds-MP-val_len-comment-SUM}{89,909}
\DefMacro{ds-MP-val_len-comment-MAX}{130}
\DefMacro{ds-MP-val_len-comment-MIN}{1}
\DefMacro{ds-MP-val_len-comment-MEDIAN}{10.00}
\DefMacro{ds-MP-val_len-comment-STDEV}{7.69}
\DefMacro{ds-MP-val_len-comment-MODE}{9}
\DefMacro{ds-MP-val_len-comment-CNT}{7,778}
\DefMacro{ds-MP-val_len-name-AVG}{2.45}
\DefMacro{ds-MP-val_len-name-SUM}{19,048}
\DefMacro{ds-MP-val_len-name-MAX}{14}
\DefMacro{ds-MP-val_len-name-MIN}{1}
\DefMacro{ds-MP-val_len-name-MEDIAN}{2.00}
\DefMacro{ds-MP-val_len-name-STDEV}{1.32}
\DefMacro{ds-MP-val_len-name-MODE}{2}
\DefMacro{ds-MP-val_len-name-CNT}{7,778}
\DefMacro{ds-CP-val_num-data}{9,460}
\DefMacro{ds-CP-val_len-code-le-100}{78.04}
\DefMacro{ds-CP-val_len-code-le-150}{87.80}
\DefMacro{ds-CP-val_len-code-le-200}{92.70}
\DefMacro{ds-CP-val_len-comment-le-20}{90.01}
\DefMacro{ds-CP-val_len-comment-le-30}{97.09}
\DefMacro{ds-CP-val_len-comment-le-50}{99.67}
\DefMacro{ds-CP-val_len-name-le-2}{69.25}
\DefMacro{ds-CP-val_len-name-le-3}{87.34}
\DefMacro{ds-CP-val_len-name-le-6}{99.21}
\DefMacro{ds-CP-val_len-code-AVG}{77.15}
\DefMacro{ds-CP-val_len-code-SUM}{729,808}
\DefMacro{ds-CP-val_len-code-MAX}{1,701}
\DefMacro{ds-CP-val_len-code-MIN}{9}
\DefMacro{ds-CP-val_len-code-MEDIAN}{46.00}
\DefMacro{ds-CP-val_len-code-STDEV}{94.11}
\DefMacro{ds-CP-val_len-code-MODE}{19}
\DefMacro{ds-CP-val_len-code-CNT}{9,460}
\DefMacro{ds-CP-val_len-comment-AVG}{11.70}
\DefMacro{ds-CP-val_len-comment-SUM}{110,647}
\DefMacro{ds-CP-val_len-comment-MAX}{93}
\DefMacro{ds-CP-val_len-comment-MIN}{1}
\DefMacro{ds-CP-val_len-comment-MEDIAN}{10.00}
\DefMacro{ds-CP-val_len-comment-STDEV}{7.53}
\DefMacro{ds-CP-val_len-comment-MODE}{8}
\DefMacro{ds-CP-val_len-comment-CNT}{9,460}
\DefMacro{ds-CP-val_len-name-AVG}{2.17}
\DefMacro{ds-CP-val_len-name-SUM}{20,575}
\DefMacro{ds-CP-val_len-name-MAX}{13}
\DefMacro{ds-CP-val_len-name-MIN}{1}
\DefMacro{ds-CP-val_len-name-MEDIAN}{2.00}
\DefMacro{ds-CP-val_len-name-STDEV}{1.27}
\DefMacro{ds-CP-val_len-name-MODE}{1}
\DefMacro{ds-CP-val_len-name-CNT}{9,460}
\DefMacro{ds-T-val_num-data}{11,958}
\DefMacro{ds-T-val_len-code-le-100}{77.33}
\DefMacro{ds-T-val_len-code-le-150}{86.06}
\DefMacro{ds-T-val_len-code-le-200}{90.94}
\DefMacro{ds-T-val_len-comment-le-20}{90.24}
\DefMacro{ds-T-val_len-comment-le-30}{97.44}
\DefMacro{ds-T-val_len-comment-le-50}{99.69}
\DefMacro{ds-T-val_len-name-le-2}{58.93}
\DefMacro{ds-T-val_len-name-le-3}{80.85}
\DefMacro{ds-T-val_len-name-le-6}{98.08}
\DefMacro{ds-T-val_len-code-AVG}{81.83}
\DefMacro{ds-T-val_len-code-SUM}{978,558}
\DefMacro{ds-T-val_len-code-MAX}{1,817}
\DefMacro{ds-T-val_len-code-MIN}{9}
\DefMacro{ds-T-val_len-code-MEDIAN}{42.00}
\DefMacro{ds-T-val_len-code-STDEV}{117.53}
\DefMacro{ds-T-val_len-code-MODE}{10}
\DefMacro{ds-T-val_len-code-CNT}{11,958}
\DefMacro{ds-T-val_len-comment-AVG}{11.44}
\DefMacro{ds-T-val_len-comment-SUM}{136,836}
\DefMacro{ds-T-val_len-comment-MAX}{168}
\DefMacro{ds-T-val_len-comment-MIN}{1}
\DefMacro{ds-T-val_len-comment-MEDIAN}{10.00}
\DefMacro{ds-T-val_len-comment-STDEV}{7.85}
\DefMacro{ds-T-val_len-comment-MODE}{5}
\DefMacro{ds-T-val_len-comment-CNT}{11,958}
\DefMacro{ds-T-val_len-name-AVG}{2.51}
\DefMacro{ds-T-val_len-name-SUM}{30,018}
\DefMacro{ds-T-val_len-name-MAX}{14}
\DefMacro{ds-T-val_len-name-MIN}{1}
\DefMacro{ds-T-val_len-name-MEDIAN}{2.00}
\DefMacro{ds-T-val_len-name-STDEV}{1.46}
\DefMacro{ds-T-val_len-name-MODE}{2}
\DefMacro{ds-T-val_len-name-CNT}{11,958}
\DefMacro{ds-MP-test_standard_num-data}{15,556}
\DefMacro{ds-MP-test_standard_len-code-le-100}{79.20}
\DefMacro{ds-MP-test_standard_len-code-le-150}{87.66}
\DefMacro{ds-MP-test_standard_len-code-le-200}{92.30}
\DefMacro{ds-MP-test_standard_len-comment-le-20}{90.58}
\DefMacro{ds-MP-test_standard_len-comment-le-30}{97.82}
\DefMacro{ds-MP-test_standard_len-comment-le-50}{99.72}
\DefMacro{ds-MP-test_standard_len-name-le-2}{58.09}
\DefMacro{ds-MP-test_standard_len-name-le-3}{82.35}
\DefMacro{ds-MP-test_standard_len-name-le-6}{98.91}
\DefMacro{ds-MP-test_standard_len-code-AVG}{75.47}
\DefMacro{ds-MP-test_standard_len-code-SUM}{1,174,057}
\DefMacro{ds-MP-test_standard_len-code-MAX}{1,819}
\DefMacro{ds-MP-test_standard_len-code-MIN}{9}
\DefMacro{ds-MP-test_standard_len-code-MEDIAN}{39.00}
\DefMacro{ds-MP-test_standard_len-code-STDEV}{109.41}
\DefMacro{ds-MP-test_standard_len-code-MODE}{10}
\DefMacro{ds-MP-test_standard_len-code-CNT}{15,556}
\DefMacro{ds-MP-test_standard_len-comment-AVG}{11.51}
\DefMacro{ds-MP-test_standard_len-comment-SUM}{179,105}
\DefMacro{ds-MP-test_standard_len-comment-MAX}{172}
\DefMacro{ds-MP-test_standard_len-comment-MIN}{1}
\DefMacro{ds-MP-test_standard_len-comment-MEDIAN}{10.00}
\DefMacro{ds-MP-test_standard_len-comment-STDEV}{7.52}
\DefMacro{ds-MP-test_standard_len-comment-MODE}{9}
\DefMacro{ds-MP-test_standard_len-comment-CNT}{15,556}
\DefMacro{ds-MP-test_standard_len-name-AVG}{2.48}
\DefMacro{ds-MP-test_standard_len-name-SUM}{38,548}
\DefMacro{ds-MP-test_standard_len-name-MAX}{15}
\DefMacro{ds-MP-test_standard_len-name-MIN}{1}
\DefMacro{ds-MP-test_standard_len-name-MEDIAN}{2.00}
\DefMacro{ds-MP-test_standard_len-name-STDEV}{1.33}
\DefMacro{ds-MP-test_standard_len-name-MODE}{2}
\DefMacro{ds-MP-test_standard_len-name-CNT}{15,556}
\DefMacro{ds-CP-test_standard_num-data}{17,136}
\DefMacro{ds-CP-test_standard_len-code-le-100}{76.18}
\DefMacro{ds-CP-test_standard_len-code-le-150}{86.15}
\DefMacro{ds-CP-test_standard_len-code-le-200}{90.94}
\DefMacro{ds-CP-test_standard_len-comment-le-20}{90.87}
\DefMacro{ds-CP-test_standard_len-comment-le-30}{97.96}
\DefMacro{ds-CP-test_standard_len-comment-le-50}{99.75}
\DefMacro{ds-CP-test_standard_len-name-le-2}{52.49}
\DefMacro{ds-CP-test_standard_len-name-le-3}{79.54}
\DefMacro{ds-CP-test_standard_len-name-le-6}{99.11}
\DefMacro{ds-CP-test_standard_len-code-AVG}{82.48}
\DefMacro{ds-CP-test_standard_len-code-SUM}{1,413,346}
\DefMacro{ds-CP-test_standard_len-code-MAX}{1,815}
\DefMacro{ds-CP-test_standard_len-code-MIN}{9}
\DefMacro{ds-CP-test_standard_len-code-MEDIAN}{42.00}
\DefMacro{ds-CP-test_standard_len-code-STDEV}{119.11}
\DefMacro{ds-CP-test_standard_len-code-MODE}{10}
\DefMacro{ds-CP-test_standard_len-code-CNT}{17,136}
\DefMacro{ds-CP-test_standard_len-comment-AVG}{11.59}
\DefMacro{ds-CP-test_standard_len-comment-SUM}{198,561}
\DefMacro{ds-CP-test_standard_len-comment-MAX}{168}
\DefMacro{ds-CP-test_standard_len-comment-MIN}{1}
\DefMacro{ds-CP-test_standard_len-comment-MEDIAN}{10.00}
\DefMacro{ds-CP-test_standard_len-comment-STDEV}{7.27}
\DefMacro{ds-CP-test_standard_len-comment-MODE}{8}
\DefMacro{ds-CP-test_standard_len-comment-CNT}{17,136}
\DefMacro{ds-CP-test_standard_len-name-AVG}{2.65}
\DefMacro{ds-CP-test_standard_len-name-SUM}{45,349}
\DefMacro{ds-CP-test_standard_len-name-MAX}{15}
\DefMacro{ds-CP-test_standard_len-name-MIN}{1}
\DefMacro{ds-CP-test_standard_len-name-MEDIAN}{2.00}
\DefMacro{ds-CP-test_standard_len-name-STDEV}{1.30}
\DefMacro{ds-CP-test_standard_len-name-MODE}{2}
\DefMacro{ds-CP-test_standard_len-name-CNT}{17,136}
\DefMacro{ds-T-test_standard_num-data}{10,057}
\DefMacro{ds-T-test_standard_len-code-le-100}{74.48}
\DefMacro{ds-T-test_standard_len-code-le-150}{85.35}
\DefMacro{ds-T-test_standard_len-code-le-200}{90.51}
\DefMacro{ds-T-test_standard_len-comment-le-20}{90.35}
\DefMacro{ds-T-test_standard_len-comment-le-30}{97.16}
\DefMacro{ds-T-test_standard_len-comment-le-50}{99.44}
\DefMacro{ds-T-test_standard_len-name-le-2}{55.75}
\DefMacro{ds-T-test_standard_len-name-le-3}{80.27}
\DefMacro{ds-T-test_standard_len-name-le-6}{98.51}
\DefMacro{ds-T-test_standard_len-code-AVG}{84.74}
\DefMacro{ds-T-test_standard_len-code-SUM}{852,214}
\DefMacro{ds-T-test_standard_len-code-MAX}{1,819}
\DefMacro{ds-T-test_standard_len-code-MIN}{9}
\DefMacro{ds-T-test_standard_len-code-MEDIAN}{45.00}
\DefMacro{ds-T-test_standard_len-code-STDEV}{113.52}
\DefMacro{ds-T-test_standard_len-code-MODE}{15}
\DefMacro{ds-T-test_standard_len-code-CNT}{10,057}
\DefMacro{ds-T-test_standard_len-comment-AVG}{11.49}
\DefMacro{ds-T-test_standard_len-comment-SUM}{115,545}
\DefMacro{ds-T-test_standard_len-comment-MAX}{136}
\DefMacro{ds-T-test_standard_len-comment-MIN}{1}
\DefMacro{ds-T-test_standard_len-comment-MEDIAN}{9.00}
\DefMacro{ds-T-test_standard_len-comment-STDEV}{8.12}
\DefMacro{ds-T-test_standard_len-comment-MODE}{5}
\DefMacro{ds-T-test_standard_len-comment-CNT}{10,057}
\DefMacro{ds-T-test_standard_len-name-AVG}{2.57}
\DefMacro{ds-T-test_standard_len-name-SUM}{25,856}
\DefMacro{ds-T-test_standard_len-name-MAX}{14}
\DefMacro{ds-T-test_standard_len-name-MIN}{1}
\DefMacro{ds-T-test_standard_len-name-MEDIAN}{2.00}
\DefMacro{ds-T-test_standard_len-name-STDEV}{1.38}
\DefMacro{ds-T-test_standard_len-name-MODE}{2}
\DefMacro{ds-T-test_standard_len-name-CNT}{10,057}
\DefMacro{ds-MP-CP-test_common_num-data}{3,438}
\DefMacro{ds-MP-CP-test_common_len-code-le-100}{75.54}
\DefMacro{ds-MP-CP-test_common_len-code-le-150}{84.79}
\DefMacro{ds-MP-CP-test_common_len-code-le-200}{89.76}
\DefMacro{ds-MP-CP-test_common_len-comment-le-20}{91.24}
\DefMacro{ds-MP-CP-test_common_len-comment-le-30}{98.28}
\DefMacro{ds-MP-CP-test_common_len-comment-le-50}{99.71}
\DefMacro{ds-MP-CP-test_common_len-name-le-2}{51.43}
\DefMacro{ds-MP-CP-test_common_len-name-le-3}{79.46}
\DefMacro{ds-MP-CP-test_common_len-name-le-6}{99.10}
\DefMacro{ds-MP-CP-test_common_len-code-AVG}{85.11}
\DefMacro{ds-MP-CP-test_common_len-code-SUM}{292,608}
\DefMacro{ds-MP-CP-test_common_len-code-MAX}{1,311}
\DefMacro{ds-MP-CP-test_common_len-code-MIN}{9}
\DefMacro{ds-MP-CP-test_common_len-code-MEDIAN}{42.00}
\DefMacro{ds-MP-CP-test_common_len-code-STDEV}{121.62}
\DefMacro{ds-MP-CP-test_common_len-code-MODE}{10}
\DefMacro{ds-MP-CP-test_common_len-code-CNT}{3,438}
\DefMacro{ds-MP-CP-test_common_len-comment-AVG}{11.67}
\DefMacro{ds-MP-CP-test_common_len-comment-SUM}{40,136}
\DefMacro{ds-MP-CP-test_common_len-comment-MAX}{128}
\DefMacro{ds-MP-CP-test_common_len-comment-MIN}{1}
\DefMacro{ds-MP-CP-test_common_len-comment-MEDIAN}{10.00}
\DefMacro{ds-MP-CP-test_common_len-comment-STDEV}{7.61}
\DefMacro{ds-MP-CP-test_common_len-comment-MODE}{9}
\DefMacro{ds-MP-CP-test_common_len-comment-CNT}{3,438}
\DefMacro{ds-MP-CP-test_common_len-name-AVG}{2.65}
\DefMacro{ds-MP-CP-test_common_len-name-SUM}{9,112}
\DefMacro{ds-MP-CP-test_common_len-name-MAX}{11}
\DefMacro{ds-MP-CP-test_common_len-name-MIN}{1}
\DefMacro{ds-MP-CP-test_common_len-name-MEDIAN}{2.00}
\DefMacro{ds-MP-CP-test_common_len-name-STDEV}{1.28}
\DefMacro{ds-MP-CP-test_common_len-name-MODE}{2}
\DefMacro{ds-MP-CP-test_common_len-name-CNT}{3,438}
\DefMacro{ds-MP-T-test_common_num-data}{2,025}
\DefMacro{ds-MP-T-test_common_len-code-le-100}{74.32}
\DefMacro{ds-MP-T-test_common_len-code-le-150}{85.09}
\DefMacro{ds-MP-T-test_common_len-code-le-200}{90.27}
\DefMacro{ds-MP-T-test_common_len-comment-le-20}{90.07}
\DefMacro{ds-MP-T-test_common_len-comment-le-30}{97.53}
\DefMacro{ds-MP-T-test_common_len-comment-le-50}{99.46}
\DefMacro{ds-MP-T-test_common_len-name-le-2}{52.44}
\DefMacro{ds-MP-T-test_common_len-name-le-3}{79.11}
\DefMacro{ds-MP-T-test_common_len-name-le-6}{98.52}
\DefMacro{ds-MP-T-test_common_len-code-AVG}{86.28}
\DefMacro{ds-MP-T-test_common_len-code-SUM}{174,727}
\DefMacro{ds-MP-T-test_common_len-code-MAX}{1,819}
\DefMacro{ds-MP-T-test_common_len-code-MIN}{9}
\DefMacro{ds-MP-T-test_common_len-code-MEDIAN}{46.00}
\DefMacro{ds-MP-T-test_common_len-code-STDEV}{119.50}
\DefMacro{ds-MP-T-test_common_len-code-MODE}{15}
\DefMacro{ds-MP-T-test_common_len-code-CNT}{2,025}
\DefMacro{ds-MP-T-test_common_len-comment-AVG}{11.56}
\DefMacro{ds-MP-T-test_common_len-comment-SUM}{23,407}
\DefMacro{ds-MP-T-test_common_len-comment-MAX}{128}
\DefMacro{ds-MP-T-test_common_len-comment-MIN}{1}
\DefMacro{ds-MP-T-test_common_len-comment-MEDIAN}{10.00}
\DefMacro{ds-MP-T-test_common_len-comment-STDEV}{8.05}
\DefMacro{ds-MP-T-test_common_len-comment-MODE}{5}
\DefMacro{ds-MP-T-test_common_len-comment-CNT}{2,025}
\DefMacro{ds-MP-T-test_common_len-name-AVG}{2.63}
\DefMacro{ds-MP-T-test_common_len-name-SUM}{5,327}
\DefMacro{ds-MP-T-test_common_len-name-MAX}{10}
\DefMacro{ds-MP-T-test_common_len-name-MIN}{1}
\DefMacro{ds-MP-T-test_common_len-name-MEDIAN}{2.00}
\DefMacro{ds-MP-T-test_common_len-name-STDEV}{1.38}
\DefMacro{ds-MP-T-test_common_len-name-MODE}{2}
\DefMacro{ds-MP-T-test_common_len-name-CNT}{2,025}
\DefMacro{ds-CP-T-test_common_num-data}{2,259}
\DefMacro{ds-CP-T-test_common_len-code-le-100}{67.29}
\DefMacro{ds-CP-T-test_common_len-code-le-150}{81.41}
\DefMacro{ds-CP-T-test_common_len-code-le-200}{88.18}
\DefMacro{ds-CP-T-test_common_len-comment-le-20}{90.22}
\DefMacro{ds-CP-T-test_common_len-comment-le-30}{97.34}
\DefMacro{ds-CP-T-test_common_len-comment-le-50}{99.38}
\DefMacro{ds-CP-T-test_common_len-name-le-2}{49.27}
\DefMacro{ds-CP-T-test_common_len-name-le-3}{75.56}
\DefMacro{ds-CP-T-test_common_len-name-le-6}{98.80}
\DefMacro{ds-CP-T-test_common_len-code-AVG}{98.25}
\DefMacro{ds-CP-T-test_common_len-code-SUM}{221,949}
\DefMacro{ds-CP-T-test_common_len-code-MAX}{1,311}
\DefMacro{ds-CP-T-test_common_len-code-MIN}{10}
\DefMacro{ds-CP-T-test_common_len-code-MEDIAN}{59.00}
\DefMacro{ds-CP-T-test_common_len-code-STDEV}{115.63}
\DefMacro{ds-CP-T-test_common_len-code-MODE}{16}
\DefMacro{ds-CP-T-test_common_len-code-CNT}{2,259}
\DefMacro{ds-CP-T-test_common_len-comment-AVG}{11.66}
\DefMacro{ds-CP-T-test_common_len-comment-SUM}{26,343}
\DefMacro{ds-CP-T-test_common_len-comment-MAX}{136}
\DefMacro{ds-CP-T-test_common_len-comment-MIN}{2}
\DefMacro{ds-CP-T-test_common_len-comment-MEDIAN}{10.00}
\DefMacro{ds-CP-T-test_common_len-comment-STDEV}{8.63}
\DefMacro{ds-CP-T-test_common_len-comment-MODE}{5}
\DefMacro{ds-CP-T-test_common_len-comment-CNT}{2,259}
\DefMacro{ds-CP-T-test_common_len-name-AVG}{2.75}
\DefMacro{ds-CP-T-test_common_len-name-SUM}{6,210}
\DefMacro{ds-CP-T-test_common_len-name-MAX}{10}
\DefMacro{ds-CP-T-test_common_len-name-MIN}{1}
\DefMacro{ds-CP-T-test_common_len-name-MEDIAN}{3.00}
\DefMacro{ds-CP-T-test_common_len-name-STDEV}{1.37}
\DefMacro{ds-CP-T-test_common_len-name-MODE}{2}
\DefMacro{ds-CP-T-test_common_len-name-CNT}{2,259}
\DefMacro{ds-CG-val-MP_num-data}{7,569}
\DefMacro{ds-CG-val-MP_len-code-le-100}{73.62}
\DefMacro{ds-CG-val-MP_len-code-le-150}{83.71}
\DefMacro{ds-CG-val-MP_len-code-le-200}{89.42}
\DefMacro{ds-CG-val-MP_len-comment-le-20}{88.73}
\DefMacro{ds-CG-val-MP_len-comment-le-30}{97.21}
\DefMacro{ds-CG-val-MP_len-comment-le-50}{99.67}
\DefMacro{ds-CG-val-MP_len-code-AVG}{92.97}
\DefMacro{ds-CG-val-MP_len-code-SUM}{703,685}
\DefMacro{ds-CG-val-MP_len-code-MAX}{2,016}
\DefMacro{ds-CG-val-MP_len-code-MIN}{9}
\DefMacro{ds-CG-val-MP_len-code-MEDIAN}{49.00}
\DefMacro{ds-CG-val-MP_len-code-STDEV}{133.35}
\DefMacro{ds-CG-val-MP_len-code-MODE}{16}
\DefMacro{ds-CG-val-MP_len-code-CNT}{7,569}
\DefMacro{ds-CG-val-MP_len-comment-AVG}{12.07}
\DefMacro{ds-CG-val-MP_len-comment-SUM}{91,350}
\DefMacro{ds-CG-val-MP_len-comment-MAX}{174}
\DefMacro{ds-CG-val-MP_len-comment-MIN}{1}
\DefMacro{ds-CG-val-MP_len-comment-MEDIAN}{10.00}
\DefMacro{ds-CG-val-MP_len-comment-STDEV}{8.28}
\DefMacro{ds-CG-val-MP_len-comment-MODE}{8}
\DefMacro{ds-CG-val-MP_len-comment-CNT}{7,569}
\DefMacro{ds-CG-test_common-MP-CP_num-data}{3,362}
\DefMacro{ds-CG-test_common-MP-CP_len-code-le-100}{70.32}
\DefMacro{ds-CG-test_common-MP-CP_len-code-le-150}{80.07}
\DefMacro{ds-CG-test_common-MP-CP_len-code-le-200}{85.81}
\DefMacro{ds-CG-test_common-MP-CP_len-comment-le-20}{90.10}
\DefMacro{ds-CG-test_common-MP-CP_len-comment-le-30}{97.95}
\DefMacro{ds-CG-test_common-MP-CP_len-comment-le-50}{99.58}
\DefMacro{ds-CG-test_common-MP-CP_len-code-AVG}{106.83}
\DefMacro{ds-CG-test_common-MP-CP_len-code-SUM}{359,170}
\DefMacro{ds-CG-test_common-MP-CP_len-code-MAX}{1,740}
\DefMacro{ds-CG-test_common-MP-CP_len-code-MIN}{10}
\DefMacro{ds-CG-test_common-MP-CP_len-code-MEDIAN}{51.00}
\DefMacro{ds-CG-test_common-MP-CP_len-code-STDEV}{152.84}
\DefMacro{ds-CG-test_common-MP-CP_len-code-MODE}{16}
\DefMacro{ds-CG-test_common-MP-CP_len-code-CNT}{3,362}
\DefMacro{ds-CG-test_common-MP-CP_len-comment-AVG}{12.16}
\DefMacro{ds-CG-test_common-MP-CP_len-comment-SUM}{40,883}
\DefMacro{ds-CG-test_common-MP-CP_len-comment-MAX}{134}
\DefMacro{ds-CG-test_common-MP-CP_len-comment-MIN}{1}
\DefMacro{ds-CG-test_common-MP-CP_len-comment-MEDIAN}{11.00}
\DefMacro{ds-CG-test_common-MP-CP_len-comment-STDEV}{8.00}
\DefMacro{ds-CG-test_common-MP-CP_len-comment-MODE}{8}
\DefMacro{ds-CG-test_common-MP-CP_len-comment-CNT}{3,362}
\DefMacro{ds-CG-test_standard-MP_num-data}{14,956}
\DefMacro{ds-CG-test_standard-MP_len-code-le-100}{74.28}
\DefMacro{ds-CG-test_standard-MP_len-code-le-150}{83.96}
\DefMacro{ds-CG-test_standard-MP_len-code-le-200}{89.30}
\DefMacro{ds-CG-test_standard-MP_len-comment-le-20}{89.05}
\DefMacro{ds-CG-test_standard-MP_len-comment-le-30}{97.33}
\DefMacro{ds-CG-test_standard-MP_len-comment-le-50}{99.64}
\DefMacro{ds-CG-test_standard-MP_len-code-AVG}{93.02}
\DefMacro{ds-CG-test_standard-MP_len-code-SUM}{1,391,195}
\DefMacro{ds-CG-test_standard-MP_len-code-MAX}{2,215}
\DefMacro{ds-CG-test_standard-MP_len-code-MIN}{9}
\DefMacro{ds-CG-test_standard-MP_len-code-MEDIAN}{48.00}
\DefMacro{ds-CG-test_standard-MP_len-code-STDEV}{135.59}
\DefMacro{ds-CG-test_standard-MP_len-code-MODE}{20}
\DefMacro{ds-CG-test_standard-MP_len-code-CNT}{14,956}
\DefMacro{ds-CG-test_standard-MP_len-comment-AVG}{12.03}
\DefMacro{ds-CG-test_standard-MP_len-comment-SUM}{179,981}
\DefMacro{ds-CG-test_standard-MP_len-comment-MAX}{206}
\DefMacro{ds-CG-test_standard-MP_len-comment-MIN}{1}
\DefMacro{ds-CG-test_standard-MP_len-comment-MEDIAN}{10.00}
\DefMacro{ds-CG-test_standard-MP_len-comment-STDEV}{7.98}
\DefMacro{ds-CG-test_standard-MP_len-comment-MODE}{8}
\DefMacro{ds-CG-test_standard-MP_len-comment-CNT}{14,956}
\DefMacro{ds-CG-train-MP_num-data}{50,879}
\DefMacro{ds-CG-train-MP_len-code-le-100}{74.97}
\DefMacro{ds-CG-train-MP_len-code-le-150}{84.47}
\DefMacro{ds-CG-train-MP_len-code-le-200}{89.86}
\DefMacro{ds-CG-train-MP_len-comment-le-20}{88.91}
\DefMacro{ds-CG-train-MP_len-comment-le-30}{97.12}
\DefMacro{ds-CG-train-MP_len-comment-le-50}{99.61}
\DefMacro{ds-CG-train-MP_len-code-AVG}{89.84}
\DefMacro{ds-CG-train-MP_len-code-SUM}{4,571,112}
\DefMacro{ds-CG-train-MP_len-code-MAX}{2,544}
\DefMacro{ds-CG-train-MP_len-code-MIN}{9}
\DefMacro{ds-CG-train-MP_len-code-MEDIAN}{47.00}
\DefMacro{ds-CG-train-MP_len-code-STDEV}{129.95}
\DefMacro{ds-CG-train-MP_len-code-MODE}{19}
\DefMacro{ds-CG-train-MP_len-code-CNT}{50,879}
\DefMacro{ds-CG-train-MP_len-comment-AVG}{11.98}
\DefMacro{ds-CG-train-MP_len-comment-SUM}{609,725}
\DefMacro{ds-CG-train-MP_len-comment-MAX}{212}
\DefMacro{ds-CG-train-MP_len-comment-MIN}{1}
\DefMacro{ds-CG-train-MP_len-comment-MEDIAN}{10.00}
\DefMacro{ds-CG-train-MP_len-comment-STDEV}{8.05}
\DefMacro{ds-CG-train-MP_len-comment-MODE}{8}
\DefMacro{ds-CG-train-MP_len-comment-CNT}{50,879}
\DefMacro{ds-CG-test_common-MP-T_num-data}{2,013}
\DefMacro{ds-CG-test_common-MP-T_len-code-le-100}{68.16}
\DefMacro{ds-CG-test_common-MP-T_len-code-le-150}{79.78}
\DefMacro{ds-CG-test_common-MP-T_len-code-le-200}{86.74}
\DefMacro{ds-CG-test_common-MP-T_len-comment-le-20}{88.72}
\DefMacro{ds-CG-test_common-MP-T_len-comment-le-30}{96.77}
\DefMacro{ds-CG-test_common-MP-T_len-comment-le-50}{99.45}
\DefMacro{ds-CG-test_common-MP-T_len-code-AVG}{106.15}
\DefMacro{ds-CG-test_common-MP-T_len-code-SUM}{213,679}
\DefMacro{ds-CG-test_common-MP-T_len-code-MAX}{1,981}
\DefMacro{ds-CG-test_common-MP-T_len-code-MIN}{10}
\DefMacro{ds-CG-test_common-MP-T_len-code-MEDIAN}{58.00}
\DefMacro{ds-CG-test_common-MP-T_len-code-STDEV}{144.12}
\DefMacro{ds-CG-test_common-MP-T_len-code-MODE}{22}
\DefMacro{ds-CG-test_common-MP-T_len-code-CNT}{2,013}
\DefMacro{ds-CG-test_common-MP-T_len-comment-AVG}{12.10}
\DefMacro{ds-CG-test_common-MP-T_len-comment-SUM}{24,350}
\DefMacro{ds-CG-test_common-MP-T_len-comment-MAX}{134}
\DefMacro{ds-CG-test_common-MP-T_len-comment-MIN}{2}
\DefMacro{ds-CG-test_common-MP-T_len-comment-MEDIAN}{10.00}
\DefMacro{ds-CG-test_common-MP-T_len-comment-STDEV}{8.50}
\DefMacro{ds-CG-test_common-MP-T_len-comment-MODE}{8}
\DefMacro{ds-CG-test_common-MP-T_len-comment-CNT}{2,013}
\DefMacro{ds-CG-val-CP_num-data}{8,938}
\DefMacro{ds-CG-val-CP_len-code-le-100}{73.42}
\DefMacro{ds-CG-val-CP_len-code-le-150}{83.42}
\DefMacro{ds-CG-val-CP_len-code-le-200}{89.24}
\DefMacro{ds-CG-val-CP_len-comment-le-20}{88.70}
\DefMacro{ds-CG-val-CP_len-comment-le-30}{96.65}
\DefMacro{ds-CG-val-CP_len-comment-le-50}{99.62}
\DefMacro{ds-CG-val-CP_len-code-AVG}{93.12}
\DefMacro{ds-CG-val-CP_len-code-SUM}{832,266}
\DefMacro{ds-CG-val-CP_len-code-MAX}{2,021}
\DefMacro{ds-CG-val-CP_len-code-MIN}{10}
\DefMacro{ds-CG-val-CP_len-code-MEDIAN}{53.00}
\DefMacro{ds-CG-val-CP_len-code-STDEV}{113.79}
\DefMacro{ds-CG-val-CP_len-code-MODE}{43}
\DefMacro{ds-CG-val-CP_len-code-CNT}{8,938}
\DefMacro{ds-CG-val-CP_len-comment-AVG}{12.09}
\DefMacro{ds-CG-val-CP_len-comment-SUM}{108,044}
\DefMacro{ds-CG-val-CP_len-comment-MAX}{114}
\DefMacro{ds-CG-val-CP_len-comment-MIN}{1}
\DefMacro{ds-CG-val-CP_len-comment-MEDIAN}{10.00}
\DefMacro{ds-CG-val-CP_len-comment-STDEV}{7.90}
\DefMacro{ds-CG-val-CP_len-comment-MODE}{8}
\DefMacro{ds-CG-val-CP_len-comment-CNT}{8,938}
\DefMacro{ds-CG-test_common-CP-T_num-data}{2,220}
\DefMacro{ds-CG-test_common-CP-T_len-code-le-100}{59.82}
\DefMacro{ds-CG-test_common-CP-T_len-code-le-150}{73.69}
\DefMacro{ds-CG-test_common-CP-T_len-code-le-200}{82.52}
\DefMacro{ds-CG-test_common-CP-T_len-comment-le-20}{88.29}
\DefMacro{ds-CG-test_common-CP-T_len-comment-le-30}{96.80}
\DefMacro{ds-CG-test_common-CP-T_len-comment-le-50}{99.37}
\DefMacro{ds-CG-test_common-CP-T_len-code-AVG}{125.01}
\DefMacro{ds-CG-test_common-CP-T_len-code-SUM}{277,528}
\DefMacro{ds-CG-test_common-CP-T_len-code-MAX}{1,311}
\DefMacro{ds-CG-test_common-CP-T_len-code-MIN}{10}
\DefMacro{ds-CG-test_common-CP-T_len-code-MEDIAN}{73.00}
\DefMacro{ds-CG-test_common-CP-T_len-code-STDEV}{147.76}
\DefMacro{ds-CG-test_common-CP-T_len-code-MODE}{16}
\DefMacro{ds-CG-test_common-CP-T_len-code-CNT}{2,220}
\DefMacro{ds-CG-test_common-CP-T_len-comment-AVG}{12.20}
\DefMacro{ds-CG-test_common-CP-T_len-comment-SUM}{27,095}
\DefMacro{ds-CG-test_common-CP-T_len-comment-MAX}{136}
\DefMacro{ds-CG-test_common-CP-T_len-comment-MIN}{2}
\DefMacro{ds-CG-test_common-CP-T_len-comment-MEDIAN}{10.00}
\DefMacro{ds-CG-test_common-CP-T_len-comment-STDEV}{9.01}
\DefMacro{ds-CG-test_common-CP-T_len-comment-MODE}{5}
\DefMacro{ds-CG-test_common-CP-T_len-comment-CNT}{2,220}
\DefMacro{ds-CG-test_standard-CP_num-data}{15,661}
\DefMacro{ds-CG-test_standard-CP_len-code-le-100}{69.82}
\DefMacro{ds-CG-test_standard-CP_len-code-le-150}{80.38}
\DefMacro{ds-CG-test_standard-CP_len-code-le-200}{86.69}
\DefMacro{ds-CG-test_standard-CP_len-comment-le-20}{89.53}
\DefMacro{ds-CG-test_standard-CP_len-comment-le-30}{97.54}
\DefMacro{ds-CG-test_standard-CP_len-comment-le-50}{99.69}
\DefMacro{ds-CG-test_standard-CP_len-code-AVG}{106.11}
\DefMacro{ds-CG-test_standard-CP_len-code-SUM}{1,661,776}
\DefMacro{ds-CG-test_standard-CP_len-code-MAX}{2,544}
\DefMacro{ds-CG-test_standard-CP_len-code-MIN}{10}
\DefMacro{ds-CG-test_standard-CP_len-code-MEDIAN}{53.00}
\DefMacro{ds-CG-test_standard-CP_len-code-STDEV}{155.61}
\DefMacro{ds-CG-test_standard-CP_len-code-MODE}{16}
\DefMacro{ds-CG-test_standard-CP_len-code-CNT}{15,661}
\DefMacro{ds-CG-test_standard-CP_len-comment-AVG}{12.13}
\DefMacro{ds-CG-test_standard-CP_len-comment-SUM}{189,979}
\DefMacro{ds-CG-test_standard-CP_len-comment-MAX}{212}
\DefMacro{ds-CG-test_standard-CP_len-comment-MIN}{1}
\DefMacro{ds-CG-test_standard-CP_len-comment-MEDIAN}{10.00}
\DefMacro{ds-CG-test_standard-CP_len-comment-STDEV}{7.69}
\DefMacro{ds-CG-test_standard-CP_len-comment-MODE}{8}
\DefMacro{ds-CG-test_standard-CP_len-comment-CNT}{15,661}
\DefMacro{ds-CG-train-CP_num-data}{50,879}
\DefMacro{ds-CG-train-CP_len-code-le-100}{76.36}
\DefMacro{ds-CG-train-CP_len-code-le-150}{85.59}
\DefMacro{ds-CG-train-CP_len-code-le-200}{90.67}
\DefMacro{ds-CG-train-CP_len-comment-le-20}{88.74}
\DefMacro{ds-CG-train-CP_len-comment-le-30}{97.12}
\DefMacro{ds-CG-train-CP_len-comment-le-50}{99.61}
\DefMacro{ds-CG-train-CP_len-code-AVG}{86.23}
\DefMacro{ds-CG-train-CP_len-code-SUM}{4,387,455}
\DefMacro{ds-CG-train-CP_len-code-MAX}{2,699}
\DefMacro{ds-CG-train-CP_len-code-MIN}{9}
\DefMacro{ds-CG-train-CP_len-code-MEDIAN}{44.00}
\DefMacro{ds-CG-train-CP_len-code-STDEV}{127.89}
\DefMacro{ds-CG-train-CP_len-code-MODE}{25}
\DefMacro{ds-CG-train-CP_len-code-CNT}{50,879}
\DefMacro{ds-CG-train-CP_len-comment-AVG}{11.96}
\DefMacro{ds-CG-train-CP_len-comment-SUM}{608,502}
\DefMacro{ds-CG-train-CP_len-comment-MAX}{206}
\DefMacro{ds-CG-train-CP_len-comment-MIN}{1}
\DefMacro{ds-CG-train-CP_len-comment-MEDIAN}{10.00}
\DefMacro{ds-CG-train-CP_len-comment-STDEV}{8.19}
\DefMacro{ds-CG-train-CP_len-comment-MODE}{9}
\DefMacro{ds-CG-train-CP_len-comment-CNT}{50,879}
\DefMacro{ds-CG-val-T_num-data}{11,312}
\DefMacro{ds-CG-val-T_len-code-le-100}{70.66}
\DefMacro{ds-CG-val-T_len-code-le-150}{81.24}
\DefMacro{ds-CG-val-T_len-code-le-200}{87.05}
\DefMacro{ds-CG-val-T_len-comment-le-20}{88.56}
\DefMacro{ds-CG-val-T_len-comment-le-30}{96.96}
\DefMacro{ds-CG-val-T_len-comment-le-50}{99.62}
\DefMacro{ds-CG-val-T_len-code-AVG}{104.88}
\DefMacro{ds-CG-val-T_len-code-SUM}{1,186,426}
\DefMacro{ds-CG-val-T_len-code-MAX}{2,215}
\DefMacro{ds-CG-val-T_len-code-MIN}{9}
\DefMacro{ds-CG-val-T_len-code-MEDIAN}{55.00}
\DefMacro{ds-CG-val-T_len-code-STDEV}{149.17}
\DefMacro{ds-CG-val-T_len-code-MODE}{20}
\DefMacro{ds-CG-val-T_len-code-CNT}{11,312}
\DefMacro{ds-CG-val-T_len-comment-AVG}{12.20}
\DefMacro{ds-CG-val-T_len-comment-SUM}{137,970}
\DefMacro{ds-CG-val-T_len-comment-MAX}{212}
\DefMacro{ds-CG-val-T_len-comment-MIN}{1}
\DefMacro{ds-CG-val-T_len-comment-MEDIAN}{10.00}
\DefMacro{ds-CG-val-T_len-comment-STDEV}{8.30}
\DefMacro{ds-CG-val-T_len-comment-MODE}{8}
\DefMacro{ds-CG-val-T_len-comment-CNT}{11,312}
\DefMacro{ds-CG-test_standard-T_num-data}{9,870}
\DefMacro{ds-CG-test_standard-T_len-code-le-100}{68.38}
\DefMacro{ds-CG-test_standard-T_len-code-le-150}{79.75}
\DefMacro{ds-CG-test_standard-T_len-code-le-200}{86.80}
\DefMacro{ds-CG-test_standard-T_len-comment-le-20}{88.99}
\DefMacro{ds-CG-test_standard-T_len-comment-le-30}{96.53}
\DefMacro{ds-CG-test_standard-T_len-comment-le-50}{99.37}
\DefMacro{ds-CG-test_standard-T_len-code-AVG}{105.36}
\DefMacro{ds-CG-test_standard-T_len-code-SUM}{1,039,901}
\DefMacro{ds-CG-test_standard-T_len-code-MAX}{2,178}
\DefMacro{ds-CG-test_standard-T_len-code-MIN}{10}
\DefMacro{ds-CG-test_standard-T_len-code-MEDIAN}{56.50}
\DefMacro{ds-CG-test_standard-T_len-code-STDEV}{140.13}
\DefMacro{ds-CG-test_standard-T_len-code-MODE}{22}
\DefMacro{ds-CG-test_standard-T_len-code-CNT}{9,870}
\DefMacro{ds-CG-test_standard-T_len-comment-AVG}{12.04}
\DefMacro{ds-CG-test_standard-T_len-comment-SUM}{118,786}
\DefMacro{ds-CG-test_standard-T_len-comment-MAX}{136}
\DefMacro{ds-CG-test_standard-T_len-comment-MIN}{1}
\DefMacro{ds-CG-test_standard-T_len-comment-MEDIAN}{10.00}
\DefMacro{ds-CG-test_standard-T_len-comment-STDEV}{8.55}
\DefMacro{ds-CG-test_standard-T_len-comment-MODE}{8}
\DefMacro{ds-CG-test_standard-T_len-comment-CNT}{9,870}
\DefMacro{ds-CG-train-T_num-data}{50,879}
\DefMacro{ds-CG-train-T_len-code-le-100}{76.55}
\DefMacro{ds-CG-train-T_len-code-le-150}{85.70}
\DefMacro{ds-CG-train-T_len-code-le-200}{90.74}
\DefMacro{ds-CG-train-T_len-comment-le-20}{88.83}
\DefMacro{ds-CG-train-T_len-comment-le-30}{97.29}
\DefMacro{ds-CG-train-T_len-comment-le-50}{99.68}
\DefMacro{ds-CG-train-T_len-code-AVG}{85.83}
\DefMacro{ds-CG-train-T_len-code-SUM}{4,367,057}
\DefMacro{ds-CG-train-T_len-code-MAX}{2,699}
\DefMacro{ds-CG-train-T_len-code-MIN}{9}
\DefMacro{ds-CG-train-T_len-code-MEDIAN}{45.00}
\DefMacro{ds-CG-train-T_len-code-STDEV}{127.04}
\DefMacro{ds-CG-train-T_len-code-MODE}{16}
\DefMacro{ds-CG-train-T_len-code-CNT}{50,879}
\DefMacro{ds-CG-train-T_len-comment-AVG}{12.00}
\DefMacro{ds-CG-train-T_len-comment-SUM}{610,737}
\DefMacro{ds-CG-train-T_len-comment-MAX}{206}
\DefMacro{ds-CG-train-T_len-comment-MIN}{1}
\DefMacro{ds-CG-train-T_len-comment-MEDIAN}{10.00}
\DefMacro{ds-CG-train-T_len-comment-STDEV}{7.88}
\DefMacro{ds-CG-train-T_len-comment-MODE}{8}
\DefMacro{ds-CG-train-T_len-comment-CNT}{50,879}
\DefMacro{ds-MN-val-MP_num-data}{7,523}
\DefMacro{ds-MN-val-MP_len-code-le-100}{74.00}
\DefMacro{ds-MN-val-MP_len-code-le-150}{83.89}
\DefMacro{ds-MN-val-MP_len-code-le-200}{89.47}
\DefMacro{ds-MN-val-MP_len-name-le-2}{56.40}
\DefMacro{ds-MN-val-MP_len-name-le-3}{81.72}
\DefMacro{ds-MN-val-MP_len-name-le-6}{98.67}
\DefMacro{ds-MN-val-MP_len-code-AVG}{91.70}
\DefMacro{ds-MN-val-MP_len-code-SUM}{689,882}
\DefMacro{ds-MN-val-MP_len-code-MAX}{2,014}
\DefMacro{ds-MN-val-MP_len-code-MIN}{9}
\DefMacro{ds-MN-val-MP_len-code-MEDIAN}{47.00}
\DefMacro{ds-MN-val-MP_len-code-STDEV}{133.57}
\DefMacro{ds-MN-val-MP_len-code-MODE}{15}
\DefMacro{ds-MN-val-MP_len-code-CNT}{7,523}
\DefMacro{ds-MN-val-MP_len-name-AVG}{2.52}
\DefMacro{ds-MN-val-MP_len-name-SUM}{18,988}
\DefMacro{ds-MN-val-MP_len-name-MAX}{20}
\DefMacro{ds-MN-val-MP_len-name-MIN}{1}
\DefMacro{ds-MN-val-MP_len-name-MEDIAN}{2.00}
\DefMacro{ds-MN-val-MP_len-name-STDEV}{1.38}
\DefMacro{ds-MN-val-MP_len-name-MODE}{2}
\DefMacro{ds-MN-val-MP_len-name-CNT}{7,523}
\DefMacro{ds-MN-test_common-MP-CP_num-data}{3,344}
\DefMacro{ds-MN-test_common-MP-CP_len-code-le-100}{70.93}
\DefMacro{ds-MN-test_common-MP-CP_len-code-le-150}{80.17}
\DefMacro{ds-MN-test_common-MP-CP_len-code-le-200}{85.94}
\DefMacro{ds-MN-test_common-MP-CP_len-name-le-2}{49.55}
\DefMacro{ds-MN-test_common-MP-CP_len-name-le-3}{78.20}
\DefMacro{ds-MN-test_common-MP-CP_len-name-le-6}{98.92}
\DefMacro{ds-MN-test_common-MP-CP_len-code-AVG}{105.42}
\DefMacro{ds-MN-test_common-MP-CP_len-code-SUM}{352,511}
\DefMacro{ds-MN-test_common-MP-CP_len-code-MAX}{1,739}
\DefMacro{ds-MN-test_common-MP-CP_len-code-MIN}{10}
\DefMacro{ds-MN-test_common-MP-CP_len-code-MEDIAN}{50.00}
\DefMacro{ds-MN-test_common-MP-CP_len-code-STDEV}{153.12}
\DefMacro{ds-MN-test_common-MP-CP_len-code-MODE}{15}
\DefMacro{ds-MN-test_common-MP-CP_len-code-CNT}{3,344}
\DefMacro{ds-MN-test_common-MP-CP_len-name-AVG}{2.70}
\DefMacro{ds-MN-test_common-MP-CP_len-name-SUM}{9,036}
\DefMacro{ds-MN-test_common-MP-CP_len-name-MAX}{12}
\DefMacro{ds-MN-test_common-MP-CP_len-name-MIN}{1}
\DefMacro{ds-MN-test_common-MP-CP_len-name-MEDIAN}{3.00}
\DefMacro{ds-MN-test_common-MP-CP_len-name-STDEV}{1.34}
\DefMacro{ds-MN-test_common-MP-CP_len-name-MODE}{2}
\DefMacro{ds-MN-test_common-MP-CP_len-name-CNT}{3,344}
\DefMacro{ds-MN-test_standard-MP_num-data}{14,796}
\DefMacro{ds-MN-test_standard-MP_len-code-le-100}{74.64}
\DefMacro{ds-MN-test_standard-MP_len-code-le-150}{84.08}
\DefMacro{ds-MN-test_standard-MP_len-code-le-200}{89.36}
\DefMacro{ds-MN-test_standard-MP_len-name-le-2}{55.52}
\DefMacro{ds-MN-test_standard-MP_len-name-le-3}{80.64}
\DefMacro{ds-MN-test_standard-MP_len-name-le-6}{98.53}
\DefMacro{ds-MN-test_standard-MP_len-code-AVG}{91.97}
\DefMacro{ds-MN-test_standard-MP_len-code-SUM}{1,360,799}
\DefMacro{ds-MN-test_standard-MP_len-code-MAX}{2,213}
\DefMacro{ds-MN-test_standard-MP_len-code-MIN}{9}
\DefMacro{ds-MN-test_standard-MP_len-code-MEDIAN}{47.00}
\DefMacro{ds-MN-test_standard-MP_len-code-STDEV}{136.02}
\DefMacro{ds-MN-test_standard-MP_len-code-MODE}{15}
\DefMacro{ds-MN-test_standard-MP_len-code-CNT}{14,796}
\DefMacro{ds-MN-test_standard-MP_len-name-AVG}{2.56}
\DefMacro{ds-MN-test_standard-MP_len-name-SUM}{37,853}
\DefMacro{ds-MN-test_standard-MP_len-name-MAX}{16}
\DefMacro{ds-MN-test_standard-MP_len-name-MIN}{1}
\DefMacro{ds-MN-test_standard-MP_len-name-MEDIAN}{2.00}
\DefMacro{ds-MN-test_standard-MP_len-name-STDEV}{1.40}
\DefMacro{ds-MN-test_standard-MP_len-name-MODE}{2}
\DefMacro{ds-MN-test_standard-MP_len-name-CNT}{14,796}
\DefMacro{ds-MN-train-MP_num-data}{50,879}
\DefMacro{ds-MN-train-MP_len-code-le-100}{75.45}
\DefMacro{ds-MN-train-MP_len-code-le-150}{84.75}
\DefMacro{ds-MN-train-MP_len-code-le-200}{90.03}
\DefMacro{ds-MN-train-MP_len-name-le-2}{57.27}
\DefMacro{ds-MN-train-MP_len-name-le-3}{81.31}
\DefMacro{ds-MN-train-MP_len-name-le-6}{98.54}
\DefMacro{ds-MN-train-MP_len-code-AVG}{88.15}
\DefMacro{ds-MN-train-MP_len-code-SUM}{4,485,088}
\DefMacro{ds-MN-train-MP_len-code-MAX}{2,537}
\DefMacro{ds-MN-train-MP_len-code-MIN}{9}
\DefMacro{ds-MN-train-MP_len-code-MEDIAN}{45.00}
\DefMacro{ds-MN-train-MP_len-code-STDEV}{129.87}
\DefMacro{ds-MN-train-MP_len-code-MODE}{19}
\DefMacro{ds-MN-train-MP_len-code-CNT}{50,879}
\DefMacro{ds-MN-train-MP_len-name-AVG}{2.52}
\DefMacro{ds-MN-train-MP_len-name-SUM}{128,308}
\DefMacro{ds-MN-train-MP_len-name-MAX}{21}
\DefMacro{ds-MN-train-MP_len-name-MIN}{1}
\DefMacro{ds-MN-train-MP_len-name-MEDIAN}{2.00}
\DefMacro{ds-MN-train-MP_len-name-STDEV}{1.38}
\DefMacro{ds-MN-train-MP_len-name-MODE}{2}
\DefMacro{ds-MN-train-MP_len-name-CNT}{50,879}
\DefMacro{ds-MN-test_common-MP-T_num-data}{2,011}
\DefMacro{ds-MN-test_common-MP-T_len-code-le-100}{68.72}
\DefMacro{ds-MN-test_common-MP-T_len-code-le-150}{80.01}
\DefMacro{ds-MN-test_common-MP-T_len-code-le-200}{86.92}
\DefMacro{ds-MN-test_common-MP-T_len-name-le-2}{50.92}
\DefMacro{ds-MN-test_common-MP-T_len-name-le-3}{77.13}
\DefMacro{ds-MN-test_common-MP-T_len-name-le-6}{98.01}
\DefMacro{ds-MN-test_common-MP-T_len-code-AVG}{104.38}
\DefMacro{ds-MN-test_common-MP-T_len-code-SUM}{209,903}
\DefMacro{ds-MN-test_common-MP-T_len-code-MAX}{1,981}
\DefMacro{ds-MN-test_common-MP-T_len-code-MIN}{10}
\DefMacro{ds-MN-test_common-MP-T_len-code-MEDIAN}{56.00}
\DefMacro{ds-MN-test_common-MP-T_len-code-STDEV}{144.06}
\DefMacro{ds-MN-test_common-MP-T_len-code-MODE}{22}
\DefMacro{ds-MN-test_common-MP-T_len-code-CNT}{2,011}
\DefMacro{ds-MN-test_common-MP-T_len-name-AVG}{2.70}
\DefMacro{ds-MN-test_common-MP-T_len-name-SUM}{5,421}
\DefMacro{ds-MN-test_common-MP-T_len-name-MAX}{11}
\DefMacro{ds-MN-test_common-MP-T_len-name-MIN}{1}
\DefMacro{ds-MN-test_common-MP-T_len-name-MEDIAN}{2.00}
\DefMacro{ds-MN-test_common-MP-T_len-name-STDEV}{1.46}
\DefMacro{ds-MN-test_common-MP-T_len-name-MODE}{2}
\DefMacro{ds-MN-test_common-MP-T_len-name-CNT}{2,011}
\DefMacro{ds-MN-val-CP_num-data}{8,811}
\DefMacro{ds-MN-val-CP_len-code-le-100}{73.57}
\DefMacro{ds-MN-val-CP_len-code-le-150}{83.52}
\DefMacro{ds-MN-val-CP_len-code-le-200}{89.32}
\DefMacro{ds-MN-val-CP_len-name-le-2}{64.60}
\DefMacro{ds-MN-val-CP_len-name-le-3}{84.67}
\DefMacro{ds-MN-val-CP_len-name-le-6}{98.91}
\DefMacro{ds-MN-val-CP_len-code-AVG}{92.28}
\DefMacro{ds-MN-val-CP_len-code-SUM}{813,037}
\DefMacro{ds-MN-val-CP_len-code-MAX}{2,021}
\DefMacro{ds-MN-val-CP_len-code-MIN}{9}
\DefMacro{ds-MN-val-CP_len-code-MEDIAN}{52.00}
\DefMacro{ds-MN-val-CP_len-code-STDEV}{113.91}
\DefMacro{ds-MN-val-CP_len-code-MODE}{27}
\DefMacro{ds-MN-val-CP_len-code-CNT}{8,811}
\DefMacro{ds-MN-val-CP_len-name-AVG}{2.31}
\DefMacro{ds-MN-val-CP_len-name-SUM}{20,312}
\DefMacro{ds-MN-val-CP_len-name-MAX}{13}
\DefMacro{ds-MN-val-CP_len-name-MIN}{1}
\DefMacro{ds-MN-val-CP_len-name-MEDIAN}{2.00}
\DefMacro{ds-MN-val-CP_len-name-STDEV}{1.35}
\DefMacro{ds-MN-val-CP_len-name-MODE}{2}
\DefMacro{ds-MN-val-CP_len-name-CNT}{8,811}
\DefMacro{ds-MN-test_common-CP-T_num-data}{2,211}
\DefMacro{ds-MN-test_common-CP-T_len-code-le-100}{60.33}
\DefMacro{ds-MN-test_common-CP-T_len-code-le-150}{74.04}
\DefMacro{ds-MN-test_common-CP-T_len-code-le-200}{82.81}
\DefMacro{ds-MN-test_common-CP-T_len-name-le-2}{46.54}
\DefMacro{ds-MN-test_common-CP-T_len-name-le-3}{73.77}
\DefMacro{ds-MN-test_common-CP-T_len-name-le-6}{98.24}
\DefMacro{ds-MN-test_common-CP-T_len-code-AVG}{123.46}
\DefMacro{ds-MN-test_common-CP-T_len-code-SUM}{272,965}
\DefMacro{ds-MN-test_common-CP-T_len-code-MAX}{1,311}
\DefMacro{ds-MN-test_common-CP-T_len-code-MIN}{10}
\DefMacro{ds-MN-test_common-CP-T_len-code-MEDIAN}{73.00}
\DefMacro{ds-MN-test_common-CP-T_len-code-STDEV}{147.82}
\DefMacro{ds-MN-test_common-CP-T_len-code-MODE}{15}
\DefMacro{ds-MN-test_common-CP-T_len-code-CNT}{2,211}
\DefMacro{ds-MN-test_common-CP-T_len-name-AVG}{2.82}
\DefMacro{ds-MN-test_common-CP-T_len-name-SUM}{6,235}
\DefMacro{ds-MN-test_common-CP-T_len-name-MAX}{11}
\DefMacro{ds-MN-test_common-CP-T_len-name-MIN}{1}
\DefMacro{ds-MN-test_common-CP-T_len-name-MEDIAN}{3.00}
\DefMacro{ds-MN-test_common-CP-T_len-name-STDEV}{1.43}
\DefMacro{ds-MN-test_common-CP-T_len-name-MODE}{2}
\DefMacro{ds-MN-test_common-CP-T_len-name-CNT}{2,211}
\DefMacro{ds-MN-test_standard-CP_num-data}{15,332}
\DefMacro{ds-MN-test_standard-CP_len-code-le-100}{69.91}
\DefMacro{ds-MN-test_standard-CP_len-code-le-150}{80.37}
\DefMacro{ds-MN-test_standard-CP_len-code-le-200}{86.63}
\DefMacro{ds-MN-test_standard-CP_len-name-le-2}{50.39}
\DefMacro{ds-MN-test_standard-CP_len-name-le-3}{78.10}
\DefMacro{ds-MN-test_standard-CP_len-name-le-6}{98.62}
\DefMacro{ds-MN-test_standard-CP_len-code-AVG}{105.87}
\DefMacro{ds-MN-test_standard-CP_len-code-SUM}{1,623,220}
\DefMacro{ds-MN-test_standard-CP_len-code-MAX}{2,537}
\DefMacro{ds-MN-test_standard-CP_len-code-MIN}{10}
\DefMacro{ds-MN-test_standard-CP_len-code-MEDIAN}{53.00}
\DefMacro{ds-MN-test_standard-CP_len-code-STDEV}{156.76}
\DefMacro{ds-MN-test_standard-CP_len-code-MODE}{15}
\DefMacro{ds-MN-test_standard-CP_len-code-CNT}{15,332}
\DefMacro{ds-MN-test_standard-CP_len-name-AVG}{2.71}
\DefMacro{ds-MN-test_standard-CP_len-name-SUM}{41,558}
\DefMacro{ds-MN-test_standard-CP_len-name-MAX}{16}
\DefMacro{ds-MN-test_standard-CP_len-name-MIN}{1}
\DefMacro{ds-MN-test_standard-CP_len-name-MEDIAN}{2.00}
\DefMacro{ds-MN-test_standard-CP_len-name-STDEV}{1.38}
\DefMacro{ds-MN-test_standard-CP_len-name-MODE}{2}
\DefMacro{ds-MN-test_standard-CP_len-name-CNT}{15,332}
\DefMacro{ds-MN-train-CP_num-data}{50,879}
\DefMacro{ds-MN-train-CP_len-code-le-100}{76.81}
\DefMacro{ds-MN-train-CP_len-code-le-150}{85.83}
\DefMacro{ds-MN-train-CP_len-code-le-200}{90.80}
\DefMacro{ds-MN-train-CP_len-name-le-2}{57.82}
\DefMacro{ds-MN-train-CP_len-name-le-3}{81.64}
\DefMacro{ds-MN-train-CP_len-name-le-6}{98.47}
\DefMacro{ds-MN-train-CP_len-code-AVG}{84.56}
\DefMacro{ds-MN-train-CP_len-code-SUM}{4,302,410}
\DefMacro{ds-MN-train-CP_len-code-MAX}{2,699}
\DefMacro{ds-MN-train-CP_len-code-MIN}{9}
\DefMacro{ds-MN-train-CP_len-code-MEDIAN}{43.00}
\DefMacro{ds-MN-train-CP_len-code-STDEV}{127.81}
\DefMacro{ds-MN-train-CP_len-code-MODE}{10}
\DefMacro{ds-MN-train-CP_len-code-CNT}{50,879}
\DefMacro{ds-MN-train-CP_len-name-AVG}{2.50}
\DefMacro{ds-MN-train-CP_len-name-SUM}{127,358}
\DefMacro{ds-MN-train-CP_len-name-MAX}{21}
\DefMacro{ds-MN-train-CP_len-name-MIN}{1}
\DefMacro{ds-MN-train-CP_len-name-MEDIAN}{2.00}
\DefMacro{ds-MN-train-CP_len-name-STDEV}{1.39}
\DefMacro{ds-MN-train-CP_len-name-MODE}{2}
\DefMacro{ds-MN-train-CP_len-name-CNT}{50,879}
\DefMacro{ds-MN-val-T_num-data}{11,223}
\DefMacro{ds-MN-val-T_len-code-le-100}{71.00}
\DefMacro{ds-MN-val-T_len-code-le-150}{81.44}
\DefMacro{ds-MN-val-T_len-code-le-200}{87.16}
\DefMacro{ds-MN-val-T_len-name-le-2}{55.67}
\DefMacro{ds-MN-val-T_len-name-le-3}{78.29}
\DefMacro{ds-MN-val-T_len-name-le-6}{97.61}
\DefMacro{ds-MN-val-T_len-code-AVG}{103.71}
\DefMacro{ds-MN-val-T_len-code-SUM}{1,163,964}
\DefMacro{ds-MN-val-T_len-code-MAX}{2,213}
\DefMacro{ds-MN-val-T_len-code-MIN}{9}
\DefMacro{ds-MN-val-T_len-code-MEDIAN}{54.00}
\DefMacro{ds-MN-val-T_len-code-STDEV}{149.45}
\DefMacro{ds-MN-val-T_len-code-MODE}{20}
\DefMacro{ds-MN-val-T_len-code-CNT}{11,223}
\DefMacro{ds-MN-val-T_len-name-AVG}{2.63}
\DefMacro{ds-MN-val-T_len-name-SUM}{29,529}
\DefMacro{ds-MN-val-T_len-name-MAX}{17}
\DefMacro{ds-MN-val-T_len-name-MIN}{1}
\DefMacro{ds-MN-val-T_len-name-MEDIAN}{2.00}
\DefMacro{ds-MN-val-T_len-name-STDEV}{1.54}
\DefMacro{ds-MN-val-T_len-name-MODE}{2}
\DefMacro{ds-MN-val-T_len-name-CNT}{11,223}
\DefMacro{ds-MN-test_standard-T_num-data}{9,807}
\DefMacro{ds-MN-test_standard-T_len-code-le-100}{68.79}
\DefMacro{ds-MN-test_standard-T_len-code-le-150}{79.95}
\DefMacro{ds-MN-test_standard-T_len-code-le-200}{86.96}
\DefMacro{ds-MN-test_standard-T_len-name-le-2}{53.99}
\DefMacro{ds-MN-test_standard-T_len-name-le-3}{78.48}
\DefMacro{ds-MN-test_standard-T_len-name-le-6}{98.00}
\DefMacro{ds-MN-test_standard-T_len-code-AVG}{104.09}
\DefMacro{ds-MN-test_standard-T_len-code-SUM}{1,020,781}
\DefMacro{ds-MN-test_standard-T_len-code-MAX}{2,176}
\DefMacro{ds-MN-test_standard-T_len-code-MIN}{9}
\DefMacro{ds-MN-test_standard-T_len-code-MEDIAN}{55.00}
\DefMacro{ds-MN-test_standard-T_len-code-STDEV}{140.30}
\DefMacro{ds-MN-test_standard-T_len-code-MODE}{23}
\DefMacro{ds-MN-test_standard-T_len-code-CNT}{9,807}
\DefMacro{ds-MN-test_standard-T_len-name-AVG}{2.64}
\DefMacro{ds-MN-test_standard-T_len-name-SUM}{25,849}
\DefMacro{ds-MN-test_standard-T_len-name-MAX}{20}
\DefMacro{ds-MN-test_standard-T_len-name-MIN}{1}
\DefMacro{ds-MN-test_standard-T_len-name-MEDIAN}{2.00}
\DefMacro{ds-MN-test_standard-T_len-name-STDEV}{1.46}
\DefMacro{ds-MN-test_standard-T_len-name-MODE}{2}
\DefMacro{ds-MN-test_standard-T_len-name-CNT}{9,807}
\DefMacro{ds-MN-train-T_num-data}{50,879}
\DefMacro{ds-MN-train-T_len-code-le-100}{77.04}
\DefMacro{ds-MN-train-T_len-code-le-150}{85.96}
\DefMacro{ds-MN-train-T_len-code-le-200}{90.87}
\DefMacro{ds-MN-train-T_len-name-le-2}{57.45}
\DefMacro{ds-MN-train-T_len-name-le-3}{82.12}
\DefMacro{ds-MN-train-T_len-name-le-6}{98.84}
\DefMacro{ds-MN-train-T_len-code-AVG}{84.17}
\DefMacro{ds-MN-train-T_len-code-SUM}{4,282,301}
\DefMacro{ds-MN-train-T_len-code-MAX}{2,699}
\DefMacro{ds-MN-train-T_len-code-MIN}{9}
\DefMacro{ds-MN-train-T_len-code-MEDIAN}{43.00}
\DefMacro{ds-MN-train-T_len-code-STDEV}{126.98}
\DefMacro{ds-MN-train-T_len-code-MODE}{15}
\DefMacro{ds-MN-train-T_len-code-CNT}{50,879}
\DefMacro{ds-MN-train-T_len-name-AVG}{2.50}
\DefMacro{ds-MN-train-T_len-name-SUM}{127,033}
\DefMacro{ds-MN-train-T_len-name-MAX}{21}
\DefMacro{ds-MN-train-T_len-name-MIN}{1}
\DefMacro{ds-MN-train-T_len-name-MEDIAN}{2.00}
\DefMacro{ds-MN-train-T_len-name-STDEV}{1.34}
\DefMacro{ds-MN-train-T_len-name-MODE}{2}
\DefMacro{ds-MN-train-T_len-name-CNT}{50,879}

\DefMacro{result-CG_CP_DeepComHybridESE19_test_common-CP-T_bleu}{11.0}
\DefMacro{result-CG_CP_DeepComHybridESE19_test_common-CP-T_exact_match}{1.3}
\DefMacro{result-CG_CP_DeepComHybridESE19_test_common-CP-T_meteor}{18.2}
\DefMacro{result-CG_CP_DeepComHybridESE19_test_common-CP-T_rouge_l_f}{22.8}
\DefMacro{result-CG_CP_DeepComHybridESE19_test_common-CP-T_rouge_l_p}{23.8}
\DefMacro{result-CG_CP_DeepComHybridESE19_test_common-CP-T_rouge_l_r}{26.1}
\DefMacro{result-CG_CP_DeepComHybridESE19_test_common-MP-CP_bleu}{11.4}
\DefMacro{result-CG_CP_DeepComHybridESE19_test_common-MP-CP_exact_match}{1.4}
\DefMacro{result-CG_CP_DeepComHybridESE19_test_common-MP-CP_meteor}{18.3}
\DefMacro{result-CG_CP_DeepComHybridESE19_test_common-MP-CP_rouge_l_f}{23.9}
\DefMacro{result-CG_CP_DeepComHybridESE19_test_common-MP-CP_rouge_l_p}{25.5}
\DefMacro{result-CG_CP_DeepComHybridESE19_test_common-MP-CP_rouge_l_r}{26.2}
\DefMacro{result-CG_CP_DeepComHybridESE19_test_standard_bleu}{11.6}
\DefMacro{result-CG_CP_DeepComHybridESE19_test_standard_exact_match}{1.6}
\DefMacro{result-CG_CP_DeepComHybridESE19_test_standard_meteor}{18.4}
\DefMacro{result-CG_CP_DeepComHybridESE19_test_standard_rouge_l_f}{24.1}
\DefMacro{result-CG_CP_DeepComHybridESE19_test_standard_rouge_l_p}{25.8}
\DefMacro{result-CG_CP_DeepComHybridESE19_test_standard_rouge_l_r}{26.4}
\DefMacro{result-CG_CP_DeepComHybridESE19_val_bleu}{11.8}
\DefMacro{result-CG_CP_DeepComHybridESE19_val_exact_match}{3.4}
\DefMacro{result-CG_CP_DeepComHybridESE19_val_meteor}{17.9}
\DefMacro{result-CG_CP_DeepComHybridESE19_val_rouge_l_f}{22.4}
\DefMacro{result-CG_CP_DeepComHybridESE19_val_rouge_l_p}{23.2}
\DefMacro{result-CG_CP_DeepComHybridESE19_val_rouge_l_r}{25.2}
\DefMacro{result-CG_CP_RNNBaseline_test_common-CP-T_bleu}{13.4}
\DefMacro{result-CG_CP_RNNBaseline_test_common-CP-T_exact_match}{1.1}
\DefMacro{result-CG_CP_RNNBaseline_test_common-CP-T_meteor}{21.2}
\DefMacro{result-CG_CP_RNNBaseline_test_common-CP-T_rouge_l_f}{28.7}
\DefMacro{result-CG_CP_RNNBaseline_test_common-CP-T_rouge_l_p}{35.2}
\DefMacro{result-CG_CP_RNNBaseline_test_common-CP-T_rouge_l_r}{27.7}
\DefMacro{result-CG_CP_RNNBaseline_test_common-MP-CP_bleu}{13.7}
\DefMacro{result-CG_CP_RNNBaseline_test_common-MP-CP_exact_match}{1.2}
\DefMacro{result-CG_CP_RNNBaseline_test_common-MP-CP_meteor}{21.2}
\DefMacro{result-CG_CP_RNNBaseline_test_common-MP-CP_rouge_l_f}{29.4}
\DefMacro{result-CG_CP_RNNBaseline_test_common-MP-CP_rouge_l_p}{36.8}
\DefMacro{result-CG_CP_RNNBaseline_test_common-MP-CP_rouge_l_r}{27.4}
\DefMacro{result-CG_CP_RNNBaseline_test_standard_bleu}{13.7}
\DefMacro{result-CG_CP_RNNBaseline_test_standard_exact_match}{1.3}
\DefMacro{result-CG_CP_RNNBaseline_test_standard_meteor}{21.2}
\DefMacro{result-CG_CP_RNNBaseline_test_standard_rouge_l_f}{29.6}
\DefMacro{result-CG_CP_RNNBaseline_test_standard_rouge_l_p}{37.1}
\DefMacro{result-CG_CP_RNNBaseline_test_standard_rouge_l_r}{27.6}
\DefMacro{result-CG_CP_RNNBaseline_val_bleu}{13.7}
\DefMacro{result-CG_CP_RNNBaseline_val_exact_match}{2.2}
\DefMacro{result-CG_CP_RNNBaseline_val_meteor}{19.9}
\DefMacro{result-CG_CP_RNNBaseline_val_rouge_l_f}{27.3}
\DefMacro{result-CG_CP_RNNBaseline_val_rouge_l_p}{33.2}
\DefMacro{result-CG_CP_RNNBaseline_val_rouge_l_r}{26.3}
\DefMacro{result-CG_CP_TransformerACL20_test_common-CP-T_bleu}{14.3}
\DefMacro{result-CG_CP_TransformerACL20_test_common-CP-T_exact_match}{1.7}
\DefMacro{result-CG_CP_TransformerACL20_test_common-CP-T_meteor}{21.6}
\DefMacro{result-CG_CP_TransformerACL20_test_common-CP-T_rouge_l_f}{30.0}
\DefMacro{result-CG_CP_TransformerACL20_test_common-CP-T_rouge_l_p}{38.2}
\DefMacro{result-CG_CP_TransformerACL20_test_common-CP-T_rouge_l_r}{27.8}
\DefMacro{result-CG_CP_TransformerACL20_test_common-MP-CP_bleu}{14.1}
\DefMacro{result-CG_CP_TransformerACL20_test_common-MP-CP_exact_match}{1.7}
\DefMacro{result-CG_CP_TransformerACL20_test_common-MP-CP_meteor}{21.4}
\DefMacro{result-CG_CP_TransformerACL20_test_common-MP-CP_rouge_l_f}{30.4}
\DefMacro{result-CG_CP_TransformerACL20_test_common-MP-CP_rouge_l_p}{39.4}
\DefMacro{result-CG_CP_TransformerACL20_test_common-MP-CP_rouge_l_r}{27.4}
\DefMacro{result-CG_CP_TransformerACL20_test_standard_bleu}{14.2}
\DefMacro{result-CG_CP_TransformerACL20_test_standard_exact_match}{1.8}
\DefMacro{result-CG_CP_TransformerACL20_test_standard_meteor}{21.1}
\DefMacro{result-CG_CP_TransformerACL20_test_standard_rouge_l_f}{30.2}
\DefMacro{result-CG_CP_TransformerACL20_test_standard_rouge_l_p}{39.4}
\DefMacro{result-CG_CP_TransformerACL20_test_standard_rouge_l_r}{27.3}
\DefMacro{result-CG_CP_TransformerACL20_val_bleu}{14.2}
\DefMacro{result-CG_CP_TransformerACL20_val_exact_match}{3.3}
\DefMacro{result-CG_CP_TransformerACL20_val_meteor}{19.5}
\DefMacro{result-CG_CP_TransformerACL20_val_rouge_l_f}{27.4}
\DefMacro{result-CG_CP_TransformerACL20_val_rouge_l_p}{35.0}
\DefMacro{result-CG_CP_TransformerACL20_val_rouge_l_r}{25.4}
\DefMacro{result-CG_MP_DeepComHybridESE19_test_common-MP-CP_bleu}{45.0}
\DefMacro{result-CG_MP_DeepComHybridESE19_test_common-MP-CP_exact_match}{30.2}
\DefMacro{result-CG_MP_DeepComHybridESE19_test_common-MP-CP_meteor}{52.9}
\DefMacro{result-CG_MP_DeepComHybridESE19_test_common-MP-CP_rouge_l_f}{57.0}
\DefMacro{result-CG_MP_DeepComHybridESE19_test_common-MP-CP_rouge_l_p}{58.4}
\DefMacro{result-CG_MP_DeepComHybridESE19_test_common-MP-CP_rouge_l_r}{58.3}
\DefMacro{result-CG_MP_DeepComHybridESE19_test_common-MP-T_bleu}{52.6}
\DefMacro{result-CG_MP_DeepComHybridESE19_test_common-MP-T_exact_match}{39.5}
\DefMacro{result-CG_MP_DeepComHybridESE19_test_common-MP-T_meteor}{59.3}
\DefMacro{result-CG_MP_DeepComHybridESE19_test_common-MP-T_rouge_l_f}{62.8}
\DefMacro{result-CG_MP_DeepComHybridESE19_test_common-MP-T_rouge_l_p}{64.1}
\DefMacro{result-CG_MP_DeepComHybridESE19_test_common-MP-T_rouge_l_r}{64.1}
\DefMacro{result-CG_MP_DeepComHybridESE19_test_standard_bleu}{49.6}
\DefMacro{result-CG_MP_DeepComHybridESE19_test_standard_exact_match}{35.2}
\DefMacro{result-CG_MP_DeepComHybridESE19_test_standard_meteor}{57.2}
\DefMacro{result-CG_MP_DeepComHybridESE19_test_standard_rouge_l_f}{60.9}
\DefMacro{result-CG_MP_DeepComHybridESE19_test_standard_rouge_l_p}{62.1}
\DefMacro{result-CG_MP_DeepComHybridESE19_test_standard_rouge_l_r}{62.4}
\DefMacro{result-CG_MP_DeepComHybridESE19_val_bleu}{48.9}
\DefMacro{result-CG_MP_DeepComHybridESE19_val_exact_match}{34.3}
\DefMacro{result-CG_MP_DeepComHybridESE19_val_meteor}{56.6}
\DefMacro{result-CG_MP_DeepComHybridESE19_val_rouge_l_f}{60.3}
\DefMacro{result-CG_MP_DeepComHybridESE19_val_rouge_l_p}{61.6}
\DefMacro{result-CG_MP_DeepComHybridESE19_val_rouge_l_r}{61.7}
\DefMacro{result-CG_MP_RNNBaseline_test_common-MP-CP_bleu}{53.8}
\DefMacro{result-CG_MP_RNNBaseline_test_common-MP-CP_exact_match}{37.3}
\DefMacro{result-CG_MP_RNNBaseline_test_common-MP-CP_meteor}{62.0}
\DefMacro{result-CG_MP_RNNBaseline_test_common-MP-CP_rouge_l_f}{66.7}
\DefMacro{result-CG_MP_RNNBaseline_test_common-MP-CP_rouge_l_p}{70.7}
\DefMacro{result-CG_MP_RNNBaseline_test_common-MP-CP_rouge_l_r}{65.7}
\DefMacro{result-CG_MP_RNNBaseline_test_common-MP-T_bleu}{61.7}
\DefMacro{result-CG_MP_RNNBaseline_test_common-MP-T_exact_match}{48.7}
\DefMacro{result-CG_MP_RNNBaseline_test_common-MP-T_meteor}{68.0}
\DefMacro{result-CG_MP_RNNBaseline_test_common-MP-T_rouge_l_f}{72.0}
\DefMacro{result-CG_MP_RNNBaseline_test_common-MP-T_rouge_l_p}{75.4}
\DefMacro{result-CG_MP_RNNBaseline_test_common-MP-T_rouge_l_r}{71.2}
\DefMacro{result-CG_MP_RNNBaseline_test_standard_bleu}{58.6}
\DefMacro{result-CG_MP_RNNBaseline_test_standard_exact_match}{43.5}
\DefMacro{result-CG_MP_RNNBaseline_test_standard_meteor}{66.0}
\DefMacro{result-CG_MP_RNNBaseline_test_standard_rouge_l_f}{70.0}
\DefMacro{result-CG_MP_RNNBaseline_test_standard_rouge_l_p}{73.4}
\DefMacro{result-CG_MP_RNNBaseline_test_standard_rouge_l_r}{69.4}
\DefMacro{result-CG_MP_RNNBaseline_val_bleu}{57.7}
\DefMacro{result-CG_MP_RNNBaseline_val_exact_match}{42.3}
\DefMacro{result-CG_MP_RNNBaseline_val_meteor}{65.1}
\DefMacro{result-CG_MP_RNNBaseline_val_rouge_l_f}{69.5}
\DefMacro{result-CG_MP_RNNBaseline_val_rouge_l_p}{73.1}
\DefMacro{result-CG_MP_RNNBaseline_val_rouge_l_r}{68.9}
\DefMacro{result-CG_MP_TransformerACL20_test_common-MP-CP_bleu}{58.1}
\DefMacro{result-CG_MP_TransformerACL20_test_common-MP-CP_exact_match}{41.1}
\DefMacro{result-CG_MP_TransformerACL20_test_common-MP-CP_meteor}{66.0}
\DefMacro{result-CG_MP_TransformerACL20_test_common-MP-CP_rouge_l_f}{70.1}
\DefMacro{result-CG_MP_TransformerACL20_test_common-MP-CP_rouge_l_p}{73.5}
\DefMacro{result-CG_MP_TransformerACL20_test_common-MP-CP_rouge_l_r}{69.3}
\DefMacro{result-CG_MP_TransformerACL20_test_common-MP-T_bleu}{65.6}
\DefMacro{result-CG_MP_TransformerACL20_test_common-MP-T_exact_match}{52.3}
\DefMacro{result-CG_MP_TransformerACL20_test_common-MP-T_meteor}{71.5}
\DefMacro{result-CG_MP_TransformerACL20_test_common-MP-T_rouge_l_f}{74.9}
\DefMacro{result-CG_MP_TransformerACL20_test_common-MP-T_rouge_l_p}{77.7}
\DefMacro{result-CG_MP_TransformerACL20_test_common-MP-T_rouge_l_r}{74.3}
\DefMacro{result-CG_MP_TransformerACL20_test_standard_bleu}{62.5}
\DefMacro{result-CG_MP_TransformerACL20_test_standard_exact_match}{46.9}
\DefMacro{result-CG_MP_TransformerACL20_test_standard_meteor}{69.8}
\DefMacro{result-CG_MP_TransformerACL20_test_standard_rouge_l_f}{73.3}
\DefMacro{result-CG_MP_TransformerACL20_test_standard_rouge_l_p}{76.1}
\DefMacro{result-CG_MP_TransformerACL20_test_standard_rouge_l_r}{72.8}
\DefMacro{result-CG_MP_TransformerACL20_val_bleu}{61.5}
\DefMacro{result-CG_MP_TransformerACL20_val_exact_match}{45.7}
\DefMacro{result-CG_MP_TransformerACL20_val_meteor}{68.9}
\DefMacro{result-CG_MP_TransformerACL20_val_rouge_l_f}{72.7}
\DefMacro{result-CG_MP_TransformerACL20_val_rouge_l_p}{75.6}
\DefMacro{result-CG_MP_TransformerACL20_val_rouge_l_r}{72.2}
\DefMacro{result-CG_T_DeepComHybridESE19_test_common-CP-T_bleu}{45.6}
\DefMacro{result-CG_T_DeepComHybridESE19_test_common-CP-T_exact_match}{35.4}
\DefMacro{result-CG_T_DeepComHybridESE19_test_common-CP-T_meteor}{52.0}
\DefMacro{result-CG_T_DeepComHybridESE19_test_common-CP-T_rouge_l_f}{55.9}
\DefMacro{result-CG_T_DeepComHybridESE19_test_common-CP-T_rouge_l_p}{57.3}
\DefMacro{result-CG_T_DeepComHybridESE19_test_common-CP-T_rouge_l_r}{57.3}
\DefMacro{result-CG_T_DeepComHybridESE19_test_common-MP-T_bleu}{43.2}
\DefMacro{result-CG_T_DeepComHybridESE19_test_common-MP-T_exact_match}{31.0}
\DefMacro{result-CG_T_DeepComHybridESE19_test_common-MP-T_meteor}{50.0}
\DefMacro{result-CG_T_DeepComHybridESE19_test_common-MP-T_rouge_l_f}{53.9}
\DefMacro{result-CG_T_DeepComHybridESE19_test_common-MP-T_rouge_l_p}{54.8}
\DefMacro{result-CG_T_DeepComHybridESE19_test_common-MP-T_rouge_l_r}{56.0}
\DefMacro{result-CG_T_DeepComHybridESE19_test_standard_bleu}{43.1}
\DefMacro{result-CG_T_DeepComHybridESE19_test_standard_exact_match}{31.0}
\DefMacro{result-CG_T_DeepComHybridESE19_test_standard_meteor}{50.1}
\DefMacro{result-CG_T_DeepComHybridESE19_test_standard_rouge_l_f}{53.9}
\DefMacro{result-CG_T_DeepComHybridESE19_test_standard_rouge_l_p}{55.0}
\DefMacro{result-CG_T_DeepComHybridESE19_test_standard_rouge_l_r}{56.0}
\DefMacro{result-CG_T_DeepComHybridESE19_val_bleu}{48.3}
\DefMacro{result-CG_T_DeepComHybridESE19_val_exact_match}{36.5}
\DefMacro{result-CG_T_DeepComHybridESE19_val_meteor}{55.0}
\DefMacro{result-CG_T_DeepComHybridESE19_val_rouge_l_f}{58.3}
\DefMacro{result-CG_T_DeepComHybridESE19_val_rouge_l_p}{58.9}
\DefMacro{result-CG_T_DeepComHybridESE19_val_rouge_l_r}{60.8}
\DefMacro{result-CG_T_RNNBaseline_test_common-CP-T_bleu}{53.3}
\DefMacro{result-CG_T_RNNBaseline_test_common-CP-T_exact_match}{42.7}
\DefMacro{result-CG_T_RNNBaseline_test_common-CP-T_meteor}{59.6}
\DefMacro{result-CG_T_RNNBaseline_test_common-CP-T_rouge_l_f}{64.3}
\DefMacro{result-CG_T_RNNBaseline_test_common-CP-T_rouge_l_p}{68.4}
\DefMacro{result-CG_T_RNNBaseline_test_common-CP-T_rouge_l_r}{63.2}
\DefMacro{result-CG_T_RNNBaseline_test_common-MP-T_bleu}{53.2}
\DefMacro{result-CG_T_RNNBaseline_test_common-MP-T_exact_match}{41.0}
\DefMacro{result-CG_T_RNNBaseline_test_common-MP-T_meteor}{59.8}
\DefMacro{result-CG_T_RNNBaseline_test_common-MP-T_rouge_l_f}{64.1}
\DefMacro{result-CG_T_RNNBaseline_test_common-MP-T_rouge_l_p}{67.6}
\DefMacro{result-CG_T_RNNBaseline_test_common-MP-T_rouge_l_r}{63.7}
\DefMacro{result-CG_T_RNNBaseline_test_standard_bleu}{53.0}
\DefMacro{result-CG_T_RNNBaseline_test_standard_exact_match}{40.8}
\DefMacro{result-CG_T_RNNBaseline_test_standard_meteor}{59.4}
\DefMacro{result-CG_T_RNNBaseline_test_standard_rouge_l_f}{63.9}
\DefMacro{result-CG_T_RNNBaseline_test_standard_rouge_l_p}{67.4}
\DefMacro{result-CG_T_RNNBaseline_test_standard_rouge_l_r}{63.4}
\DefMacro{result-CG_T_RNNBaseline_val_bleu}{58.4}
\DefMacro{result-CG_T_RNNBaseline_val_exact_match}{45.8}
\DefMacro{result-CG_T_RNNBaseline_val_meteor}{64.8}
\DefMacro{result-CG_T_RNNBaseline_val_rouge_l_f}{68.5}
\DefMacro{result-CG_T_RNNBaseline_val_rouge_l_p}{71.2}
\DefMacro{result-CG_T_RNNBaseline_val_rouge_l_r}{68.8}
\DefMacro{result-CG_T_TransformerACL20_test_common-CP-T_bleu}{56.1}
\DefMacro{result-CG_T_TransformerACL20_test_common-CP-T_exact_match}{45.8}
\DefMacro{result-CG_T_TransformerACL20_test_common-CP-T_meteor}{62.1}
\DefMacro{result-CG_T_TransformerACL20_test_common-CP-T_rouge_l_f}{66.4}
\DefMacro{result-CG_T_TransformerACL20_test_common-CP-T_rouge_l_p}{70.4}
\DefMacro{result-CG_T_TransformerACL20_test_common-CP-T_rouge_l_r}{65.3}
\DefMacro{result-CG_T_TransformerACL20_test_common-MP-T_bleu}{56.3}
\DefMacro{result-CG_T_TransformerACL20_test_common-MP-T_exact_match}{44.2}
\DefMacro{result-CG_T_TransformerACL20_test_common-MP-T_meteor}{62.7}
\DefMacro{result-CG_T_TransformerACL20_test_common-MP-T_rouge_l_f}{66.6}
\DefMacro{result-CG_T_TransformerACL20_test_common-MP-T_rouge_l_p}{69.6}
\DefMacro{result-CG_T_TransformerACL20_test_common-MP-T_rouge_l_r}{66.3}
\DefMacro{result-CG_T_TransformerACL20_test_standard_bleu}{56.4}
\DefMacro{result-CG_T_TransformerACL20_test_standard_exact_match}{44.4}
\DefMacro{result-CG_T_TransformerACL20_test_standard_meteor}{62.7}
\DefMacro{result-CG_T_TransformerACL20_test_standard_rouge_l_f}{66.7}
\DefMacro{result-CG_T_TransformerACL20_test_standard_rouge_l_p}{69.8}
\DefMacro{result-CG_T_TransformerACL20_test_standard_rouge_l_r}{66.3}
\DefMacro{result-CG_T_TransformerACL20_val_bleu}{62.2}
\DefMacro{result-CG_T_TransformerACL20_val_exact_match}{50.5}
\DefMacro{result-CG_T_TransformerACL20_val_meteor}{68.2}
\DefMacro{result-CG_T_TransformerACL20_val_rouge_l_f}{71.3}
\DefMacro{result-CG_T_TransformerACL20_val_rouge_l_p}{73.4}
\DefMacro{result-CG_T_TransformerACL20_val_rouge_l_r}{71.9}

\DefMacro{result-MN_CP_Code2SeqICLR19_test_common-CP-T_exact_match}{5.9}
\DefMacro{result-MN_CP_Code2SeqICLR19_test_common-CP-T_set_match_f}{28.8}
\DefMacro{result-MN_CP_Code2SeqICLR19_test_common-CP-T_set_match_p}{35.5}
\DefMacro{result-MN_CP_Code2SeqICLR19_test_common-CP-T_set_match_r}{26.5}
\DefMacro{result-MN_CP_Code2SeqICLR19_test_common-CP-T_token_acc}{19.9}
\DefMacro{result-MN_CP_Code2SeqICLR19_test_common-MP-CP_exact_match}{7.6}
\DefMacro{result-MN_CP_Code2SeqICLR19_test_common-MP-CP_set_match_f}{33.0}
\DefMacro{result-MN_CP_Code2SeqICLR19_test_common-MP-CP_set_match_p}{39.8}
\DefMacro{result-MN_CP_Code2SeqICLR19_test_common-MP-CP_set_match_r}{30.3}
\DefMacro{result-MN_CP_Code2SeqICLR19_test_common-MP-CP_token_acc}{24.2}
\DefMacro{result-MN_CP_Code2SeqICLR19_test_standard_exact_match}{8.0}
\DefMacro{result-MN_CP_Code2SeqICLR19_test_standard_set_match_f}{32.7}
\DefMacro{result-MN_CP_Code2SeqICLR19_test_standard_set_match_p}{39.3}
\DefMacro{result-MN_CP_Code2SeqICLR19_test_standard_set_match_r}{30.1}
\DefMacro{result-MN_CP_Code2SeqICLR19_test_standard_token_acc}{24.0}
\DefMacro{result-MN_CP_Code2SeqICLR19_val_exact_match}{9.1}
\DefMacro{result-MN_CP_Code2SeqICLR19_val_set_match_f}{23.6}
\DefMacro{result-MN_CP_Code2SeqICLR19_val_set_match_p}{27.7}
\DefMacro{result-MN_CP_Code2SeqICLR19_val_set_match_r}{22.4}
\DefMacro{result-MN_CP_Code2SeqICLR19_val_token_acc}{17.3}
\DefMacro{result-MN_CP_Code2VecPOPL19_test_common-CP-T_exact_match}{5.4}
\DefMacro{result-MN_CP_Code2VecPOPL19_test_common-CP-T_set_match_f}{13.0}
\DefMacro{result-MN_CP_Code2VecPOPL19_test_common-CP-T_set_match_p}{14.4}
\DefMacro{result-MN_CP_Code2VecPOPL19_test_common-CP-T_set_match_r}{12.9}
\DefMacro{result-MN_CP_Code2VecPOPL19_test_common-CP-T_token_acc}{9.7}
\DefMacro{result-MN_CP_Code2VecPOPL19_test_common-MP-CP_exact_match}{6.5}
\DefMacro{result-MN_CP_Code2VecPOPL19_test_common-MP-CP_set_match_f}{16.7}
\DefMacro{result-MN_CP_Code2VecPOPL19_test_common-MP-CP_set_match_p}{18.9}
\DefMacro{result-MN_CP_Code2VecPOPL19_test_common-MP-CP_set_match_r}{16.4}
\DefMacro{result-MN_CP_Code2VecPOPL19_test_common-MP-CP_token_acc}{12.9}
\DefMacro{result-MN_CP_Code2VecPOPL19_test_standard_exact_match}{6.6}
\DefMacro{result-MN_CP_Code2VecPOPL19_test_standard_set_match_f}{16.6}
\DefMacro{result-MN_CP_Code2VecPOPL19_test_standard_set_match_p}{18.6}
\DefMacro{result-MN_CP_Code2VecPOPL19_test_standard_set_match_r}{16.3}
\DefMacro{result-MN_CP_Code2VecPOPL19_test_standard_token_acc}{12.9}
\DefMacro{result-MN_CP_Code2VecPOPL19_val_exact_match}{8.9}
\DefMacro{result-MN_CP_Code2VecPOPL19_val_set_match_f}{13.7}
\DefMacro{result-MN_CP_Code2VecPOPL19_val_set_match_p}{14.6}
\DefMacro{result-MN_CP_Code2VecPOPL19_val_set_match_r}{13.7}
\DefMacro{result-MN_CP_Code2VecPOPL19_val_token_acc}{11.4}
\DefMacro{result-MN_MP_Code2SeqICLR19_test_common-MP-CP_exact_match}{17.6}
\DefMacro{result-MN_MP_Code2SeqICLR19_test_common-MP-CP_set_match_f}{46.5}
\DefMacro{result-MN_MP_Code2SeqICLR19_test_common-MP-CP_set_match_p}{52.6}
\DefMacro{result-MN_MP_Code2SeqICLR19_test_common-MP-CP_set_match_r}{44.0}
\DefMacro{result-MN_MP_Code2SeqICLR19_test_common-MP-CP_token_acc}{37.9}
\DefMacro{result-MN_MP_Code2SeqICLR19_test_common-MP-T_exact_match}{18.9}
\DefMacro{result-MN_MP_Code2SeqICLR19_test_common-MP-T_set_match_f}{46.9}
\DefMacro{result-MN_MP_Code2SeqICLR19_test_common-MP-T_set_match_p}{52.7}
\DefMacro{result-MN_MP_Code2SeqICLR19_test_common-MP-T_set_match_r}{44.5}
\DefMacro{result-MN_MP_Code2SeqICLR19_test_common-MP-T_token_acc}{37.4}
\DefMacro{result-MN_MP_Code2SeqICLR19_test_standard_exact_match}{19.7}
\DefMacro{result-MN_MP_Code2SeqICLR19_test_standard_set_match_f}{47.2}
\DefMacro{result-MN_MP_Code2SeqICLR19_test_standard_set_match_p}{53.1}
\DefMacro{result-MN_MP_Code2SeqICLR19_test_standard_set_match_r}{44.9}
\DefMacro{result-MN_MP_Code2SeqICLR19_test_standard_token_acc}{38.2}
\DefMacro{result-MN_MP_Code2SeqICLR19_val_exact_match}{20.2}
\DefMacro{result-MN_MP_Code2SeqICLR19_val_set_match_f}{47.2}
\DefMacro{result-MN_MP_Code2SeqICLR19_val_set_match_p}{53.2}
\DefMacro{result-MN_MP_Code2SeqICLR19_val_set_match_r}{44.9}
\DefMacro{result-MN_MP_Code2SeqICLR19_val_token_acc}{38.5}
\DefMacro{result-MN_MP_Code2VecPOPL19_test_common-MP-CP_exact_match}{42.7}
\DefMacro{result-MN_MP_Code2VecPOPL19_test_common-MP-CP_set_match_f}{57.9}
\DefMacro{result-MN_MP_Code2VecPOPL19_test_common-MP-CP_set_match_p}{59.3}
\DefMacro{result-MN_MP_Code2VecPOPL19_test_common-MP-CP_set_match_r}{57.7}
\DefMacro{result-MN_MP_Code2VecPOPL19_test_common-MP-CP_token_acc}{54.2}
\DefMacro{result-MN_MP_Code2VecPOPL19_test_common-MP-T_exact_match}{50.5}
\DefMacro{result-MN_MP_Code2VecPOPL19_test_common-MP-T_set_match_f}{63.7}
\DefMacro{result-MN_MP_Code2VecPOPL19_test_common-MP-T_set_match_p}{65.1}
\DefMacro{result-MN_MP_Code2VecPOPL19_test_common-MP-T_set_match_r}{63.5}
\DefMacro{result-MN_MP_Code2VecPOPL19_test_common-MP-T_token_acc}{59.5}
\DefMacro{result-MN_MP_Code2VecPOPL19_test_standard_exact_match}{47.1}
\DefMacro{result-MN_MP_Code2VecPOPL19_test_standard_set_match_f}{60.9}
\DefMacro{result-MN_MP_Code2VecPOPL19_test_standard_set_match_p}{62.3}
\DefMacro{result-MN_MP_Code2VecPOPL19_test_standard_set_match_r}{60.7}
\DefMacro{result-MN_MP_Code2VecPOPL19_test_standard_token_acc}{56.7}
\DefMacro{result-MN_MP_Code2VecPOPL19_val_exact_match}{46.9}
\DefMacro{result-MN_MP_Code2VecPOPL19_val_set_match_f}{60.5}
\DefMacro{result-MN_MP_Code2VecPOPL19_val_set_match_p}{61.9}
\DefMacro{result-MN_MP_Code2VecPOPL19_val_set_match_r}{60.4}
\DefMacro{result-MN_MP_Code2VecPOPL19_val_token_acc}{56.5}
\DefMacro{result-MN_T_Code2SeqICLR19_test_common-CP-T_exact_match}{13.3}
\DefMacro{result-MN_T_Code2SeqICLR19_test_common-CP-T_set_match_f}{40.6}
\DefMacro{result-MN_T_Code2SeqICLR19_test_common-CP-T_set_match_p}{46.2}
\DefMacro{result-MN_T_Code2SeqICLR19_test_common-CP-T_set_match_r}{38.4}
\DefMacro{result-MN_T_Code2SeqICLR19_test_common-CP-T_token_acc}{31.8}
\DefMacro{result-MN_T_Code2SeqICLR19_test_common-MP-T_exact_match}{16.0}
\DefMacro{result-MN_T_Code2SeqICLR19_test_common-MP-T_set_match_f}{42.9}
\DefMacro{result-MN_T_Code2SeqICLR19_test_common-MP-T_set_match_p}{49.2}
\DefMacro{result-MN_T_Code2SeqICLR19_test_common-MP-T_set_match_r}{40.3}
\DefMacro{result-MN_T_Code2SeqICLR19_test_common-MP-T_token_acc}{33.7}
\DefMacro{result-MN_T_Code2SeqICLR19_test_standard_exact_match}{15.8}
\DefMacro{result-MN_T_Code2SeqICLR19_test_standard_set_match_f}{42.0}
\DefMacro{result-MN_T_Code2SeqICLR19_test_standard_set_match_p}{47.9}
\DefMacro{result-MN_T_Code2SeqICLR19_test_standard_set_match_r}{39.8}
\DefMacro{result-MN_T_Code2SeqICLR19_test_standard_token_acc}{33.1}
\DefMacro{result-MN_T_Code2SeqICLR19_val_exact_match}{17.0}
\DefMacro{result-MN_T_Code2SeqICLR19_val_set_match_f}{41.8}
\DefMacro{result-MN_T_Code2SeqICLR19_val_set_match_p}{47.8}
\DefMacro{result-MN_T_Code2SeqICLR19_val_set_match_r}{39.5}
\DefMacro{result-MN_T_Code2SeqICLR19_val_token_acc}{33.4}
\DefMacro{result-MN_T_Code2VecPOPL19_test_common-CP-T_exact_match}{43.9}
\DefMacro{result-MN_T_Code2VecPOPL19_test_common-CP-T_set_match_f}{53.9}
\DefMacro{result-MN_T_Code2VecPOPL19_test_common-CP-T_set_match_p}{55.3}
\DefMacro{result-MN_T_Code2VecPOPL19_test_common-CP-T_set_match_r}{53.8}
\DefMacro{result-MN_T_Code2VecPOPL19_test_common-CP-T_token_acc}{50.4}
\DefMacro{result-MN_T_Code2VecPOPL19_test_common-MP-T_exact_match}{46.9}
\DefMacro{result-MN_T_Code2VecPOPL19_test_common-MP-T_set_match_f}{56.2}
\DefMacro{result-MN_T_Code2VecPOPL19_test_common-MP-T_set_match_p}{57.8}
\DefMacro{result-MN_T_Code2VecPOPL19_test_common-MP-T_set_match_r}{55.8}
\DefMacro{result-MN_T_Code2VecPOPL19_test_common-MP-T_token_acc}{52.6}
\DefMacro{result-MN_T_Code2VecPOPL19_test_standard_exact_match}{47.3}
\DefMacro{result-MN_T_Code2VecPOPL19_test_standard_set_match_f}{56.3}
\DefMacro{result-MN_T_Code2VecPOPL19_test_standard_set_match_p}{57.7}
\DefMacro{result-MN_T_Code2VecPOPL19_test_standard_set_match_r}{56.0}
\DefMacro{result-MN_T_Code2VecPOPL19_test_standard_token_acc}{52.9}
\DefMacro{result-MN_T_Code2VecPOPL19_val_exact_match}{53.4}
\DefMacro{result-MN_T_Code2VecPOPL19_val_set_match_f}{61.6}
\DefMacro{result-MN_T_Code2VecPOPL19_val_set_match_p}{63.1}
\DefMacro{result-MN_T_Code2VecPOPL19_val_set_match_r}{61.3}
\DefMacro{result-MN_T_Code2VecPOPL19_val_token_acc}{58.4}

\DefMacro{ds-CG-MP-test_standard-original_data_same_code}{14,956}
\DefMacro{ds-CG-MP-test_standard-filtered_data_same_code}{12,225}
\DefMacro{ds-CG-MP-test_standard-filter_same_code}{2,731}
\DefMacro{ds-CG-MP-test_standard-filter_pct_same_code}{18.3\%}
\DefMacro{ds-CG-CP-test_standard-original_data_same_code}{15,661}
\DefMacro{ds-CG-CP-test_standard-filtered_data_same_code}{15,394}
\DefMacro{ds-CG-CP-test_standard-filter_same_code}{267}
\DefMacro{ds-CG-CP-test_standard-filter_pct_same_code}{1.7\%}
\DefMacro{ds-CG-T-test_standard-original_data_same_code}{9,870}
\DefMacro{ds-CG-T-test_standard-filtered_data_same_code}{8,586}
\DefMacro{ds-CG-T-test_standard-filter_same_code}{1,284}
\DefMacro{ds-CG-T-test_standard-filter_pct_same_code}{13.0\%}
\DefMacro{ds-CG-MP-test_common-MP-CP-original_data_same_code}{3,362}
\DefMacro{ds-CG-MP-test_common-MP-CP-filtered_data_same_code}{2,830}
\DefMacro{ds-CG-MP-test_common-MP-CP-filter_same_code}{532}
\DefMacro{ds-CG-MP-test_common-MP-CP-filter_pct_same_code}{15.8\%}
\DefMacro{ds-CG-MP-test_common-MP-T-original_data_same_code}{2,013}
\DefMacro{ds-CG-MP-test_common-MP-T-filtered_data_same_code}{1,692}
\DefMacro{ds-CG-MP-test_common-MP-T-filter_same_code}{321}
\DefMacro{ds-CG-MP-test_common-MP-T-filter_pct_same_code}{15.9\%}
\DefMacro{ds-CG-CP-test_common-CP-T-original_data_same_code}{2,220}
\DefMacro{ds-CG-CP-test_common-CP-T-filtered_data_same_code}{1,819}
\DefMacro{ds-CG-CP-test_common-CP-T-filter_same_code}{401}
\DefMacro{ds-CG-CP-test_common-CP-T-filter_pct_same_code}{18.1\%}
\DefMacro{ds-CG-total-original_data_same_code}{48,082}
\DefMacro{ds-CG-total-filtered_data_same_code}{42,546}
\DefMacro{ds-CG-total-filter_same_code}{5,536}
\DefMacro{ds-CG-total-filter_pct_same_code}{11.5\%}
\DefMacro{ds-CG-test_standard-original_data_same_code}{40,487}
\DefMacro{ds-CG-test_standard-filtered_data_same_code}{36,205}
\DefMacro{ds-CG-test_standard-filter_same_code}{4,282}
\DefMacro{ds-CG-test_standard-filter_pct_same_code}{10.6\%}
\DefMacro{ds-CG-test_common-original_data_same_code}{7,595}
\DefMacro{ds-CG-test_common-filtered_data_same_code}{6,341}
\DefMacro{ds-CG-test_common-filter_same_code}{1,254}
\DefMacro{ds-CG-test_common-filter_pct_same_code}{16.5\%}

\DefMacro{ds-CG-MP-test_standard-original_data_same_nl}{14,956}
\DefMacro{ds-CG-MP-test_standard-filtered_data_same_nl}{7,054}
\DefMacro{ds-CG-MP-test_standard-filter_same_nl}{7,902}
\DefMacro{ds-CG-MP-test_standard-filter_pct_same_nl}{52.8\%}
\DefMacro{ds-CG-CP-test_standard-original_data_same_nl}{15,661}
\DefMacro{ds-CG-CP-test_standard-filtered_data_same_nl}{15,277}
\DefMacro{ds-CG-CP-test_standard-filter_same_nl}{384}
\DefMacro{ds-CG-CP-test_standard-filter_pct_same_nl}{2.5\%}
\DefMacro{ds-CG-T-test_standard-original_data_same_nl}{9,870}
\DefMacro{ds-CG-T-test_standard-filtered_data_same_nl}{3,999}
\DefMacro{ds-CG-T-test_standard-filter_same_nl}{5,871}
\DefMacro{ds-CG-T-test_standard-filter_pct_same_nl}{59.5\%}
\DefMacro{ds-CG-MP-test_common-MP-CP-original_data_same_nl}{3,362}
\DefMacro{ds-CG-MP-test_common-MP-CP-filtered_data_same_nl}{1,781}
\DefMacro{ds-CG-MP-test_common-MP-CP-filter_same_nl}{1,581}
\DefMacro{ds-CG-MP-test_common-MP-CP-filter_pct_same_nl}{47.0\%}
\DefMacro{ds-CG-MP-test_common-MP-T-original_data_same_nl}{2,013}
\DefMacro{ds-CG-MP-test_common-MP-T-filtered_data_same_nl}{626}
\DefMacro{ds-CG-MP-test_common-MP-T-filter_same_nl}{1,387}
\DefMacro{ds-CG-MP-test_common-MP-T-filter_pct_same_nl}{68.9\%}
\DefMacro{ds-CG-CP-test_common-CP-T-original_data_same_nl}{2,220}
\DefMacro{ds-CG-CP-test_common-CP-T-filtered_data_same_nl}{912}
\DefMacro{ds-CG-CP-test_common-CP-T-filter_same_nl}{1,308}
\DefMacro{ds-CG-CP-test_common-CP-T-filter_pct_same_nl}{58.9\%}
\DefMacro{ds-CG-total-original_data_same_nl}{48,082}
\DefMacro{ds-CG-total-filtered_data_same_nl}{29,649}
\DefMacro{ds-CG-total-filter_same_nl}{18,433}
\DefMacro{ds-CG-total-filter_pct_same_nl}{38.3\%}
\DefMacro{ds-CG-test_standard-original_data_same_nl}{40,487}
\DefMacro{ds-CG-test_standard-filtered_data_same_nl}{26,330}
\DefMacro{ds-CG-test_standard-filter_same_nl}{14,157}
\DefMacro{ds-CG-test_standard-filter_pct_same_nl}{35.0\%}
\DefMacro{ds-CG-test_common-original_data_same_nl}{7,595}
\DefMacro{ds-CG-test_common-filtered_data_same_nl}{3,319}
\DefMacro{ds-CG-test_common-filter_same_nl}{4,276}
\DefMacro{ds-CG-test_common-filter_pct_same_nl}{56.3\%}

\DefMacro{ds-CG-MP-test_standard-original_data_sim_90}{14,956}
\DefMacro{ds-CG-MP-test_standard-filtered_data_sim_90}{5,514}
\DefMacro{ds-CG-MP-test_standard-filter_sim_90}{9,442}
\DefMacro{ds-CG-MP-test_standard-filter_pct_sim_90}{63.1\%}
\DefMacro{ds-CG-CP-test_standard-original_data_sim_90}{15,661}
\DefMacro{ds-CG-CP-test_standard-filtered_data_sim_90}{15,000}
\DefMacro{ds-CG-CP-test_standard-filter_sim_90}{661}
\DefMacro{ds-CG-CP-test_standard-filter_pct_sim_90}{4.2\%}
\DefMacro{ds-CG-T-test_standard-original_data_sim_90}{9,870}
\DefMacro{ds-CG-T-test_standard-filtered_data_sim_90}{3,467}
\DefMacro{ds-CG-T-test_standard-filter_sim_90}{6,403}
\DefMacro{ds-CG-T-test_standard-filter_pct_sim_90}{64.9\%}
\DefMacro{ds-CG-MP-test_common-MP-CP-original_data_sim_90}{3,362}
\DefMacro{ds-CG-MP-test_common-MP-CP-filtered_data_sim_90}{1,380}
\DefMacro{ds-CG-MP-test_common-MP-CP-filter_sim_90}{1,982}
\DefMacro{ds-CG-MP-test_common-MP-CP-filter_pct_sim_90}{59.0\%}
\DefMacro{ds-CG-MP-test_common-MP-T-original_data_sim_90}{2,013}
\DefMacro{ds-CG-MP-test_common-MP-T-filtered_data_sim_90}{457}
\DefMacro{ds-CG-MP-test_common-MP-T-filter_sim_90}{1,556}
\DefMacro{ds-CG-MP-test_common-MP-T-filter_pct_sim_90}{77.3\%}
\DefMacro{ds-CG-CP-test_common-CP-T-original_data_sim_90}{2,220}
\DefMacro{ds-CG-CP-test_common-CP-T-filtered_data_sim_90}{811}
\DefMacro{ds-CG-CP-test_common-CP-T-filter_sim_90}{1,409}
\DefMacro{ds-CG-CP-test_common-CP-T-filter_pct_sim_90}{63.5\%}
\DefMacro{ds-CG-total-original_data_sim_90}{48,082}
\DefMacro{ds-CG-total-filtered_data_sim_90}{26,629}
\DefMacro{ds-CG-total-filter_sim_90}{21,453}
\DefMacro{ds-CG-total-filter_pct_sim_90}{44.6\%}
\DefMacro{ds-CG-test_standard-original_data_sim_90}{40,487}
\DefMacro{ds-CG-test_standard-filtered_data_sim_90}{23,981}
\DefMacro{ds-CG-test_standard-filter_sim_90}{16,506}
\DefMacro{ds-CG-test_standard-filter_pct_sim_90}{40.8\%}
\DefMacro{ds-CG-test_common-original_data_sim_90}{7,595}
\DefMacro{ds-CG-test_common-filtered_data_sim_90}{2,648}
\DefMacro{ds-CG-test_common-filter_sim_90}{4,947}
\DefMacro{ds-CG-test_common-filter_pct_sim_90}{65.1\%}

\DefMacro{ds-MN-MP-test_standard-original_data_same_code}{14,796}
\DefMacro{ds-MN-MP-test_standard-filtered_data_same_code}{11,853}
\DefMacro{ds-MN-MP-test_standard-filter_same_code}{2,943}
\DefMacro{ds-MN-MP-test_standard-filter_pct_same_code}{19.9\%}
\DefMacro{ds-MN-CP-test_standard-original_data_same_code}{15,332}
\DefMacro{ds-MN-CP-test_standard-filtered_data_same_code}{14,990}
\DefMacro{ds-MN-CP-test_standard-filter_same_code}{342}
\DefMacro{ds-MN-CP-test_standard-filter_pct_same_code}{2.2\%}
\DefMacro{ds-MN-T-test_standard-original_data_same_code}{9,807}
\DefMacro{ds-MN-T-test_standard-filtered_data_same_code}{8,463}
\DefMacro{ds-MN-T-test_standard-filter_same_code}{1,344}
\DefMacro{ds-MN-T-test_standard-filter_pct_same_code}{13.7\%}
\DefMacro{ds-MN-MP-test_common-MP-CP-original_data_same_code}{3,344}
\DefMacro{ds-MN-MP-test_common-MP-CP-filtered_data_same_code}{2,725}
\DefMacro{ds-MN-MP-test_common-MP-CP-filter_same_code}{619}
\DefMacro{ds-MN-MP-test_common-MP-CP-filter_pct_same_code}{18.5\%}
\DefMacro{ds-MN-MP-test_common-MP-T-original_data_same_code}{2,011}
\DefMacro{ds-MN-MP-test_common-MP-T-filtered_data_same_code}{1,635}
\DefMacro{ds-MN-MP-test_common-MP-T-filter_same_code}{376}
\DefMacro{ds-MN-MP-test_common-MP-T-filter_pct_same_code}{18.7\%}
\DefMacro{ds-MN-CP-test_common-CP-T-original_data_same_code}{2,211}
\DefMacro{ds-MN-CP-test_common-CP-T-filtered_data_same_code}{1,772}
\DefMacro{ds-MN-CP-test_common-CP-T-filter_same_code}{439}
\DefMacro{ds-MN-CP-test_common-CP-T-filter_pct_same_code}{19.9\%}
\DefMacro{ds-MN-total-original_data_same_code}{47,501}
\DefMacro{ds-MN-total-filtered_data_same_code}{41,438}
\DefMacro{ds-MN-total-filter_same_code}{6,063}
\DefMacro{ds-MN-total-filter_pct_same_code}{12.8\%}
\DefMacro{ds-MN-test_standard-original_data_same_code}{39,935}
\DefMacro{ds-MN-test_standard-filtered_data_same_code}{35,306}
\DefMacro{ds-MN-test_standard-filter_same_code}{4,629}
\DefMacro{ds-MN-test_standard-filter_pct_same_code}{11.6\%}
\DefMacro{ds-MN-test_common-original_data_same_code}{7,566}
\DefMacro{ds-MN-test_common-filtered_data_same_code}{6,132}
\DefMacro{ds-MN-test_common-filter_same_code}{1,434}
\DefMacro{ds-MN-test_common-filter_pct_same_code}{19.0\%}

\DefMacro{ds-MN-MP-test_standard-original_data_same_nl}{14,796}
\DefMacro{ds-MN-MP-test_standard-filtered_data_same_nl}{4,037}
\DefMacro{ds-MN-MP-test_standard-filter_same_nl}{10,759}
\DefMacro{ds-MN-MP-test_standard-filter_pct_same_nl}{72.7\%}
\DefMacro{ds-MN-CP-test_standard-original_data_same_nl}{15,332}
\DefMacro{ds-MN-CP-test_standard-filtered_data_same_nl}{10,631}
\DefMacro{ds-MN-CP-test_standard-filter_same_nl}{4,701}
\DefMacro{ds-MN-CP-test_standard-filter_pct_same_nl}{30.7\%}
\DefMacro{ds-MN-T-test_standard-original_data_same_nl}{9,807}
\DefMacro{ds-MN-T-test_standard-filtered_data_same_nl}{2,335}
\DefMacro{ds-MN-T-test_standard-filter_same_nl}{7,472}
\DefMacro{ds-MN-T-test_standard-filter_pct_same_nl}{76.2\%}
\DefMacro{ds-MN-MP-test_common-MP-CP-original_data_same_nl}{3,344}
\DefMacro{ds-MN-MP-test_common-MP-CP-filtered_data_same_nl}{1,107}
\DefMacro{ds-MN-MP-test_common-MP-CP-filter_same_nl}{2,237}
\DefMacro{ds-MN-MP-test_common-MP-CP-filter_pct_same_nl}{66.9\%}
\DefMacro{ds-MN-MP-test_common-MP-T-original_data_same_nl}{2,011}
\DefMacro{ds-MN-MP-test_common-MP-T-filtered_data_same_nl}{378}
\DefMacro{ds-MN-MP-test_common-MP-T-filter_same_nl}{1,633}
\DefMacro{ds-MN-MP-test_common-MP-T-filter_pct_same_nl}{81.2\%}
\DefMacro{ds-MN-CP-test_common-CP-T-original_data_same_nl}{2,211}
\DefMacro{ds-MN-CP-test_common-CP-T-filtered_data_same_nl}{592}
\DefMacro{ds-MN-CP-test_common-CP-T-filter_same_nl}{1,619}
\DefMacro{ds-MN-CP-test_common-CP-T-filter_pct_same_nl}{73.2\%}
\DefMacro{ds-MN-total-original_data_same_nl}{47,501}
\DefMacro{ds-MN-total-filtered_data_same_nl}{19,080}
\DefMacro{ds-MN-total-filter_same_nl}{28,421}
\DefMacro{ds-MN-total-filter_pct_same_nl}{59.8\%}
\DefMacro{ds-MN-test_standard-original_data_same_nl}{39,935}
\DefMacro{ds-MN-test_standard-filtered_data_same_nl}{17,003}
\DefMacro{ds-MN-test_standard-filter_same_nl}{22,932}
\DefMacro{ds-MN-test_standard-filter_pct_same_nl}{57.4\%}
\DefMacro{ds-MN-test_common-original_data_same_nl}{7,566}
\DefMacro{ds-MN-test_common-filtered_data_same_nl}{2,077}
\DefMacro{ds-MN-test_common-filter_same_nl}{5,489}
\DefMacro{ds-MN-test_common-filter_pct_same_nl}{72.5\%}

\DefMacro{ds-MN-MP-test_standard-original_data_sim_90}{14,796}
\DefMacro{ds-MN-MP-test_standard-filtered_data_sim_90}{3,260}
\DefMacro{ds-MN-MP-test_standard-filter_sim_90}{11,536}
\DefMacro{ds-MN-MP-test_standard-filter_pct_sim_90}{78.0\%}
\DefMacro{ds-MN-CP-test_standard-original_data_sim_90}{15,332}
\DefMacro{ds-MN-CP-test_standard-filtered_data_sim_90}{10,234}
\DefMacro{ds-MN-CP-test_standard-filter_sim_90}{5,098}
\DefMacro{ds-MN-CP-test_standard-filter_pct_sim_90}{33.3\%}
\DefMacro{ds-MN-T-test_standard-original_data_sim_90}{9,807}
\DefMacro{ds-MN-T-test_standard-filtered_data_sim_90}{2,134}
\DefMacro{ds-MN-T-test_standard-filter_sim_90}{7,673}
\DefMacro{ds-MN-T-test_standard-filter_pct_sim_90}{78.2\%}
\DefMacro{ds-MN-MP-test_common-MP-CP-original_data_sim_90}{3,344}
\DefMacro{ds-MN-MP-test_common-MP-CP-filtered_data_sim_90}{880}
\DefMacro{ds-MN-MP-test_common-MP-CP-filter_sim_90}{2,464}
\DefMacro{ds-MN-MP-test_common-MP-CP-filter_pct_sim_90}{73.7\%}
\DefMacro{ds-MN-MP-test_common-MP-T-original_data_sim_90}{2,011}
\DefMacro{ds-MN-MP-test_common-MP-T-filtered_data_sim_90}{315}
\DefMacro{ds-MN-MP-test_common-MP-T-filter_sim_90}{1,696}
\DefMacro{ds-MN-MP-test_common-MP-T-filter_pct_sim_90}{84.3\%}
\DefMacro{ds-MN-CP-test_common-CP-T-original_data_sim_90}{2,211}
\DefMacro{ds-MN-CP-test_common-CP-T-filtered_data_sim_90}{537}
\DefMacro{ds-MN-CP-test_common-CP-T-filter_sim_90}{1,674}
\DefMacro{ds-MN-CP-test_common-CP-T-filter_pct_sim_90}{75.7\%}
\DefMacro{ds-MN-total-original_data_sim_90}{47,501}
\DefMacro{ds-MN-total-filtered_data_sim_90}{17,360}
\DefMacro{ds-MN-total-filter_sim_90}{30,141}
\DefMacro{ds-MN-total-filter_pct_sim_90}{63.5\%}
\DefMacro{ds-MN-test_standard-original_data_sim_90}{39,935}
\DefMacro{ds-MN-test_standard-filtered_data_sim_90}{15,628}
\DefMacro{ds-MN-test_standard-filter_sim_90}{24,307}
\DefMacro{ds-MN-test_standard-filter_pct_sim_90}{60.9\%}
\DefMacro{ds-MN-test_common-original_data_sim_90}{7,566}
\DefMacro{ds-MN-test_common-filtered_data_sim_90}{1,732}
\DefMacro{ds-MN-test_common-filter_sim_90}{5,834}
\DefMacro{ds-MN-test_common-filter_pct_sim_90}{77.1\%}

\newcommand{\NumCodeSumAll}{18\xspace}
\newcommand{\NumCodeSumMP}{15\xspace}
\newcommand{\NumCodeSumCP}{4\xspace}
\newcommand{\NumCodeSumT}{0\xspace}

\newcommand{\NumProjectPlanned}{1,793\xspace}
\newcommand{\NumProjectPlannedStars}{1,000\xspace}
\newcommand{\NumProjectPlannedTrending}{793\xspace}

\newcommand{\NumProject}{\UseMacro{ds-num-proj}\xspace}

\newcommand{\NumTrial}{three\xspace}

\title{\Title}

\author{
Pengyu Nie,
Jiyang Zhang,
Junyi Jessy Li,
Raymond J. Mooney,
Milos Gligoric\\
The University of Texas at Austin\\
\texttt{\{pynie@, jiyang.zhang@, jessy@austin.,}\\\texttt{mooney@cs., gligoric@\}utexas.edu}
}

\begin{document}
\maketitle

\begin{abstract}

There has been a growing interest in developing machine learning (ML)
models for code summarization tasks, e.g., \comgen and \metnam.
Despite substantial increase in the effectiveness of ML models, the
evaluation \methodologies, i.e., the way people split datasets into
\train, \val, and \test sets, were not well studied.  Specifically, no
prior work on code summarization considered the timestamps of code and
comments during evaluation.
This may lead to evaluations that are inconsistent with the intended
use cases.
In this paper, we introduce the \temporal
\emethodology, which is novel to the code summarization research
community,{ and compare it with} the \mixedproj
and \crossproj \methodologies that have been commonly used.
Each \methodology can be mapped to some use cases, and the \temporal
\methodology should be adopted in the evaluation of ML models for code
summarization.
To assess the impact of \methodologies, we collect a dataset of
\codecom pairs with timestamps to train and evaluate several
recent ML models for code summarization.  
Our experiments show that different \methodologies lead to conflicting
evaluation results.
We invite the community to expand the set of \methodologies used in
evaluations.

\end{abstract}

\section{Introduction}
\label{sec:intro}

Over the last several years, there has been a growing interest in
applying machine learning (ML) models to code summarization tasks,
such as \comgen~\cite{ IyerETAL16Summarizing, HuETAL18Deep,
WanETAL18SummarizationRL, LiangAndZhu18Automatic, HuETAL18Summarizing,
LeClairETAL19Neural, FernandesETAL19Summarization, XuETAL19Method,
LeClairAndMcMillan19CodeSum, LeClairETAL20Summarization,
HuETAL20DeepComHybrid, AhmadETAL20Transformer, CaiETAL20TAG,
GrosETAL20CodeCommentTranslation} and
\metnam~\cite{AllamanisETAL16Convolutional, AlonETAL19Code2seq,
AlonETAL19Code2vec, FernandesETAL19Summarization,
NguyenETAL20MethodNames}.
Substantial progress has been reported over years, usually measured in
terms of automatic metrics~\cite{RoyETAL21EvalMetrics}.

Despite a solid progress in generating more accurate summaries, the
\emph{\emethodology}, i.e., the way we obtain \train, \val, and \test
sets, is solely based on
conventional ML practices in natural language summarization, without
taking into account the domain
knowledge of software engineering and software evolution.
\CRNew{For example, temporal relations among \examples in the dataset
are important because the style of newer code summaries can be
affected by older code summaries; however, they are not explicitly
modeled in the evaluation of code summarization in prior work, which
assumed the \examples in the dataset are independent and
identically distributed.}
This gap could lead to inflated values
for automatic metrics reported in papers
and misunderstanding if a model might actually be useful once adopted.

\begin{figure}[t]
  \centering
  \scalebox{0.85}{




\begin{tikzpicture}[
  line width=0.4pt,
  node distance=0ex and 0em,
  every node/.style={scale=0.9},
  gridBox/.style={rectangle, opacity=0, draw=red},
  box/.style={rectangle, draw=black, inner sep=2pt, font=\small},
  rounded box/.style={rectangle, rounded corners, draw=black, inner sep=2pt, font=\small},
  anno/.style={font=\footnotesize},
]

  \node (b-l-train) at (0, -4ex) [sExample, dsTrain] {};
  \node (b-l-train-text) [right = .1em of b-l-train] [anno] {\Train};
  \node (b-l-val) [right = .2em of b-l-train-text] [sExample, dsVal] {};
  \node (b-l-val-text) [right = .1em of b-l-val] [anno] {\Val};
  \node (b-l-test) [right = .2em of b-l-val-text] [sExample, dsTest] {};
  \node (b-l-test-text) [right = .1em of b-l-test] [anno] {\Test};

  \tikzset{sEvoProject/.style={sProject, minimum height=5.5em, minimum width=3em}}

  \node (b-p1) at (0,0) [anchor=south,sEvoProject] {};
  \node (b-anno1) [below = 0 of b-p1.south] [anno] {project 1};
  \node (b-p2) [right = \wSepProject of b-p1.east] [sEvoProject] {};
  \node (b-anno2) [below = 0 of b-p2.south] [anno] {project 2};
  \node (b-p3) [right = \wSepProject of b-p2.east] [sEvoProject] {};
  \node (b-anno3) [below = 0 of b-p3.south] [anno] {project 3};
  \node (b-p4) [right = 2 * \wSepProject of b-p3.east] [sEvoProject] {};
  \node (b-anno4) [below = 0 of b-p4.south] [anno] {project n-1};
  \node (b-p5) [right = \wSepProject of b-p4.east] [sEvoProject] {};
  \node (b-anno5) [below = 0 of b-p5.south] [anno] {project n};

  \node (b-ellipsis0) at ($(b-p3.east)!0.5!(b-p4.west)$) [] {\dots};

  \node[coordinate] (c-time-beg) [below left = 1ex and 1em of b-p1.north west] {};
  \node[coordinate] (c-time-end) [above left = 0ex and 1em of b-p1.south west] {};
  \draw[->] (c-time-beg) -- (c-time-end);
  \node (b-time) [above left = -.5ex and .3em of c-time-end] [anno] {time};

  \foreach \pi in {1,2,3,4,5} {
    \foreach \ex/\ey/\ds [count=\ei] in {
      -.8em/-1.5em/dsTrain, 0em/-1.5em/dsTrain, .8em/-1.5em/dsTrain,
      -.8em/0/dsVal,  0em/0/dsVal, .8em/0/dsVal,
      -.8em/1.5em/dsTest, 0em/1.5em/dsTest, .8em/1.5em/dsTest} {
      \node (b-p\pi-\ei) [below right = \ey and \ex of b-p\pi.center, anchor=center] [sExample, \ds] {};
    }
  };

  \node (b-t1) at (c-time-beg |- b-p1-1) [circle, minimum size=.2em, inner sep=0, draw=black, fill=black] {};
  \node (b-t1-text) [left = .2em of b-t1] [anno] {\atimepp};
  \node (b-t2) at (c-time-beg |- b-p1-4) [circle, minimum size=.2em, inner sep=0, draw=black, fill=black] {};
  \node (b-t2-text) [left = .2em of b-t2] [anno] {\atimep};
  \node (b-t3) at (c-time-beg |- b-p1-7) [circle, minimum size=.2em, inner sep=0, draw=black, fill=black] {};
  \node (b-t3-text) [left = .2em of b-t3] [anno] {\atime};

\end{tikzpicture}

  }
  \vspace{-0pt}
  \caption{\Cmode use case that can be evaluated with the proposed \temporal
    \methodology. \label{fig:method-evoaware}}
  \vspace{-2pt}
\end{figure}
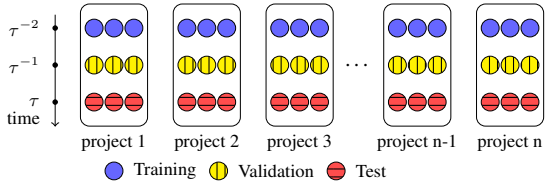

The key missing piece in prior work is the description of the targeted
\emph{use cases} for their ML models.  Prior work has implicitly
targeted only the \bmode use case: applying the model to existing code
regardless of \emph{when} the code is written.  However, a more
\realistic scenario could be the \cmode use case: training the model
with code available at a timestamp, and using the model on new code
after that timestamp (as illustrated in
Figure~\ref{fig:method-evoaware}).  Considering that programming
languages evolve and coding styles are constantly revised, results
obtained in \bmode could be very different from those obtained in
\cmode.  Thus, it is insufficient to only report the task being
targeted in a paper, and \emph{it is necessary to explain intended use
cases for the ML models}.  Once the task and use cases are clearly
defined, an appropriate \emethodology (or potentially several
\methodologies) should be used.

In this paper, we study recent literature on ML models for code
summarization.  By
reasoning about their \emethodologies (which we call \mixedproj and
\crossproj), we define two use cases that could be evaluated by
these \methodologies.
Next, we define a more \CRNew{practical} use case when a developer uses a
fixed model continuously over some period of time.
We describe an appropriate \emethodology for this use case: \temporal.
Finally, we evaluate several existing ML models using the three
\methodologies.

We highlight two key findings.  First, depending on the employed
\methodology we end up with conflicting conclusions, i.e., using one
\methodology, model A is better than model B, and using another
\methodology, model B is better than model A.
Second, our results show that the absolute values for automatic
metrics vary widely across the three \methodologies, which indicates
that models might be useful only for some use cases but not others.
Thus, it is imperative that future work describes what use case is
being targeted and use the appropriate \emethodology.  

\emph{In summary, this paper argues that we need to more diligently
  choose \emethodology and report results of ML models.}  Regardless of whether or not the
conclusions of prior work hold across \methodologies, we should always
choose the \methodology appropriate for the targeted task and use
case.  We hope the community will join us in the effort to define
the most \realistic use cases and the evaluation \methodology
for each use case.

We hope that our work will inspire others to design and formalize use
cases and methodologies for other tasks.  Only a few \CRNew{research studies} on
defect prediction~\cite{DAmbrosETAL12DefectPrediction,
  TanETAL15Online, WangETAL16DefectPrediction,
  KameiETAL16JITDefectPrediction}, program
repair~\cite{LutellierETAL20CoCoNut}, and bug
localization~\cite{PradelETAL20Scaffle} took into consideration
software evolution when evaluating ML models.
Taking software evolution into account in those
\CRNew{tasks} appears more natural, but \CRNew{is not}
more important than \CRNew{in} code summarization.
Moreover, \CRNew{for the first time, we present}
an extensive list of potential use cases and evaluation \methodologies
side-by-side, as well as the impact of choosing various \methodologies
on the performance of ML models.

\CRNew{Our code and data are available at \RPURL.}

\section{\Methodologies}
\label{sec:methodologies}

We first summarize two commonly used \methodologies: \mixedproj
(\S\ref{sec:methodologies:mixedproj}) and \crossproj
(\S\ref{sec:methodologies:crossproj}).  Then, we introduce a novel
\temporal \methodology (\S\ref{sec:methodologies:temporal}).  We will
use $\atimepp < \atimep < \atime$ to denote specific points in time
(i.e., timestamps).

{\def\arraystretch{1.03}

\begin{table*}[t]
\begin{footnotesize}
  \centering
  \begin{tabular}{@{}l | l@{\hspace{5pt}}l@{\hspace{6pt}}l@{\hspace{5pt}}>{\scriptsize}l c@{\hspace{5pt}}c@{\hspace{3pt}}c}
    \toprule
    & & & & & \multicolumn{3}{c}{\textbf{\Methodology}} \\\cline{6-8}
    \multirow{-2}{*}{\textbf{Task}} &
    \multirow{-2}{*}{\textbf{Reference}} &
    \multirow{-2}{*}{\textbf{Published at}} &
    \multirow{-2}{*}{\textbf{Language}} &
    \multirow{-2}{*}{\footnotesize\textbf{Automatic Metrics}} &
    \textbf{\Amp} & \textbf{\Acp} & \textbf{\At} \\
    \midrule

    & \citet{IyerETAL16Summarizing} & ACL'16 & \Csharp, SQL & \BLEU, \METEOR & \mycheckmark & \mycross & \mycross \\
    & \citet{WanETAL18SummarizationRL} & ASE'18 & Python & \BLEU, \METEOR, \ROUGEf, CIDER & \mycheckmark & \mycross & \mycross \\ %
    & \citet{XuETAL18Graph2Seq} & EMNLP'18 & SQL & \BLEU & \mycheckmark & \mycross & \mycross \\ %
    & \citet{FernandesETAL19Summarization} & ICLR'19 & \Csharp & \BLEU, \ROUGEf, ROUGE-2, \Fone & \mycheckmark & \mycross & \mycross \\ %
    & \citet{LeClairETAL19Neural} & ICSE'19 & Java & \BLEU & \mycross & \mycheckmark & \mycross \\ %
    \rowcolor{blue!10} \cellcolor{white} & \citet{HuETAL18Deep,HuETAL20DeepComHybrid} & ICPC'18, ESE'20 & Java & \BLEU, \METEOR, \Precision, \Recall, \Fone & \mycheckmark & \mycross & \mycross \\ %
    & \citet{LeClairETAL20Summarization} & ICPC'20 & Java & \BLEU, \ROUGEf & \mycross & \mycheckmark & \mycross \\ %
    & \citet{CaiETAL20TAG} & ACL'20 & SQL, Python & \BLEU, \ROUGEf, ROUGE-2 & \mycheckmark & \mycross & \mycross \\ %
    \rowcolor{blue!10} \cellcolor{white} & \citet{AhmadETAL20Transformer} & ACL'20 & Java, Python & \BLEU, \METEOR, \ROUGEf & \mycheckmark & \mycross & \mycross \\ %
    & \citet{FengETAL20CodeBERT} & EMNLP'20 & Java, Python, etc. & \BLEU & \mycheckmark & \mycross & \mycross \\ %
    \multirow{-11}{*}{\rotatebox[origin=c]{90}{\makecell[c]{Comment\\Generation}}} & \citet{AhmadETAL21PLBART} & NAACL'21 & Java, Python, etc. & \BLEU & \mycheckmark & \mycross & \mycross \\ %
    \midrule
    & \citet{AllamanisETAL16Convolutional} & ICML'16 & Java & \Precision, \Recall, \Fone, \XMatch & \mycheckmark & \mycross & \mycross \\
    & \citet{FernandesETAL19Summarization} & ICLR'19 & Java, \Csharp & \ROUGEf, ROUGE-2, \Fone & \mycheckmark & \mycross & \mycross \\ %
    & \citet{XuETAL19Method} & PEPM'19 & Java & \Precision, \Recall, \Fone, \XMatch & \mycheckmark & \mycross & \mycross \\ %
    \rowcolor{blue!10} \cellcolor{white} & \citet{AlonETAL19Code2vec} & POPL'19 & Java & \Precision, \Recall, \Fone & \mycheckmark & \mycross & \mycross \\ %
    \rowcolor{blue!10} \cellcolor{white} & \citet{AlonETAL19Code2seq} & ICLR'19 & Java & \Precision, \Recall, \Fone & \mycross & \mycheckmark & \mycross \\ %
    & \citet{YonaiETAL19Mercem} & APSEC'19 & Java & Top-10 Accuracy & \mycheckmark & \mycross & \mycross \\ %
    \multirow{-7}{*}{\rotatebox[origin=c]{90}{\makecell[c]{Method\\Naming}}} & \citet{NguyenETAL20MethodNames} & ICSE'20 & Java & \Precision, \Recall, \Fone & \mycheckmark & \mycheckmark & \mycross \\ %
    \midrule
    Sum & \na & \na & \na & {\footnotesize\na} & \NumCodeSumMP & \NumCodeSumCP & \NumCodeSumT \\
    \bottomrule
  \end{tabular}
  \caption{\TCPriorWorkCodeSum}
  \vspace{\TVPriorWorkCodeSum}
\end{footnotesize}
\end{table*}

}

Table~\ref{table:prior-work-code-sum} lists prior work on developing
new ML models for code summarization.
The last three columns show which \methodology/\methodologies were
used in the evaluation in each work (\Amp: \mixedproj, \Acp:
\crossproj, \At: \temporal).
Out of \NumCodeSumAll \CRNew{papers} we found, \NumCodeSumMP used the
\mixedproj \methodology and \NumCodeSumCP used the \crossproj
\methodology.  
No prior work used the \temporal \methodology.

\begin{figure}[t]
  \centering
  \scalebox{0.85}{




\begin{tikzpicture}[
  line width=0.4pt,
  node distance=0ex and 0em,
  every node/.style={scale=0.9},
  gridBox/.style={rectangle, opacity=0, draw=red},
  box/.style={rectangle, draw=black, inner sep=2pt, font=\small},
  rounded box/.style={rectangle, rounded corners, draw=black, inner sep=2pt, font=\small},
  anno/.style={font=\footnotesize},
]

  \node (b-l-train) at (0, -4ex) [sExample, dsTrain] {};
  \node (b-l-train-text) [right = .1em of b-l-train] [anno] {\Train};
  \node (b-l-val) [right = .2em of b-l-train-text] [sExample, dsVal] {};
  \node (b-l-val-text) [right = .1em of b-l-val] [anno] {\Val};
  \node (b-l-test) [right = .2em of b-l-val-text] [sExample, dsTest] {};
  \node (b-l-test-text) [right = .1em of b-l-test] [anno] {\Test};

  \node (b-p1) at (0,0) [anchor=south, sProject] {};
  \node (b-anno1) [below = 0 of b-p1.south] [anno] {project 1};
  \node (b-p2) [right = \wSepProject of b-p1.east] [sProject] {};
  \node (b-anno2) [below = 0 of b-p2.south] [anno] {project 2};
  \node (b-p3) [right = \wSepProject of b-p2.east] [sProject] {};
  \node (b-anno3) [below = 0 of b-p3.south] [anno] {project 3};
  \node (b-p4) [right = 2 * \wSepProject of b-p3.east] [sProject] {};
  \node (b-anno4) [below = 0 of b-p4.south] [anno] {project n-1};
  \node (b-p5) [right = \wSepProject of b-p4.east] [sProject] {};
  \node (b-anno5) [below = 0 of b-p5.south] [anno] {project n};

  \node (b-ellipsis) at ($(b-p3.east)!0.5!(b-p4.west)$) [] {\dots};

  \foreach \pi in {1,2,3,4,5} {
    \foreach [evaluate={
        \hash = Mod((\ei*17+19) - \pi, 5)
    }] \ex/\ey [count=\ei] in {
      -.8em/-1em, 0em/-1em, .8em/-1em,
      -1.2em/0, -.4em/0, .4em/0, 1.2em/0,
      -.8em/1em, 0em/1em, .8em/1em
    } {
      \tikzset{ds/.code={\ifcase\hash \tikzset{dsVal} \or \tikzset{dsTest} \else \tikzset{dsTrain} \fi}}
      \node (b-p\pi-\ei) [below right = \ey and \ex of b-p\pi.center, anchor=center] [sExample, ds] {};
    }
  };

\end{tikzpicture}

  }
  \vspace{-14pt}
  \caption{\Mixedproj \methodology. \label{fig:method-mixedproj}}
  \vspace{-2pt}
\end{figure}

\begin{figure}[t]
  \centering
  \scalebox{0.85}{




\begin{tikzpicture}[
  line width=0.4pt,
  node distance=0ex and 0em,
  every node/.style={scale=0.9},
  gridBox/.style={rectangle, opacity=0, draw=red},
  box/.style={rectangle, draw=black, inner sep=2pt, font=\small},
  rounded box/.style={rectangle, rounded corners, draw=black, inner sep=2pt, font=\small},
  anno/.style={font=\footnotesize},
]

  \node (b-l-train) at (0, -4ex) [sExample, dsTrain] {};
  \node (b-l-train-text) [right = .1em of b-l-train] [anno] {\Train};
  \node (b-l-val) [right = .2em of b-l-train-text] [sExample, dsVal] {};
  \node (b-l-val-text) [right = .1em of b-l-val] [anno] {\Val};
  \node (b-l-test) [right = .2em of b-l-val-text] [sExample, dsTest] {};
  \node (b-l-test-text) [right = .1em of b-l-test] [anno] {\Test};

  \node (b-p1) at (0,0) [anchor=south, sProject] {};
  \node (b-anno1) [below = 0 of b-p1.south] [anno] {project 1};
  \node (b-p2) [right = \wSepProject of b-p1.east] [sProject] {};
  \node (b-anno2) [below = 0 of b-p2.south] [anno] {project 2};
  \node (b-p3) [right = \wSepProject of b-p2.east] [sProject] {};
  \node (b-anno3) [below = 0 of b-p3.south] [anno] {project 3};
  \node (b-p4) [right = 1.5 * \wSepProject of b-p3.east] [sProject] {};
  \node (b-anno4) [below = 0 of b-p4.south] [anno] {project m};
  \node (b-p5) [right = 1.5 * \wSepProject of b-p4.east] [sProject] {};
  \node (b-anno5) [below = 0 of b-p5.south] [anno] {project n};

  \node (b-ellipsis1) at ($(b-p3.east)!0.5!(b-p4.west)$) [] {\dots};
  \node (b-ellipsis2) at ($(b-p4.east)!0.5!(b-p5.west)$) [] {\dots};

  \foreach \pi/\ds in {1/dsTrain,2/dsTrain,3/dsTrain,4/dsVal,5/dsTest} {
    \foreach \ex/\ey [count=\ei] in {
      -.8em/-1em, 0em/-1em, .8em/-1em,
      -1.2em/0, -.4em/0, .4em/0, 1.2em/0,
      -.8em/1em, 0em/1em, .8em/1em
    } {
      \node (b-p\pi-\ei) [below right = \ey and \ex of b-p\pi.center, anchor=center] [sExample, \ds] {};
    }
  };

\end{tikzpicture}

  }
  \vspace{-14pt}
  \caption{\Crossproj \methodology. \label{fig:method-crossproj}}
  \vspace{-0pt}
\end{figure}

\subsection{\MixedProj}
\label{sec:methodologies:mixedproj}

The \emph{\mixedproj} \methodology, which is the most commonly used
\methodology in prior work, extracts \examples (code and
comments) at a single timestamp (\atime) from various projects, then
randomly shuffles the \examples and splits them into \train, \val, and
\test sets.

Figure~\ref{fig:method-mixedproj} illustrates this \methodology, where
each box represents a project and each circle represents a \example.
This \methodology is \emph{time-unaware}, i.e., it does not consider
if \examples in the \test sets are committed into a project before or
after \examples in the \train or \val sets.

\subsection{\CrossProj}
\label{sec:methodologies:crossproj}

The \emph{\crossproj} \methodology, also commonly used in prior work,
extracts \examples at a single timestamp (\atime) from various
projects as well.  Unlike the \mixedproj \methodology, the \crossproj
\methodology splits the \emph{set of projects} into three disjoint
sets for \train, \val, and \test.  Thus, the \examples from one
project are contained in only one of the \train, \val, and \test sets.

Figure~\ref{fig:method-crossproj} illustrates this methodology.
The \crossproj methodology is explicitly evaluating the ability
to{ generalize} a model to new projects.
However, \crossproj is also time-unaware, i.e., it does not consider
if the \examples from a project in the \test set come before or after
the \examples from the projects in the \train set.

\subsection{\TemporalX}
\label{sec:methodologies:temporal}

We introduce a novel \methodology: \emph{\temporal}.  Unlike the
\methodologies explained earlier, the \temporal \methodology is
\emph{time-aware}, i.e., the \examples in the \train set were
available in the projects \emph{before} the \examples in the \val set,
which were in turn available \emph{before} the \examples in the \test
set.

Figure~\ref{fig:method-evoaware} illustrates this \methodology.  The
\examples available before \atimepp (i.e., their timestamps are
earlier than \atimepp) are assigned to the \train set.  The \examples
available after \atimepp and before \atimep are assigned to the \val
set.  And finally, the \examples available after \atimep and before
\atime (which is the time when the dataset is collected) are assigned
to the \test set.  This assignment may not be the only approach to
satisfy the definition of the \temporal \methodology, but is one
approach that utilizes all \examples collected at \atime.  Alternative
assignments, e.g., excluding \examples available before \atimeppp (a
timestamp earlier than \atimepp) from the \train set, may have other
benefits, which we leave for future work to study.

\section{Use Cases}
\label{sec:use-cases}

\Methodologies are used to set up experiments and obtain an
appropriate dataset split for the evaluation.  However, they do
\emph{not} describe the envisioned usage of an ML model.
Prior work picked 
\CRNew{a \methodology in order to set up experiments,} but we argue
that ML models should be described with respect to \emph{use cases},
i.e., how will the developers use the models eventually.
Once a use case is chosen, an appropriate \methodology can be selected
to evaluate the model.  

In this section, we define three use cases via examples of the \comgen
task.  The first two use cases are ``extracted'' from prior work.
Namely, we reason about the \mixedproj and the \crossproj
\methodologies used in prior work and try to link each to a (somewhat)
\realistic use case.
The third use case is inspired by our own development and can be
evaluated using the \temporal \methodology.  Note that we do not try
to provide an exhaustive list of use cases, but rather to start off
this important discussion on the distinction between a use case and an
\emethodology.  
For the simplicity of our discussion, we only focus on the \train and
\test sets (since the \val set can be regarded as the ``open'' \test
set for tuning).

\subsection{In-Project Batch-Mode Use Case}
\label{sec:use-cases:ipmode}

Consider Alice, a developer at a large software company.  Alice has
been developing several software features in her project over an
extended period of time (since \atimep), but
she only wrote comments for a part of her code.  At one point
(\atime), she decides it is time to add documentations for the
\methods without comments, with the help of an ML model.
Alice decides to train a model using already existing \examples (i.e.,
\codecom pairs for the \methods with comments) in her code, and since
this may provide only a small number of \train \examples, she also
uses the \examples (available at time \atime) from
other projects.  We call this \emph{\ipmode use
case}, because Alice trains a new model every time she wants to use
the model, and she applies it to a large amount of \methods that may
be added before or after the \methods in the \train set.  This use
case can be evaluated using the \mixedproj \methodology
(\S\ref{sec:methodologies:mixedproj}).

Because prior work using the \mixedproj \methodology did not set any
limit on the timestamps for \examples in \train and \test sets, the
time difference between \examples in the two sets
can be arbitrarily large.  Moreover,
the model is applied on all projects that it has been trained on.
These two facts make the \ipmode use case less \realistic, for example, a \example from project A available at time
\atime may be used to predict a \example from project B available at
time \atimep, and a \example from project B available at time \atime
may be used to predict a \example from project A available at time
\atimep, simultaneously.

\subsection{Cross-Project Batch-Mode Use Case}
\label{sec:use-cases:pmode}

In this case, we assume that Alice works on a project (since \atimep)
without writing any documentation for her code.
At some point (\atime), Alice decides to document all her \methods,
again with the help of an ML model.
Since Alice does not have any comments in her code, she decides to
only train on the \examples (i.e., \codecom pairs) from other
projects (at time \atime).
Once the model is trained, she uses it to generate comments for all
the \methods in her project.  We call this \emph{\pmode use case},
because Alice trains a new model at a specific timestamp and applies
it to all the \methods on a new project.
(Note that once she integrates the comments that she likes, she can
use them in the future for training a new ML model, which matches
\ipmode use case, or potentially she could decide to ignore those
comments and always generates new comments, but this is
unlikely.)  This use case can be evaluated using the \crossproj
\methodology (\S\ref{sec:methodologies:crossproj}).

While the \crossproj \methodology is reasonable for evaluating model
generalizability, the \pmode use case does make \CRNew{strong} assumptions
(e.g., 
\CRNew{no documentation exists for any \method in the targeted projects}).

\subsection{Continuous-Mode Use Case}
\label{sec:use-cases:cmode}

In this case, Alice writes comments for each \method around the same
time as the \method itself.  For example, Alice might integrate a
model for \comgen into her IDE that would suggest comments once Alice
indicates that a \method is complete.  (Updating and maintaining
comments as code evolves~\cite{PanthaplackelETAL20CommentUpdate,
  LiuETAL20CommentUpdate, LinETAL21CommentUpdate} is an important
topic, but orthogonal to our work.)
Suppose at \atimep, Alice downloads the latest model trained on the
data available in her project and other projects before \atimep; such
model could be trained by her company and retrained every once in a
while (finding an appropriate frequency at which to retrain the model
is a topic worth exploring in the future).
She can keep using the same model until \atime when she decides to use a new model.  We call this
\emph{\cmode}, because the only \examples that can be used to train
the model are the \examples from the past.  This use case can be
evaluated using the \temporal \methodology
(\S\ref{sec:methodologies:temporal}).

\section{Application of \Methodologies}
\label{sec:application}

\begin{figure*}[t]
  \centering
  \scalebox{0.86}{
    \input{tikz/eval-settings}
  }
  \vspace{-0pt}
  \caption{Steps of processing a dataset into \train, \val, \tests,
    and \testc sets. \label{fig:eval-settings}}
  \vspace{-0pt}
\end{figure*}

We describe the steps to apply the \methodologies following their
definitions (\S\ref{sec:methodologies}) with a given dataset, as
illustrated in Figure~\ref{fig:eval-settings}.  The input dataset
contains \examples with timestamps, and the outputs include: a \train
and \val set for each \methodology to train models; a \emph{\tests}
set for each \methodology to evaluate the models for this \methodology
only; and a \emph{\testc} set for each pair of \methodologies to
\emph{compare} the same models on the two \methodologies.  Appendix~\ref{sec:appendix:formulas}
presents the formulas of each step.

\MyPara{Step~1: time-segment}  See Figure~\ref{fig:eval-settings} top
left part. A project is horizontally segmented into three parts by
timestamps \atimepp and \atimep.

\MyPara{Step~2: in-project split}  See Figure~\ref{fig:eval-settings}
top right part. A project is further vertically segmented into three
parts randomly, which is orthogonal to the time segments in step~1.

\MyPara{Step~3: cross-project split}  See
Figure~\ref{fig:eval-settings} middle part. Projects are assigned to
\train, \val, and \test sets randomly, which is orthogonal to the time
segments and in-project splits in step~1~and~2.

\MyPara{Step~4: grouping}  Now that the dataset is broken down to
small segments across three dimensions (time, in-project, and
cross-project), this step groups the appropriate segments to obtain
the \train (\ATrain), \val (\AVal), and \tests (\ATestS) sets for each
\methodology.  This is visualized in Figure~\ref{fig:eval-settings}
bottom left part.

\MyPara{Step~5: intersection}  The \testc (\ATestC) set of two
\methodologies is the intersection of their \ATestS sets.  This is
visualized in Figure~\ref{fig:eval-settings} bottom right part.

In theory, we could compare all three \methodologies on the
intersection of the three \ATestS sets, but in practice, this set is
too small (far less than 4\% of all \examples when we assign 20\%
projects and 20\% \examples in each project into \test set).

\MyPara{Step~6: postprocessing} To avoid being impacted by the
differences in the number of \train \examples for different
\methodologies, we (randomly) downsample their \ATrain sets to the
same size (i.e., the size of the smallest \ATrain set).\footnote{This
is not required if training ML models under a specific \methodology
without comparing to other \methodologies.}

The evaluation (\AVal, \ATestS, \ATestC) sets may contain \examples
that are duplicates of some \examples in the \ATrain set, due to code
clones~\cite{SajnaniETAL16SourcererCC, RoyETAL09CodeClone} and
software evolution~\cite{FluriETAL07CodeCommentCoEvolve,
ZaidmanETAL11CodeTestCoEvolve}.  We remove those \examples as they
induce noise to the evaluation of ML
models~\cite{Allamanis19CodeDuplication}.  We present the results of
removing exact-duplicates in the main paper, but we also perform
experiments of removing near-duplicates to further reduce this noise
and report their results in
Appendix~\ref{sec:appendix:near-duplicates} (which do not affect our
main findings).

\section{Experiments}
\label{sec:eval}

We run several existing ML models using different \methodologies to
understand their impact on automatic metrics, which are commonly used
to judge the performance of models.

\subsection{Tasks}
\label{sec:eval:tasks}

We focus on two most studied code summarization tasks: \comgen and
\metnam.  
We gave our best to select well-studied, representative,
publicly-available models for each task; adding more models may reveal
other interesting observations but is computationally costly, which we
leave for future work.

\MyPara{\Comgen} Developers frequently write comments in natural
language together with their code to describe APIs, deliver messages
to users, and to communicate among
themselves~\cite{PadioleauETAL09Listening, NieETAL18Natural,
PascarellaETAL19Classifying}.
Maintaining comments is tedious and error-prone, and incorrect or
outdated comments could lead to bugs~\cite{TanETAL07Icomment,
TanETAL12TComment, RatolAndRobillard17Detecting,
PanthaplackelETAL21InconsistencyDetection}.
\Comgen tries to automatically generate comments from code.  Prior
work mostly focused on generating an API comment (e.g., \javadoc
summary) given a \method.

We used three models: \UseMacro{model-DeepComHybridESE19} model from
\citet{HuETAL18Deep, HuETAL20DeepComHybrid},
\UseMacro{model-TransformerACL20} model and
\UseMacro{model-RNNBaseline} baseline from
\citet{AhmadETAL20Transformer}.
We used four automatic metrics that are frequently reported in prior
work: \BLEU~\cite{PapineniETAL02BLEU} (average sentence-level \bleu-4
with smoothing~\cite{LinAndOch04ORANGE}),
\METEOR~\cite{BanerjeeAndLavie05METEOR},
\ROUGEf~\cite{LinAndOch04Automatic}, and \Xmatch (exact match
accuracy).

\MyPara{\Metnam} Descriptive names for code elements (variables,
\methods, classes, etc.) are a vital part of readable and maintainable
code~\cite{HostAndOstvold09MethodNames, AllamanisETAL15Names}.  Naming
\methods is particularly important and challenging, because the names
need to be both concise---usually containing only a few tokens---and
comprehensible---such that they describe the key functionality of the
code~\cite{LawrieETAL06Whats}.

We used two models: \UseMacro{model-Code2VecPOPL19}
from~\citet{AlonETAL19Code2vec} and \UseMacro{model-Code2SeqICLR19}
from~\citet{AlonETAL19Code2seq}.
We used four automatic metrics that are frequently reported in prior
work: \Precision, \Recall, \Fone, and \Xmatch (exact match accuracy).

\subsection{Data}
\label{sec:eval:data}

We could not easily reuse existing datasets from prior work because
the timestamps of \examples are \emph{not} available.  We extracted
\examples with timestamps from \NumProject popular and active
open-source Java projects \CRNew{using English for summaries (comments
and names)} from GitHub.  We collected \examples before \atime = 2021
\ajanone, and we time-segmented \examples by \atimepp = 2019 \ajanone
and \atimep = 2020 \ajanone.  The splitting ratios for in-project and
cross-project splits are 70\%, 10\%, 20\%.

\begin{table}[t]
\begin{footnotesize}
\begin{center}
\begin{tabular}{ @{\hspace{2pt}} l @{\hspace{2pt}} | @{\hspace{3pt}} c @{\hspace{5pt}} r @{\hspace{5pt}} r @{\hspace{5pt}} r @{\hspace{3pt}}c@{\hspace{3pt}} c @{\hspace{5pt}} r@{\hspace{2pt}} }
\toprule
\textbf{Task} & & \textbf{\ATrain} & \textbf{\AVal} & \textbf{\ATestS} & & & \textbf{\ATestC} \\
\midrule
 & \UseMacro{TH-ds-MP}
 & \UseMacro{ds-CG-train-MP_num-data}
 & \UseMacro{ds-CG-val-MP_num-data}
 & \UseMacro{ds-CG-test_standard-MP_num-data}
 & \tikz[remember picture, baseline] \node[inner sep=2pt, outer sep=0, yshift=1ex] (CGMP-base) {\phantom{XX}};
 & \UseMacro{TH-ds-MP-CP}
 & \UseMacro{ds-CG-test_common-MP-CP_num-data}
\\
 & \UseMacro{TH-ds-CP}
 & \UseMacro{ds-CG-train-CP_num-data}
 & \UseMacro{ds-CG-val-CP_num-data}
 & \UseMacro{ds-CG-test_standard-CP_num-data}
 & \tikz[remember picture, baseline] \node[inner sep=2pt, outer sep=0, yshift=1ex] (CGCP-base) {\phantom{XX}};
 & \UseMacro{TH-ds-MP-T}
 & \UseMacro{ds-CG-test_common-MP-T_num-data}
\\
\multirow{-3.5}{*}{\rotatebox[origin=c]{90}{\UseMacro{TaskM_CG}}}
 & \UseMacro{TH-ds-T}
 & \UseMacro{ds-CG-train-T_num-data}
 & \UseMacro{ds-CG-val-T_num-data}
 & \UseMacro{ds-CG-test_standard-T_num-data}
 & \tikz[remember picture, baseline] \node[inner sep=2pt, outer sep=0, yshift=1ex] (CGT-base) {\phantom{XX}};
 & \UseMacro{TH-ds-CP-T}
 & \UseMacro{ds-CG-test_common-CP-T_num-data}
\\
\midrule
 & \UseMacro{TH-ds-MP}
 & \UseMacro{ds-MN-train-MP_num-data}
 & \UseMacro{ds-MN-val-MP_num-data}
 & \UseMacro{ds-MN-test_standard-MP_num-data}
 & \tikz[remember picture, baseline] \node[inner sep=2pt, outer sep=0, yshift=1ex] (MNMP-base) {\phantom{XX}};
 & \UseMacro{TH-ds-MP-CP}
 & \UseMacro{ds-MN-test_common-MP-CP_num-data}
\\
 & \UseMacro{TH-ds-CP}
 & \UseMacro{ds-MN-train-CP_num-data}
 & \UseMacro{ds-MN-val-CP_num-data}
 & \UseMacro{ds-MN-test_standard-CP_num-data}
 & \tikz[remember picture, baseline] \node[inner sep=2pt, outer sep=0, yshift=1ex] (MNCP-base) {\phantom{XX}};
 & \UseMacro{TH-ds-MP-T}
 & \UseMacro{ds-MN-test_common-MP-T_num-data}
\\
\multirow{-3.5}{*}{\rotatebox[origin=c]{90}{\UseMacro{TaskM_MN}}}
 & \UseMacro{TH-ds-T}
 & \UseMacro{ds-MN-train-T_num-data}
 & \UseMacro{ds-MN-val-T_num-data}
 & \UseMacro{ds-MN-test_standard-T_num-data}
 & \tikz[remember picture, baseline] \node[inner sep=2pt, outer sep=0, yshift=1ex] (MNT-base) {\phantom{XX}};
 & \UseMacro{TH-ds-CP-T}
 & \UseMacro{ds-MN-test_common-CP-T_num-data}
\\
\bottomrule
\end{tabular}
\begin{tikzpicture}[remember picture, overlay, thick]
\draw[->] (CGMP-base.west) .. controls ($(CGMP-base.east) - (1em,0)$) .. (CGMP-base.east);
\draw (CGCP-base.west) .. controls ($(CGMP-base.east) - (1em,0)$) .. (CGMP-base.east);
\draw[->] (CGMP-base.west) .. controls ($(CGCP-base.east) - (1em,0)$) .. (CGCP-base.east);
\draw (CGT-base.west) .. controls ($(CGCP-base.east) - (1em,0)$) .. (CGCP-base.east);
\draw[->] (CGCP-base.west) .. controls ($(CGT-base.east) - (1em,0)$) .. (CGT-base.east);
\draw (CGT-base.west) .. controls ($(CGT-base.east) - (1em,0)$) .. (CGT-base.east);
\draw[->] (MNMP-base.west) .. controls ($(MNMP-base.east) - (1em,0)$) .. (MNMP-base.east);
\draw (MNCP-base.west) .. controls ($(MNMP-base.east) - (1em,0)$) .. (MNMP-base.east);
\draw[->] (MNMP-base.west) .. controls ($(MNCP-base.east) - (1em,0)$) .. (MNCP-base.east);
\draw (MNT-base.west) .. controls ($(MNCP-base.east) - (1em,0)$) .. (MNCP-base.east);
\draw[->] (MNCP-base.west) .. controls ($(MNT-base.east) - (1em,0)$) .. (MNT-base.east);
\draw (MNT-base.west) .. controls ($(MNT-base.east) - (1em,0)$) .. (MNT-base.east);
\end{tikzpicture}
\end{center}
\end{footnotesize}
\vspace{\UseMacro{TV-dataset-metrics-small}}
\caption{\UseMacro{TC-dataset-metrics-small}}
\end{table}

Table~\ref{table:dataset-metrics-small} presents the number of
\examples in each set for each \methodology.  We present more details
and metrics of data collection in Appendix~\ref{sec:appendix:data}.

\subsection{Results}
\label{sec:eval:results}

We use the hyper-parameters provided in the original papers.  \Val
sets are used for early-stopping if needed by the model.  We run each
model \NumTrial times with different random seeds.
Appendix~\ref{sec:appendix:repro} presents more details of our
experiments to support their reproducibility.

\begin{table}[t]
\begin{footnotesize}
\begin{center}
\begin{tabular}{@{\hspace{0pt}} l @{\hspace{0pt}} | r@{\hspace{3pt}}r | r@{\hspace{3pt}}r | r@{\hspace{3pt}}r @{\hspace{0pt}}}
\toprule
\makecell[c]{\UseMacro{TH-train-on}}
 & \makecell[c]{\UseMacro{TH-MP}}
 & \makecell[c]{\UseMacro{TH-CP}}
 & \makecell[c]{\UseMacro{TH-MP}}
 & \makecell[c]{\UseMacro{TH-T}}
 & \makecell[c]{\UseMacro{TH-CP}}
 & \makecell[c]{\UseMacro{TH-T}}
\\ \cline{2-3} \cline{4-5} \cline{6-7}
\makecell[c]{\UseMacro{TH-test-on}}
 & \multicolumn{2}{c|}{\UseMacro{TH-MP-CP}}
 & \multicolumn{2}{c|}{\UseMacro{TH-MP-T}}
 & \multicolumn{2}{c}{\UseMacro{TH-CP-T}}
\\
\midrule
\midrule
\multicolumn{7}{c}{\UseMacro{TH-metric-table-bleu}} \\
\midrule
\UseMacro{TH-model-DeepComHybridESE19}
 & \UseMacro{result-CG_MP_DeepComHybridESE19_test_common-MP-CP_bleu}
 & \UseMacro{result-CG_CP_DeepComHybridESE19_test_common-MP-CP_bleu}
 & $^{\beta}$\UseMacro{result-CG_MP_DeepComHybridESE19_test_common-MP-T_bleu}
 & \UseMacro{result-CG_T_DeepComHybridESE19_test_common-MP-T_bleu}
 & \UseMacro{result-CG_CP_DeepComHybridESE19_test_common-CP-T_bleu}
 & \UseMacro{result-CG_T_DeepComHybridESE19_test_common-CP-T_bleu}
\\
\UseMacro{TH-model-RNNBaseline}
 & \UseMacro{result-CG_MP_RNNBaseline_test_common-MP-CP_bleu}
 & $^{\alpha}$\UseMacro{result-CG_CP_RNNBaseline_test_common-MP-CP_bleu}
 & \UseMacro{result-CG_MP_RNNBaseline_test_common-MP-T_bleu}
 & $^{\beta}$\UseMacro{result-CG_T_RNNBaseline_test_common-MP-T_bleu}
 & \UseMacro{result-CG_CP_RNNBaseline_test_common-CP-T_bleu}
 & \UseMacro{result-CG_T_RNNBaseline_test_common-CP-T_bleu}
\\
\UseMacro{TH-model-TransformerACL20}
 & \textbf{\UseMacro{result-CG_MP_TransformerACL20_test_common-MP-CP_bleu}}
 & $^{\alpha}$\textbf{\UseMacro{result-CG_CP_TransformerACL20_test_common-MP-CP_bleu}}
 & \textbf{\UseMacro{result-CG_MP_TransformerACL20_test_common-MP-T_bleu}}
 & \textbf{\UseMacro{result-CG_T_TransformerACL20_test_common-MP-T_bleu}}
 & \textbf{\UseMacro{result-CG_CP_TransformerACL20_test_common-CP-T_bleu}}
 & \textbf{\UseMacro{result-CG_T_TransformerACL20_test_common-CP-T_bleu}}
\\
\midrule
\midrule
\multicolumn{7}{c}{\UseMacro{TH-metric-table-meteor}} \\
\midrule
\UseMacro{TH-model-DeepComHybridESE19}
 & \UseMacro{result-CG_MP_DeepComHybridESE19_test_common-MP-CP_meteor}
 & \UseMacro{result-CG_CP_DeepComHybridESE19_test_common-MP-CP_meteor}
 & $^{\delta}$\UseMacro{result-CG_MP_DeepComHybridESE19_test_common-MP-T_meteor}
 & \UseMacro{result-CG_T_DeepComHybridESE19_test_common-MP-T_meteor}
 & \UseMacro{result-CG_CP_DeepComHybridESE19_test_common-CP-T_meteor}
 & \UseMacro{result-CG_T_DeepComHybridESE19_test_common-CP-T_meteor}
\\
\UseMacro{TH-model-RNNBaseline}
 & \UseMacro{result-CG_MP_RNNBaseline_test_common-MP-CP_meteor}
 & $^{\gamma}$\UseMacro{result-CG_CP_RNNBaseline_test_common-MP-CP_meteor}
 & \UseMacro{result-CG_MP_RNNBaseline_test_common-MP-T_meteor}
 & $^{\delta}$\UseMacro{result-CG_T_RNNBaseline_test_common-MP-T_meteor}
 & $^{\epsilon}$\UseMacro{result-CG_CP_RNNBaseline_test_common-CP-T_meteor}
 & \UseMacro{result-CG_T_RNNBaseline_test_common-CP-T_meteor}
\\
\UseMacro{TH-model-TransformerACL20}
 & \textbf{\UseMacro{result-CG_MP_TransformerACL20_test_common-MP-CP_meteor}}
 & $^{\gamma}$\textbf{\UseMacro{result-CG_CP_TransformerACL20_test_common-MP-CP_meteor}}
 & \textbf{\UseMacro{result-CG_MP_TransformerACL20_test_common-MP-T_meteor}}
 & \textbf{\UseMacro{result-CG_T_TransformerACL20_test_common-MP-T_meteor}}
 & $^{\epsilon}$\textbf{\UseMacro{result-CG_CP_TransformerACL20_test_common-CP-T_meteor}}
 & \textbf{\UseMacro{result-CG_T_TransformerACL20_test_common-CP-T_meteor}}
\\
\midrule
\midrule
\multicolumn{7}{c}{\UseMacro{TH-metric-table-rouge_l_f}} \\
\midrule
\UseMacro{TH-model-DeepComHybridESE19}
 & \UseMacro{result-CG_MP_DeepComHybridESE19_test_common-MP-CP_rouge_l_f}
 & \UseMacro{result-CG_CP_DeepComHybridESE19_test_common-MP-CP_rouge_l_f}
 & $^{\zeta}$\UseMacro{result-CG_MP_DeepComHybridESE19_test_common-MP-T_rouge_l_f}
 & \UseMacro{result-CG_T_DeepComHybridESE19_test_common-MP-T_rouge_l_f}
 & \UseMacro{result-CG_CP_DeepComHybridESE19_test_common-CP-T_rouge_l_f}
 & \UseMacro{result-CG_T_DeepComHybridESE19_test_common-CP-T_rouge_l_f}
\\
\UseMacro{TH-model-RNNBaseline}
 & \UseMacro{result-CG_MP_RNNBaseline_test_common-MP-CP_rouge_l_f}
 & \UseMacro{result-CG_CP_RNNBaseline_test_common-MP-CP_rouge_l_f}
 & \UseMacro{result-CG_MP_RNNBaseline_test_common-MP-T_rouge_l_f}
 & $^{\zeta}$\UseMacro{result-CG_T_RNNBaseline_test_common-MP-T_rouge_l_f}
 & \UseMacro{result-CG_CP_RNNBaseline_test_common-CP-T_rouge_l_f}
 & \UseMacro{result-CG_T_RNNBaseline_test_common-CP-T_rouge_l_f}
\\
\UseMacro{TH-model-TransformerACL20}
 & \textbf{\UseMacro{result-CG_MP_TransformerACL20_test_common-MP-CP_rouge_l_f}}
 & \textbf{\UseMacro{result-CG_CP_TransformerACL20_test_common-MP-CP_rouge_l_f}}
 & \textbf{\UseMacro{result-CG_MP_TransformerACL20_test_common-MP-T_rouge_l_f}}
 & \textbf{\UseMacro{result-CG_T_TransformerACL20_test_common-MP-T_rouge_l_f}}
 & \textbf{\UseMacro{result-CG_CP_TransformerACL20_test_common-CP-T_rouge_l_f}}
 & \textbf{\UseMacro{result-CG_T_TransformerACL20_test_common-CP-T_rouge_l_f}}
\\
\midrule
\midrule
\multicolumn{7}{c}{\UseMacro{TH-metric-table-exact_match}} \\
\midrule
\UseMacro{TH-model-DeepComHybridESE19}
 & \UseMacro{result-CG_MP_DeepComHybridESE19_test_common-MP-CP_exact_match}
 & $^{\eta}$$^{\theta}$\UseMacro{result-CG_CP_DeepComHybridESE19_test_common-MP-CP_exact_match}
 & $^{\kappa}$\UseMacro{result-CG_MP_DeepComHybridESE19_test_common-MP-T_exact_match}
 & \UseMacro{result-CG_T_DeepComHybridESE19_test_common-MP-T_exact_match}
 & $^{\lambda}$$^{\mu}$\UseMacro{result-CG_CP_DeepComHybridESE19_test_common-CP-T_exact_match}
 & \UseMacro{result-CG_T_DeepComHybridESE19_test_common-CP-T_exact_match}
\\
\UseMacro{TH-model-RNNBaseline}
 & \UseMacro{result-CG_MP_RNNBaseline_test_common-MP-CP_exact_match}
 & $^{\eta}$$^{\iota}$\UseMacro{result-CG_CP_RNNBaseline_test_common-MP-CP_exact_match}
 & \UseMacro{result-CG_MP_RNNBaseline_test_common-MP-T_exact_match}
 & $^{\kappa}$\UseMacro{result-CG_T_RNNBaseline_test_common-MP-T_exact_match}
 & $^{\lambda}$\UseMacro{result-CG_CP_RNNBaseline_test_common-CP-T_exact_match}
 & \UseMacro{result-CG_T_RNNBaseline_test_common-CP-T_exact_match}
\\
\UseMacro{TH-model-TransformerACL20}
 & \textbf{\UseMacro{result-CG_MP_TransformerACL20_test_common-MP-CP_exact_match}}
 & $^{\theta}$$^{\iota}$\textbf{\UseMacro{result-CG_CP_TransformerACL20_test_common-MP-CP_exact_match}}
 & \textbf{\UseMacro{result-CG_MP_TransformerACL20_test_common-MP-T_exact_match}}
 & \textbf{\UseMacro{result-CG_T_TransformerACL20_test_common-MP-T_exact_match}}
 & $^{\mu}$\textbf{\UseMacro{result-CG_CP_TransformerACL20_test_common-CP-T_exact_match}}
 & \textbf{\UseMacro{result-CG_T_TransformerACL20_test_common-CP-T_exact_match}}
\\
\bottomrule
\end{tabular}
\end{center}
\end{footnotesize}
\vspace{\UseMacro{TV-results-CG}}
\caption{\UseMacro{TC-results-CG}}
\end{table}

\begin{table}[t]
\begin{footnotesize}
\begin{center}
\begin{tabular}{@{\hspace{0pt}} l @{\hspace{0pt}} | r@{\hspace{3pt}}r | r@{\hspace{3pt}}r | r@{\hspace{3pt}}r @{\hspace{0pt}}}
\toprule
\makecell[c]{\UseMacro{TH-train-on}}
 & \makecell[c]{\UseMacro{TH-MP}}
 & \makecell[c]{\UseMacro{TH-CP}}
 & \makecell[c]{\UseMacro{TH-MP}}
 & \makecell[c]{\UseMacro{TH-T}}
 & \makecell[c]{\UseMacro{TH-CP}}
 & \makecell[c]{\UseMacro{TH-T}}
\\ \cline{2-3} \cline{4-5} \cline{6-7}
\makecell[c]{\UseMacro{TH-test-on}}
 & \multicolumn{2}{c|}{\UseMacro{TH-MP-CP}}
 & \multicolumn{2}{c|}{\UseMacro{TH-MP-T}}
 & \multicolumn{2}{c}{\UseMacro{TH-CP-T}}
\\
\midrule
\midrule
\multicolumn{7}{c}{\UseMacro{TH-metric-table-set_match_p}} \\
\midrule
\UseMacro{TH-model-Code2VecPOPL19}
 & \textbf{\UseMacro{result-MN_MP_Code2VecPOPL19_test_common-MP-CP_set_match_p}}
 & \UseMacro{result-MN_CP_Code2VecPOPL19_test_common-MP-CP_set_match_p}
 & \textbf{\UseMacro{result-MN_MP_Code2VecPOPL19_test_common-MP-T_set_match_p}}
 & \textbf{\UseMacro{result-MN_T_Code2VecPOPL19_test_common-MP-T_set_match_p}}
 & \UseMacro{result-MN_CP_Code2VecPOPL19_test_common-CP-T_set_match_p}
 & \textbf{\UseMacro{result-MN_T_Code2VecPOPL19_test_common-CP-T_set_match_p}}
\\
\UseMacro{TH-model-Code2SeqICLR19}
 & \UseMacro{result-MN_MP_Code2SeqICLR19_test_common-MP-CP_set_match_p}
 & \textbf{\UseMacro{result-MN_CP_Code2SeqICLR19_test_common-MP-CP_set_match_p}}
 & \UseMacro{result-MN_MP_Code2SeqICLR19_test_common-MP-T_set_match_p}
 & \UseMacro{result-MN_T_Code2SeqICLR19_test_common-MP-T_set_match_p}
 & \textbf{\UseMacro{result-MN_CP_Code2SeqICLR19_test_common-CP-T_set_match_p}}
 & \UseMacro{result-MN_T_Code2SeqICLR19_test_common-CP-T_set_match_p}
\\
\midrule
\midrule
\multicolumn{7}{c}{\UseMacro{TH-metric-table-set_match_r}} \\
\midrule
\UseMacro{TH-model-Code2VecPOPL19}
 & \textbf{\UseMacro{result-MN_MP_Code2VecPOPL19_test_common-MP-CP_set_match_r}}
 & \UseMacro{result-MN_CP_Code2VecPOPL19_test_common-MP-CP_set_match_r}
 & \textbf{\UseMacro{result-MN_MP_Code2VecPOPL19_test_common-MP-T_set_match_r}}
 & \textbf{\UseMacro{result-MN_T_Code2VecPOPL19_test_common-MP-T_set_match_r}}
 & \UseMacro{result-MN_CP_Code2VecPOPL19_test_common-CP-T_set_match_r}
 & \textbf{\UseMacro{result-MN_T_Code2VecPOPL19_test_common-CP-T_set_match_r}}
\\
\UseMacro{TH-model-Code2SeqICLR19}
 & \UseMacro{result-MN_MP_Code2SeqICLR19_test_common-MP-CP_set_match_r}
 & \textbf{\UseMacro{result-MN_CP_Code2SeqICLR19_test_common-MP-CP_set_match_r}}
 & \UseMacro{result-MN_MP_Code2SeqICLR19_test_common-MP-T_set_match_r}
 & \UseMacro{result-MN_T_Code2SeqICLR19_test_common-MP-T_set_match_r}
 & \textbf{\UseMacro{result-MN_CP_Code2SeqICLR19_test_common-CP-T_set_match_r}}
 & \UseMacro{result-MN_T_Code2SeqICLR19_test_common-CP-T_set_match_r}
\\
\midrule
\midrule
\multicolumn{7}{c}{\UseMacro{TH-metric-table-set_match_f}} \\
\midrule
\UseMacro{TH-model-Code2VecPOPL19}
 & \textbf{\UseMacro{result-MN_MP_Code2VecPOPL19_test_common-MP-CP_set_match_f}}
 & \UseMacro{result-MN_CP_Code2VecPOPL19_test_common-MP-CP_set_match_f}
 & \textbf{\UseMacro{result-MN_MP_Code2VecPOPL19_test_common-MP-T_set_match_f}}
 & \textbf{\UseMacro{result-MN_T_Code2VecPOPL19_test_common-MP-T_set_match_f}}
 & \UseMacro{result-MN_CP_Code2VecPOPL19_test_common-CP-T_set_match_f}
 & \textbf{\UseMacro{result-MN_T_Code2VecPOPL19_test_common-CP-T_set_match_f}}
\\
\UseMacro{TH-model-Code2SeqICLR19}
 & \UseMacro{result-MN_MP_Code2SeqICLR19_test_common-MP-CP_set_match_f}
 & \textbf{\UseMacro{result-MN_CP_Code2SeqICLR19_test_common-MP-CP_set_match_f}}
 & \UseMacro{result-MN_MP_Code2SeqICLR19_test_common-MP-T_set_match_f}
 & \UseMacro{result-MN_T_Code2SeqICLR19_test_common-MP-T_set_match_f}
 & \textbf{\UseMacro{result-MN_CP_Code2SeqICLR19_test_common-CP-T_set_match_f}}
 & \UseMacro{result-MN_T_Code2SeqICLR19_test_common-CP-T_set_match_f}
\\
\midrule
\midrule
\multicolumn{7}{c}{\UseMacro{TH-metric-table-exact_match}} \\
\midrule
\UseMacro{TH-model-Code2VecPOPL19}
 & \textbf{\UseMacro{result-MN_MP_Code2VecPOPL19_test_common-MP-CP_exact_match}}
 & $^{\alpha}$\UseMacro{result-MN_CP_Code2VecPOPL19_test_common-MP-CP_exact_match}
 & \textbf{\UseMacro{result-MN_MP_Code2VecPOPL19_test_common-MP-T_exact_match}}
 & \textbf{\UseMacro{result-MN_T_Code2VecPOPL19_test_common-MP-T_exact_match}}
 & $^{\beta}$\UseMacro{result-MN_CP_Code2VecPOPL19_test_common-CP-T_exact_match}
 & \textbf{\UseMacro{result-MN_T_Code2VecPOPL19_test_common-CP-T_exact_match}}
\\
\UseMacro{TH-model-Code2SeqICLR19}
 & \UseMacro{result-MN_MP_Code2SeqICLR19_test_common-MP-CP_exact_match}
 & $^{\alpha}$\textbf{\UseMacro{result-MN_CP_Code2SeqICLR19_test_common-MP-CP_exact_match}}
 & \UseMacro{result-MN_MP_Code2SeqICLR19_test_common-MP-T_exact_match}
 & \UseMacro{result-MN_T_Code2SeqICLR19_test_common-MP-T_exact_match}
 & $^{\beta}$\textbf{\UseMacro{result-MN_CP_Code2SeqICLR19_test_common-CP-T_exact_match}}
 & \UseMacro{result-MN_T_Code2SeqICLR19_test_common-CP-T_exact_match}
\\
\bottomrule
\end{tabular}
\end{center}
\end{footnotesize}
\vspace{\UseMacro{TV-results-MN}}
\caption{\UseMacro{TC-results-MN}}
\end{table}

Tables~\ref{table:results-cg}~and~\ref{table:results-mn} present the
results for \comgen and \metnam, respectively.  Each table has four
parts and each part contains the results for one metric.  Each number
is the metric of a model (name at column~1) trained on the \ATrain set
of a \methodology (name at row~1) and evaluated on a \ATestC set
involving that \methodology (name at row~2).  The best
results are in bold
text.  The results marked with the same Greek letter are
\emph{not} statistically significantly different.\footnote{We
conducted statistical significance tests using bootstrap
tests~\cite{Berg-KirkpatrickETAL12Empirical} with confidence level
95\%.}  Appendix~\ref{sec:appendix:more-results} presents the results
on \AVal and \ATestS sets, and bar plots visualizing the results.

\subsection{Findings}
\label{sec:eval:findings}

\noindent
\textbf{Depending on the \methodology, one model can perform better or
  worse than another.}  On \metnam task, we found that
\UseMacro{model-Code2SeqICLR19} outperforms
\UseMacro{model-Code2VecPOPL19} only in \crossproj \methodology but
not the other \methodologies, consistently on all metrics.
Our observation aligns with the finding in the original
paper~\cite{AlonETAL19Code2seq} that \UseMacro{model-Code2SeqICLR19}
outperforms \UseMacro{model-Code2VecPOPL19} when using the \crossproj
\methodology.
\CRNew{The reason is that in contrary to \UseMacro{model-Code2SeqICLR19}
which generates a name as a sequence of \subtoks,
\UseMacro{model-Code2VecPOPL19} generates a name by retrieving a name
in the \ATrain set, and thus has better chances to generate correct
names under the \mixedproj and \temporal \methodologies where the
names in the \ATest set are similar to the names in the \ATrain set.}

This finding suggests that a model may work better for one use case
but not another---in this case, \UseMacro{model-Code2SeqICLR19}
performs better in the \pmode use case, but
\UseMacro{model-Code2VecPOPL19} performs better in the \ipmode and the
\cmode use case.

\noindent
\textbf{Depending on the \methodology, the differences between models
may or may not be observable.}  For example, for \comgen, on the
\ATestC set of \crossproj and \temporal \methodologies when using the
\meteor metric (Table~\ref{table:results-cg}, columns 6--7),
\UseMacro{model-TransformerACL20} significantly outperforms
\UseMacro{model-RNNBaseline} when trained on the \temporal \ATrain
set, but does not when trained on the \crossproj \ATrain set.  Similar
observations can be made on the \bleu and
\xmatch metrics for \comgen, and the \xmatch
metric for \metnam.

Two models' results being not statistically significantly different
indicates that their difference is 
\CRNew{not reliable}.
We could not find reference points for this finding in prior work
(unfortunately, \citet{AhmadETAL20Transformer} did not compare
\UseMacro{model-RNNBaseline} with \UseMacro{model-TransformerACL20}
though both were provided in their replication package).

\noindent
\textbf{Results under the \mixedproj \methodology are inflated.}  We
found that the results under the \mixedproj \methodology are always
higher than the other two \methodologies.  
This is not surprising as ML models have difficulty in generalizing to
\examples that are different from the \ATrain set.

Considering that the \mixedproj \methodology represents a less
\realistic use case than the other two \methodologies, the \mixedproj
\methodology always over-estimates the models' usefulness.
As such, we suggest that the \mixedproj \methodology should never be
used unless the model is targeted specially for the \ipmode use case
(\S\ref{sec:use-cases}).

\noindent
\textbf{Results under the \crossproj \methodology may be an
under-estimation of the more \realistic \cmode use case.}  We found
that the results under the \crossproj \methodology are always lower
than the results under the \temporal \methodology, consistently on all
metrics in both tasks.
We have{ discussed} that the \cmode use
case is more \realistic than others
(\S\ref{sec:use-cases}).  This suggests that the usefulness of
the models in prior work using the \crossproj \methodology may have
been under-estimated.

\noindent
\textbf{Findings in prior work may not hold when using a different
  \methodology or a different dataset.}  We found that the findings
reported by prior work may not hold in our experiment:
for example, the finding ``\UseMacro{model-DeepComHybridESE19}
outperforms \UseMacro{model-RNNBaseline}'' from
\citet{HuETAL20DeepComHybrid} does not hold on our dataset (one reason
could be the \UseMacro{model-RNNBaseline} code we used is more recent
than the version that \UseMacro{model-DeepComHybridESE19} based on).
This indicates that researchers should specify the targeted use case,
the employed \methodology, and the used dataset when reporting
findings, and expect that the findings may not generalize to a
different use case or dataset.

\section{Future Work}
\label{sec:future}

\subsection{Methodologies for Other SE Areas Using ML Models}
\label{sec:future:area}

\CRNew{We studied the impact of different \emethodologies in the
context of code summarization, and future work can study their impacts
on other software engineering (SE) areas using ML models.} We briefly
discuss the potential ways and challenges of transferring our
\methodologies from code summarization to \CRNew{ML models for other
SE tasks, including generation tasks (e.g., commit message generation
and code synthesis) and non-generation tasks (e.g., defect prediction
and bug localization).}
The key is to modify the application steps of the
\methodologies based on the format of \examples (inputs and
outputs) in the targeted task.

\CRNew{For most tasks where inputs and outputs are software-related
artifacts with timestamps, the \methodologies, use cases, and
application steps defined by us should still apply.} 
\CRNew{For example, transferring our \methodologies from the code
summarization task to the commit message generation task only requires
replacing ``\codecom pairs'' to ``(code change, commit message)
pairs''.}

\CRNew{For some tasks, the input or output of one \example may change
when observed at different timestamps.}
For example, in defect prediction (pointed out by
\citet{TanETAL15Online}), suppose a commit at \atimepp was discovered
to be buggy at \atime, then when training the model at \atimep, that
commit should be labeled as not buggy.  The correct version of the
\example should be used according to its timestamp.

\subsection{Other Use Cases and \Methodologies}
\label{sec:future:method}

Out of many other use cases and \methodologies, we discuss two that
are closely related to the \cmode use case and the \temporal
methodology.  Future work can expand our study and perform experiments
on them.

\MyPara{Cross-project continuous-mode use case} Compared to the \cmode
use case, when training the model at \atime, instead of using all
projects' \examples before \atime, we only use other projects'
\examples.  The corresponding \methodology is a combination of the
\crossproj and \temporal \methodologies.  From the ML model users'
perspective, this use case is less \realistic than the \cmode use
case, because using \examples from the targeted projects can improve
the model's performance.  However, from ML
model researchers' perspective, this methodology may be used to better
evaluate the model's effectiveness on unseen \examples (while
considering software evolution).

\MyPara{Online continuous-mode use case} Compared to the \cmode use
case, when we train a new model at \atime, instead of discarding the
previous model trained at \atimep and training from scratch, we
continue training the previous model using the \examples between
\atimep and \atime, e.g., using online learning
algorithms~\cite{ShalevShwartzETAL12OnlineLearning}.  The
corresponding \methodology is similar to the \temporal \methodology,
but with multiple training and evaluation steps.  Compared to the
\temporal \methodology, the model trained using this \methodology may
have better performance as it is continuously tuned on the latest
\examples (e.g., with the latest language features).

\subsection{\CRNew{Applications of Our Study in Industry}}
\label{sec:future:industry}

\CRNew{We provide generic definitions to several representative use
cases (\ipmode, \pmode, and \cmode).  We believe these three use
cases, plus some variants of the \cmode use case
(\S\ref{sec:future:method}), should cover most use cases of ML models
in the SE industry.  In practice, it may not always be possible to
determine the target use cases in advance of deploying ML models, in
which case performing a set of experiments (similar to the one in our
study) to compare between different \methodologies and use cases can
guide the switching of use cases.  We leave studying the usages of ML
models in the SE industry and deploying the findings of our study as
techniques to benefit the SE industry as future work.}

\section{Related Work}
\label{sec:related}

\subsection{Evaluation \Methodologies}

To our best knowledge, ours is the first work to study the
\emethodologies of code summarization ML models and use the \temporal
\methodology in this area.  Outside of the code summarization area, a
couple of work on defect
prediction~\cite{DAmbrosETAL12DefectPrediction, TanETAL15Online,
WangETAL16DefectPrediction, KameiETAL16JITDefectPrediction}, one work
on program repair~\cite{LutellierETAL20CoCoNut}, and one work on bug
localization~\cite{PradelETAL20Scaffle} have taken into account the
timestamps during evaluation, specifically for their task.  The
\methodologies we proposed in this paper may also be extended to those
areas.  Moreover, 
\CRNew{our work is the first to study the impact of} the
\mixedproj, \crossproj, and \temporal \methodologies \CRNew{side-by-side}.

\citet{TuETAL18Careful} revealed the data leakage problem when using
issue tracking data caused by the
unawareness of the evolution of issue attributes.  We revealed that a
similar problem (unawareness of the timestamps of \examples in the
dataset) exists in the evaluation of code summarization tasks, and we
propose a \temporal \methodology that can be used in future research.

\citet{BenderETAL21StochasticParrots}
pointed out a similar issue in NLP, that the ML models evaluated in standard
cross-validation \methodology may incur significant bias in realistic use cases, as the models cannot adapt to the
new norms, language, and ways of communicating produced by social
movements.

\subsection{Code Summarization}

Code summarization studies the problem of
summarizing a code snippet into a natural language sentence or phrase.
The two most studied tasks in code summarization are \comgen and
\metnam (\S\ref{sec:eval:tasks}).
Table~\ref{table:prior-work-code-sum} already listed the prior work on
these two tasks.  Here, we
briefly discuss their history.

The first work for \comgen~\cite{IyerETAL16Summarizing} and
\metnam~\cite{AllamanisETAL16Convolutional} were developed based on
encoder-decoder neural networks and attention mechanism.  Other prior
work extended this basic framework in many directions: by
incorporating tree-like code context such as
AST~\cite{WanETAL18SummarizationRL, XuETAL19Method,
LeClairETAL19Neural, HuETAL18Deep, HuETAL20DeepComHybrid}; by
incorporating graph-like code context such as call graphs and data
flow graphs~\cite{XuETAL18Graph2Seq, FernandesETAL19Summarization,
YonaiETAL19Mercem, LeClairETAL20Summarization}; by incorporating
path-like code context such as paths in AST~\cite{AlonETAL19Code2vec,
AlonETAL19Code2seq}; by incorporating environment context, e.g., class
name when generating \method names~\cite{NguyenETAL20MethodNames}; by
incorporating type information~\cite{CaiETAL20TAG}; or by using more
advanced neural architecture such as
transformers~\cite{AhmadETAL20Transformer}.

Recently, pre-trained models for code
learning~\cite{FengETAL20CodeBERT, GuoETAL21GraphCodeBERT,
AhmadETAL21PLBART, WangETAL21CodeT5, ChenETAL21Codex} were built on
large datasets using general tasks (e.g., masked language
modeling), and these models can be fine-tuned on specific code learning
tasks, including \comgen and \metnam.
Evaluating pre-trained models involves a pre-training set, in addition
to the regular \train, \val, and \test sets.  Our
\methodologies can be extended for pre-trained models; for example, in
the \temporal \methodology, the pre-training set contains \examples
that are available before the \examples in
all other sets.
No prior work on pre-trained models has considered the timestamps of
\examples during evaluation.

\section{Conclusion}
\label{sec:conclusion}

We highlighted the importance of specifying targeted use cases and
adopting the correct \emethodologies during the development of ML
models for code summarization tasks (and for other software
engineering tasks).  We revealed the importance of the \realistic
\cmode use case, and introduced the \temporal
\methodology which is novel to code summarization.  Our experiments of
comparing ML models using the \temporal \methodology and using the
\mixedproj and \crossproj \methodologies (which are prevalent in the
literature) showed that the choice of \methodology impacts the results
and findings of the evaluation.  We found that \mixedproj tends to
over-estimate the effectiveness of ML models, while the \crossproj may
under-estimate it.
We hope that future work on ML models for software engineering will
dedicate extra space to document intended use cases and report
findings using various \methodologies.

\section*{Acknowledgments}
We thank Nader Al Awar, Kush Jain, Yu Liu, Darko Marinov, Sheena
Panthaplackel, August Shi, Zhiqiang Zang, and the anonymous reviewers
for their comments and feedback.  This work is partially supported by
a Google Faculty Research Award, the US National Science Foundation
under Grant Nos. CCF-1652517, IIS-1850153, and IIS-2107524, and the
University of Texas at Austin Continuing Fellowship.

\section*{Ethics Statement}
\label{sec:ethical}

Our dataset has been collected in a manner that is consistent with the
licenses provided from the sources (i.e., GitHub repositories).

The \emethodologies described in our study is expected to assist
researchers in evaluating and reporting ML models for code
summarization, and assist software developers (i.e., users of those
models) in understanding the reported metrics and choosing the correct
model that fits their use case.  Our work can be directly deployed in
code summarization research, and can potentially generalize to other
software engineering areas using ML models (\S\ref{sec:future:area}).
We expect our work to help researchers build ML models for code
summarization (and other SE areas) that are more applicable to their
intended use cases.  

We do not claim that the \methodologies and use cases described in our
study are the most \realistic ones, nor do we try to provide an
exhaustive list of them.  In particular, the \cmode use case
(\S\ref{sec:use-cases:cmode}) is inspired by our own observations
during using and developing ML models for code summarization.  We try
our best to design this use case to reflect the most common and
\realistic scenarios, but other use cases may be more valid in certain
scenarios (\S\ref{sec:future:method}).

We conducted experiments involving computation time/power, but we have
carefully chosen the number of times to repeat the experiments to both
ensure reproducibility of our research and avoid consuming excessive
energy.  We provided details of our computing platform and running
time in Appendix~\ref{sec:appendix:repro}.

\bibliography{bib}
\bibliographystyle{acl_natbib}

\clearpage
\newpage
\appendix

\section{Formulas of Application of \Methodologies}
\label{sec:appendix:formulas}

\S\ref{sec:application} described the steps to apply the
\methodologies on a given dataset.  In this section, we present the
formulas used in those steps.

Table~\ref{table:prelim} lists the symbols and functions that we use.
Recall that Figure~\ref{fig:eval-settings} visualizes all the steps.
In the following discussion, we use these abbreviations: \Amp =
\mixedproj; \Acp = \crossproj; \At = \temporal; \ATrain = \train;
\AVal = \val; \ATestS = \tests; \ATestC = \testc.

\MyPara{Step~1: time-segment} We first obtain the \examples in each
project on three selected timestamps \atimepp, \atimep, \atime:
$\aexamples^{\atimepp, \aproject}, \aexamples^{\atimep, \aproject},
\aexamples^{\atime, \aproject}$.  Then, we compute the difference of
the sets to get: the \examples \emph{after} \atimepp and \emph{before}
\atimep, denoted as $\aexamples^{\atimep\setminus\atimepp, \aproject}
= \aexamples^{\atimep, \aproject}\setminus\aexamples^{\atimepp,
\aproject}$; and the \examples \emph{after} \atimep and \emph{before}
\atime, denoted as $\aexamples^{\atime\setminus\atimep, \aproject} =
\aexamples^{\atime, \aproject}\setminus\aexamples^{\atimep,
\aproject}$.

\MyPara{Step~2: in-project split} We perform the split with the
following formula ($\apcttrain, \apctval, \apcttest$ are the splitting
ratios, and following ML practices,
$\apcttrain \gg \apcttest \gtrapprox \apctval$):
\begin{flalign*}
  &\aexamples_{train}^{\atimepp, \aproject},
  \aexamples_{val}^{\atimepp, \aproject},
  \aexamples_{test}^{\atimepp, \aproject}\\
  &\qquad= \asplit(\aexamples_{train}^{\atimepp, \aproject}, \apcttrain, \apctval, \apcttest)\\
  &\aexamples_{train}^{\atimep\setminus\atimepp, \aproject},
  \aexamples_{val}^{\atimep\setminus\atimepp, \aproject},
  \aexamples_{test}^{\atimep\setminus\atimepp, \aproject}\\
  &\qquad= \asplit(\aexamples_{train}^{\atimep\setminus\atimepp, \aproject}, \apcttrain, \apctval, \apcttest)\\
  &\aexamples_{train}^{\atime\setminus\atimep, \aproject},
  \aexamples_{val}^{\atime\setminus\atimep, \aproject},
  \aexamples_{test}^{\atime\setminus\atimep, \aproject}\\
  &\qquad= \asplit(\aexamples_{train}^{\atime\setminus\atimep, \aproject}, \apcttrain, \apctval, \apcttest)
\end{flalign*}

\MyPara{Step~3: cross-project split} Given the set of projects
\aprojects and the splitting ratios $\apcttrain, \apctval, \apcttest$,
we perform the split with the following formula:
\begin{flalign*}
  \aprojecttrain, \aprojectval, \aprojecttest &= \asplit(\ashuffle(\aprojects), \apcttrain, \apctval, \apcttest)
\end{flalign*}

\MyPara{Step~4: grouping} Table~\ref{table:eval-settings} left part
lists the formulas used in this step.

\MyPara{Step~5: intersection} Table~\ref{table:eval-settings} right
part lists the formulas used in this step.

\begin{table}[t]
\begin{footnotesize}
  \centering
  \begin{tabular}{@{\hspace{2pt}} l @{\hspace{5pt}} p{5.7cm} @{\hspace{2pt}}}
    \toprule
    \textbf{Symbol} &
    \textbf{Definition} \\
    \midrule

    \atime, \atimep, \atimepp &

    Timestamps, i.e., specific points in time.  \atimepp is earlier
    than \atimep, and \atimep is earlier than \atime ($\atimepp <
    \atimep < \atime$).\\

    \midrule

    \aprojects &

    A set of projects, from which \examples are derived.\\

    \aproject &

    A project.\\

    \midrule

    \aexamples &

    A set of \examples.\\  %

    $\aexamples^{\atime, \aproject}$ &

    The set of \examples extracted from project \aproject at timestamp
    \atime.\\

    $\aexamples^{\atime\setminus\atimep, \aproject}$ &

    $= \aexamples^{\atime, \aproject} \setminus \aexamples^{\atimep,
      \aproject}$ (where $\setminus$ is the set difference operator),
    i.e., the \examples extracted from project \aproject at timestamp
    \atime that were not available at timestamp \atimep.\\

    \midrule

    $\ashuffle(l)$ &

    Given a set (of \examples or projects) $s$, returns a set with
    the same items after random shuffling.\\

    \midrule

    \makecell[lt]{$\asplit(\aexamples,$\\$\apcttrain, \apctval, \apcttest)$} &

    Given a set of \examples $\aexamples$, splits the set into three
    sets $\aexamples_x, \aexamples_y, \aexamples_z$ such that $\left|
    \aexamples_x \right| : \left| \aexamples_y \right| : \left|
    \aexamples_z \right| \approx \apcttrain:\apctval:\apcttest$;
    requires that $\apcttrain + \apctval + \apcttest = 1$.\\

    \midrule

    \makecell[lt]{$\asplitprj(\aprojects,$\\$\atime, \apcttrain, \apctval, \apcttest)$} &

    Given a set of projects $\aprojects$, splits the set into three
    sets $\aprojects_x, \aprojects_y, \aprojects_z$ such that

    $ \left| \bigcup_{p\in\aprojects_x}{\aexamples^{\atime, p}} \right|
    : \left| \bigcup_{p\in\aprojects_y}{\aexamples^{\atime, p}} \right|
    : \left| \bigcup_{p\in\aprojects_z}{\aexamples^{\atime, p}} \right|
    \approx \apcttrain:\apctval:\apcttest$; requires that $\apcttrain +
    \apctval + \apcttest = 1$.\\

    \bottomrule
  \end{tabular}
  \caption{\TCPrelim}
  \vspace{\TVPrelim}
\end{footnotesize}
\end{table}

{\def\arraystretch{2}
\begin{table*}[t]
  \begin{scriptsize}
  \begin{center}
  \begin{tabular}{ | @{\hspace{2pt}}c@{\hspace{2pt}} | @{\hspace{2pt}}c@{\hspace{2pt}} | @{\hspace{3pt}}c@{\hspace{3pt}} |@{}c@{}| @{\hspace{2pt}}c@{\hspace{2pt}} | @{\hspace{2pt}}c@{\hspace{2pt}} | @{\hspace{3pt}}c@{\hspace{3pt}} | }
    \cline{1-3} \cline{5-7}
    \textbf{\makecell[c]{Metho-\\dology}} & \textbf{Set} & \textbf{Formula}
    & &
    \textbf{Pair} & \textbf{Set} & \textbf{Formula} \\
    \cline{1-3} \cline{5-7}
    & & & & & & \\[-4ex]
    \cline{1-3} \cline{5-7}

    \rule{0pt}{3ex}
    & \textbf{\ATrain}
    & $\bigcup_{\aproject \in \aprojects} (\aexamples_{train}^{\atimepp, \aproject} \sunion \aexamples_{train}^{\atimep\setminus\atimepp, \aproject} \sunion \aexamples_{train}^{\atime\setminus\atimep, \aproject})$
    &
    &
    &
    & \\
    \cline{2-3}

    \rule{0pt}{3ex}
    & \textbf{\AVal}
    & $\bigcup_{\aproject \in \aprojects} (\aexamples_{val}^{\atimepp, \aproject} \sunion \aexamples_{val}^{\atimep\setminus\atimepp, \aproject} \sunion \aexamples_{val}^{\atime\setminus\atimep, \aproject})$
    & \tikz[remember picture, baseline] \node[inner sep=2pt, outer sep=0] (mpxcp-base) {\phantom{XX}};
    &
    &
    & \\
    \cline{2-3}

    \rule{0pt}{3ex}
    \multirow{-3}{*}{\textbf{\Amp}}
    & \textbf{\ATestS}
    & $\bigcup_{\aproject \in \aprojects} (\aexamples_{test}^{\atimepp, \aproject} \sunion \aexamples_{test}^{\atimep\setminus\atimepp, \aproject} \sunion \aexamples_{test}^{\atime\setminus\atimep, \aproject})$
    & \tikz[remember picture, baseline] \node[inner sep=2pt, outer sep=0] (mp-base) {\phantom{XX}};
    & \multirow{-3}{*}{\textbf{\mpxcp}}
    & \multirow{-3}{*}{\textbf{\ATestC}}
    & \multirow{-3}{*}{$\bigcup_{\aproject \in \aprojects_{test}} (\aexamples_{test}^{\atimepp, \aproject} \sunion \aexamples_{test}^{\atimep\setminus\atimepp, \aproject} \sunion \aexamples_{test}^{\atime\setminus\atimep, \aproject})$} \\
    \cline{1-3} \cline{5-7}
    & & & & & & \\[-4ex]
    \cline{1-3} \cline{5-7}

    \rule{0pt}{3ex}
    & \textbf{\ATrain}
    & $\bigcup_{\aproject \in \aprojects_{train}} (\aexamples^{\atimepp, \aproject} \sunion \aexamples^{\atimep\setminus\atimepp, \aproject} \sunion \aexamples^{\atime\setminus\atimep, \aproject})$
    &
    &
    &
    & \\
    \cline{2-3}

    \rule{0pt}{3ex}
    & \textbf{\AVal}
    & $\bigcup_{\aproject \in \aprojects_{val}} (\aexamples^{\atimepp, \aproject} \sunion \aexamples^{\atimep\setminus\atimepp, \aproject} \sunion \aexamples^{\atime\setminus\atimep, \aproject})$
    & \tikz[remember picture, baseline] \node[inner sep=2pt, outer sep=0] (mpxt-base) {\phantom{XX}};
    &
    &
    & \\
    \cline{2-3}

    \rule{0pt}{3ex}
    \multirow{-3}{*}{\textbf{\Acp}}
    & \textbf{\ATestS}
    & $\bigcup_{\aproject \in \aprojects_{test}} (\aexamples^{\atimepp, \aproject} \sunion \aexamples^{\atimep\setminus\atimepp, \aproject} \sunion \aexamples^{\atime\setminus\atimep, \aproject})$
    & \tikz[remember picture, baseline] \node[inner sep=2pt, outer sep=0] (cp-base) {\phantom{XX}};
    & \multirow{-3}{*}{\textbf{\mpxt}}
    & \multirow{-3}{*}{\textbf{\ATestC}}
    & \multirow{-3}{*}{$\bigcup_{\aproject \in \aprojects} \aexamples_{test}^{\atime\setminus\atimep, \aproject}$} \\
    \cline{1-3} \cline{5-7}
    & & & & & & \\[-4ex]
    \cline{1-3} \cline{5-7}

    \rule{0pt}{3ex}
    & \textbf{\ATrain}
    & $\bigcup_{\aproject \in \aprojects} \aexamples^{\atimepp, \aproject}$
    &
    &
    &
    & \\
    \cline{2-3}

    \rule{0pt}{3ex}
    & \textbf{\AVal}
    & $\bigcup_{\aproject \in \aprojects} \aexamples^{\atimep\setminus\atimepp, \aproject}$
    & \tikz[remember picture, baseline] \node[inner sep=2pt, outer sep=0] (cpxt-base) {\phantom{XX}};
    &
    &
    & \\
    \cline{2-3}

    \rule{0pt}{3ex}
    \multirow{-3}{*}{\textbf{\At}}
    & \textbf{\ATestS}
    & $\bigcup_{\aproject \in \aprojects} \aexamples^{\atime\setminus\atimep, \aproject}$
    & \tikz[remember picture, baseline] \node[inner sep=2pt, outer sep=0] (t-base) {\phantom{XX}};
    & \multirow{-3}{*}{\textbf{\cpxt}}
    & \multirow{-3}{*}{\textbf{\ATestC}}
    & \multirow{-3}{*}{$\bigcup_{\aproject \in \aprojects_{test}} \aexamples^{\atime\setminus\atimep, \aproject}$} \\
    \cline{1-3} \cline{5-7}

  \end{tabular}

  \begin{tikzpicture}[remember picture, overlay, thick]
    \draw[->] (mp-base.west) .. controls ($(mpxcp-base.east) - (1em,0)$) .. (mpxcp-base.east);
    \draw[] (cp-base.west) .. controls ($(mpxcp-base.east) - (1em,0)$) .. (mpxcp-base.east);
    \draw[->] (mp-base.west) .. controls ($(mpxt-base.east) - (1em,0)$) .. (mpxt-base.east);
    \draw[] (t-base.west) .. controls ($(mpxt-base.east) - (1em,0)$) .. (mpxt-base.east);
    \draw[->] (cp-base.west) .. controls ($(cpxt-base.east) - (1em,0)$) .. (cpxt-base.east);
    \draw[] (t-base.west) .. controls ($(cpxt-base.east) - (1em,0)$) .. (cpxt-base.east);
  \end{tikzpicture}

  \end{center}
  \end{scriptsize}
  \caption{\TCEvalSettings}
  \vspace{\TVEvalSettings}
\end{table*}
}

\MyPara{Step~6: postprocessing} The formulas for downsampling the
\ATrain sets are:
\begin{flalign*}
&size = \min_{m\in\{\text{\Amp}, \text{\Acp}, \text{\At}\}} \left| \aexamples_{\text{\ATrain}(m)} \right|\\
&\text{for}\ m \in \{\text{\Amp}, \text{\Acp}, \text{\At}\}\text{:}\\
&\quad
\aexamples_{\text{\ATrain}(m)} \leftarrow \ashuffle(\aexamples_{\text{\ATrain}(m)})[0:size]
\end{flalign*}

To formalize the filtering of exact-duplicates and near-duplicates
(which we will further discuss in
Appendix~\ref{sec:appendix:near-duplicates}), we define
$\aclean(\aexamples_{eval}, \aexamples_{train})$ which is
task-specific.  It takes two inputs: the \examples in the evaluation
set that needs to be cleaned, and the \examples used for training; and
returns the cleaned evaluation set.  Note that when the evaluation set
is the \ATestS or \ATestC set, we also consider \examples in the \AVal
set as used for training (because they are used for hyper-parameter
tuning or early-stopping).  The formulas for this step are:
\begin{flalign*}
&\text{for}\ m \in \{\text{\Amp}, \text{\Acp}, \text{\At}\}\text{:}\\
&\quad\aexamples_{\text{\AVal}(m)} \leftarrow \aclean(\aexamples_{\text{\AVal}(m)}, \aexamples_{\text{\ATrain}(m)})\\
&\quad\aexamples_{\text{\ATestS}(m)} \leftarrow \aclean(\aexamples_{\text{\ATestS}(m)}, \aexamples_{\text{\ATrain}(m)} \sunion \aexamples_{\text{\AVal}(m)})\\
&\text{for}\ m, m' \in \{(\text{\Amp}, \text{\Acp}), (\text{\Amp}, \text{\At}), (\text{\Acp}, \text{\At})\}\text{:}\\
&\quad\aexamples_{\text{\ATestC}(m, m')} \leftarrow \aclean(\aexamples_{\text{\ATestC}(m, m')}, \\
&\qquad\aexamples_{\text{\ATrain}(m)} \sunion \aexamples_{\text{\AVal}(m)} \sunion \aexamples_{\text{\ATrain}(m')} \sunion \aexamples_{\text{\AVal}(m')})
\end{flalign*}

\section{Filtering Near-Duplicates}
\label{sec:appendix:near-duplicates}

We experimented if filtering near-duplicates can lead to any change to
our findings.  We used the following
three configurations to define near-duplicates
(there are many other ways to define
near-duplicates, which we leave for future work).  The numbers in
parentheses are the percentages of \examples considered as
near-duplicates in \ATestC sets of all pairs of \methodologies for
\comgen{} / \metnam.
\begin{itemize}[topsep=2pt,itemsep=2pt,partopsep=0ex,parsep=0ex,leftmargin=*]
\item \UseMacro{nd_same_code}: a \example is near-duplicate if any
  \example in the \train set has identical code with
  it. (\UseMacro{ds-CG-test_common-filter_pct_same_code} /
  \UseMacro{ds-MN-test_common-filter_pct_same_code})
\item \UseMacro{nd_same_nl}: a \example is near-duplicate if any
  \example in the \train set has identical summary (comment for
  \comgen or name for \metnam) with
  it. (\UseMacro{ds-CG-test_common-filter_pct_same_nl} /
  \UseMacro{ds-MN-test_common-filter_pct_same_nl})
\item \UseMacro{nd_sim_90}: a \example is near-duplicate if any
  \example in the \train set has more than 90\% similarity with it in
  terms of both code and summary; the similarity is measured by
  \subtok-level accuracy which is fast to
  compute. (\UseMacro{ds-CG-test_common-filter_pct_sim_90} /
  \UseMacro{ds-MN-test_common-filter_pct_sim_90})
\end{itemize}

The experiment results are presented in the following tables and
plots:
\begin{itemize}[topsep=3pt,itemsep=3pt,partopsep=0ex,parsep=0ex,leftmargin=*]
\foreach \ndconfig in {same_code, same_nl, sim_90}{
\item Using \UseMacro{nd_\ndconfig} configuration:
  \begin{itemize}
    \item \comgen: Table~\ref{table:nd-results:CG-\ndconfig}, Figure~\ref{fig:nd-results:CG-\ndconfig}.
    \item \metnam: Table~\ref{table:nd-results:MN-\ndconfig}, Figure~\ref{fig:nd-results:MN-\ndconfig}.
  \end{itemize}
}
\end{itemize}

We can draw several conclusions.  First of all, our findings in
Section~\ref{sec:eval:findings} still hold.  The metrics of
\UseMacro{nd_same_code} are closest to the metrics of not filtering
near-duplicates, which indicates that this filtering configuration has
little impact on evaluation results.
On the contrary, the metrics of \UseMacro{nd_same_nl} and
\UseMacro{nd_sim_90} are lower than the metrics of not filtering
near-duplicates, which means the models become less effective.  This
indicates that current ML models for code summarization are better
at following the \examples in the \train set
than generating novel summaries.

\section{Data Collection Details}
\label{sec:appendix:data}

\begin{table*}[t]
\begin{footnotesize}
\begin{center}
\begin{tabular}{ l@{\hspace{2pt}}c@{\hspace{2pt}} | r@{\hspace{4pt}}r@{\hspace{4pt}}r @{\hspace{3pt}}c@{\hspace{3pt}} r@{\hspace{4pt}}r@{\hspace{4pt}}r @{\hspace{3pt}}c@{\hspace{3pt}} r@{\hspace{4pt}}r@{\hspace{4pt}}r @{\hspace{3pt}}c@{\hspace{3pt}} r@{\hspace{4pt}}r@{\hspace{4pt}}r }
\toprule
\multicolumn{2}{c|}{} & \multicolumn{3}{c}{\UseMacro{TH-ds-MP}} & & \multicolumn{3}{c}{\UseMacro{TH-ds-CP}} & & \multicolumn{3}{c}{\UseMacro{TH-ds-T}} & & \UseMacro{TH-ds-MP-CP} & \UseMacro{TH-ds-MP-T} & \UseMacro{TH-ds-CP-T} \\\cline{3-5}\cline{7-9}\cline{11-13}\cline{15-17}
\multicolumn{2}{c|}{\multirow{-2}{*}{\THDSStat}} & \UseMacro{TH-ds-train} & \UseMacro{TH-ds-val} & \UseMacro{TH-ds-test_standard} & & \UseMacro{TH-ds-train} & \UseMacro{TH-ds-val} & \UseMacro{TH-ds-test_standard} & & \UseMacro{TH-ds-train} & \UseMacro{TH-ds-val} & \UseMacro{TH-ds-test_standard} & & \multicolumn{3}{c}{\UseMacro{TH-ds-test_common}} \\
\midrule
\multicolumn{2}{c|}{\UseMacro{TH-ds-num-data}}
 & \UseMacro{ds-CG-train-MP_num-data}
 & \UseMacro{ds-CG-val-MP_num-data}
 & \UseMacro{ds-CG-test_standard-MP_num-data}
 & 
 & \UseMacro{ds-CG-train-CP_num-data}
 & \UseMacro{ds-CG-val-CP_num-data}
 & \UseMacro{ds-CG-test_standard-CP_num-data}
 & 
 & \UseMacro{ds-CG-train-T_num-data}
 & \UseMacro{ds-CG-val-T_num-data}
 & \UseMacro{ds-CG-test_standard-T_num-data}
 & 
 & \UseMacro{ds-CG-test_common-MP-CP_num-data}
 & \UseMacro{ds-CG-test_common-MP-T_num-data}
 & \UseMacro{ds-CG-test_common-CP-T_num-data}
\\
\midrule
& \UseMacro{TH-ds-len-code-avg}
 & \UseMacro{ds-CG-train-MP_len-code-AVG}
 & \UseMacro{ds-CG-val-MP_len-code-AVG}
 & \UseMacro{ds-CG-test_standard-MP_len-code-AVG}
 & 
 & \UseMacro{ds-CG-train-CP_len-code-AVG}
 & \UseMacro{ds-CG-val-CP_len-code-AVG}
 & \UseMacro{ds-CG-test_standard-CP_len-code-AVG}
 & 
 & \UseMacro{ds-CG-train-T_len-code-AVG}
 & \UseMacro{ds-CG-val-T_len-code-AVG}
 & \UseMacro{ds-CG-test_standard-T_len-code-AVG}
 & 
 & \UseMacro{ds-CG-test_common-MP-CP_len-code-AVG}
 & \UseMacro{ds-CG-test_common-MP-T_len-code-AVG}
 & \UseMacro{ds-CG-test_common-CP-T_len-code-AVG}
\\
& \UseMacro{TH-ds-len-code-le100}
 & \UseMacro{ds-CG-train-MP_len-code-le-100}
 & \UseMacro{ds-CG-val-MP_len-code-le-100}
 & \UseMacro{ds-CG-test_standard-MP_len-code-le-100}
 & 
 & \UseMacro{ds-CG-train-CP_len-code-le-100}
 & \UseMacro{ds-CG-val-CP_len-code-le-100}
 & \UseMacro{ds-CG-test_standard-CP_len-code-le-100}
 & 
 & \UseMacro{ds-CG-train-T_len-code-le-100}
 & \UseMacro{ds-CG-val-T_len-code-le-100}
 & \UseMacro{ds-CG-test_standard-T_len-code-le-100}
 & 
 & \UseMacro{ds-CG-test_common-MP-CP_len-code-le-100}
 & \UseMacro{ds-CG-test_common-MP-T_len-code-le-100}
 & \UseMacro{ds-CG-test_common-CP-T_len-code-le-100}
\\
& \UseMacro{TH-ds-len-code-le150}
 & \UseMacro{ds-CG-train-MP_len-code-le-150}
 & \UseMacro{ds-CG-val-MP_len-code-le-150}
 & \UseMacro{ds-CG-test_standard-MP_len-code-le-150}
 & 
 & \UseMacro{ds-CG-train-CP_len-code-le-150}
 & \UseMacro{ds-CG-val-CP_len-code-le-150}
 & \UseMacro{ds-CG-test_standard-CP_len-code-le-150}
 & 
 & \UseMacro{ds-CG-train-T_len-code-le-150}
 & \UseMacro{ds-CG-val-T_len-code-le-150}
 & \UseMacro{ds-CG-test_standard-T_len-code-le-150}
 & 
 & \UseMacro{ds-CG-test_common-MP-CP_len-code-le-150}
 & \UseMacro{ds-CG-test_common-MP-T_len-code-le-150}
 & \UseMacro{ds-CG-test_common-CP-T_len-code-le-150}
\\
\multirow{-4}{*}{\UseMacro{TH-ds-len-code}} & \UseMacro{TH-ds-len-code-le200}
 & \UseMacro{ds-CG-train-MP_len-code-le-200}
 & \UseMacro{ds-CG-val-MP_len-code-le-200}
 & \UseMacro{ds-CG-test_standard-MP_len-code-le-200}
 & 
 & \UseMacro{ds-CG-train-CP_len-code-le-200}
 & \UseMacro{ds-CG-val-CP_len-code-le-200}
 & \UseMacro{ds-CG-test_standard-CP_len-code-le-200}
 & 
 & \UseMacro{ds-CG-train-T_len-code-le-200}
 & \UseMacro{ds-CG-val-T_len-code-le-200}
 & \UseMacro{ds-CG-test_standard-T_len-code-le-200}
 & 
 & \UseMacro{ds-CG-test_common-MP-CP_len-code-le-200}
 & \UseMacro{ds-CG-test_common-MP-T_len-code-le-200}
 & \UseMacro{ds-CG-test_common-CP-T_len-code-le-200}
\\
\midrule
& \UseMacro{TH-ds-len-comment-avg}
 & \UseMacro{ds-CG-train-MP_len-comment-AVG}
 & \UseMacro{ds-CG-val-MP_len-comment-AVG}
 & \UseMacro{ds-CG-test_standard-MP_len-comment-AVG}
 & 
 & \UseMacro{ds-CG-train-CP_len-comment-AVG}
 & \UseMacro{ds-CG-val-CP_len-comment-AVG}
 & \UseMacro{ds-CG-test_standard-CP_len-comment-AVG}
 & 
 & \UseMacro{ds-CG-train-T_len-comment-AVG}
 & \UseMacro{ds-CG-val-T_len-comment-AVG}
 & \UseMacro{ds-CG-test_standard-T_len-comment-AVG}
 & 
 & \UseMacro{ds-CG-test_common-MP-CP_len-comment-AVG}
 & \UseMacro{ds-CG-test_common-MP-T_len-comment-AVG}
 & \UseMacro{ds-CG-test_common-CP-T_len-comment-AVG}
\\
& \UseMacro{TH-ds-len-comment-le20}
 & \UseMacro{ds-CG-train-MP_len-comment-le-20}
 & \UseMacro{ds-CG-val-MP_len-comment-le-20}
 & \UseMacro{ds-CG-test_standard-MP_len-comment-le-20}
 & 
 & \UseMacro{ds-CG-train-CP_len-comment-le-20}
 & \UseMacro{ds-CG-val-CP_len-comment-le-20}
 & \UseMacro{ds-CG-test_standard-CP_len-comment-le-20}
 & 
 & \UseMacro{ds-CG-train-T_len-comment-le-20}
 & \UseMacro{ds-CG-val-T_len-comment-le-20}
 & \UseMacro{ds-CG-test_standard-T_len-comment-le-20}
 & 
 & \UseMacro{ds-CG-test_common-MP-CP_len-comment-le-20}
 & \UseMacro{ds-CG-test_common-MP-T_len-comment-le-20}
 & \UseMacro{ds-CG-test_common-CP-T_len-comment-le-20}
\\
& \UseMacro{TH-ds-len-comment-le30}
 & \UseMacro{ds-CG-train-MP_len-comment-le-30}
 & \UseMacro{ds-CG-val-MP_len-comment-le-30}
 & \UseMacro{ds-CG-test_standard-MP_len-comment-le-30}
 & 
 & \UseMacro{ds-CG-train-CP_len-comment-le-30}
 & \UseMacro{ds-CG-val-CP_len-comment-le-30}
 & \UseMacro{ds-CG-test_standard-CP_len-comment-le-30}
 & 
 & \UseMacro{ds-CG-train-T_len-comment-le-30}
 & \UseMacro{ds-CG-val-T_len-comment-le-30}
 & \UseMacro{ds-CG-test_standard-T_len-comment-le-30}
 & 
 & \UseMacro{ds-CG-test_common-MP-CP_len-comment-le-30}
 & \UseMacro{ds-CG-test_common-MP-T_len-comment-le-30}
 & \UseMacro{ds-CG-test_common-CP-T_len-comment-le-30}
\\
\multirow{-4}{*}{\UseMacro{TH-ds-len-comment}} & \UseMacro{TH-ds-len-comment-le50}
 & \UseMacro{ds-CG-train-MP_len-comment-le-50}
 & \UseMacro{ds-CG-val-MP_len-comment-le-50}
 & \UseMacro{ds-CG-test_standard-MP_len-comment-le-50}
 & 
 & \UseMacro{ds-CG-train-CP_len-comment-le-50}
 & \UseMacro{ds-CG-val-CP_len-comment-le-50}
 & \UseMacro{ds-CG-test_standard-CP_len-comment-le-50}
 & 
 & \UseMacro{ds-CG-train-T_len-comment-le-50}
 & \UseMacro{ds-CG-val-T_len-comment-le-50}
 & \UseMacro{ds-CG-test_standard-T_len-comment-le-50}
 & 
 & \UseMacro{ds-CG-test_common-MP-CP_len-comment-le-50}
 & \UseMacro{ds-CG-test_common-MP-T_len-comment-le-50}
 & \UseMacro{ds-CG-test_common-CP-T_len-comment-le-50}
\\
\bottomrule
\end{tabular}
\end{center}
\end{footnotesize}
\vspace{\UseMacro{TV-dataset-metrics-CG}}
\caption{\UseMacro{TC-dataset-metrics-CG}}
\end{table*}

\begin{table*}[t]
\begin{footnotesize}
\begin{center}
\begin{tabular}{ l@{\hspace{2pt}}c@{\hspace{2pt}} | r@{\hspace{4pt}}r@{\hspace{4pt}}r @{\hspace{3pt}}c@{\hspace{3pt}} r@{\hspace{4pt}}r@{\hspace{4pt}}r @{\hspace{3pt}}c@{\hspace{3pt}} r@{\hspace{4pt}}r@{\hspace{4pt}}r @{\hspace{3pt}}c@{\hspace{3pt}} r@{\hspace{4pt}}r@{\hspace{4pt}}r }
\toprule
\multicolumn{2}{c|}{} & \multicolumn{3}{c}{\UseMacro{TH-ds-MP}} & & \multicolumn{3}{c}{\UseMacro{TH-ds-CP}} & & \multicolumn{3}{c}{\UseMacro{TH-ds-T}} & & \UseMacro{TH-ds-MP-CP} & \UseMacro{TH-ds-MP-T} & \UseMacro{TH-ds-CP-T} \\\cline{3-5}\cline{7-9}\cline{11-13}\cline{15-17}
\multicolumn{2}{c|}{\multirow{-2}{*}{\THDSStat}} & \UseMacro{TH-ds-train} & \UseMacro{TH-ds-val} & \UseMacro{TH-ds-test_standard} & & \UseMacro{TH-ds-train} & \UseMacro{TH-ds-val} & \UseMacro{TH-ds-test_standard} & & \UseMacro{TH-ds-train} & \UseMacro{TH-ds-val} & \UseMacro{TH-ds-test_standard} & & \multicolumn{3}{c}{\UseMacro{TH-ds-test_common}} \\
\midrule
\multicolumn{2}{c|}{\UseMacro{TH-ds-num-data}}
 & \UseMacro{ds-MN-train-MP_num-data}
 & \UseMacro{ds-MN-val-MP_num-data}
 & \UseMacro{ds-MN-test_standard-MP_num-data}
 & 
 & \UseMacro{ds-MN-train-CP_num-data}
 & \UseMacro{ds-MN-val-CP_num-data}
 & \UseMacro{ds-MN-test_standard-CP_num-data}
 & 
 & \UseMacro{ds-MN-train-T_num-data}
 & \UseMacro{ds-MN-val-T_num-data}
 & \UseMacro{ds-MN-test_standard-T_num-data}
 & 
 & \UseMacro{ds-MN-test_common-MP-CP_num-data}
 & \UseMacro{ds-MN-test_common-MP-T_num-data}
 & \UseMacro{ds-MN-test_common-CP-T_num-data}
\\
\midrule
& \UseMacro{TH-ds-len-code-avg}
 & \UseMacro{ds-MN-train-MP_len-code-AVG}
 & \UseMacro{ds-MN-val-MP_len-code-AVG}
 & \UseMacro{ds-MN-test_standard-MP_len-code-AVG}
 & 
 & \UseMacro{ds-MN-train-CP_len-code-AVG}
 & \UseMacro{ds-MN-val-CP_len-code-AVG}
 & \UseMacro{ds-MN-test_standard-CP_len-code-AVG}
 & 
 & \UseMacro{ds-MN-train-T_len-code-AVG}
 & \UseMacro{ds-MN-val-T_len-code-AVG}
 & \UseMacro{ds-MN-test_standard-T_len-code-AVG}
 & 
 & \UseMacro{ds-MN-test_common-MP-CP_len-code-AVG}
 & \UseMacro{ds-MN-test_common-MP-T_len-code-AVG}
 & \UseMacro{ds-MN-test_common-CP-T_len-code-AVG}
\\
& \UseMacro{TH-ds-len-code-le100}
 & \UseMacro{ds-MN-train-MP_len-code-le-100}
 & \UseMacro{ds-MN-val-MP_len-code-le-100}
 & \UseMacro{ds-MN-test_standard-MP_len-code-le-100}
 & 
 & \UseMacro{ds-MN-train-CP_len-code-le-100}
 & \UseMacro{ds-MN-val-CP_len-code-le-100}
 & \UseMacro{ds-MN-test_standard-CP_len-code-le-100}
 & 
 & \UseMacro{ds-MN-train-T_len-code-le-100}
 & \UseMacro{ds-MN-val-T_len-code-le-100}
 & \UseMacro{ds-MN-test_standard-T_len-code-le-100}
 & 
 & \UseMacro{ds-MN-test_common-MP-CP_len-code-le-100}
 & \UseMacro{ds-MN-test_common-MP-T_len-code-le-100}
 & \UseMacro{ds-MN-test_common-CP-T_len-code-le-100}
\\
& \UseMacro{TH-ds-len-code-le150}
 & \UseMacro{ds-MN-train-MP_len-code-le-150}
 & \UseMacro{ds-MN-val-MP_len-code-le-150}
 & \UseMacro{ds-MN-test_standard-MP_len-code-le-150}
 & 
 & \UseMacro{ds-MN-train-CP_len-code-le-150}
 & \UseMacro{ds-MN-val-CP_len-code-le-150}
 & \UseMacro{ds-MN-test_standard-CP_len-code-le-150}
 & 
 & \UseMacro{ds-MN-train-T_len-code-le-150}
 & \UseMacro{ds-MN-val-T_len-code-le-150}
 & \UseMacro{ds-MN-test_standard-T_len-code-le-150}
 & 
 & \UseMacro{ds-MN-test_common-MP-CP_len-code-le-150}
 & \UseMacro{ds-MN-test_common-MP-T_len-code-le-150}
 & \UseMacro{ds-MN-test_common-CP-T_len-code-le-150}
\\
\multirow{-4}{*}{\UseMacro{TH-ds-len-code}} & \UseMacro{TH-ds-len-code-le200}
 & \UseMacro{ds-MN-train-MP_len-code-le-200}
 & \UseMacro{ds-MN-val-MP_len-code-le-200}
 & \UseMacro{ds-MN-test_standard-MP_len-code-le-200}
 & 
 & \UseMacro{ds-MN-train-CP_len-code-le-200}
 & \UseMacro{ds-MN-val-CP_len-code-le-200}
 & \UseMacro{ds-MN-test_standard-CP_len-code-le-200}
 & 
 & \UseMacro{ds-MN-train-T_len-code-le-200}
 & \UseMacro{ds-MN-val-T_len-code-le-200}
 & \UseMacro{ds-MN-test_standard-T_len-code-le-200}
 & 
 & \UseMacro{ds-MN-test_common-MP-CP_len-code-le-200}
 & \UseMacro{ds-MN-test_common-MP-T_len-code-le-200}
 & \UseMacro{ds-MN-test_common-CP-T_len-code-le-200}
\\
\midrule
& \UseMacro{TH-ds-len-name-avg}
 & \UseMacro{ds-MN-train-MP_len-name-AVG}
 & \UseMacro{ds-MN-val-MP_len-name-AVG}
 & \UseMacro{ds-MN-test_standard-MP_len-name-AVG}
 & 
 & \UseMacro{ds-MN-train-CP_len-name-AVG}
 & \UseMacro{ds-MN-val-CP_len-name-AVG}
 & \UseMacro{ds-MN-test_standard-CP_len-name-AVG}
 & 
 & \UseMacro{ds-MN-train-T_len-name-AVG}
 & \UseMacro{ds-MN-val-T_len-name-AVG}
 & \UseMacro{ds-MN-test_standard-T_len-name-AVG}
 & 
 & \UseMacro{ds-MN-test_common-MP-CP_len-name-AVG}
 & \UseMacro{ds-MN-test_common-MP-T_len-name-AVG}
 & \UseMacro{ds-MN-test_common-CP-T_len-name-AVG}
\\
& \UseMacro{TH-ds-len-name-le2}
 & \UseMacro{ds-MN-train-MP_len-name-le-2}
 & \UseMacro{ds-MN-val-MP_len-name-le-2}
 & \UseMacro{ds-MN-test_standard-MP_len-name-le-2}
 & 
 & \UseMacro{ds-MN-train-CP_len-name-le-2}
 & \UseMacro{ds-MN-val-CP_len-name-le-2}
 & \UseMacro{ds-MN-test_standard-CP_len-name-le-2}
 & 
 & \UseMacro{ds-MN-train-T_len-name-le-2}
 & \UseMacro{ds-MN-val-T_len-name-le-2}
 & \UseMacro{ds-MN-test_standard-T_len-name-le-2}
 & 
 & \UseMacro{ds-MN-test_common-MP-CP_len-name-le-2}
 & \UseMacro{ds-MN-test_common-MP-T_len-name-le-2}
 & \UseMacro{ds-MN-test_common-CP-T_len-name-le-2}
\\
& \UseMacro{TH-ds-len-name-le3}
 & \UseMacro{ds-MN-train-MP_len-name-le-3}
 & \UseMacro{ds-MN-val-MP_len-name-le-3}
 & \UseMacro{ds-MN-test_standard-MP_len-name-le-3}
 & 
 & \UseMacro{ds-MN-train-CP_len-name-le-3}
 & \UseMacro{ds-MN-val-CP_len-name-le-3}
 & \UseMacro{ds-MN-test_standard-CP_len-name-le-3}
 & 
 & \UseMacro{ds-MN-train-T_len-name-le-3}
 & \UseMacro{ds-MN-val-T_len-name-le-3}
 & \UseMacro{ds-MN-test_standard-T_len-name-le-3}
 & 
 & \UseMacro{ds-MN-test_common-MP-CP_len-name-le-3}
 & \UseMacro{ds-MN-test_common-MP-T_len-name-le-3}
 & \UseMacro{ds-MN-test_common-CP-T_len-name-le-3}
\\
\multirow{-4}{*}{\UseMacro{TH-ds-len-name}} & \UseMacro{TH-ds-len-name-le6}
 & \UseMacro{ds-MN-train-MP_len-name-le-6}
 & \UseMacro{ds-MN-val-MP_len-name-le-6}
 & \UseMacro{ds-MN-test_standard-MP_len-name-le-6}
 & 
 & \UseMacro{ds-MN-train-CP_len-name-le-6}
 & \UseMacro{ds-MN-val-CP_len-name-le-6}
 & \UseMacro{ds-MN-test_standard-CP_len-name-le-6}
 & 
 & \UseMacro{ds-MN-train-T_len-name-le-6}
 & \UseMacro{ds-MN-val-T_len-name-le-6}
 & \UseMacro{ds-MN-test_standard-T_len-name-le-6}
 & 
 & \UseMacro{ds-MN-test_common-MP-CP_len-name-le-6}
 & \UseMacro{ds-MN-test_common-MP-T_len-name-le-6}
 & \UseMacro{ds-MN-test_common-CP-T_len-name-le-6}
\\
\bottomrule
\end{tabular}
\end{center}
\end{footnotesize}
\vspace{\UseMacro{TV-dataset-metrics-MN}}
\caption{\UseMacro{TC-dataset-metrics-MN}}
\end{table*}

This section extends \S\ref{sec:eval:data} and describes our data
collection process in details.  Overall, our datasets are collected
and processed following the steps in \S\ref{sec:application} and
Appendix~\ref{sec:appendix:formulas}.  We started by collecting
\examples of \methods with comments from open-source GitHub projects,
and then performed task-specific processing to get the dataset for
each task.

\MyPara{Selecting projects} We initially chose \NumProjectPlanned
popular Java projects on GitHub: \NumProjectPlannedStars Java projects
with the highest number of stars (indicating how many GitHub users
bookmarked a project) and another \NumProjectPlannedTrending Java
projects whose owner is one of the famous open-source organizations on
GitHub\footnote{\url{https://github.com/collections/open-source-organizations}}.
We chose to use only Java projects
because most prior work focused on this language (see
Table~\ref{table:prior-work-code-sum}).
Then, we only kept the projects meeting the following criteria:
(1)~the number of stars should be larger than 20; (2)~the lines of
code of the project (as reported by GitHub
API\footnote{\url{https://docs.github.com/en/rest}}) should be in the
range of $[10^6, 2\times 10^6]$, to keep the number of \examples
balanced across projects; (3)~the project should have at least one
commit after \ajanone 2018.  \UseMacro{ds-num_prj_all} projects
satisfied all the criteria.

\MyPara{Collecting the raw dataset}
We set the timestamps \atimepp to 2019 \ajanone, \atimep to 2020
\ajanone, and \atime to 2021 \ajanone. For each project and for each
year in [2019, 2020, 2021], we identified the last commit in the
project before \ajanone of that year, checked-out to that commit, used
JavaParser\footnote{\url{https://javaparser.org/}} to parse all Java
files, and collected \examples of Java \methods in the form of (code,
comment, name, project, year) tuples, where the comment is the first
sentence in the \javadoc of the \method.  We discarded the \examples
where: (1)~the code or the comment contains non-English characters
(\UseMacro{ds-filter-code_non_english} and
\UseMacro{ds-filter-comment_non_english} cases respectively); (2)~the
code is longer than 10,000 characters
(\UseMacro{ds-filter-code_too_long} cases); (3)~the \method body is
empty, i.e., abstract \method (\UseMacro{ds-filter-empty_body} cases);
(4)~the comment is empty after removing tags such as
\CodeIn{@inheritDoc} (\UseMacro{ds-filter-empty_comment_summary}
cases).  If two \examples are identical except for the ``year'' label,
we would keep the one with the earliest year (e.g., two \examples from
2018 and 2019 years have identical code, comment, name, and project,
so we only keep the 2018 one).  We ended up with
\UseMacro{ds-all_num-data} \examples in the raw dataset.

Then, we follow the steps described in \S\ref{sec:application} to
split the raw dataset into \ATrain, \AVal, \ATestS sets for each
\methodology and \ATestC set for each pair of \methodologies.  The
splitting ratios (for in-project and cross-project splits) are:
$\apcttrain=70\%, \apctval=10\%, \apcttest=20\%$.

\begin{figure*}

\begin{center}
\begin{footnotesize}
\begin{tabular}{|l|c|c|c || l|c|}
\hline
& \textbf{ATrain} & \textbf{\AVal} & \textbf{\ATestS} & & \textbf{\ATestC} \\
\hline
\UseMacro{TH-ds-MP}
 & \begin{minipage}{.18\textwidth}\includegraphics[width=\textwidth]{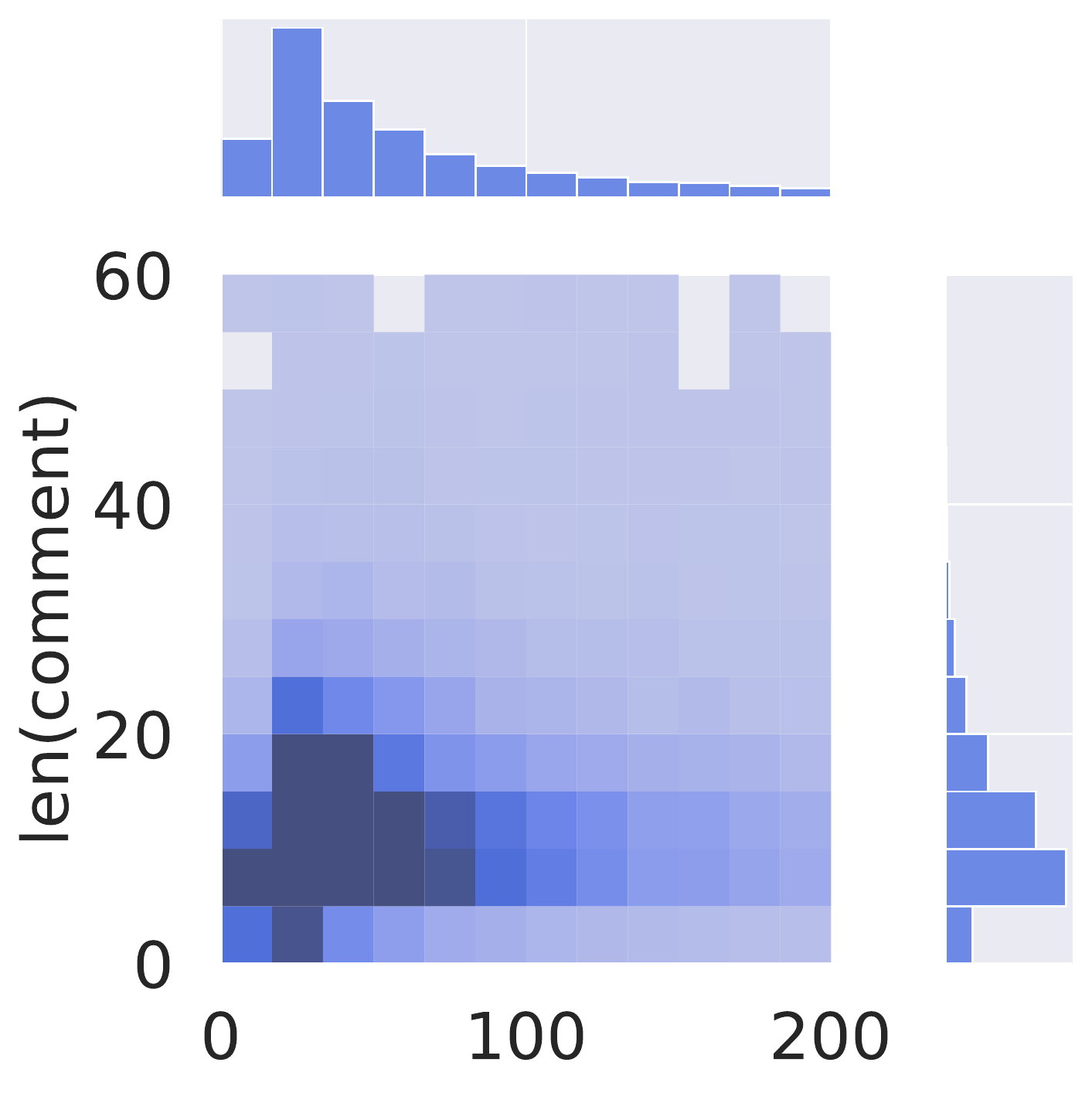}\end{minipage}
 & \begin{minipage}{.18\textwidth}\includegraphics[width=\textwidth]{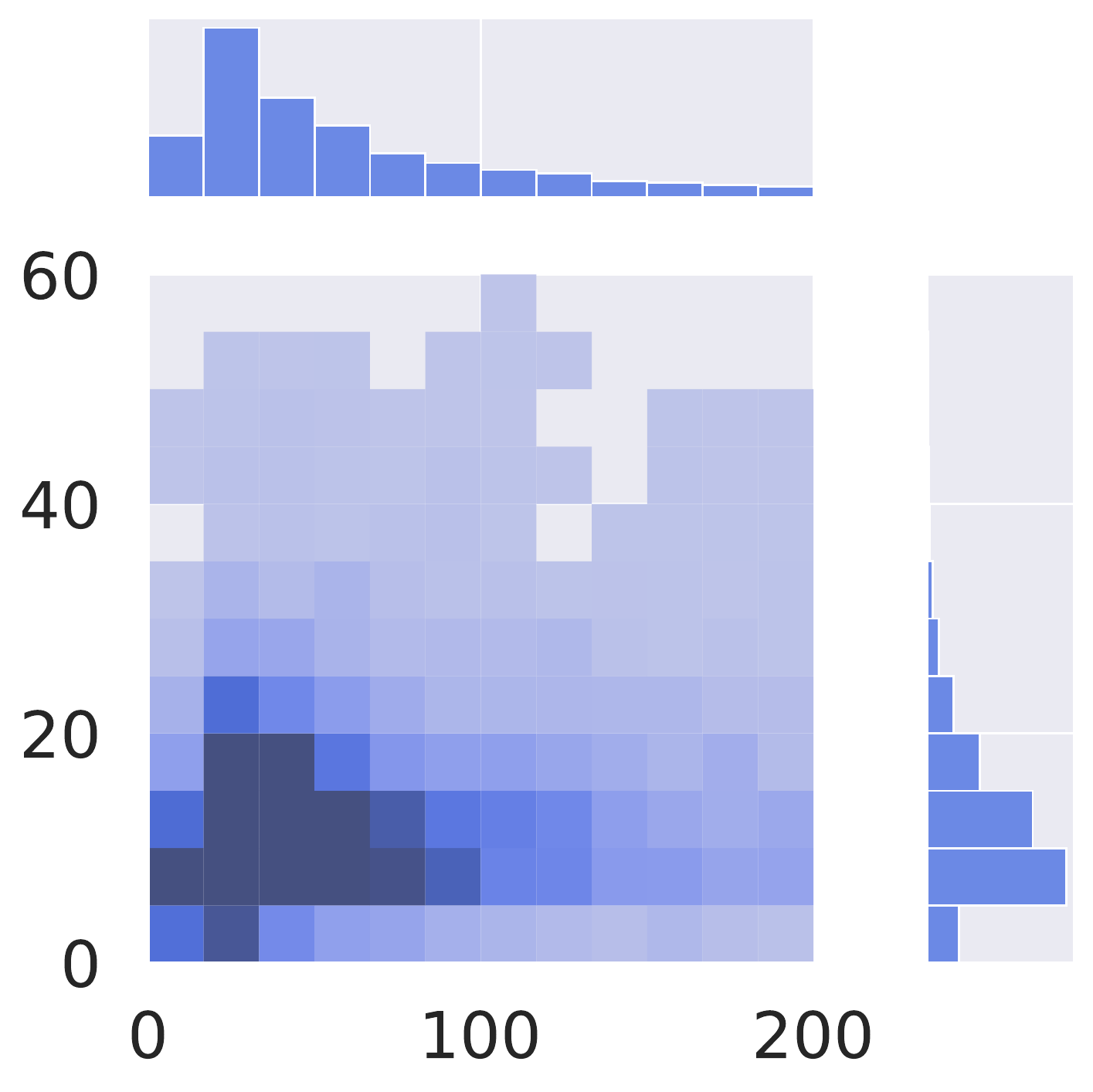}\end{minipage}
 & \begin{minipage}{.18\textwidth}\includegraphics[width=\textwidth]{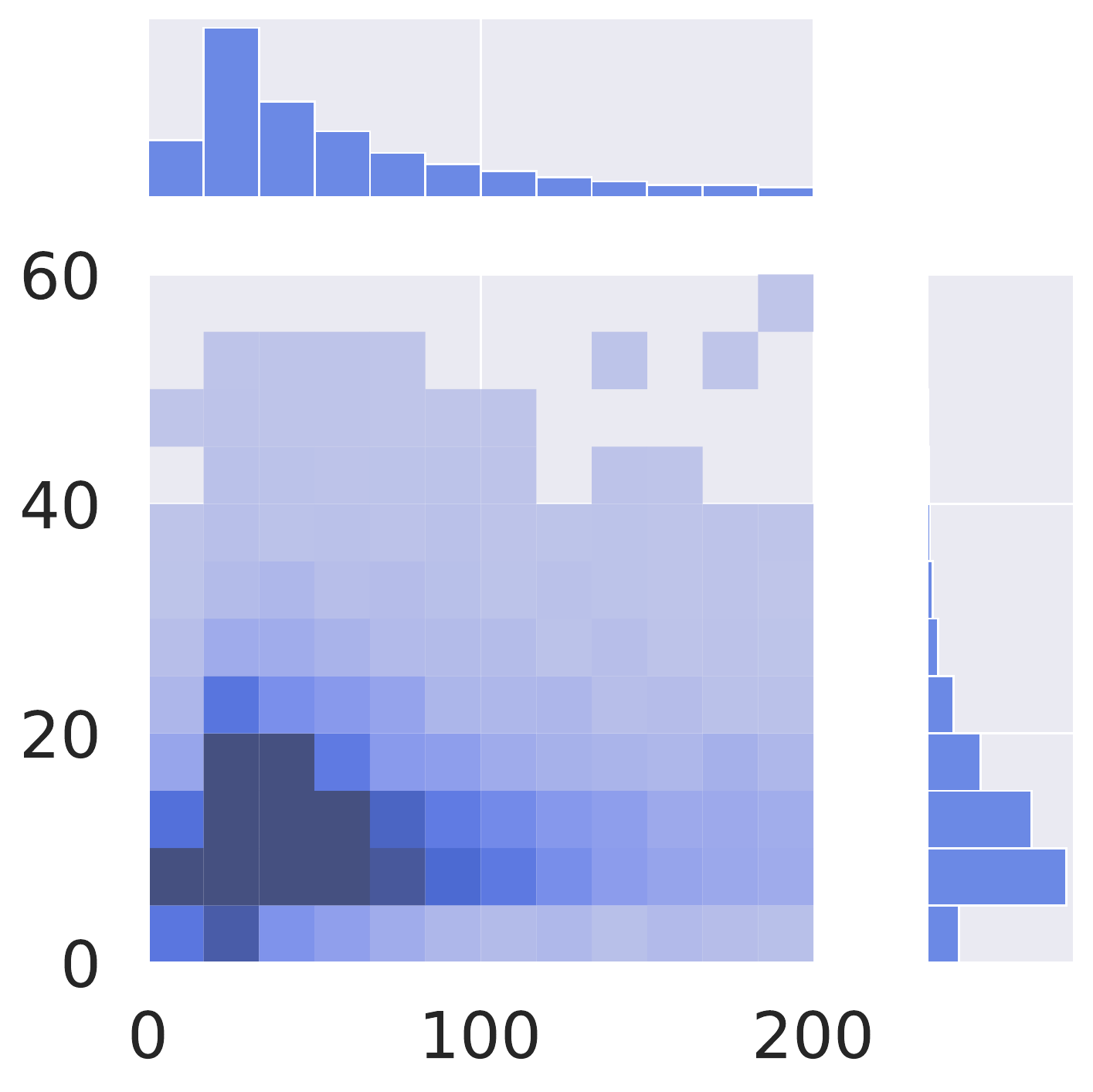}\end{minipage}
 & \UseMacro{TH-ds-MP-CP}
 & \begin{minipage}{.18\textwidth}\includegraphics[width=\textwidth]{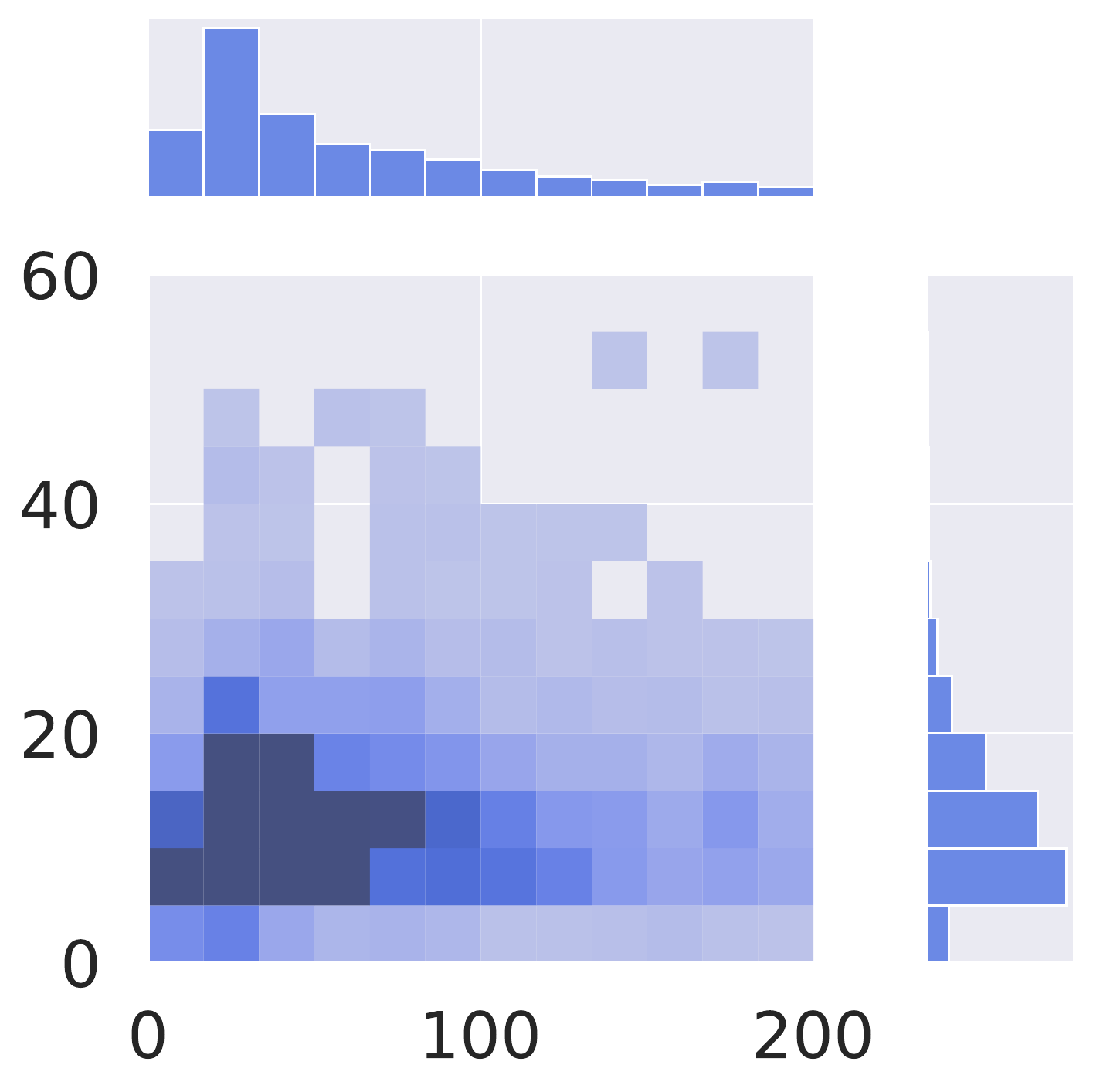}\end{minipage}
\\
\hline
\UseMacro{TH-ds-CP}
 & \begin{minipage}{.18\textwidth}\includegraphics[width=\textwidth]{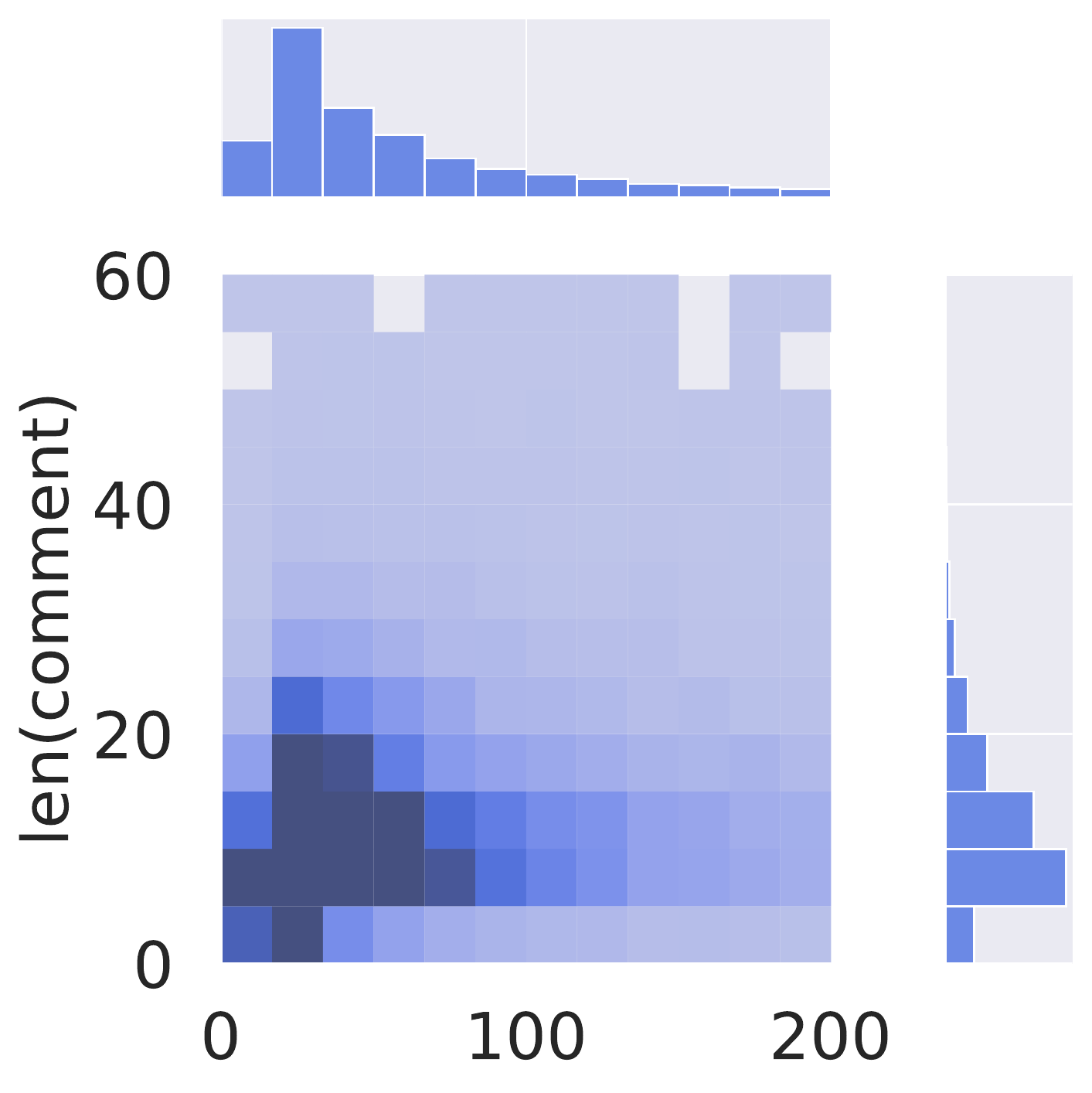}\end{minipage}
 & \begin{minipage}{.18\textwidth}\includegraphics[width=\textwidth]{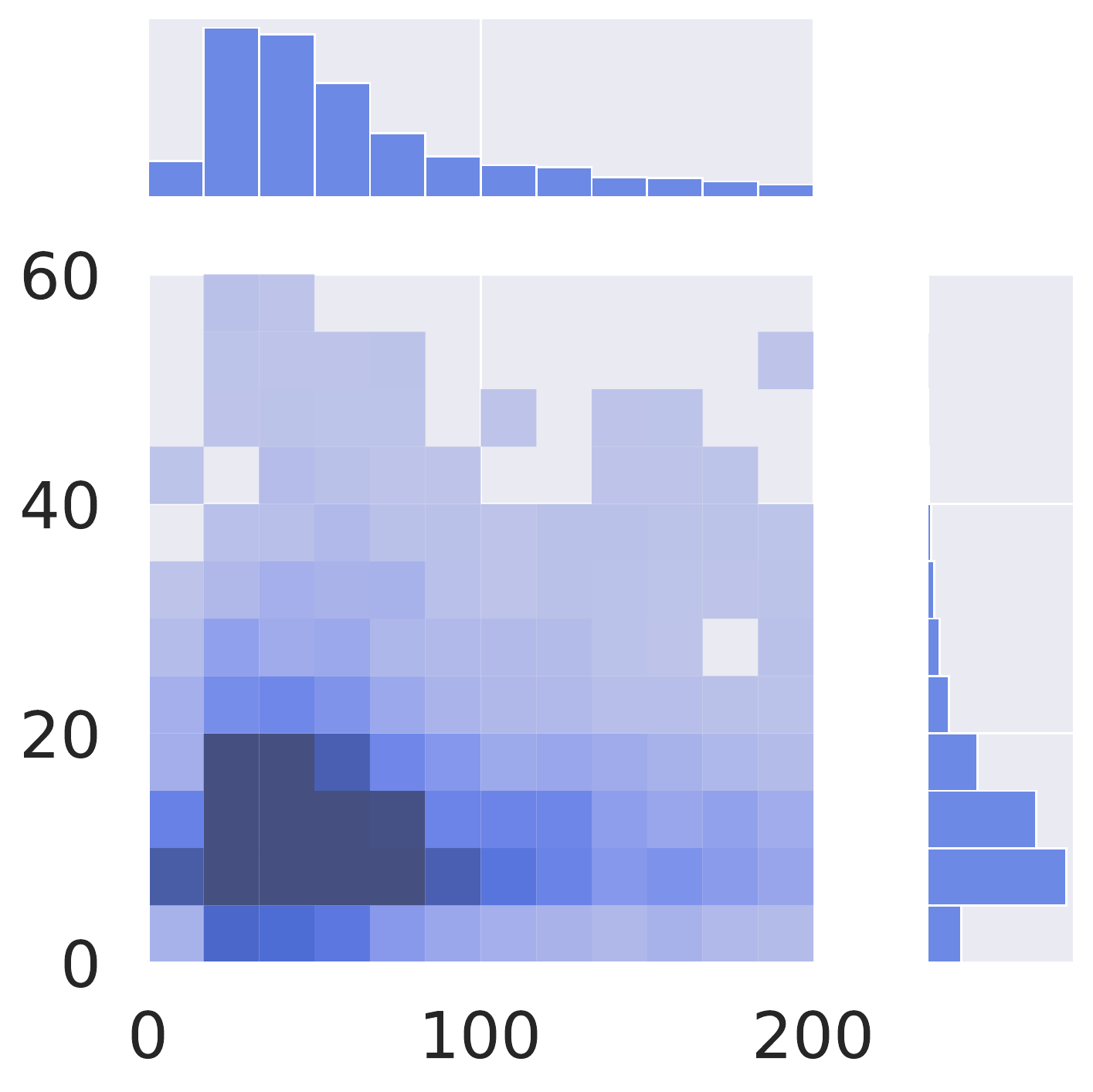}\end{minipage}
 & \begin{minipage}{.18\textwidth}\includegraphics[width=\textwidth]{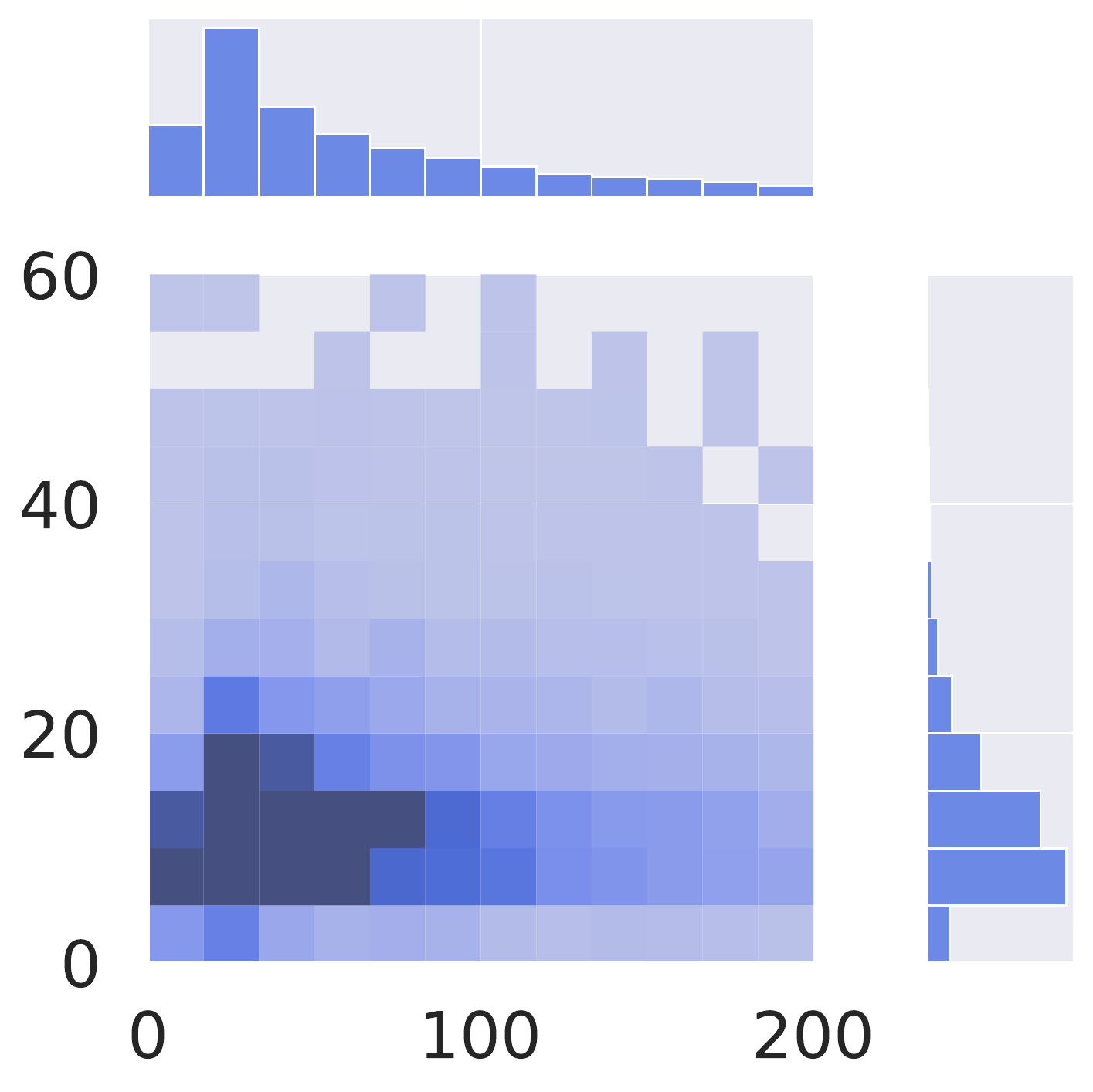}\end{minipage}
 & \UseMacro{TH-ds-MP-T}
 & \begin{minipage}{.18\textwidth}\includegraphics[width=\textwidth]{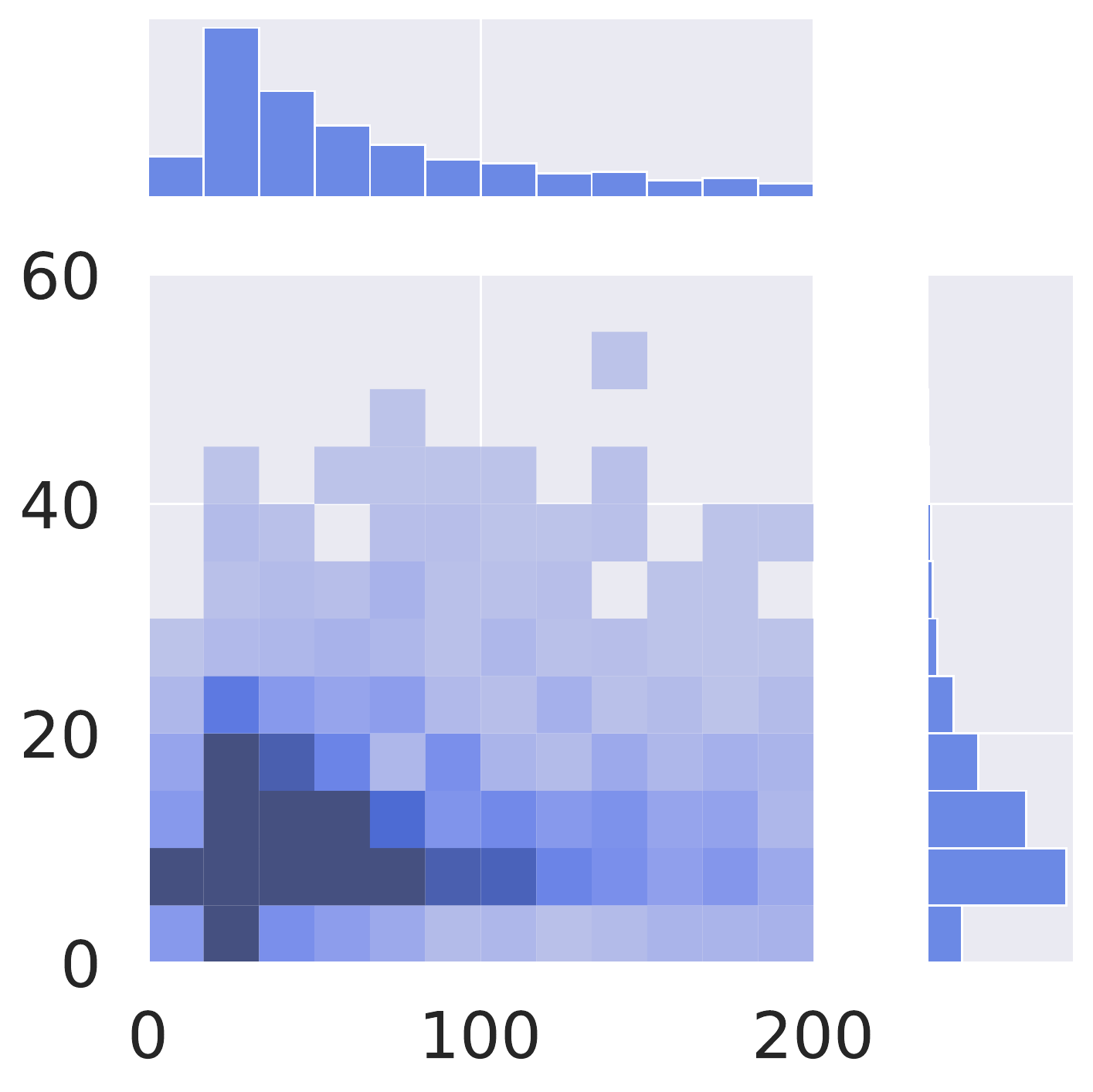}\end{minipage}
\\
\hline
\UseMacro{TH-ds-T}
 & \begin{minipage}{.18\textwidth}\includegraphics[width=\textwidth]{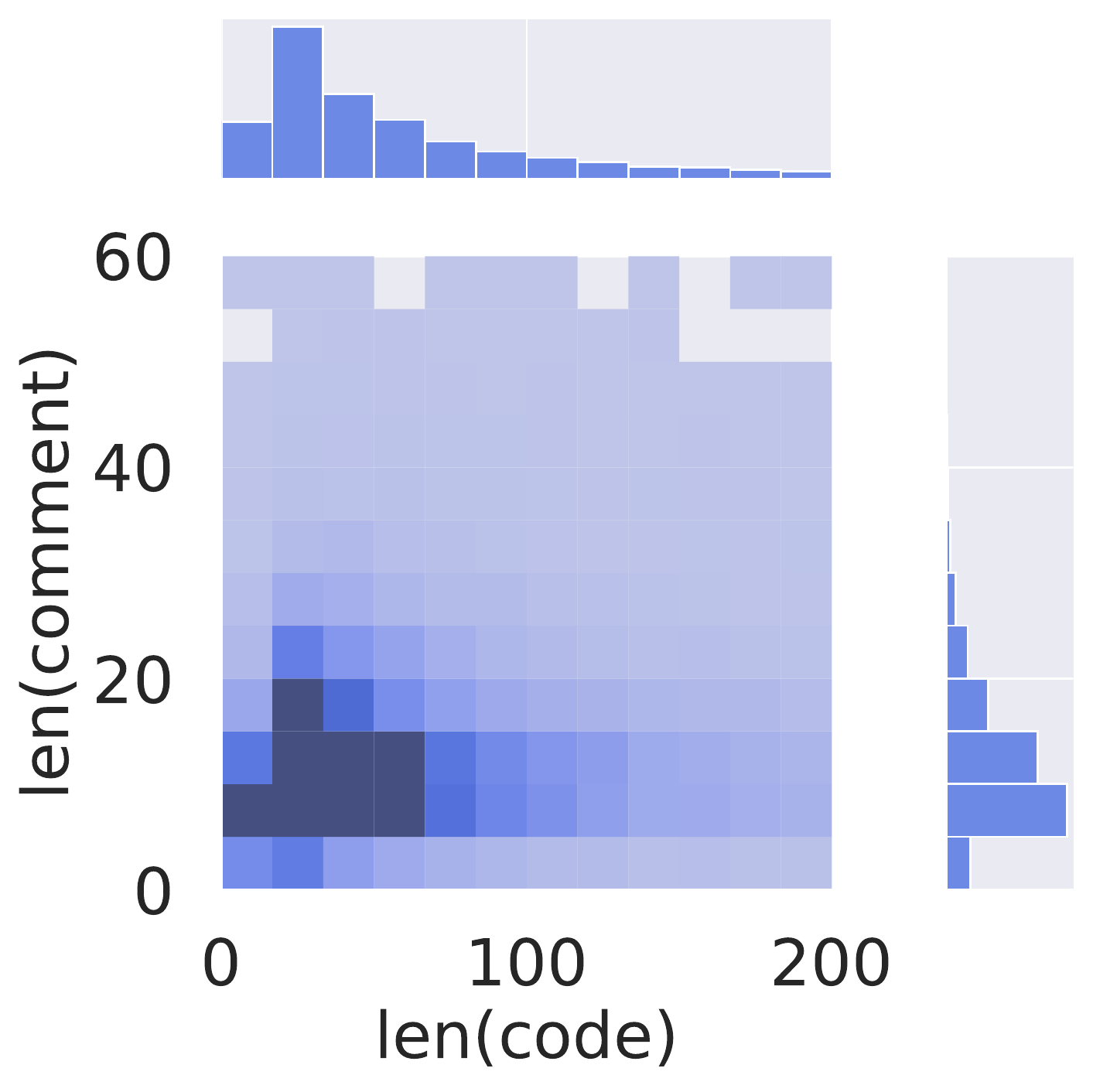}\end{minipage}
 & \begin{minipage}{.18\textwidth}\includegraphics[width=\textwidth]{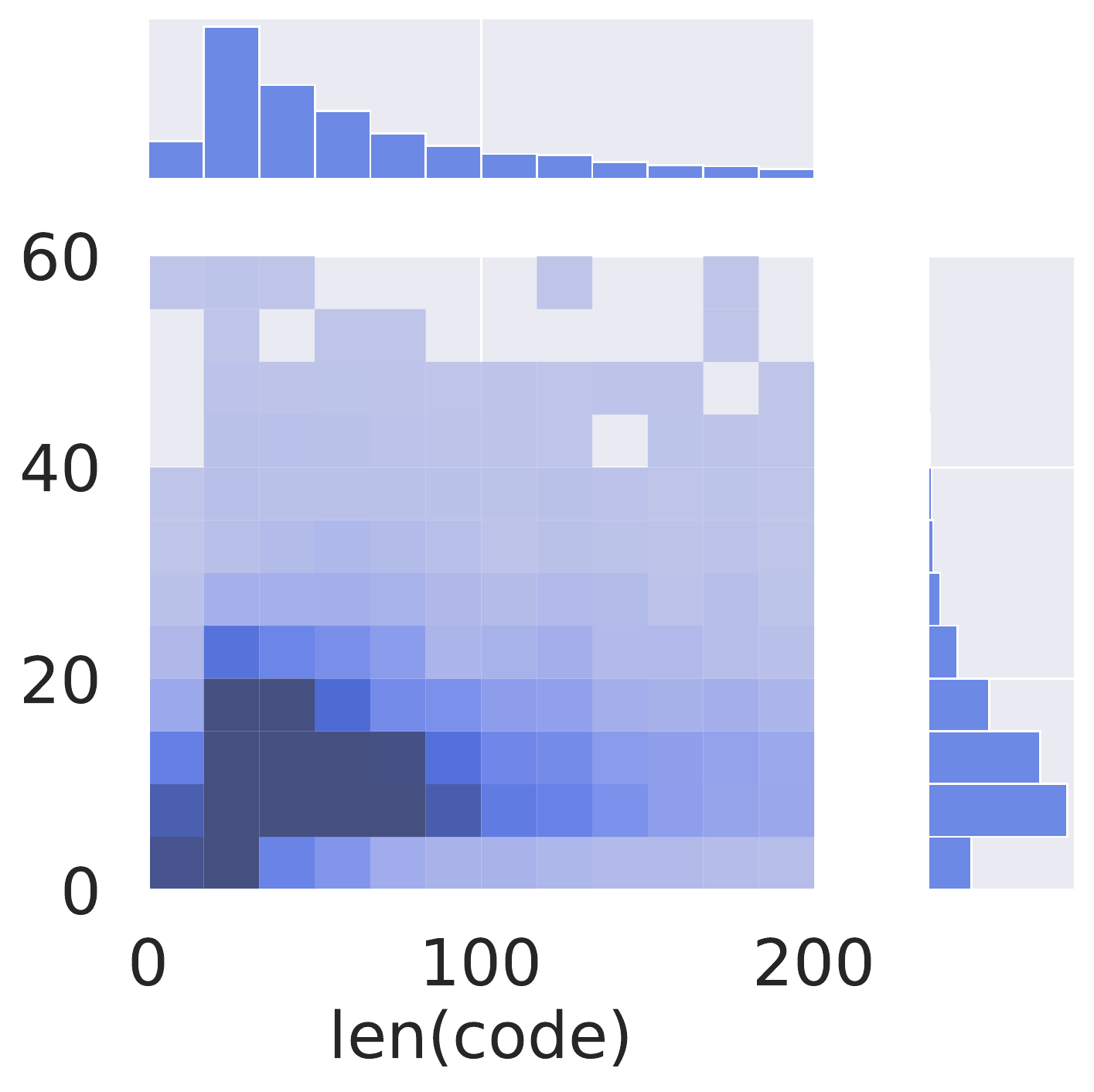}\end{minipage}
 & \begin{minipage}{.18\textwidth}\includegraphics[width=\textwidth]{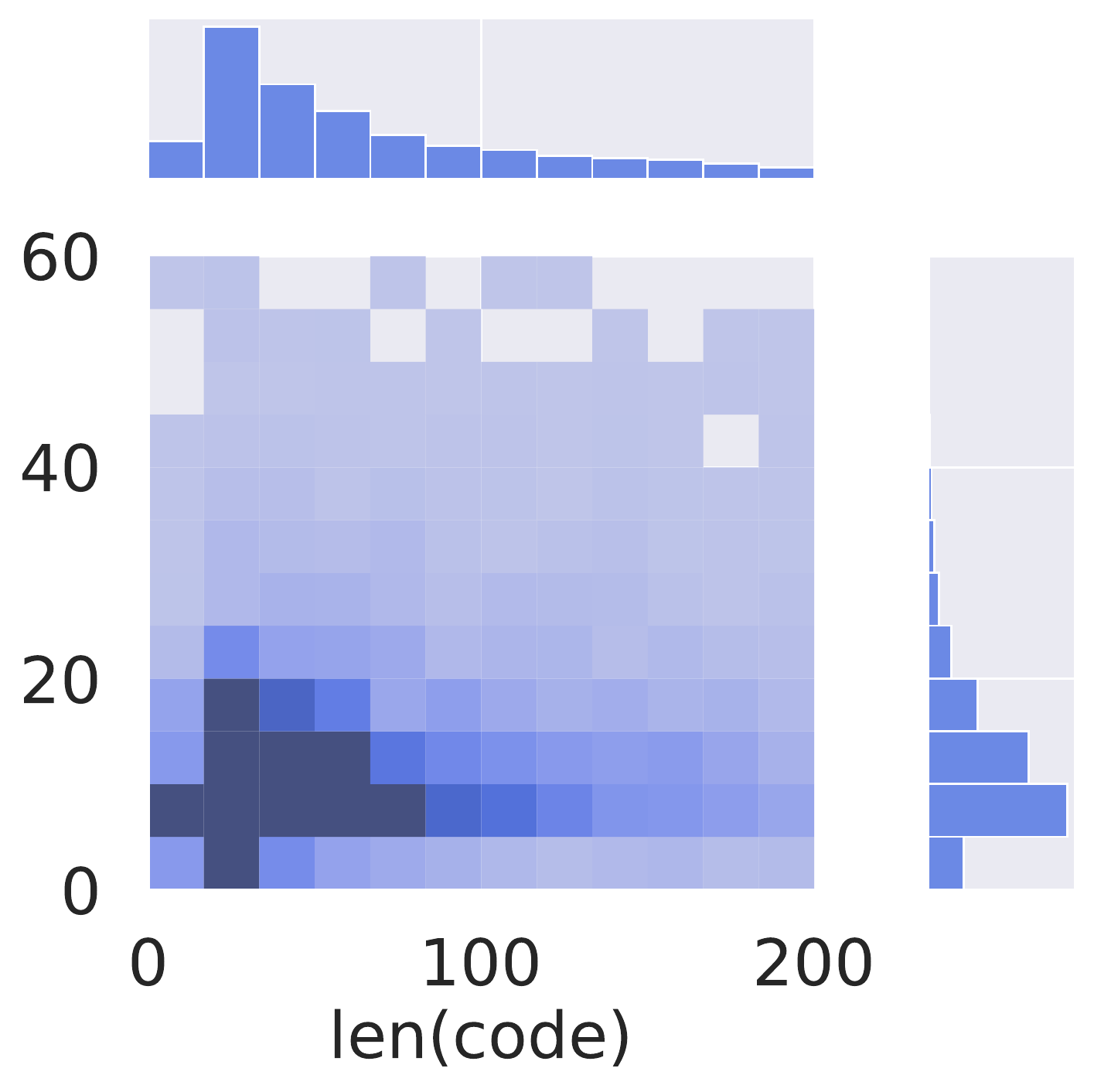}\end{minipage}
 & \UseMacro{TH-ds-CP-T}
 & \begin{minipage}{.18\textwidth}\includegraphics[width=\textwidth]{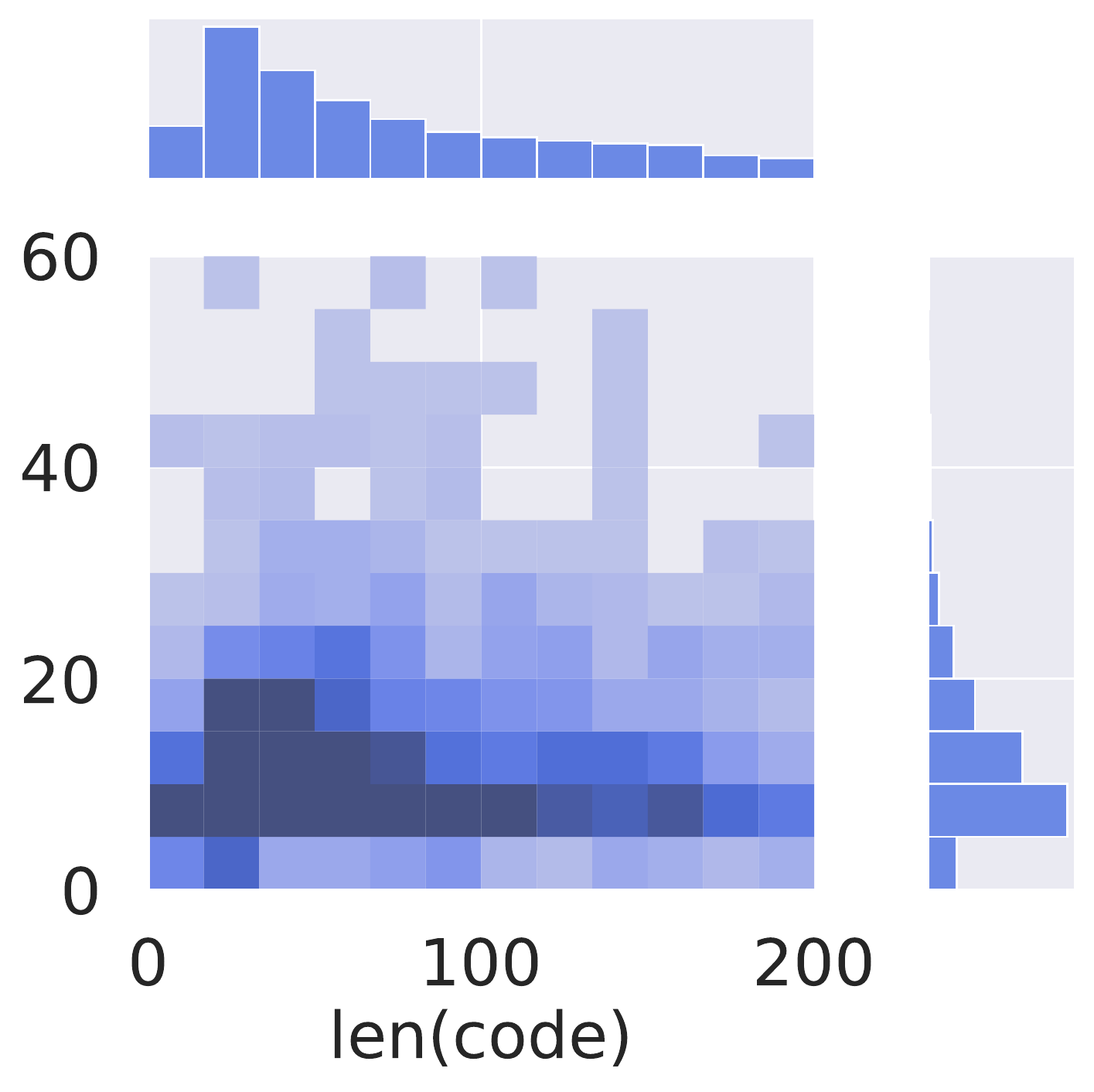}\end{minipage}
\\
\hline
\end{tabular}
\end{footnotesize}
\end{center}

\caption{Distributions of the number of \subtoks in code (x-axis)
    and the number of \subtoks in comments (y-axis) in our \comgen
    dataset.\label{fig:dataset-metrics-dist-cg}}
\end{figure*}

\MyPara{\Comgen} Table~\ref{table:dataset-metrics-cg} shows the
statistics of the \comgen dataset.  
The rows, from top to bottom, are: the number of \examples; the
average number of \subtoks in code; the percentage of \examples whose
number of \subtoks in the code is less than 100, 150, 200; the average
number of \subtoks in comments; the percentage of \examples whose
number of \subtoks in the comment is less than 20, 30, 50.
Figure~\ref{fig:dataset-metrics-dist-cg} visualizes the distributions
of the number of \subtoks in code (x-axis) and the number of \subtoks
in comments (y-axis).

\begin{figure*}

\begin{center}
\begin{footnotesize}
\begin{tabular}{|l|c|c|c || l|c|}
\hline
& \textbf{ATrain} & \textbf{\AVal} & \textbf{\ATestS} & & \textbf{\ATestC} \\
\hline
\UseMacro{TH-ds-MP}
 & \begin{minipage}{.18\textwidth}\includegraphics[width=\textwidth]{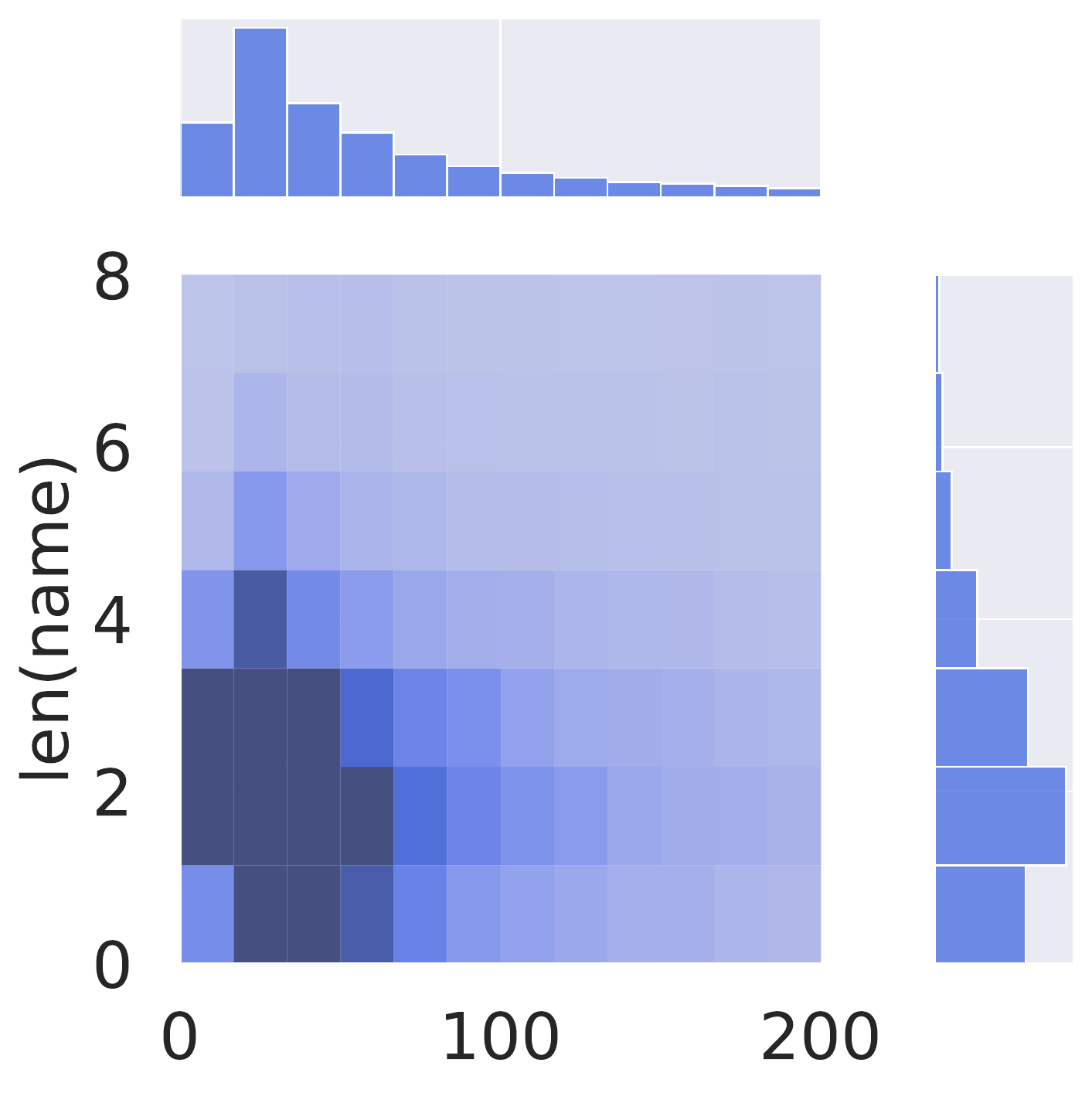}\end{minipage}
 & \begin{minipage}{.18\textwidth}\includegraphics[width=\textwidth]{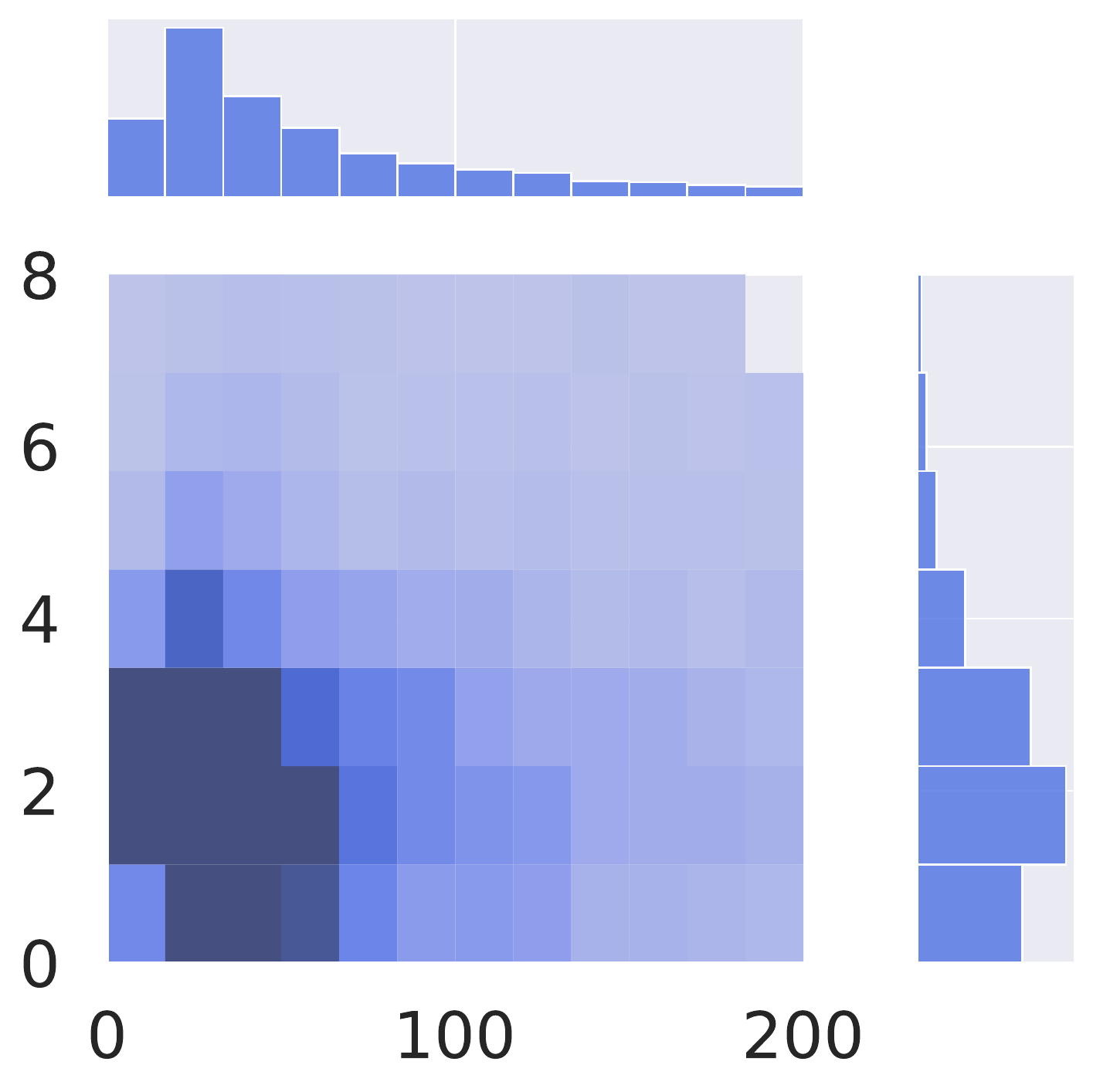}\end{minipage}
 & \begin{minipage}{.18\textwidth}\includegraphics[width=\textwidth]{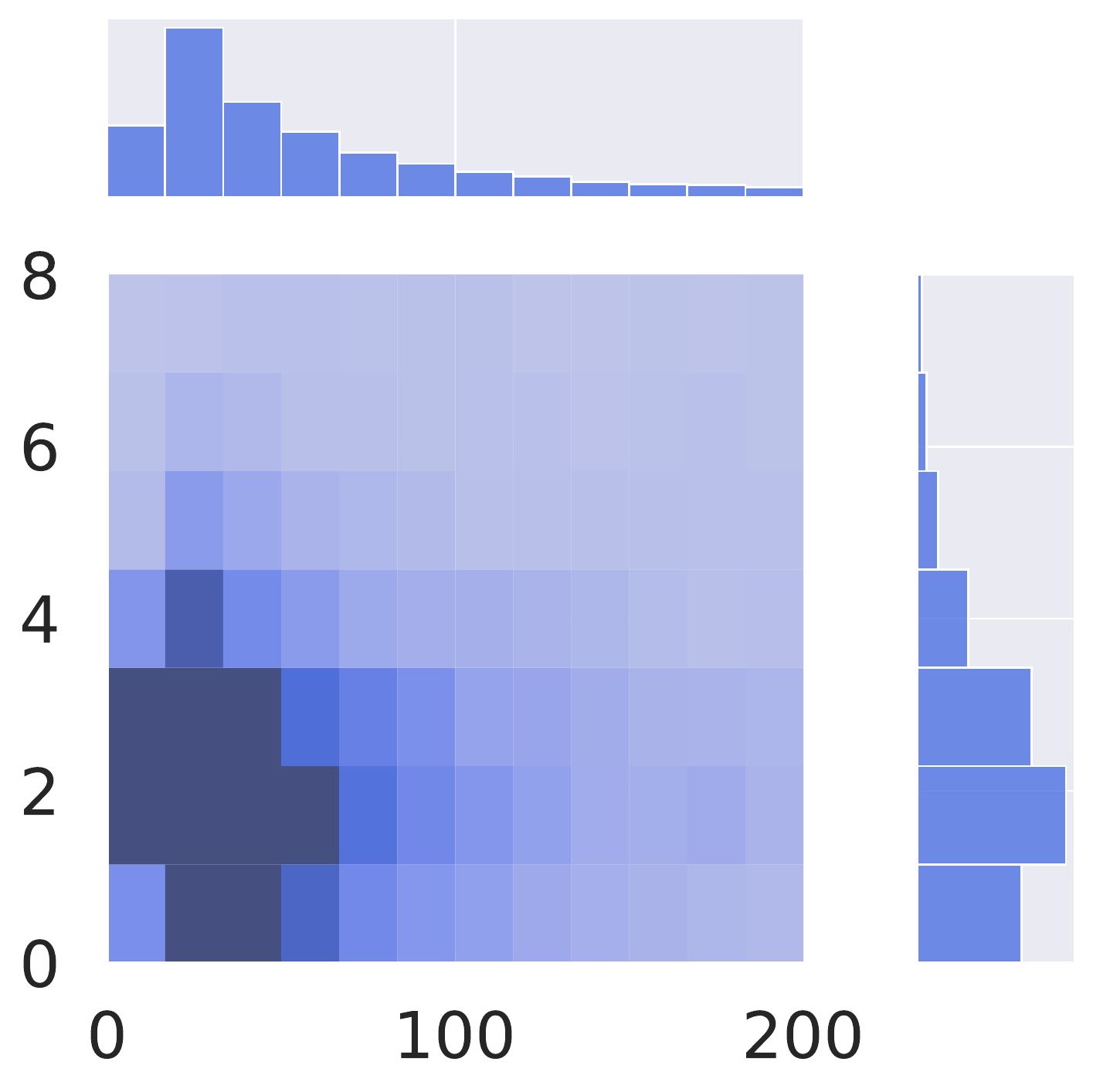}\end{minipage}
 & \UseMacro{TH-ds-MP-CP}
 & \begin{minipage}{.18\textwidth}\includegraphics[width=\textwidth]{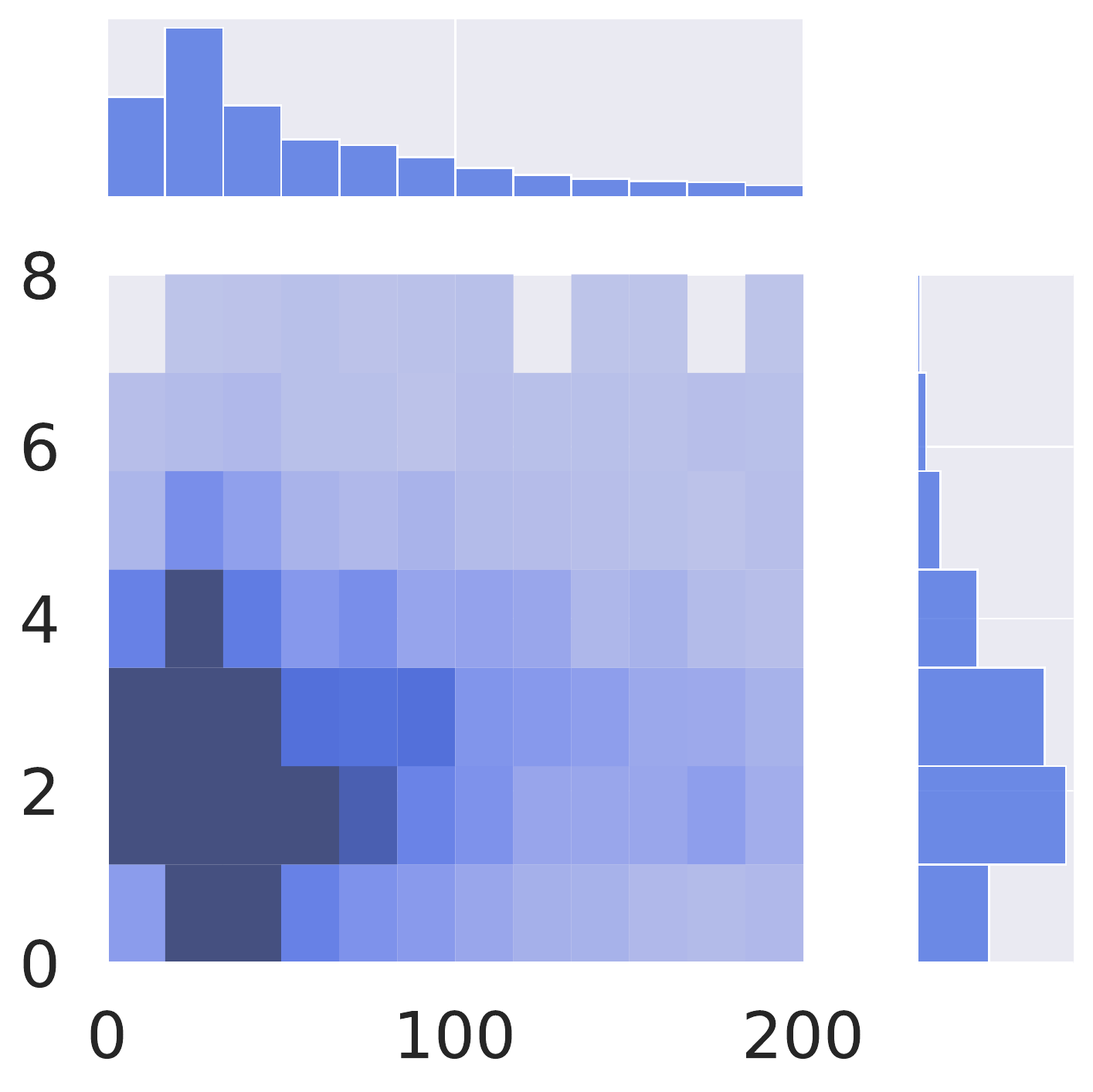}\end{minipage}
\\
\hline
\UseMacro{TH-ds-CP}
 & \begin{minipage}{.18\textwidth}\includegraphics[width=\textwidth]{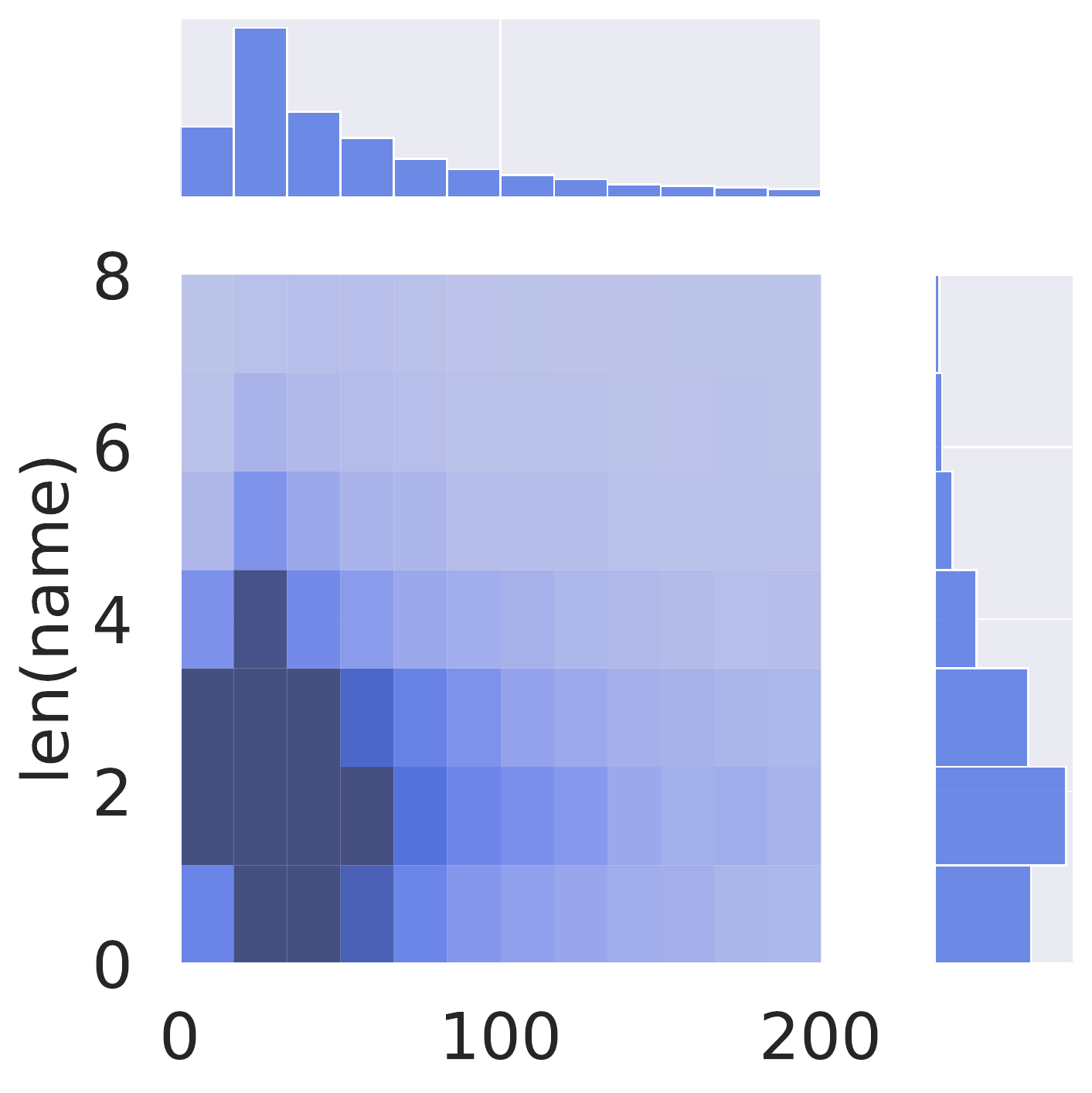}\end{minipage}
 & \begin{minipage}{.18\textwidth}\includegraphics[width=\textwidth]{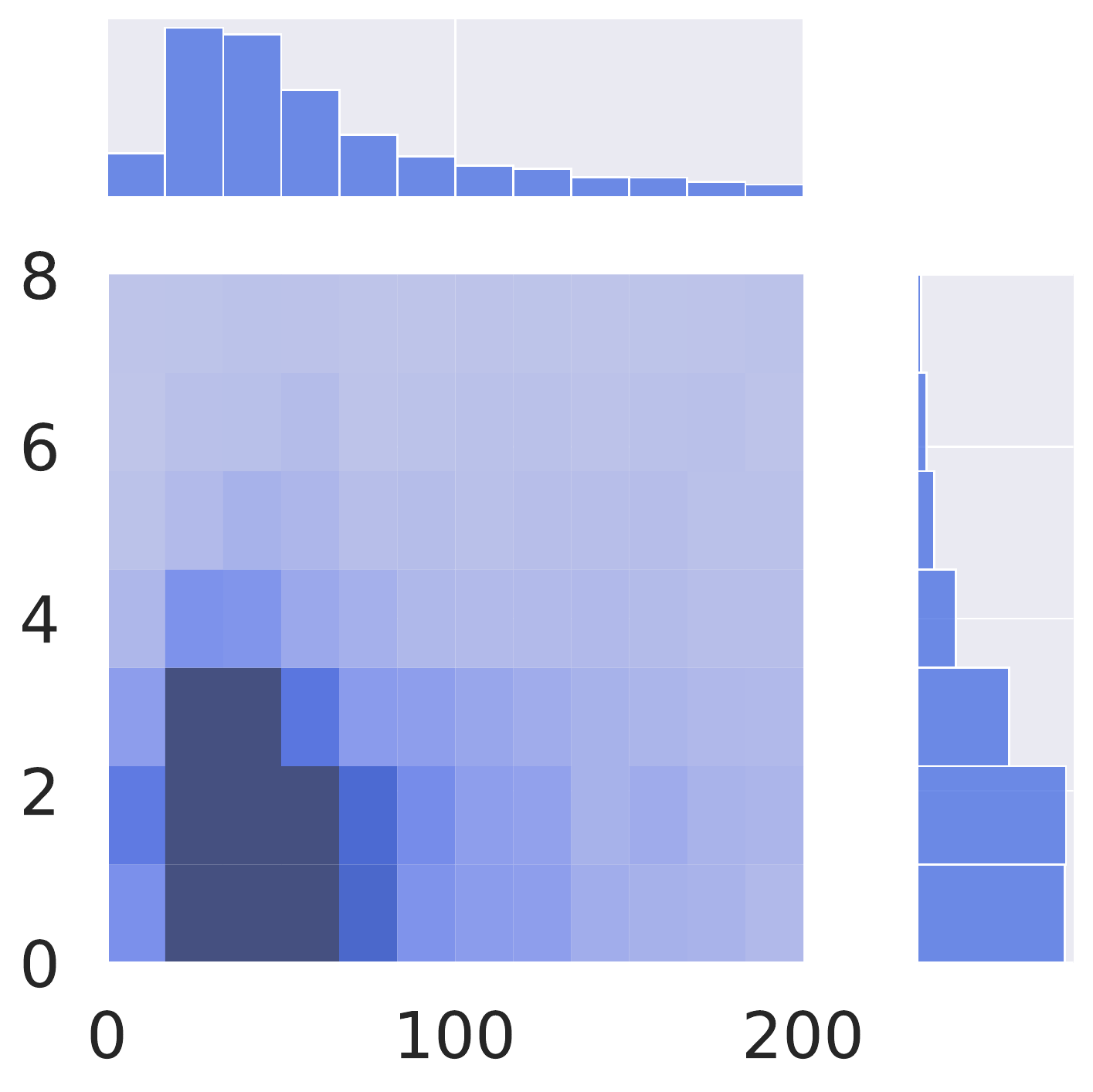}\end{minipage}
 & \begin{minipage}{.18\textwidth}\includegraphics[width=\textwidth]{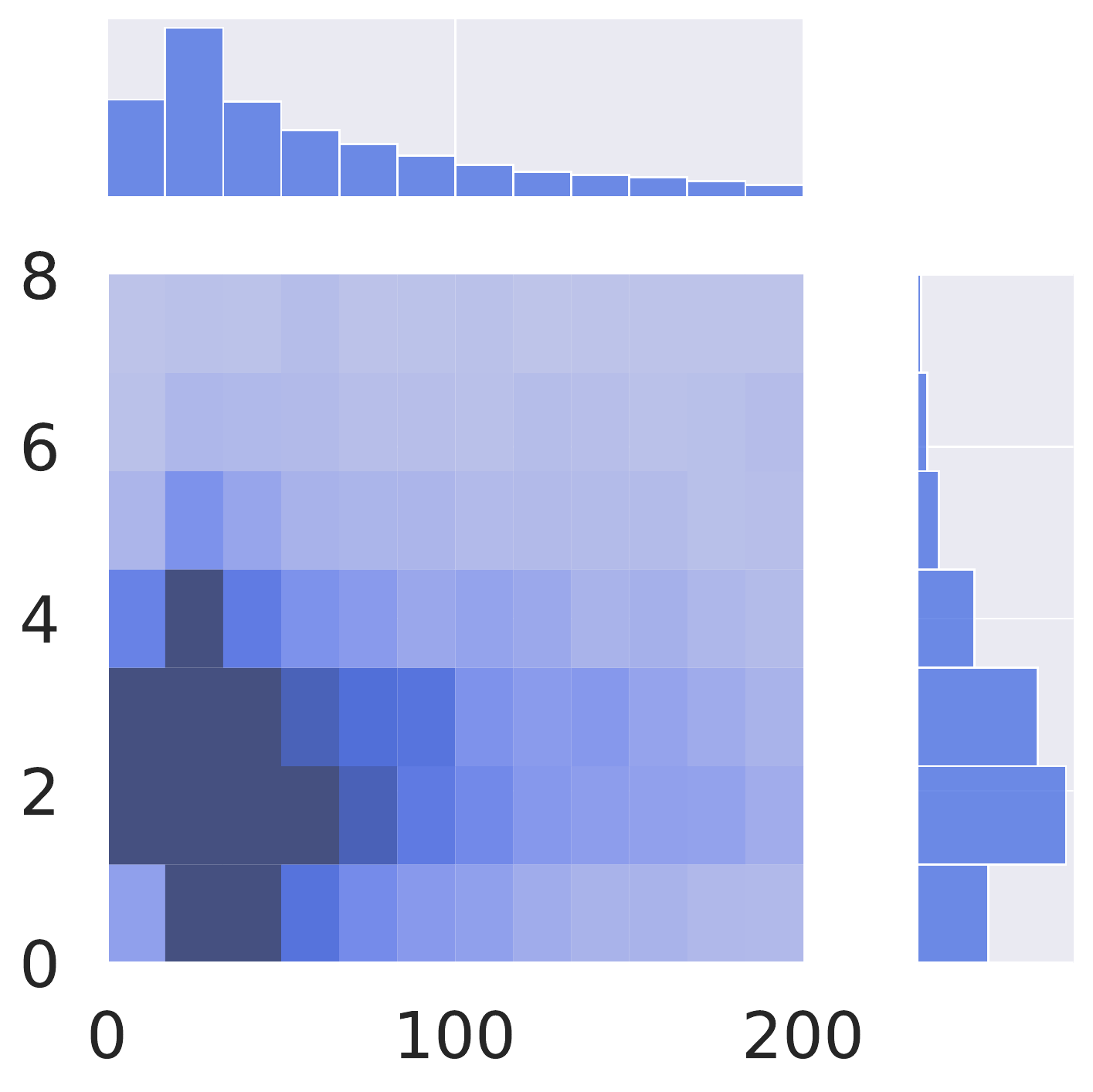}\end{minipage}
 & \UseMacro{TH-ds-MP-T}
 & \begin{minipage}{.18\textwidth}\includegraphics[width=\textwidth]{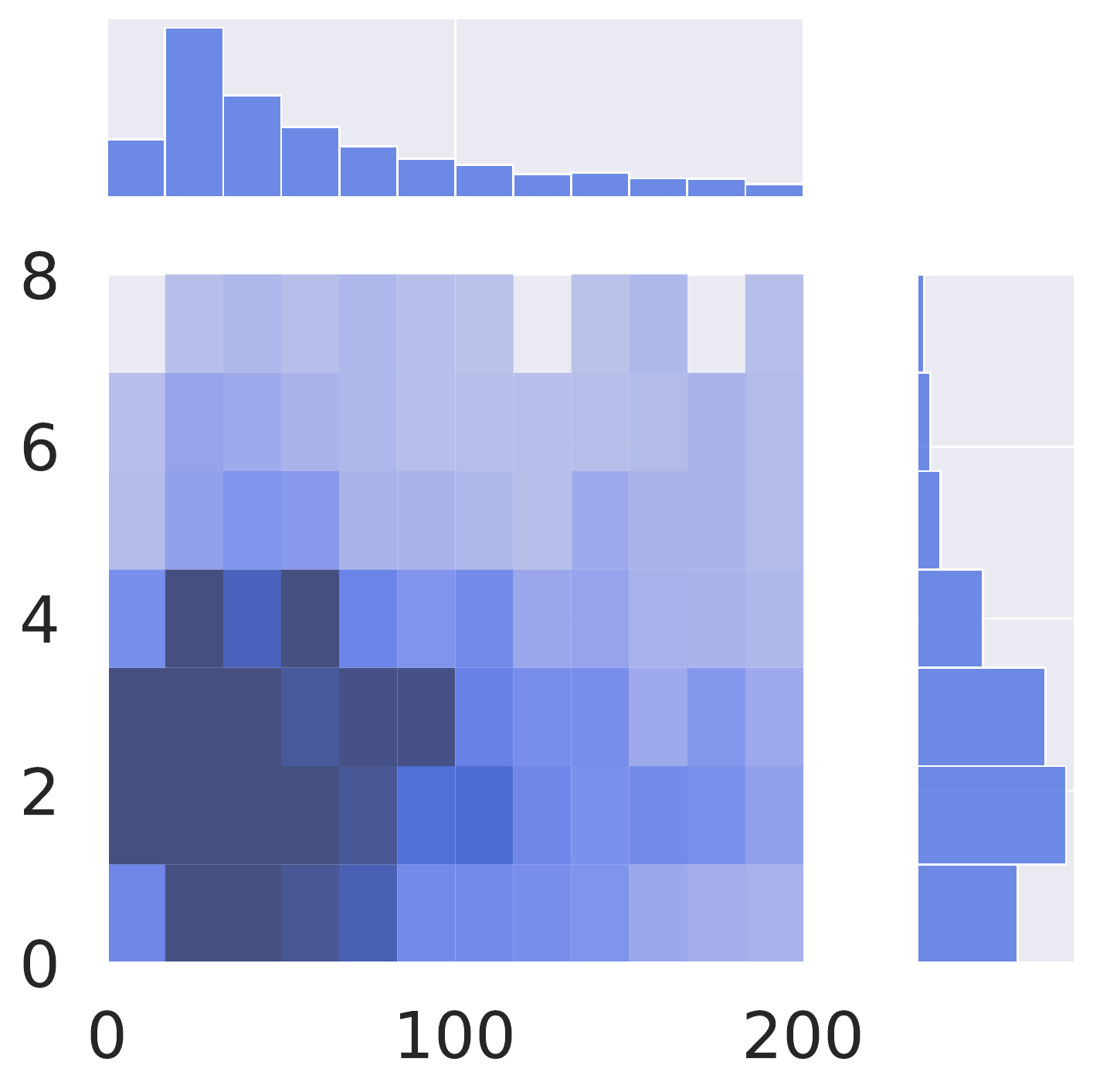}\end{minipage}
\\
\hline
\UseMacro{TH-ds-T}
 & \begin{minipage}{.18\textwidth}\includegraphics[width=\textwidth]{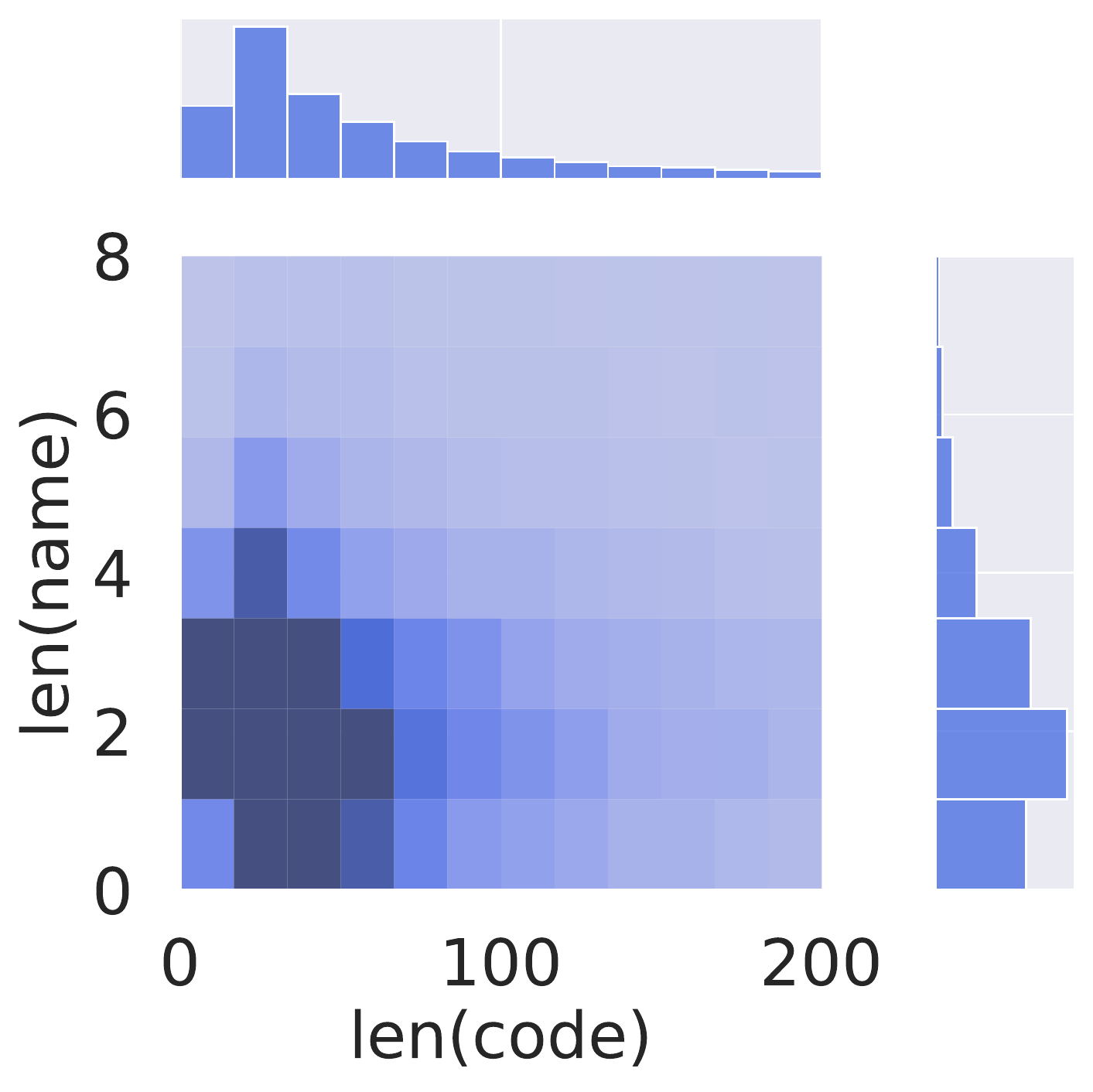}\end{minipage}
 & \begin{minipage}{.18\textwidth}\includegraphics[width=\textwidth]{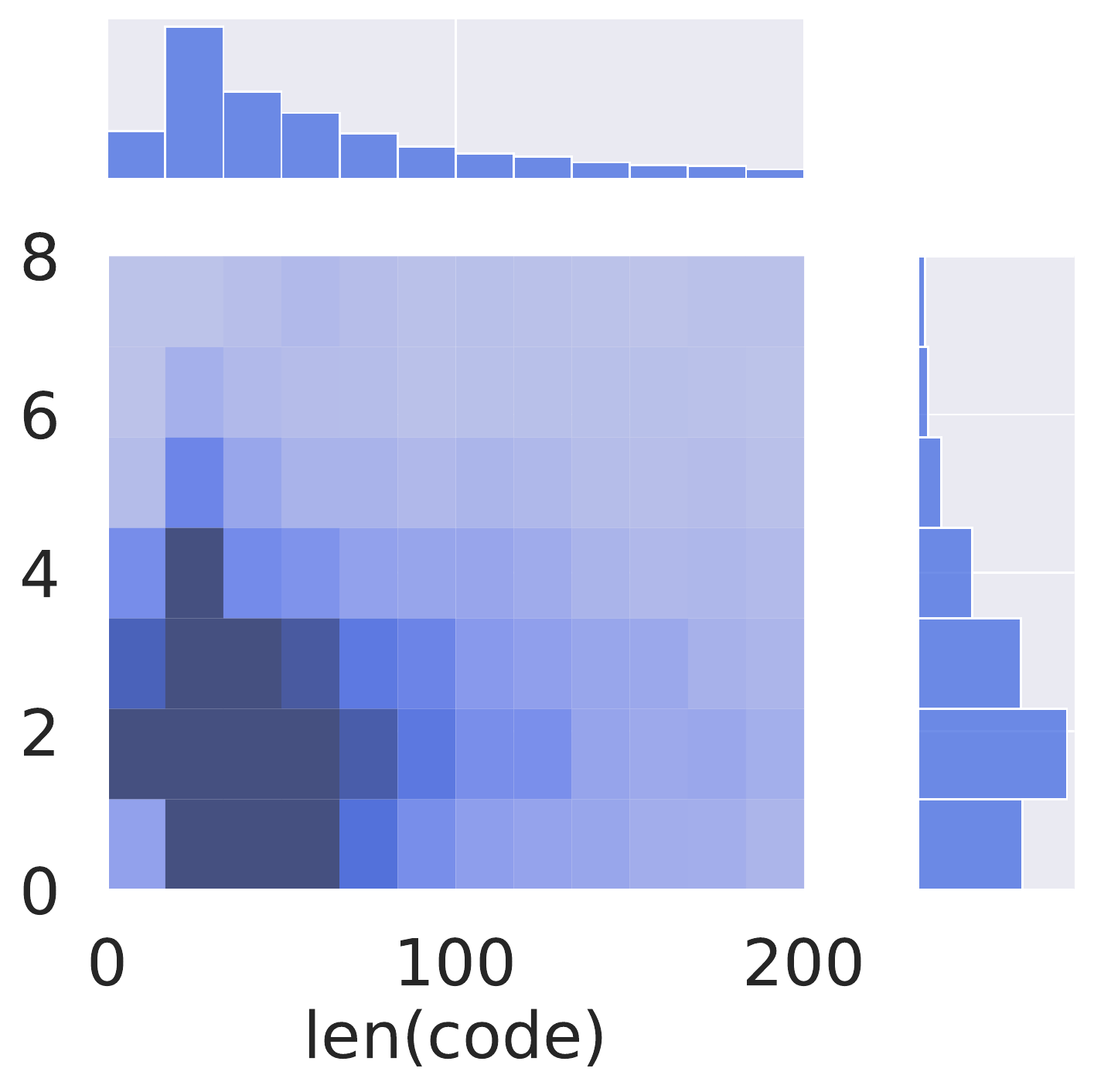}\end{minipage}
 & \begin{minipage}{.18\textwidth}\includegraphics[width=\textwidth]{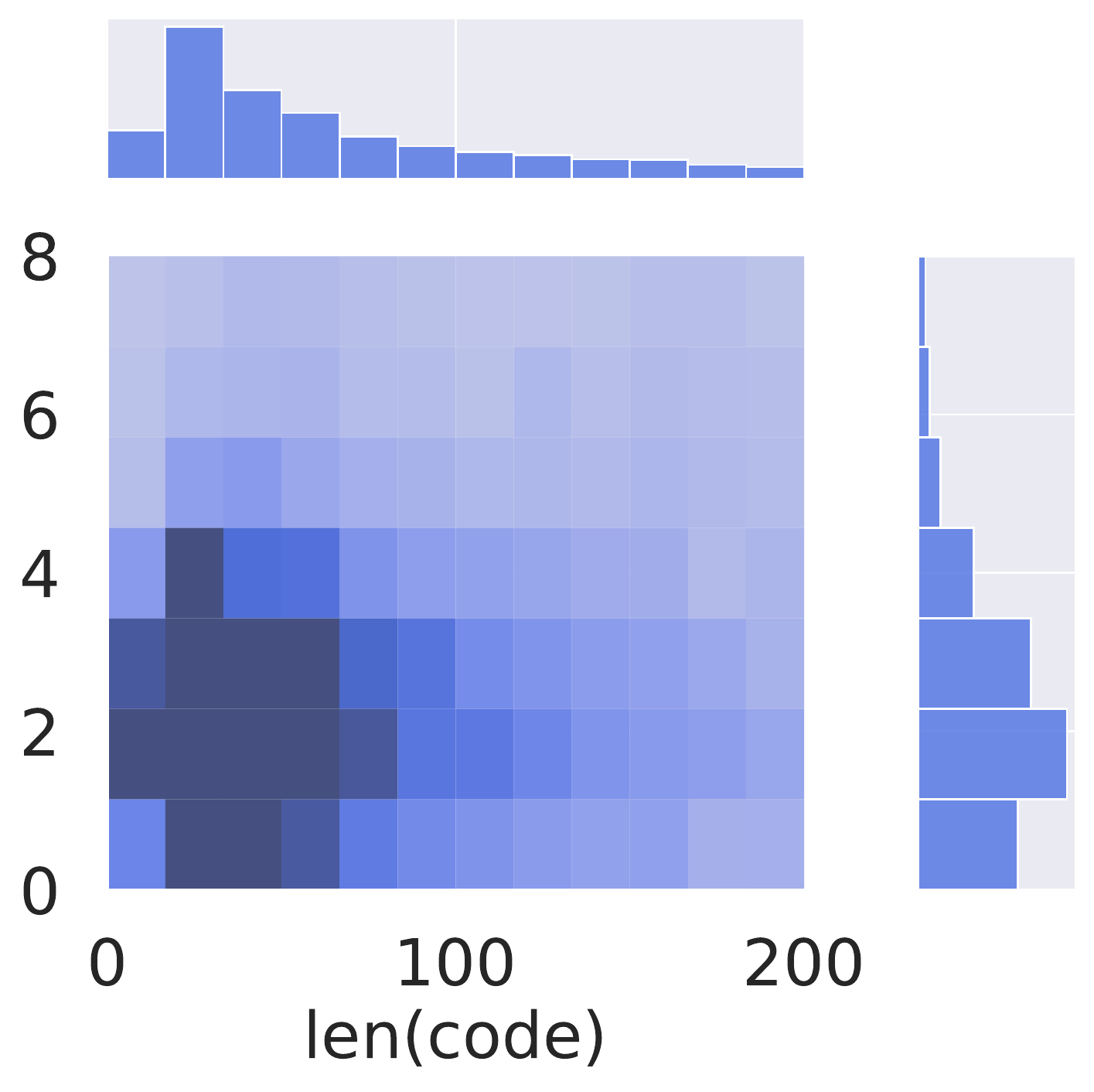}\end{minipage}
 & \UseMacro{TH-ds-CP-T}
 & \begin{minipage}{.18\textwidth}\includegraphics[width=\textwidth]{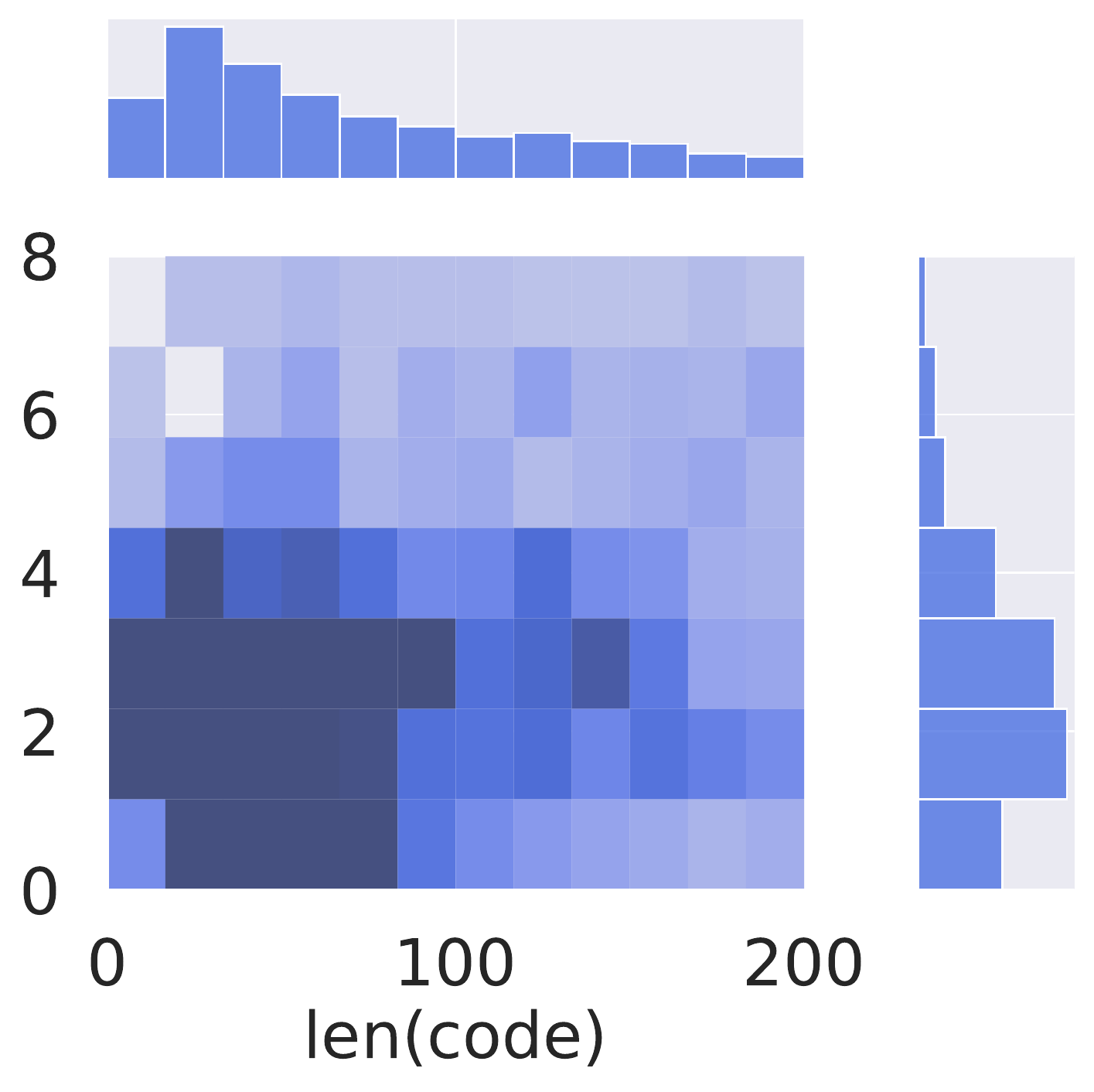}\end{minipage}
\\
\hline
\end{tabular}
\end{footnotesize}
\end{center}

\caption{Distributions of the number of \subtoks in code (x-axis)
    and the number of \subtoks in names (y-axis) in our \metnam
    dataset.\label{fig:dataset-metrics-dist-mn}}
\end{figure*}

\MyPara{\Metnam} For each \example, we replaced the appearances of its
name from its code to the special token ``METHODNAMEMASK'' such that
the models cannot cheat by looking for the name in the signature line
or in the \method body of recursive \methods.
Table~\ref{table:dataset-metrics-mn} shows the statistics of the
\metnam dataset.  
The rows, from top to bottom, are: the number of \examples; the
average number of \subtoks in code; the percentage of \examples whose
number of \subtoks in the code is less than 100, 150, 200; the average
number of \subtoks in names; the percentage of \examples whose number
of \subtoks in the name is less than 2, 3, 6.
Figure~\ref{fig:dataset-metrics-dist-mn} visualizes the distributions
of the number of \subtoks in code (x-axis) and the number of \subtoks
in names (y-axis).

\section{Experiments Details}
\label{sec:appendix:repro}

This section presents details of our experiments to support their
reproducibility.

\MyPara{Computing infrastructure} We run our experiments on a machine
with four NVIDIA 1080-TI GPUs and two Intel Xeon E5-2620 v4 CPUs.

\MyPara{Estimated runtime of models} The approximate model training
time are: \UseMacro{model-DeepComHybridESE19} 7 days;
\UseMacro{model-RNNBaseline} 4 hours;
\UseMacro{model-TransformerACL20} 10 hours;
\UseMacro{model-Code2SeqICLR19} 4 hours;
\UseMacro{model-Code2VecPOPL19} 15 minutes.  The evaluation time is
around 1--10 minutes per model per evaluation set.

\MyPara{Number of parameters} The number of parameters in each model
are: \UseMacro{model-DeepComHybridESE19} 15.6M;
\UseMacro{model-RNNBaseline} 31.3M; \UseMacro{model-TransformerACL20}
68.2M; \UseMacro{model-Code2SeqICLR19} 5.7M;
\UseMacro{model-Code2VecPOPL19} 33.1M.

\MyPara{Random seeds} The random seed used for preparing the dataset
(performing in-project and cross-project splits) is: 7.  The random
seeds used for the three times of training are: 4182, 99243, 3705.

\MyPara{Reproducibility of prior work} We used the replication
packages provided in the original papers of the models when possible.
We made (small) updates to all models to: (1)~upgrade outdated data
processing code (because of our dataset contains \examples with new
programming language features that were not considered in the past);
(2)~export evaluation results in the format compatible with our
scripts.  We integrated these updates in our replication package.

\section{Additional Experiment Results}
\label{sec:appendix:more-results}

\begin{figure*}

\begin{center}
\begin{footnotesize}
\begin{tabular}{|r|@{}c@{}c@{}|@{}c@{}c@{}|@{}c@{}c@{}|}
\hline
\makecell[c]{\UseMacro{TH-train-on}}
 & \makecell[c]{\UseMacro{TH-MP}}
 & \makecell[c]{\UseMacro{TH-CP}}
 & \makecell[c]{\UseMacro{TH-MP}}
 & \makecell[c]{\UseMacro{TH-T}}
 & \makecell[c]{\UseMacro{TH-CP}}
 & \makecell[c]{\UseMacro{TH-T}}
\\
\hline
\makecell[c]{\UseMacro{TH-test-on}}
 & \multicolumn{2}{c|}{\UseMacro{TH-MP-CP}}
 & \multicolumn{2}{c|}{\UseMacro{TH-MP-T}}
 & \multicolumn{2}{c|}{\UseMacro{TH-CP-T}}
\\
\hline
\UseMacro{TH-metric-bleu}
 & \begin{minipage}{.12\textwidth}\includegraphics[width=\textwidth]{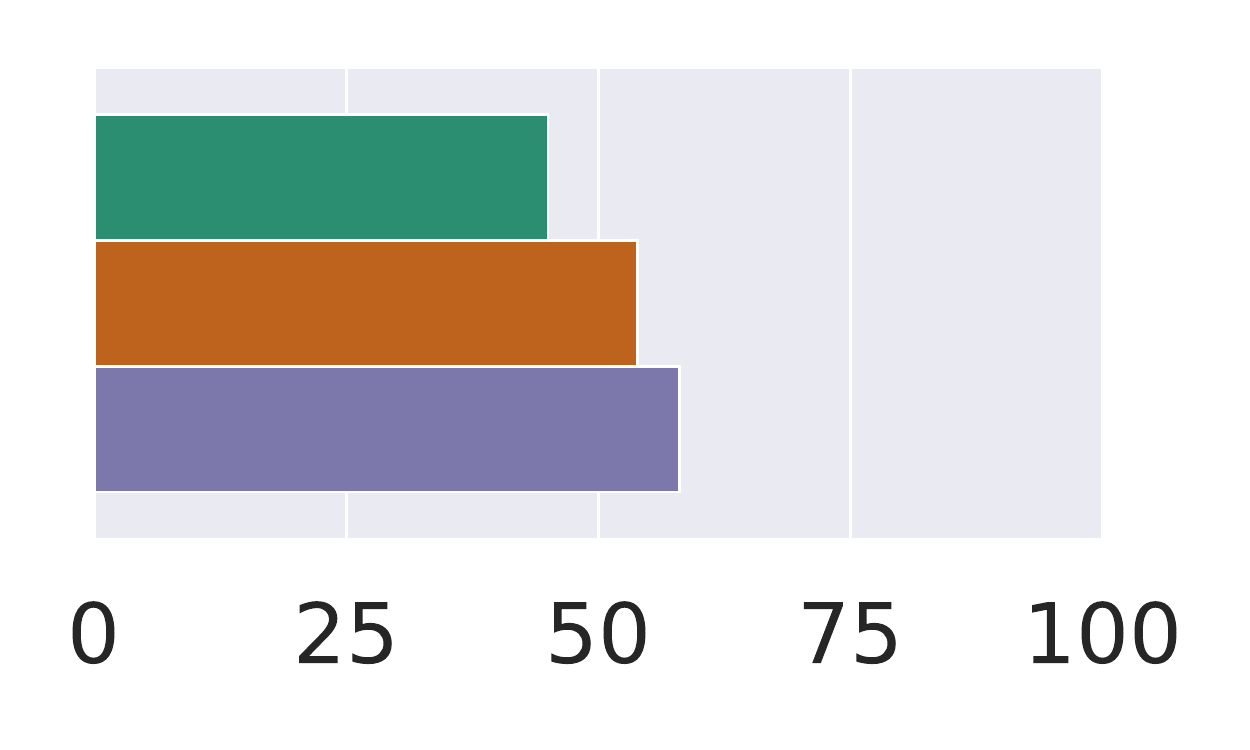}\end{minipage}
 & \begin{minipage}{.12\textwidth}\includegraphics[width=\textwidth]{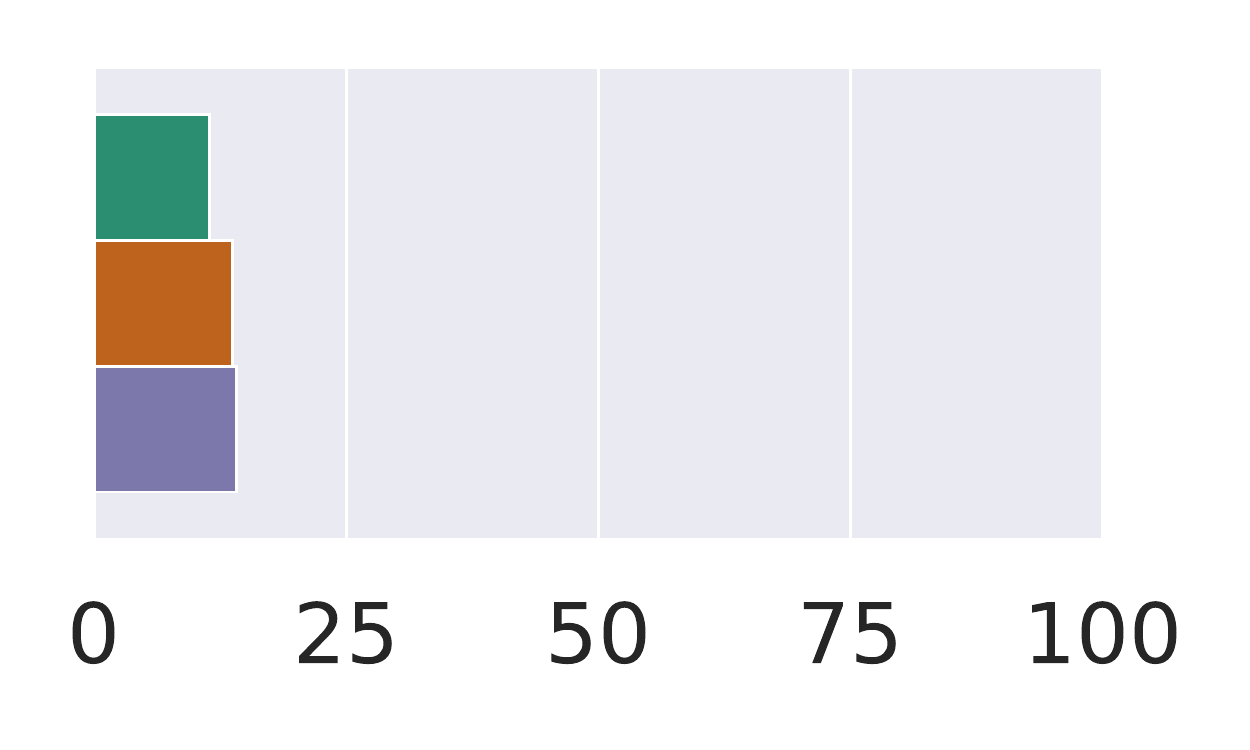}\end{minipage}
 & \begin{minipage}{.12\textwidth}\includegraphics[width=\textwidth]{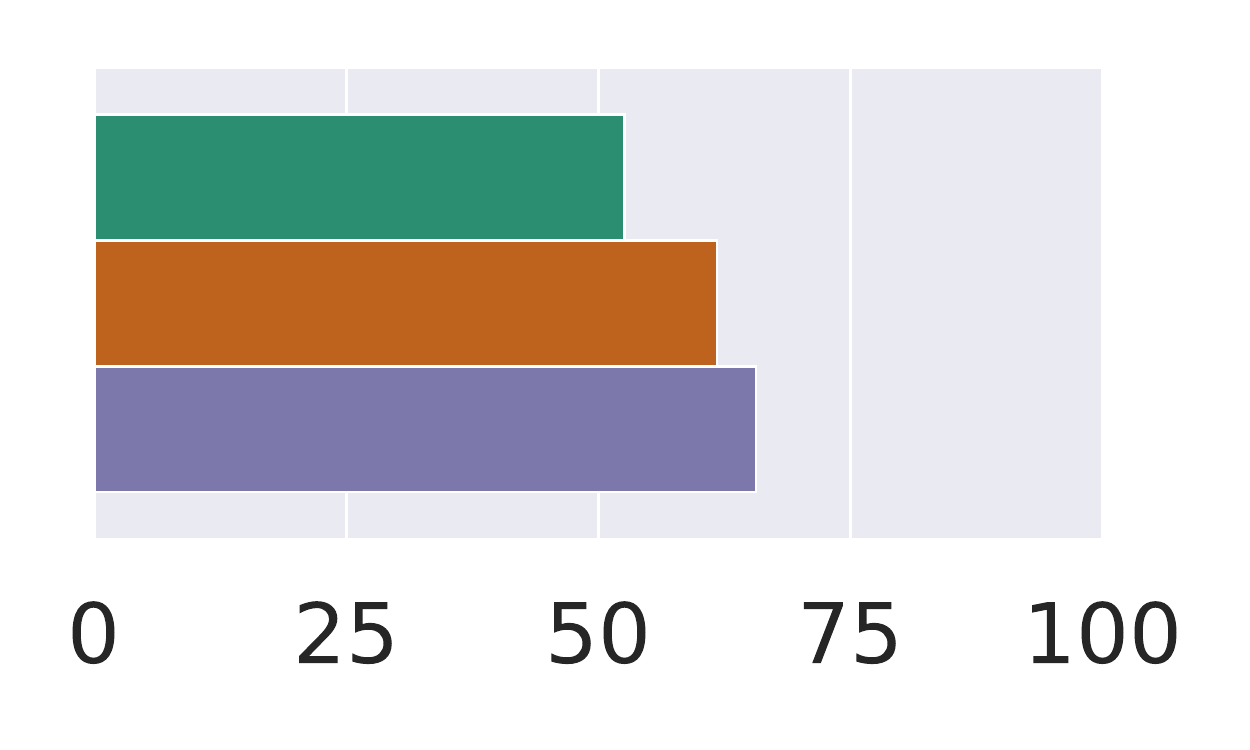}\end{minipage}
 & \begin{minipage}{.12\textwidth}\includegraphics[width=\textwidth]{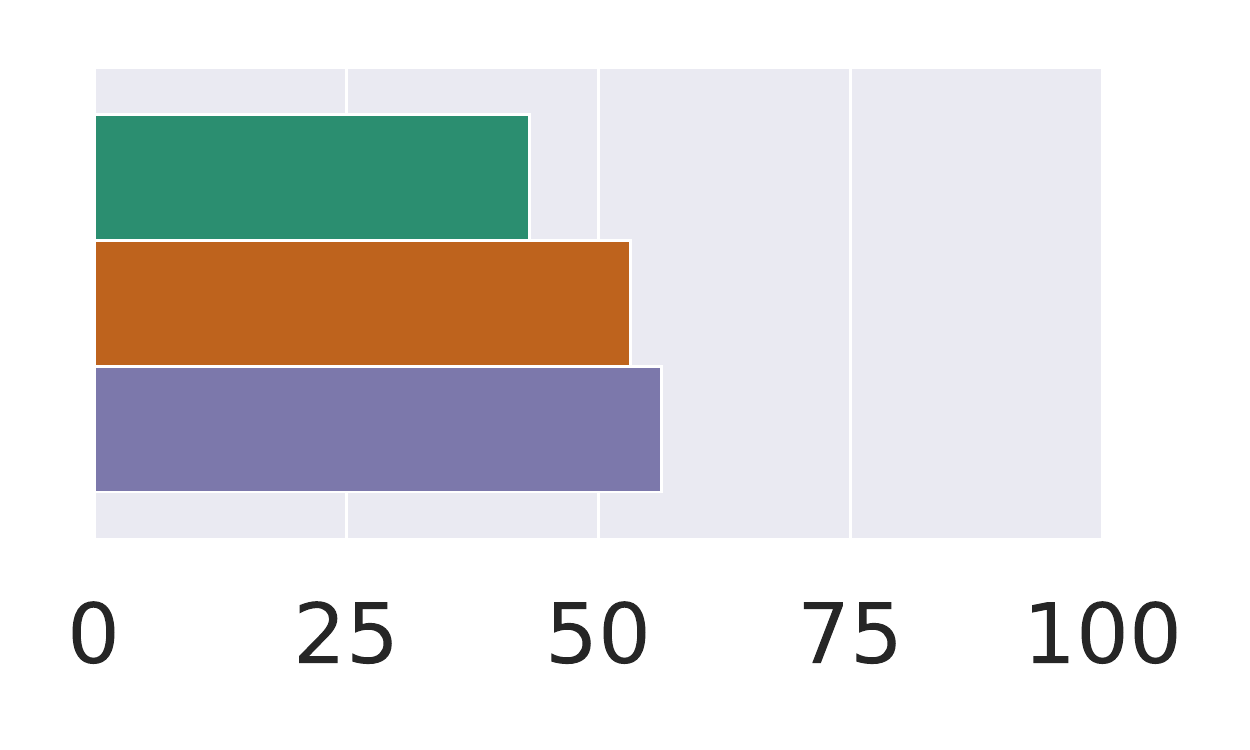}\end{minipage}
 & \begin{minipage}{.12\textwidth}\includegraphics[width=\textwidth]{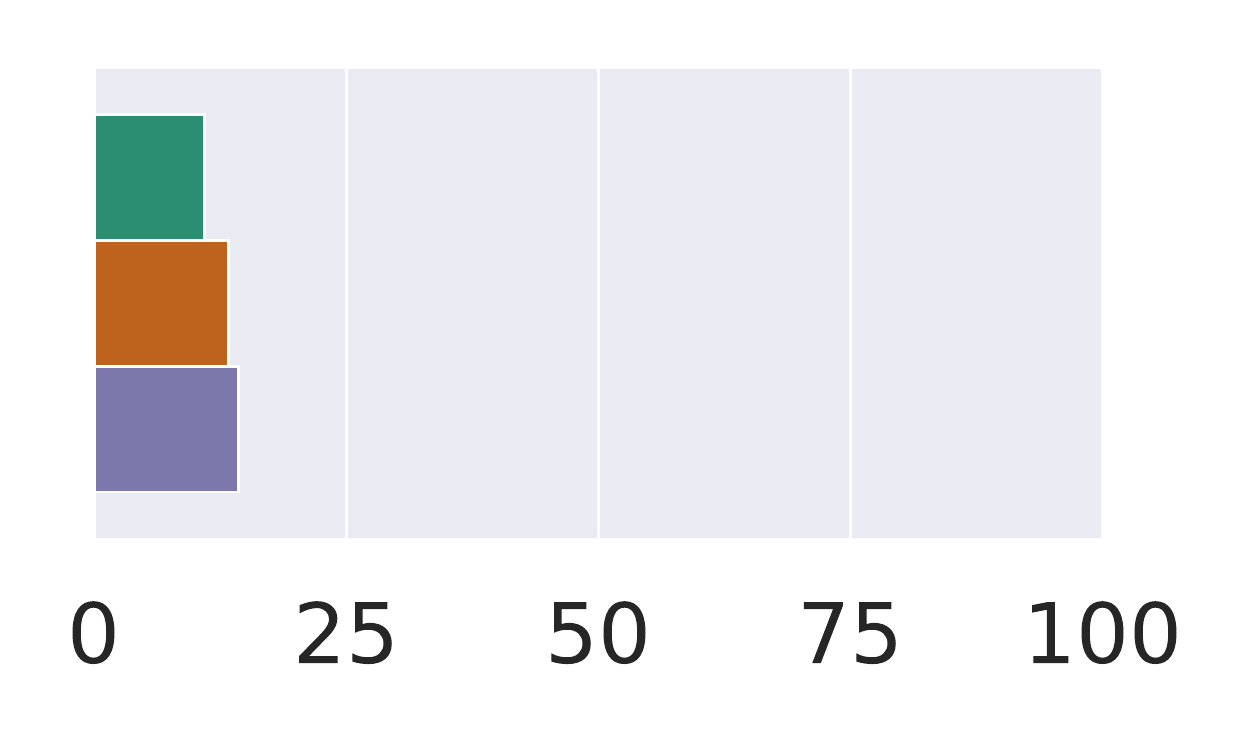}\end{minipage}
 & \begin{minipage}{.12\textwidth}\includegraphics[width=\textwidth]{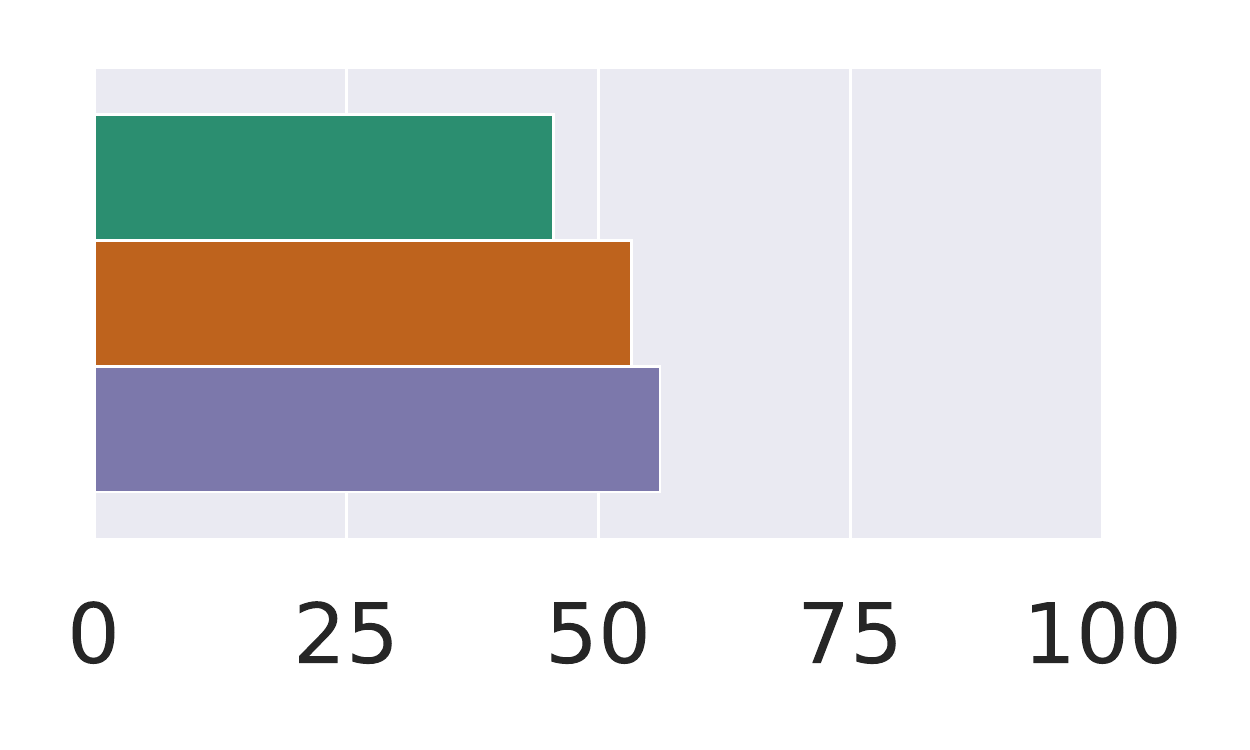}\end{minipage}
\\
\UseMacro{TH-metric-meteor}
 & \begin{minipage}{.12\textwidth}\includegraphics[width=\textwidth]{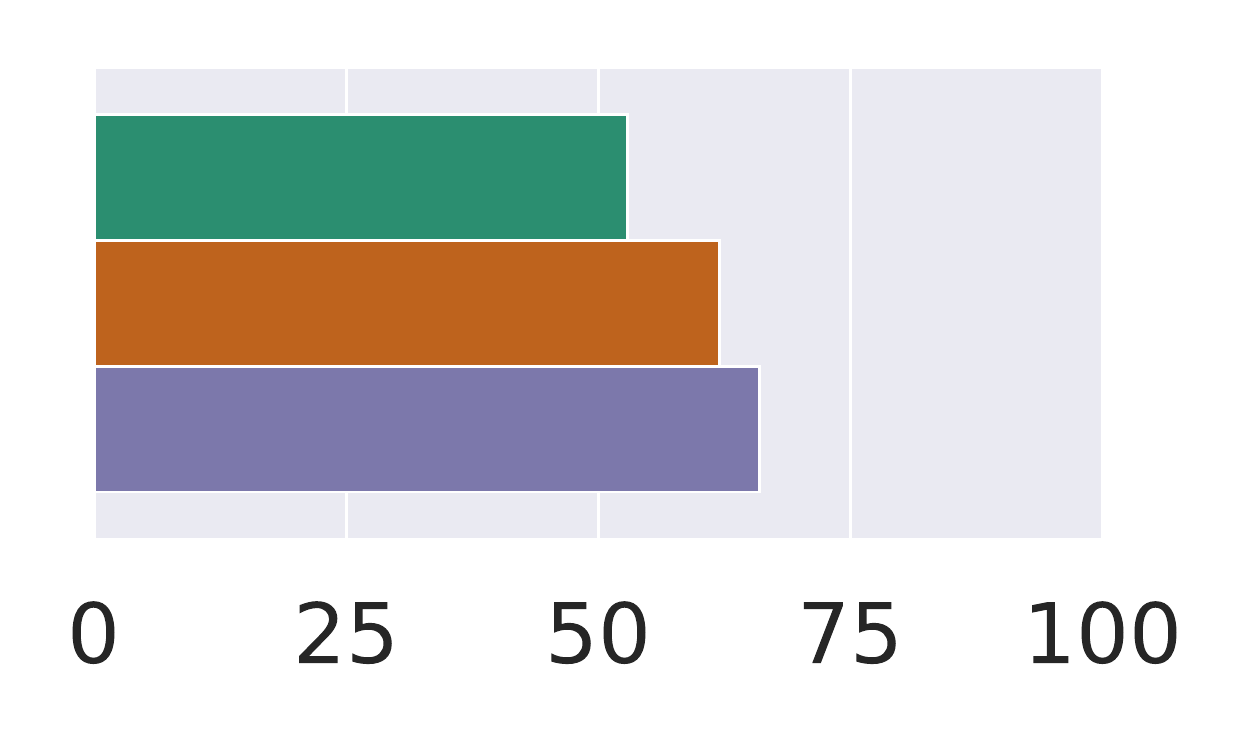}\end{minipage}
 & \begin{minipage}{.12\textwidth}\includegraphics[width=\textwidth]{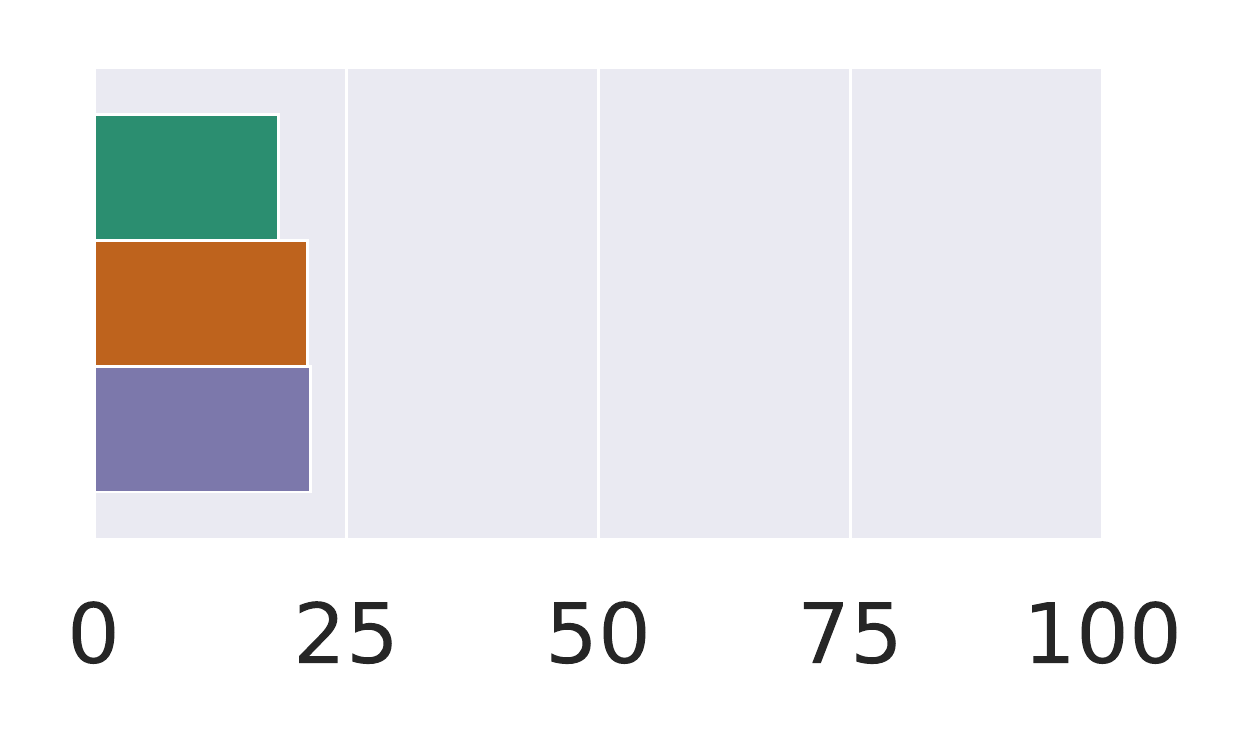}\end{minipage}
 & \begin{minipage}{.12\textwidth}\includegraphics[width=\textwidth]{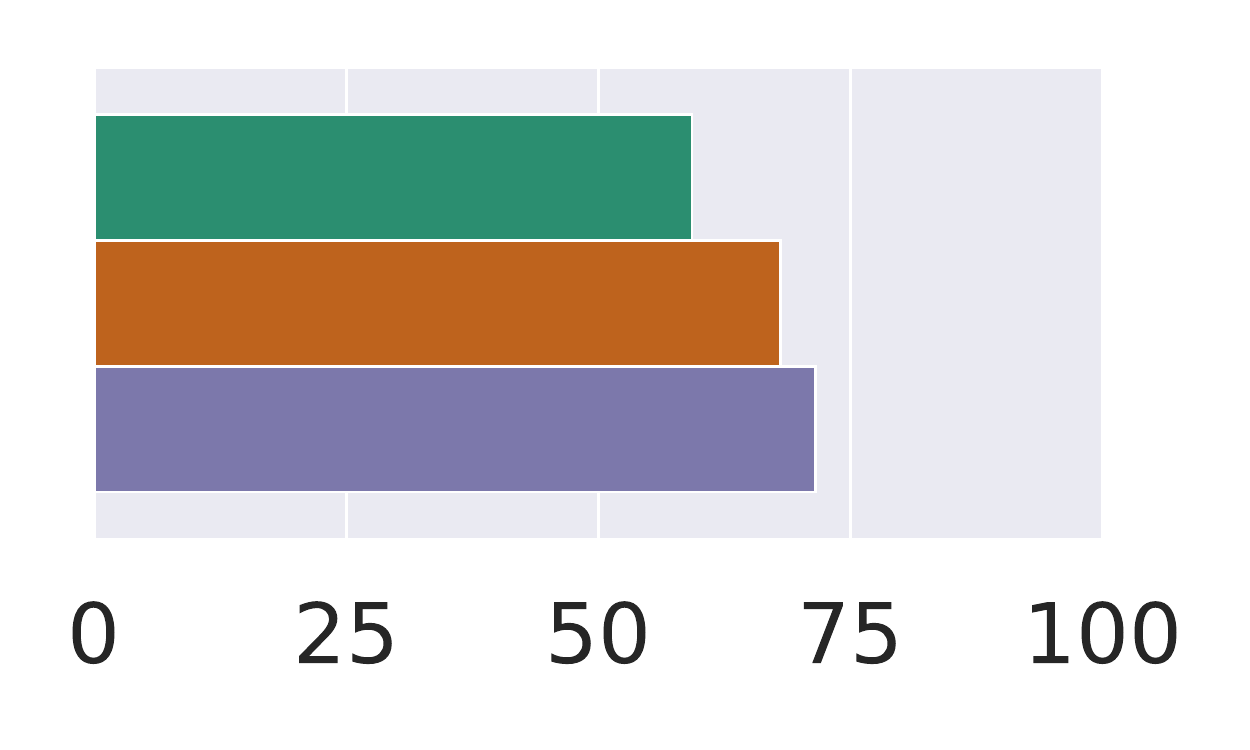}\end{minipage}
 & \begin{minipage}{.12\textwidth}\includegraphics[width=\textwidth]{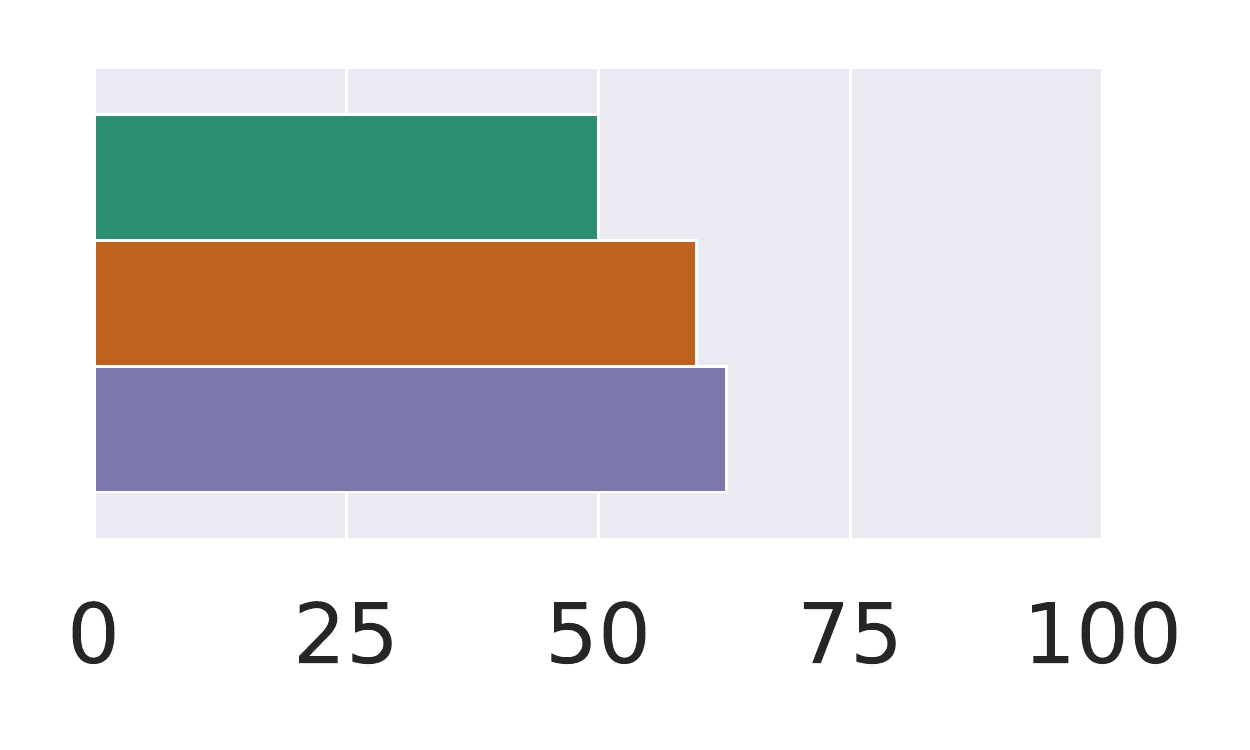}\end{minipage}
 & \begin{minipage}{.12\textwidth}\includegraphics[width=\textwidth]{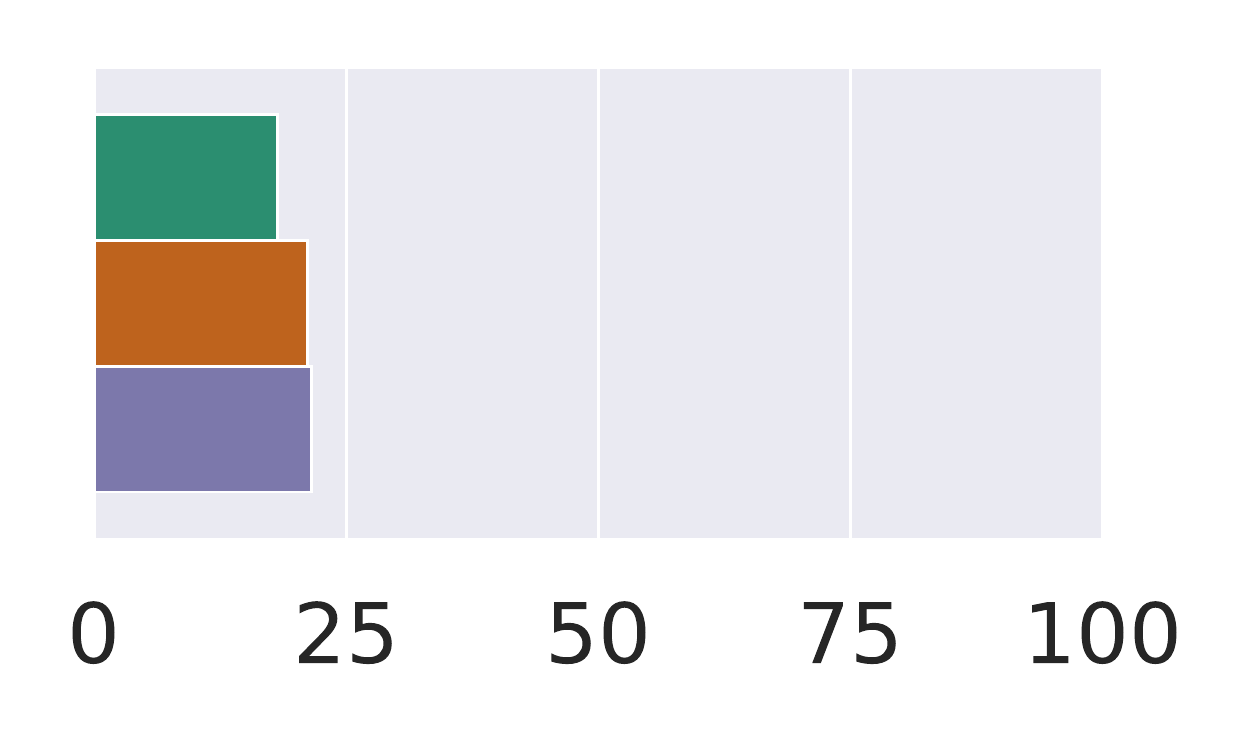}\end{minipage}
 & \begin{minipage}{.12\textwidth}\includegraphics[width=\textwidth]{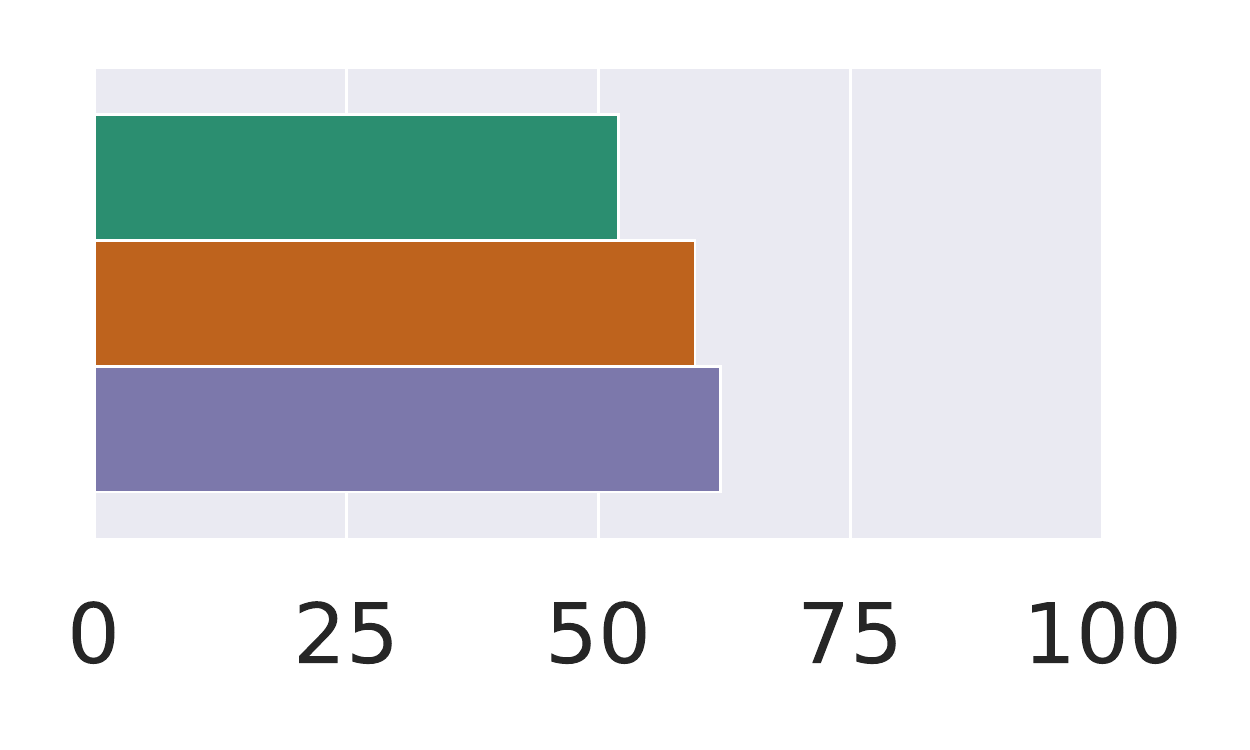}\end{minipage}
\\
\UseMacro{TH-metric-rouge_l_f}
 & \begin{minipage}{.12\textwidth}\includegraphics[width=\textwidth]{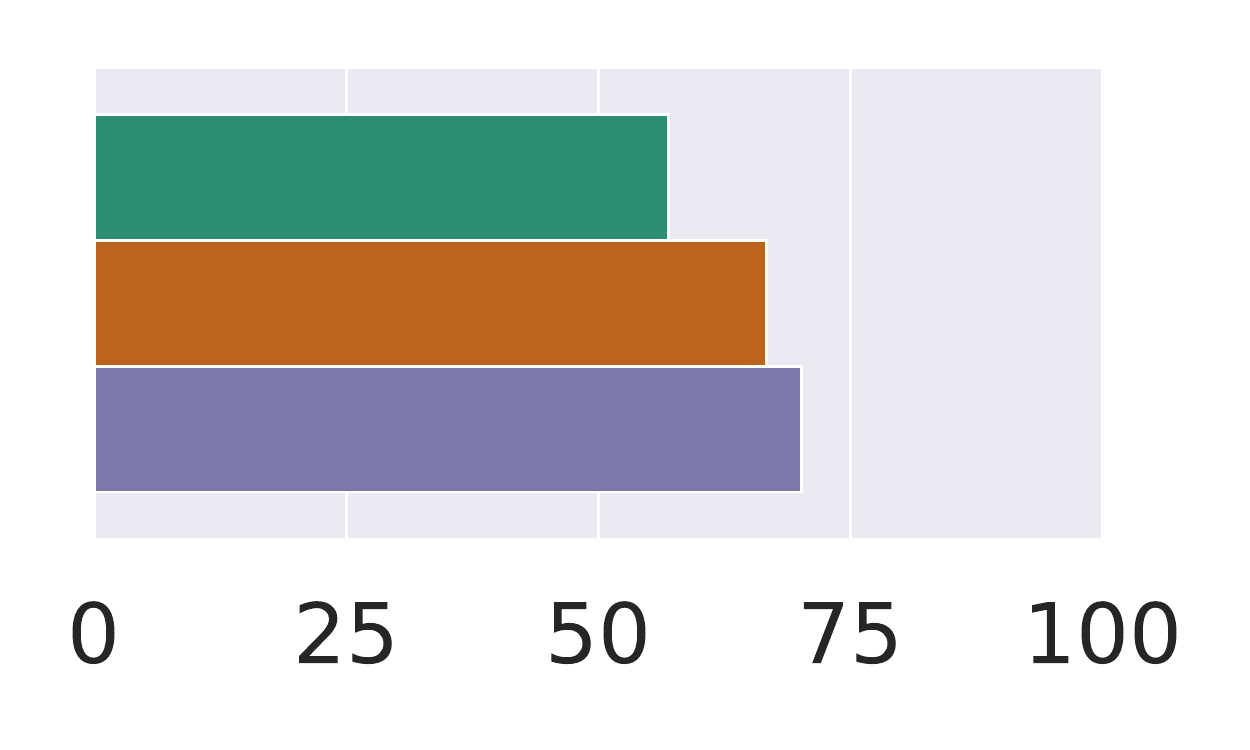}\end{minipage}
 & \begin{minipage}{.12\textwidth}\includegraphics[width=\textwidth]{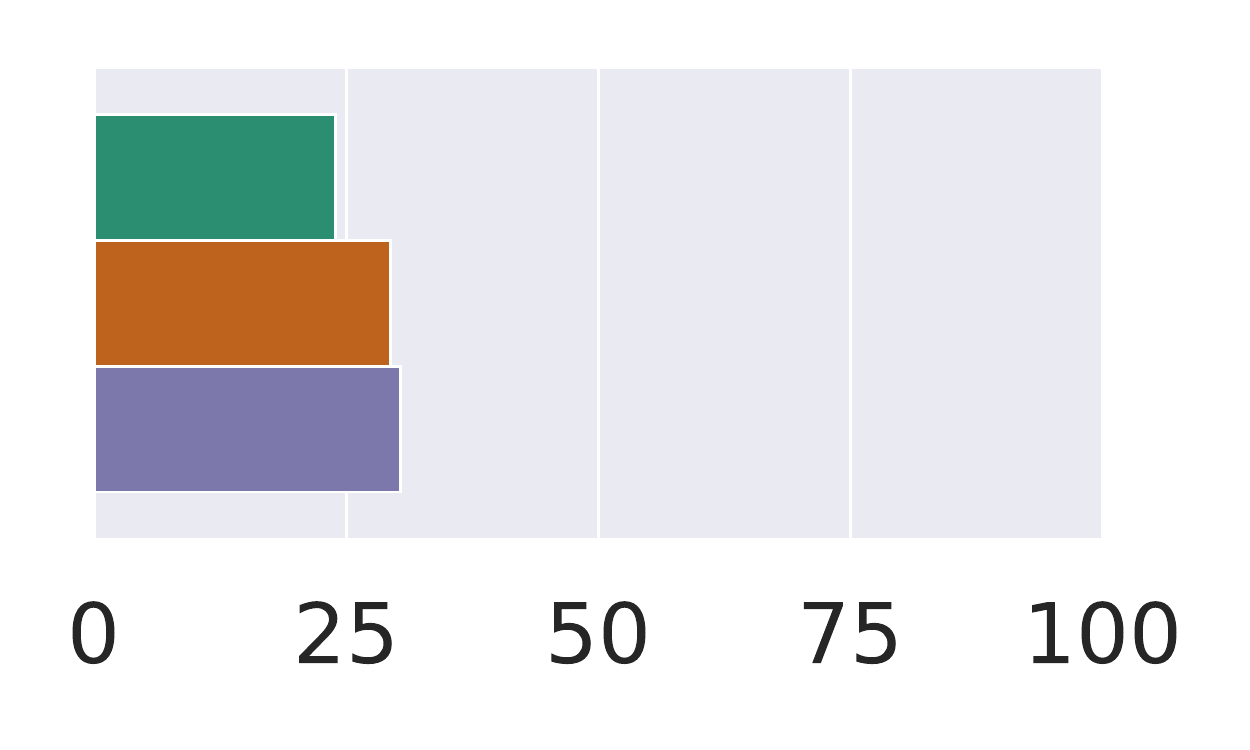}\end{minipage}
 & \begin{minipage}{.12\textwidth}\includegraphics[width=\textwidth]{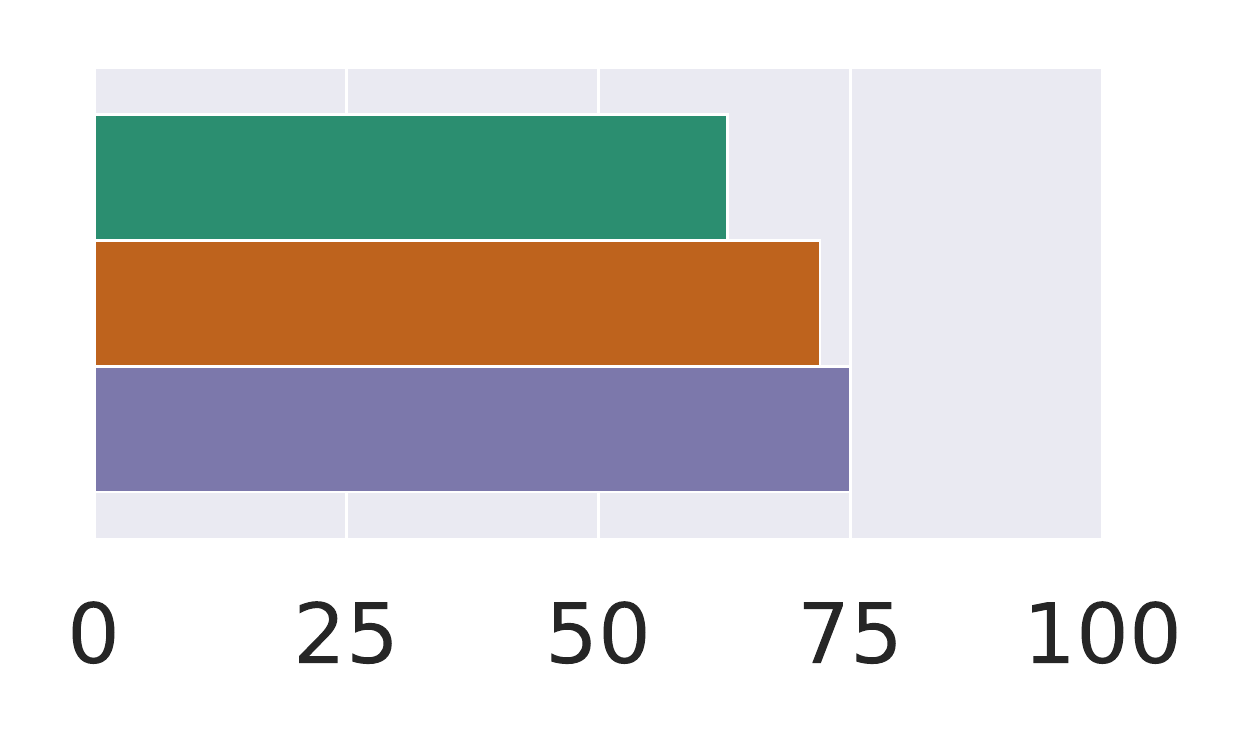}\end{minipage}
 & \begin{minipage}{.12\textwidth}\includegraphics[width=\textwidth]{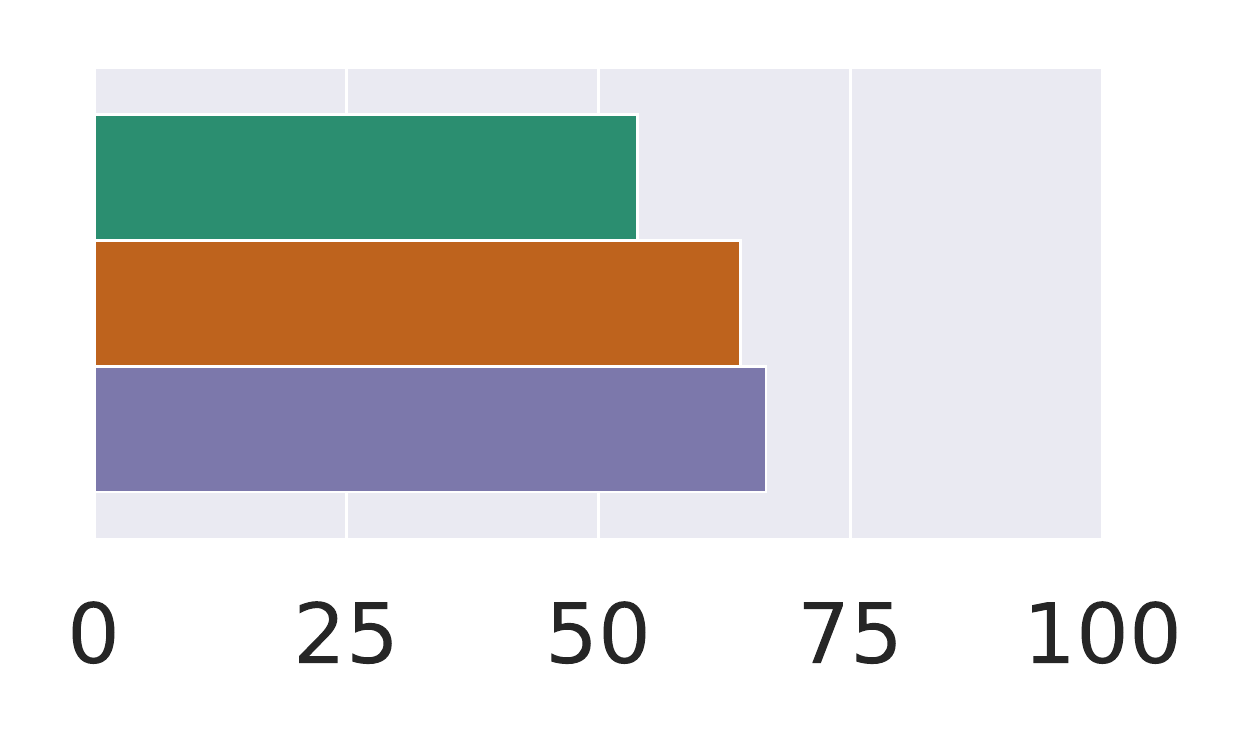}\end{minipage}
 & \begin{minipage}{.12\textwidth}\includegraphics[width=\textwidth]{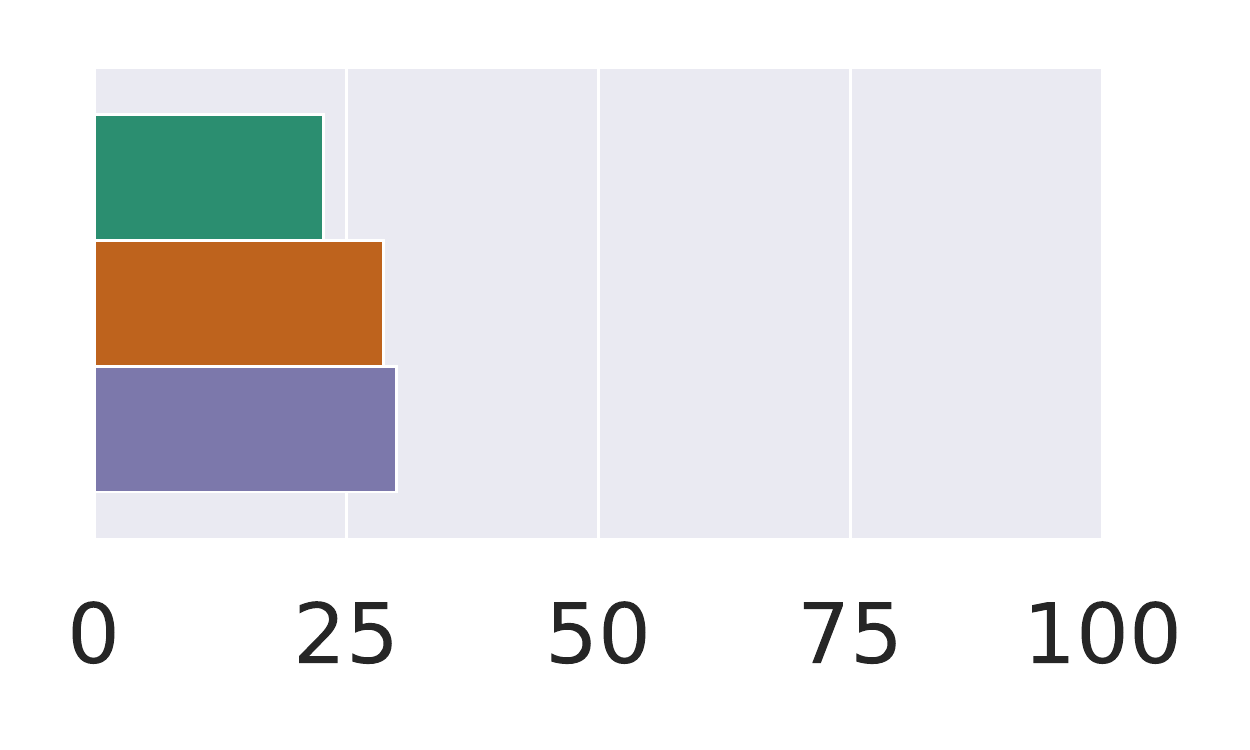}\end{minipage}
 & \begin{minipage}{.12\textwidth}\includegraphics[width=\textwidth]{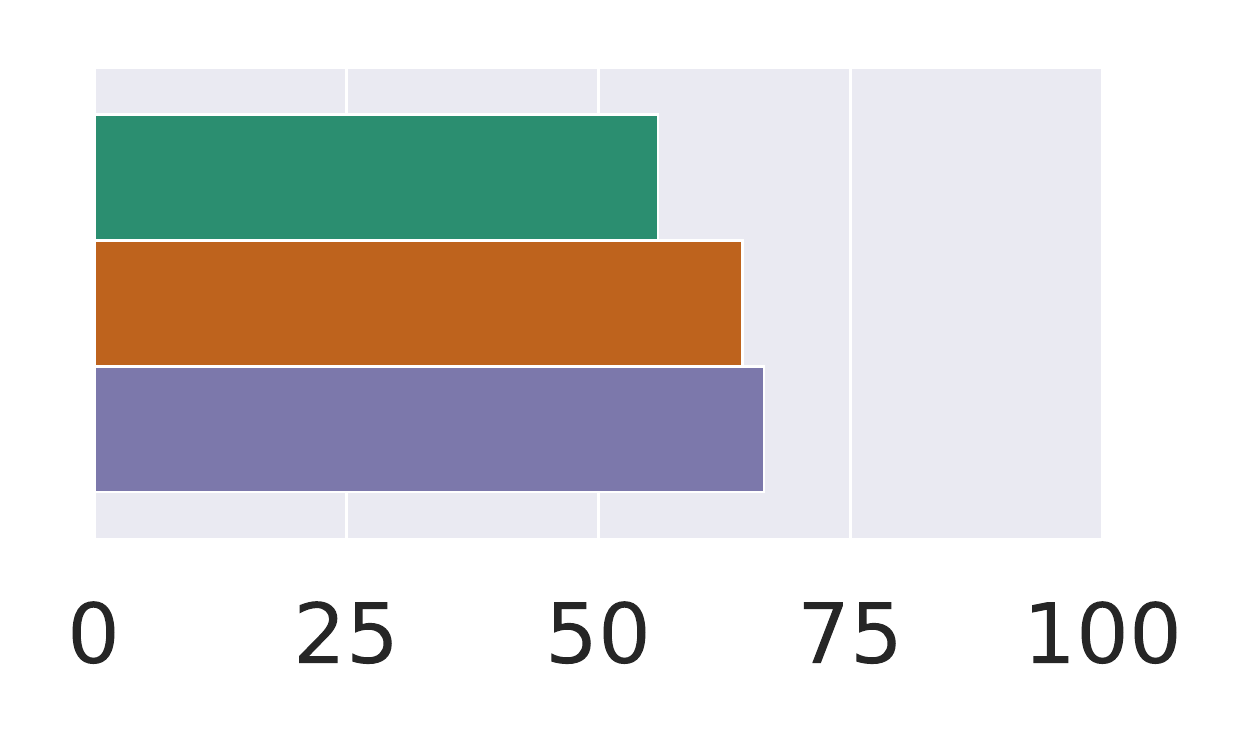}\end{minipage}
\\
\UseMacro{TH-metric-exact_match}
 & \begin{minipage}{.12\textwidth}\includegraphics[width=\textwidth]{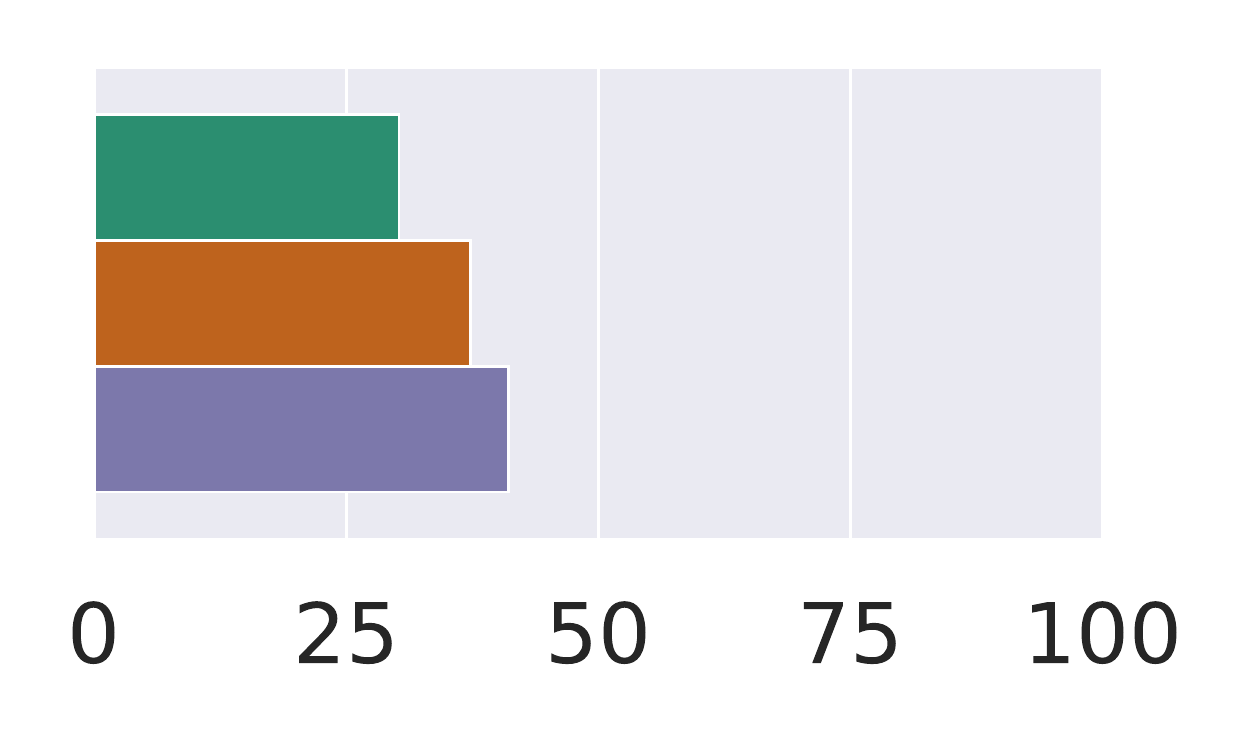}\end{minipage}
 & \begin{minipage}{.12\textwidth}\includegraphics[width=\textwidth]{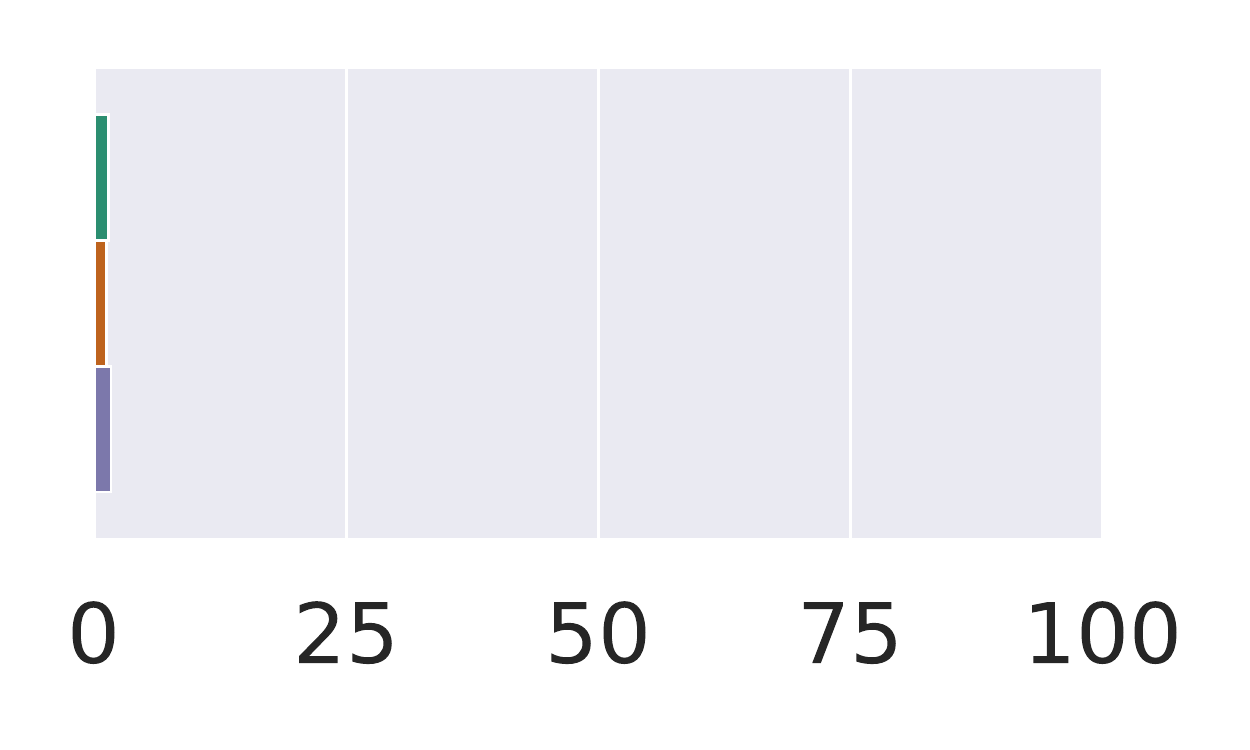}\end{minipage}
 & \begin{minipage}{.12\textwidth}\includegraphics[width=\textwidth]{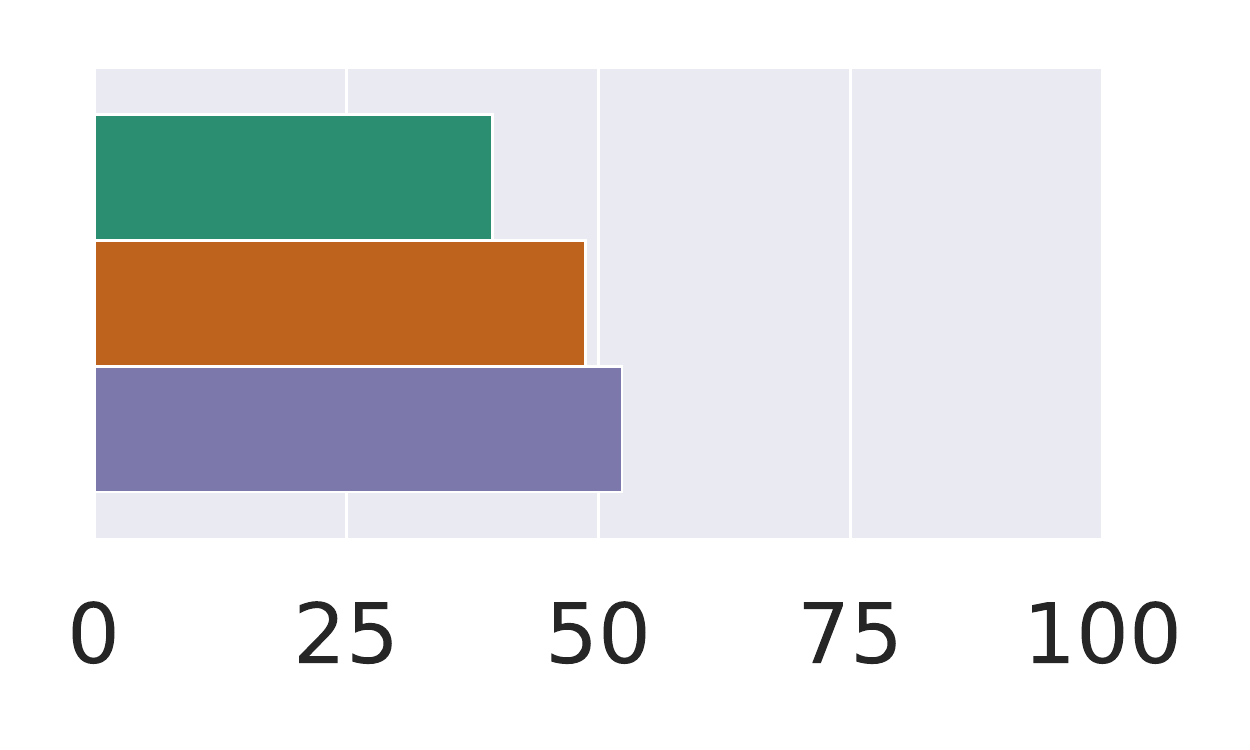}\end{minipage}
 & \begin{minipage}{.12\textwidth}\includegraphics[width=\textwidth]{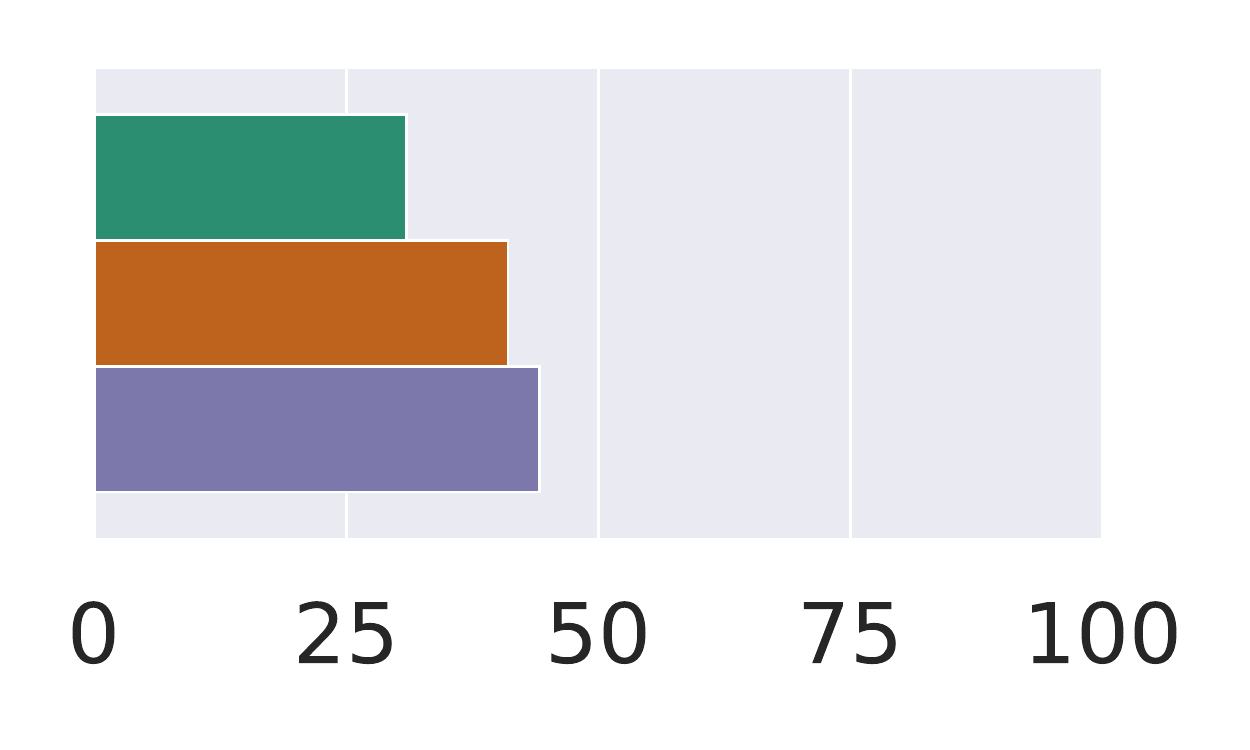}\end{minipage}
 & \begin{minipage}{.12\textwidth}\includegraphics[width=\textwidth]{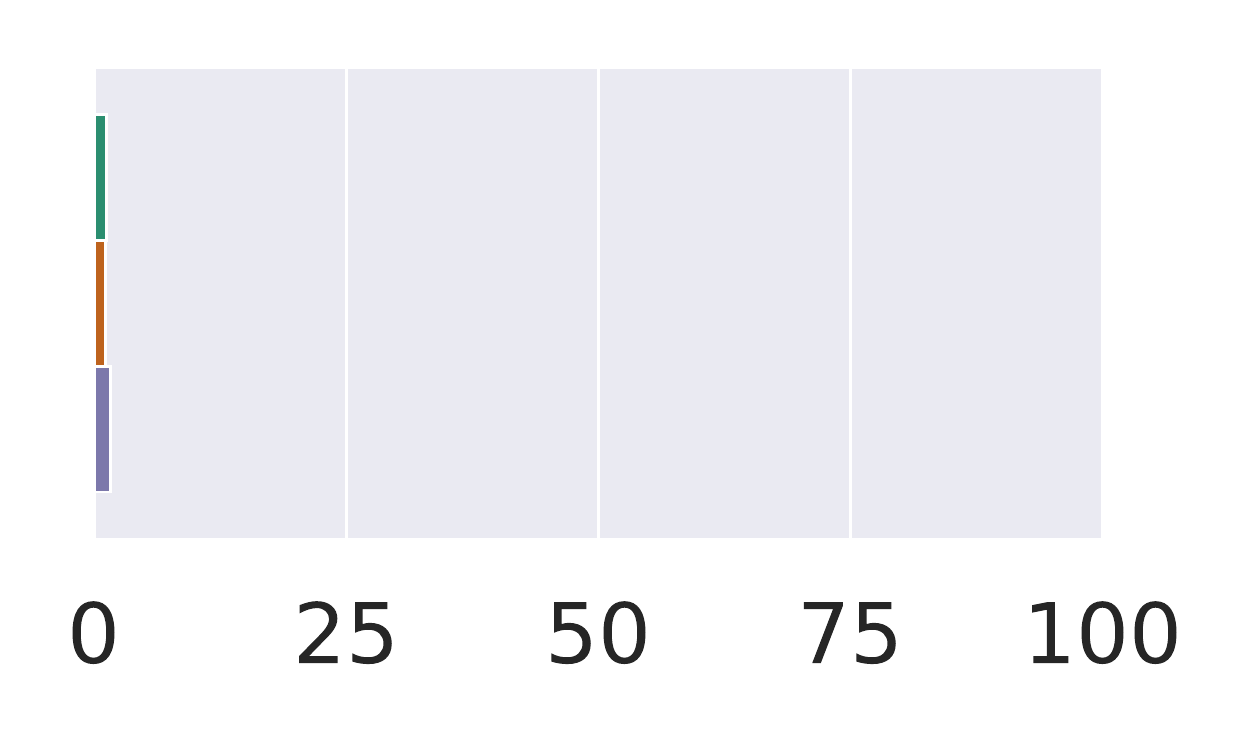}\end{minipage}
 & \begin{minipage}{.12\textwidth}\includegraphics[width=\textwidth]{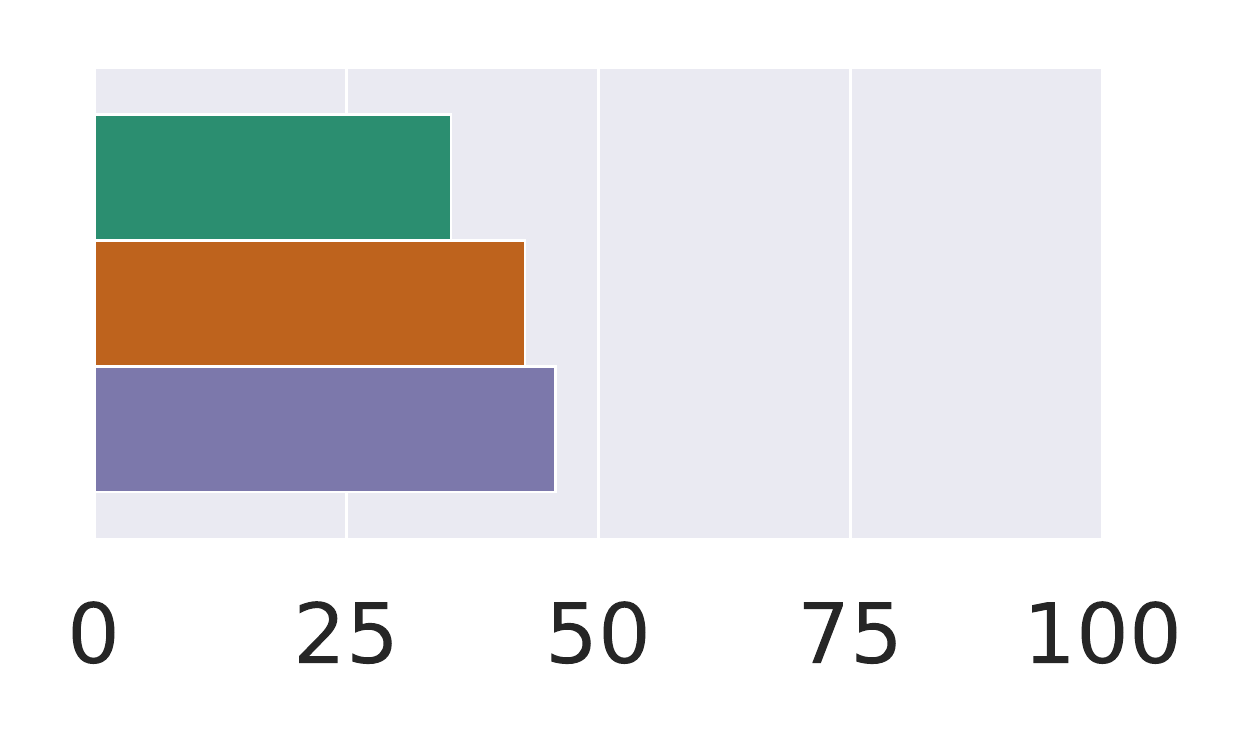}\end{minipage}
\\
\hline
\multicolumn{7}{c}{\includegraphics[scale=0.3]{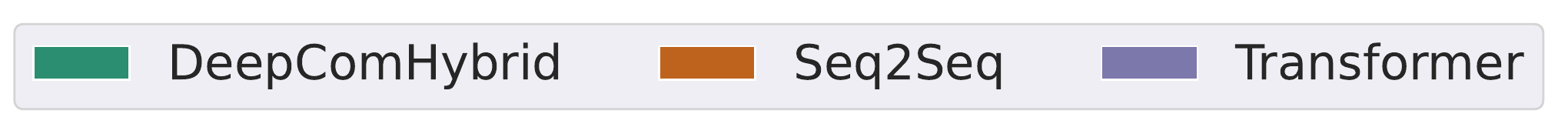}} \\
\end{tabular}
\end{footnotesize}
\end{center}

\vspace{-15pt}
\caption{Results of \comgen models on \testc sets.\label{fig:results-cg}}
\end{figure*}

\begin{figure*}

\begin{center}
\begin{footnotesize}
\begin{tabular}{|r|c|c|c|c|c|c|}
\hline
\makecell[c]{\UseMacro{TH-train-on}}
 & \multicolumn{2}{c|}{\UseMacro{TH-MP}}
 & \multicolumn{2}{c|}{\UseMacro{TH-CP}}
 & \multicolumn{2}{c|}{\UseMacro{TH-T}}
\\
\hline
\makecell[c]{\UseMacro{TH-test-on}}
 & \UseMacro{TH-val}
 & \UseMacro{TH-test_standard}
 & \UseMacro{TH-val}
 & \UseMacro{TH-test_standard}
 & \UseMacro{TH-val}
 & \UseMacro{TH-test_standard}
\\
\hline
\UseMacro{TH-metric-bleu}
 & \begin{minipage}{.12\textwidth}\includegraphics[width=\textwidth]{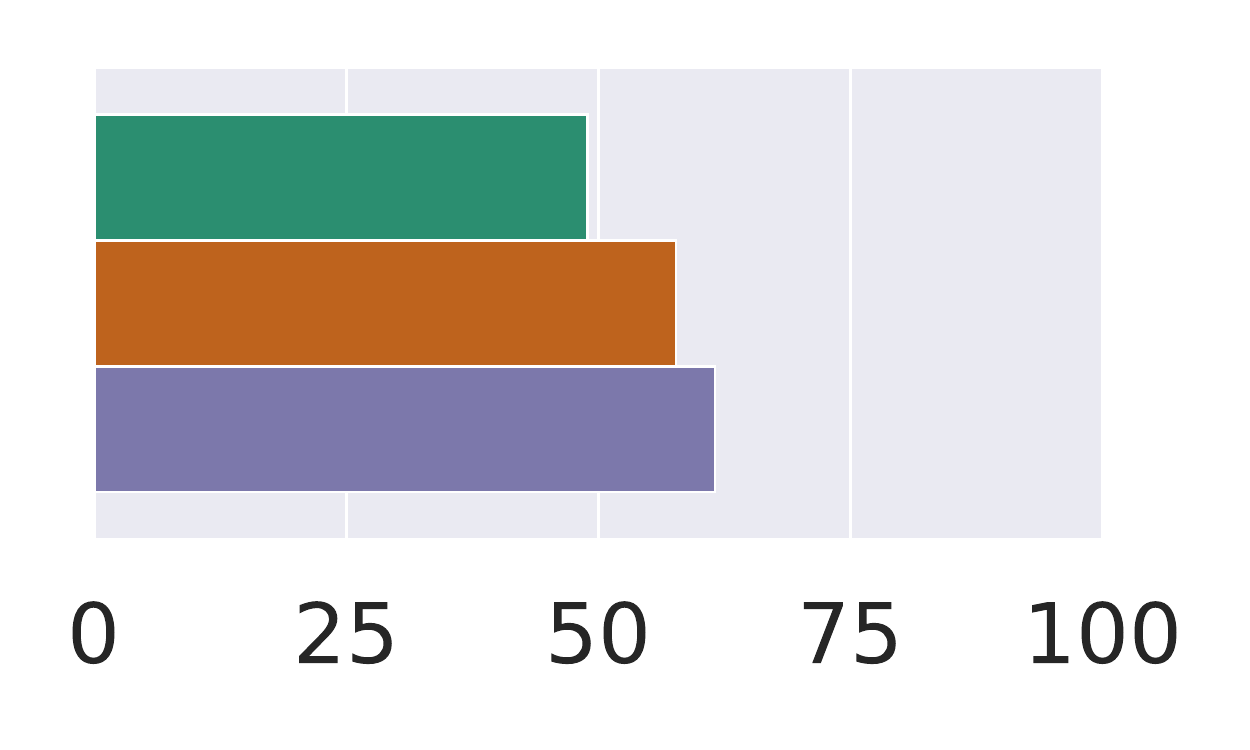}\end{minipage}
 & \begin{minipage}{.12\textwidth}\includegraphics[width=\textwidth]{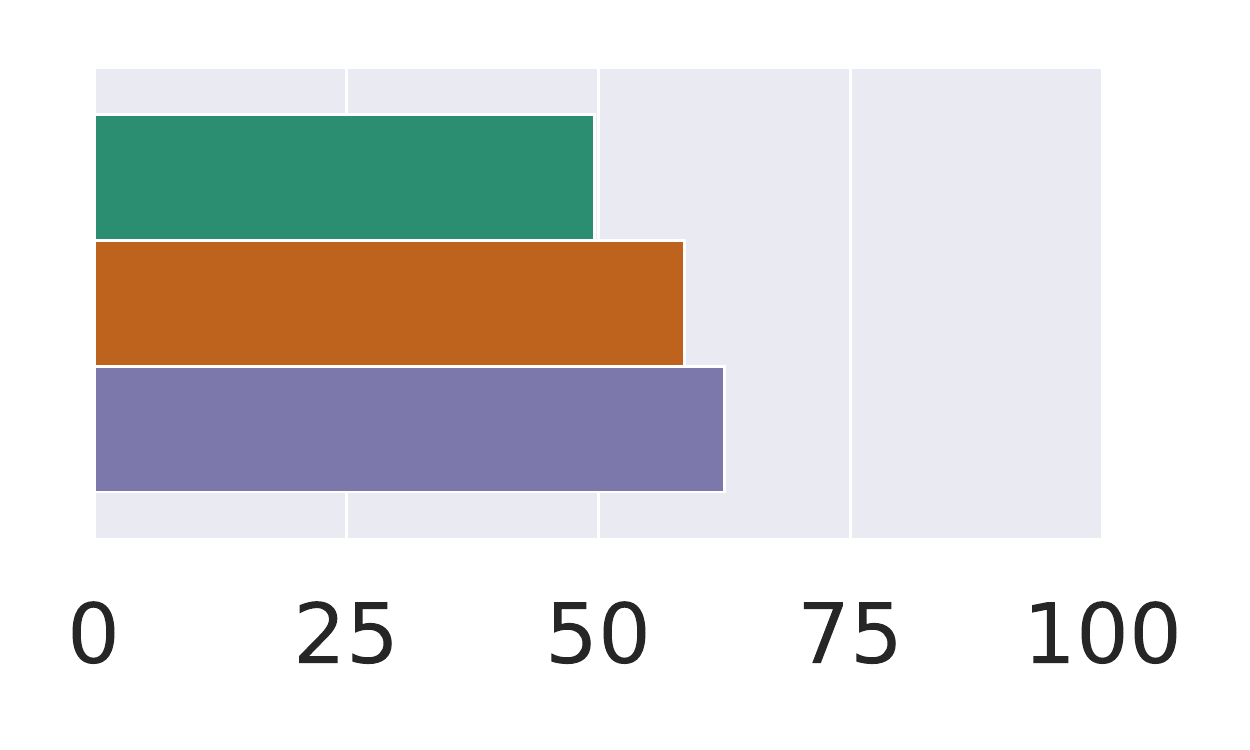}\end{minipage}
 & \begin{minipage}{.12\textwidth}\includegraphics[width=\textwidth]{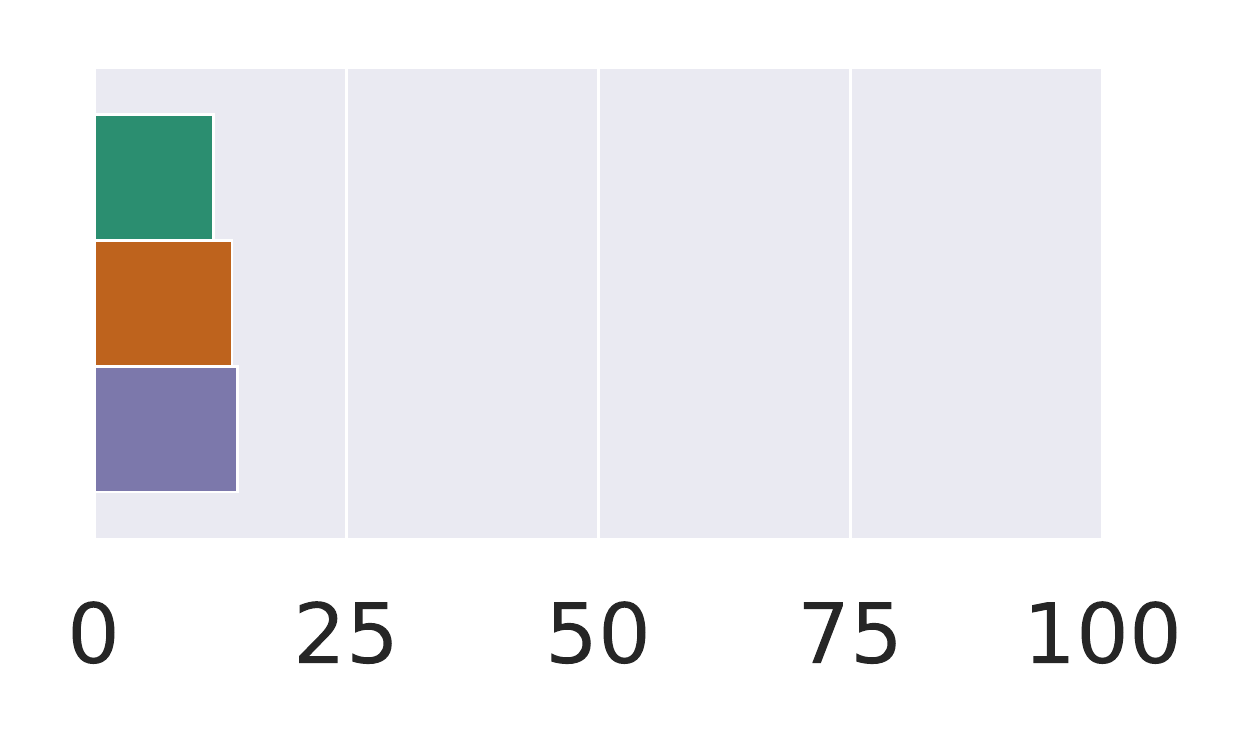}\end{minipage}
 & \begin{minipage}{.12\textwidth}\includegraphics[width=\textwidth]{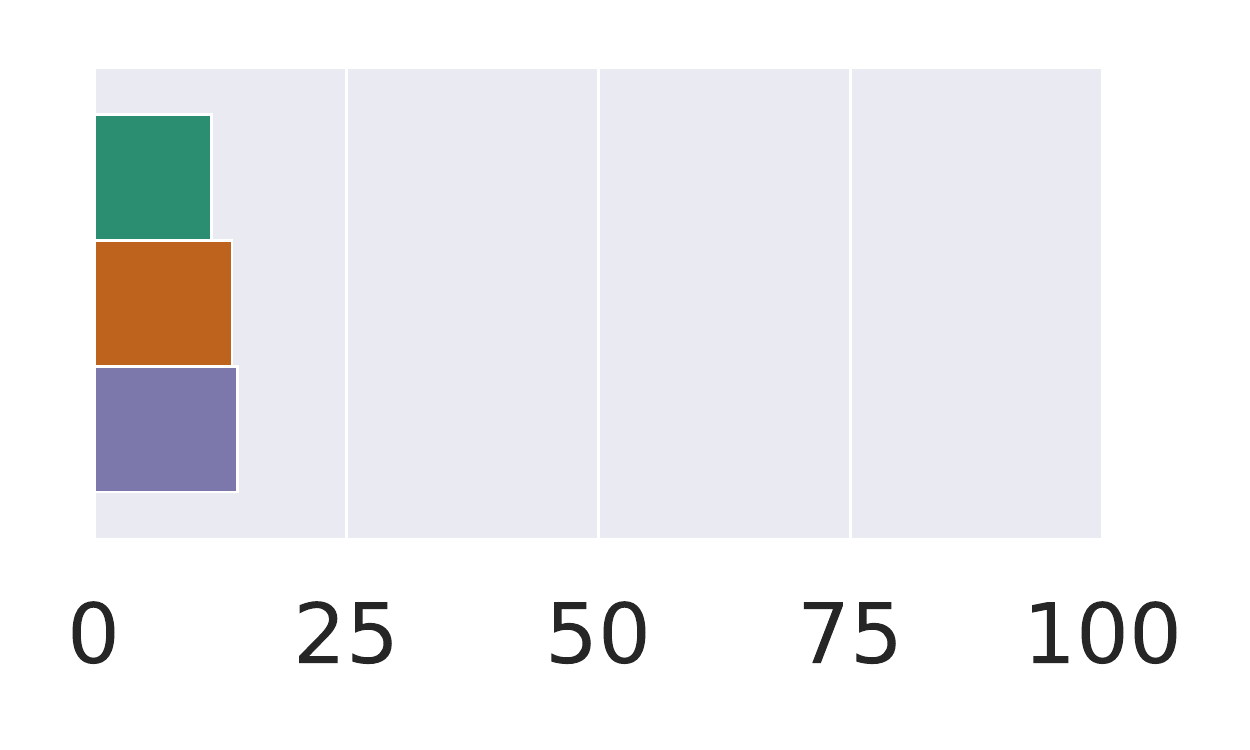}\end{minipage}
 & \begin{minipage}{.12\textwidth}\includegraphics[width=\textwidth]{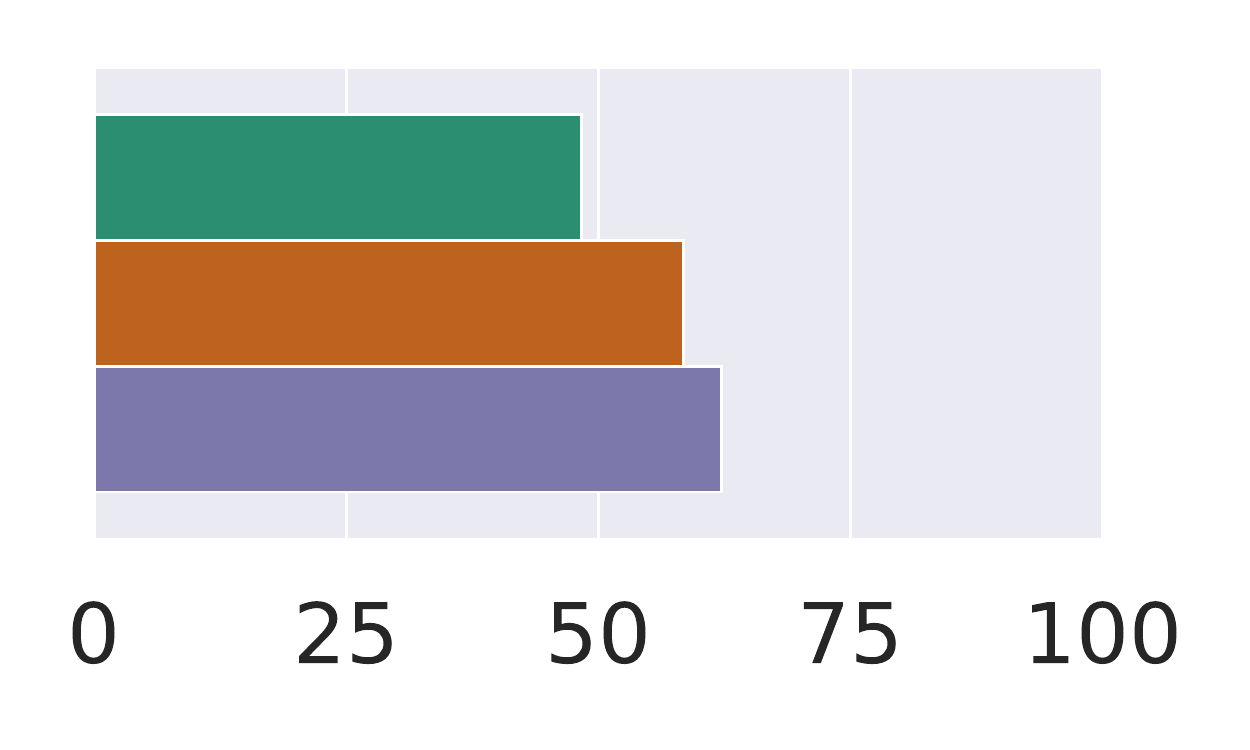}\end{minipage}
 & \begin{minipage}{.12\textwidth}\includegraphics[width=\textwidth]{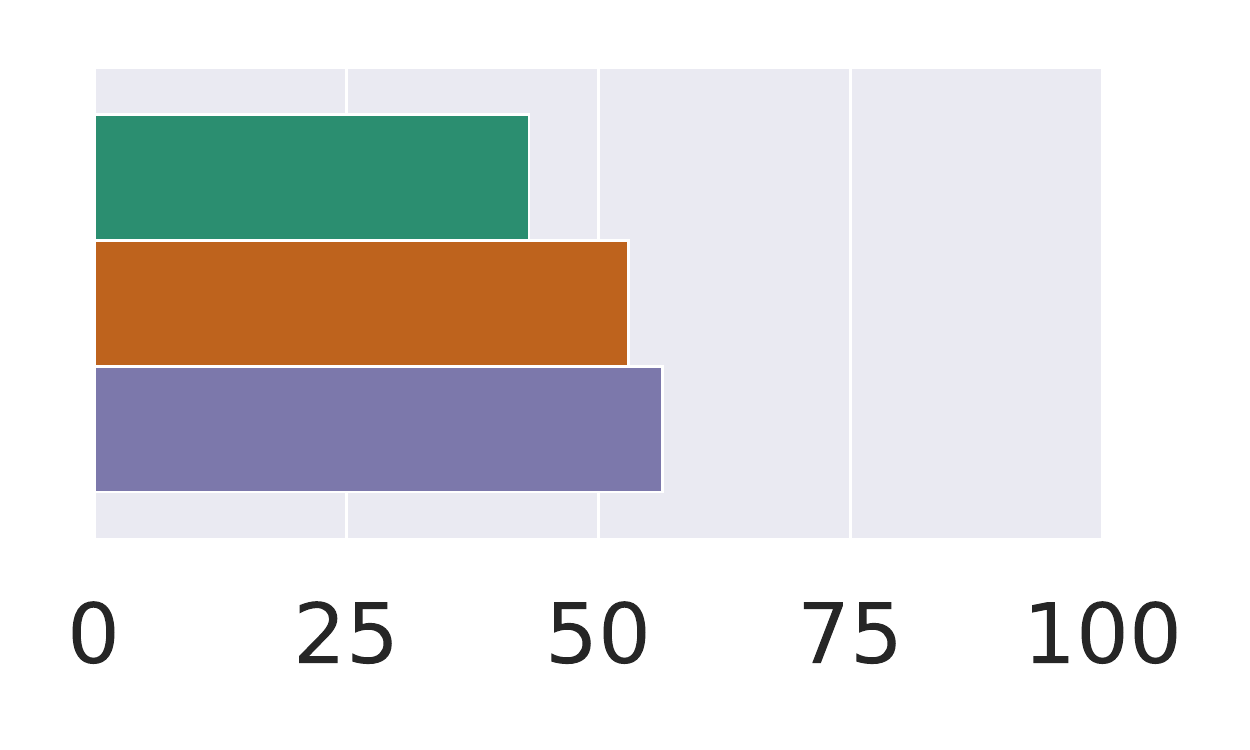}\end{minipage}
\\
\UseMacro{TH-metric-meteor}
 & \begin{minipage}{.12\textwidth}\includegraphics[width=\textwidth]{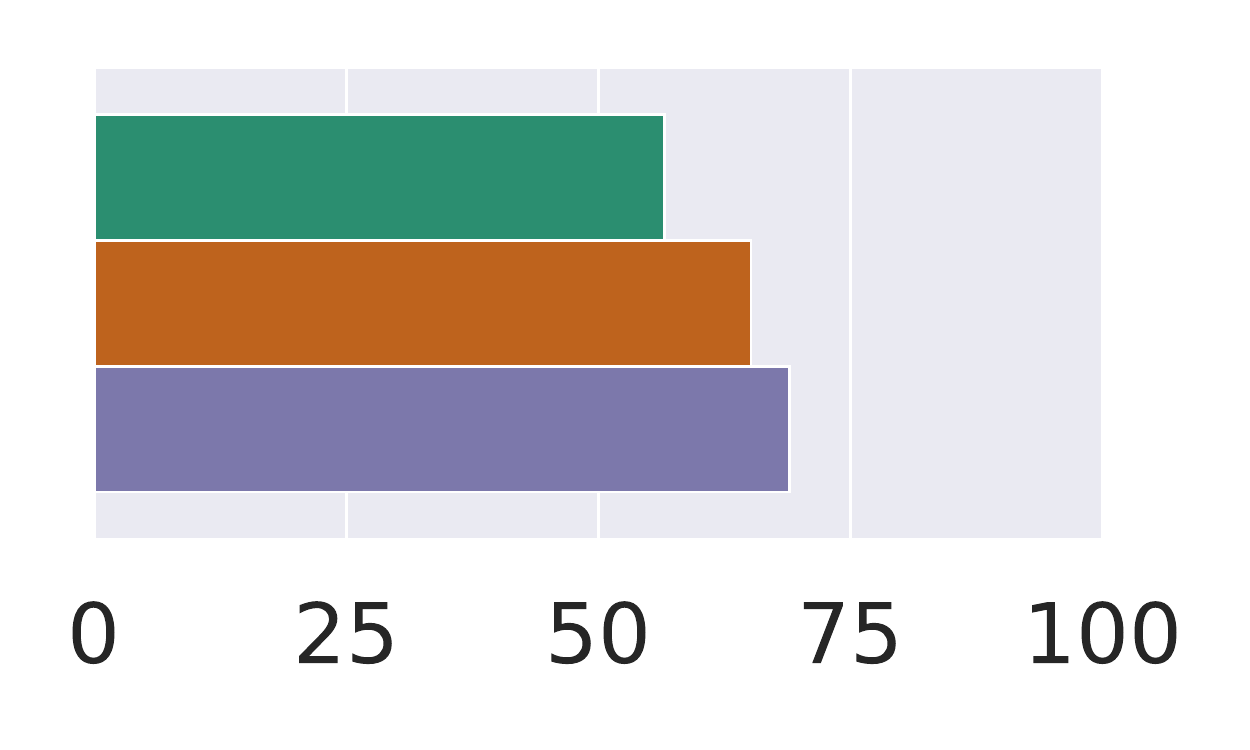}\end{minipage}
 & \begin{minipage}{.12\textwidth}\includegraphics[width=\textwidth]{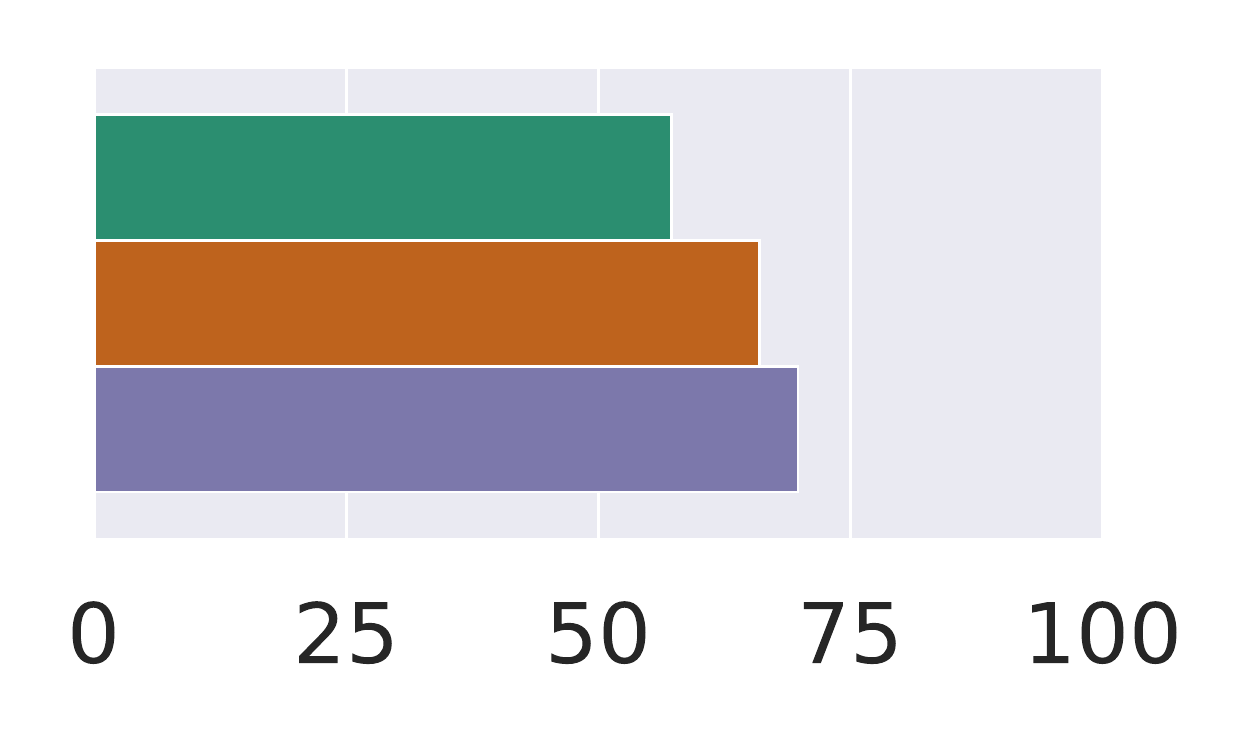}\end{minipage}
 & \begin{minipage}{.12\textwidth}\includegraphics[width=\textwidth]{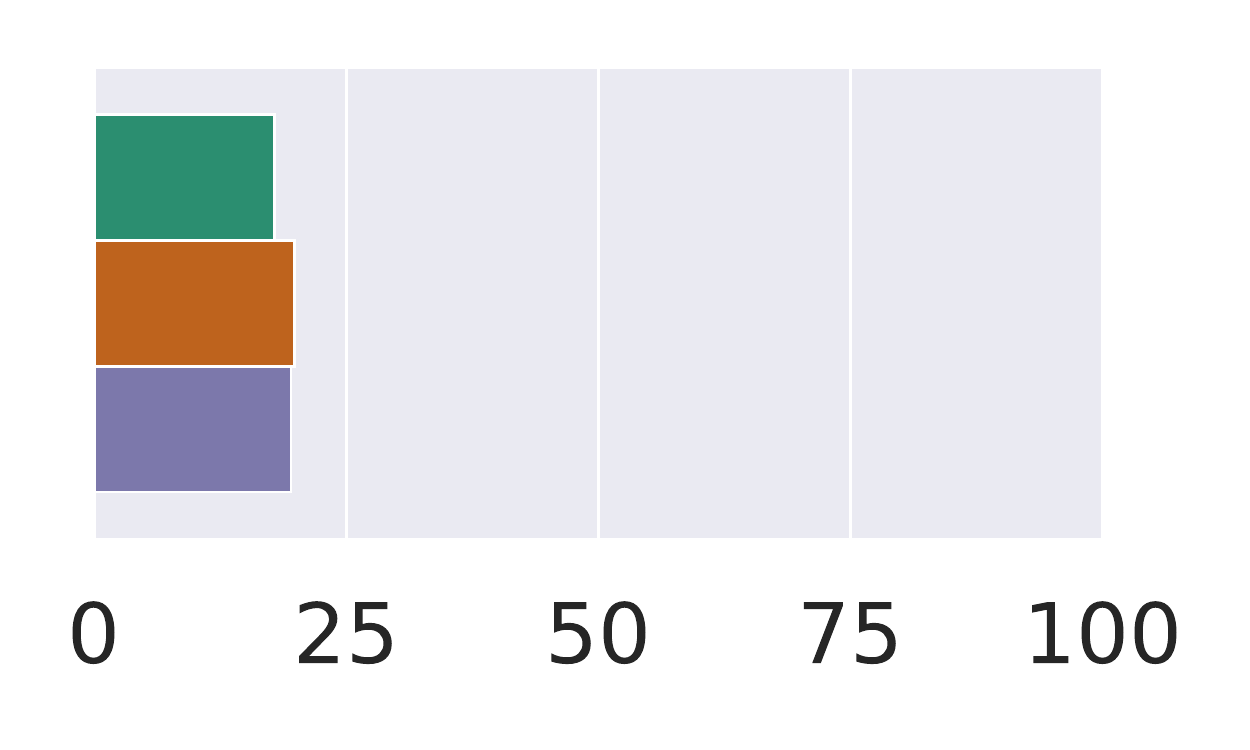}\end{minipage}
 & \begin{minipage}{.12\textwidth}\includegraphics[width=\textwidth]{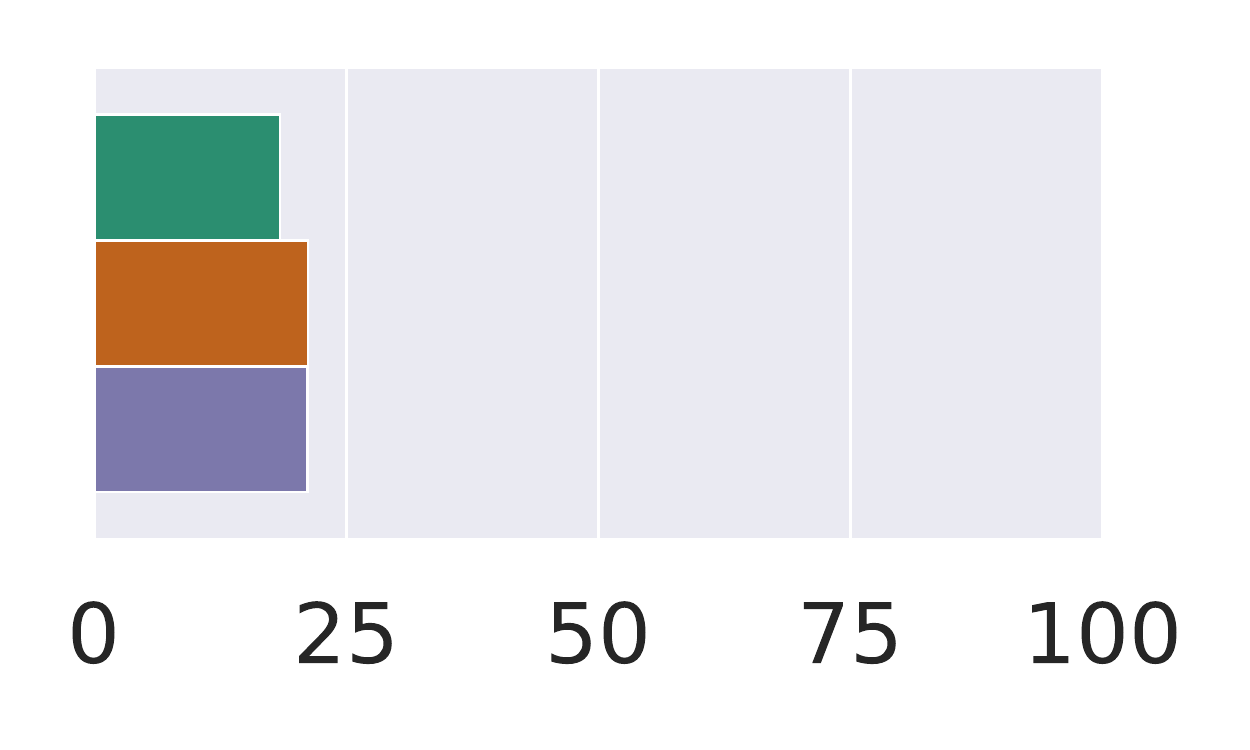}\end{minipage}
 & \begin{minipage}{.12\textwidth}\includegraphics[width=\textwidth]{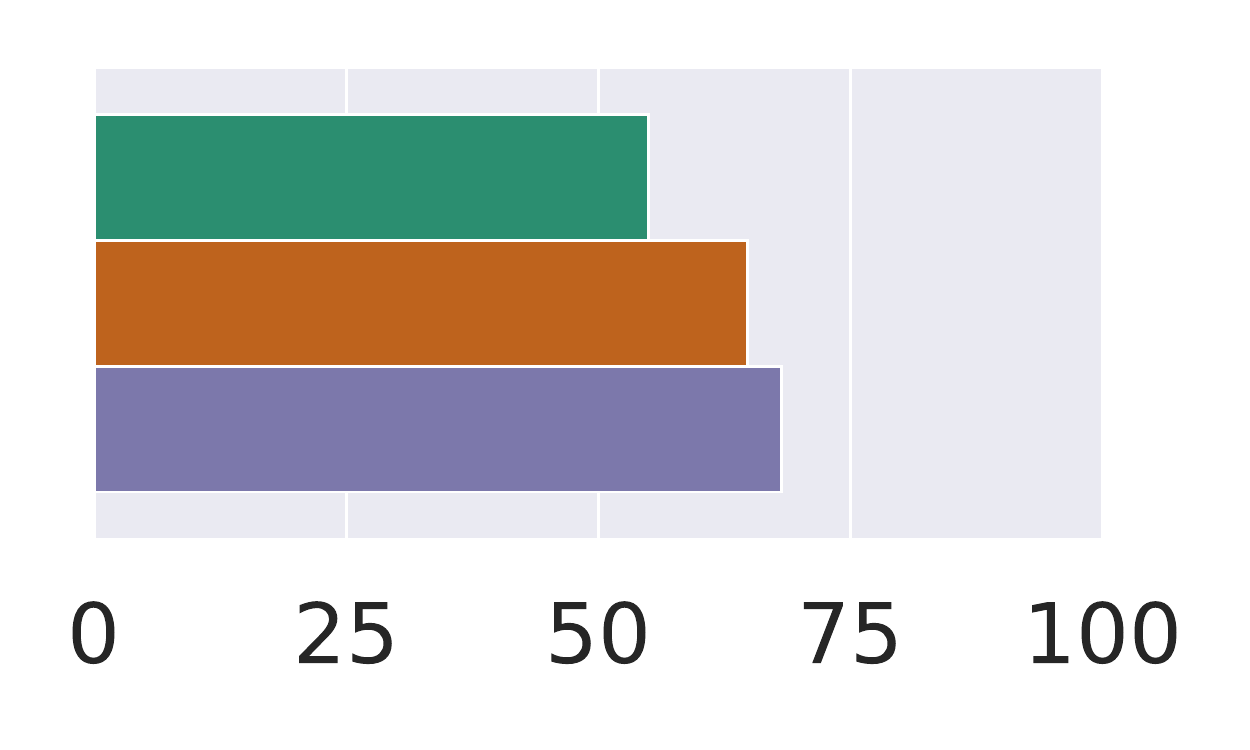}\end{minipage}
 & \begin{minipage}{.12\textwidth}\includegraphics[width=\textwidth]{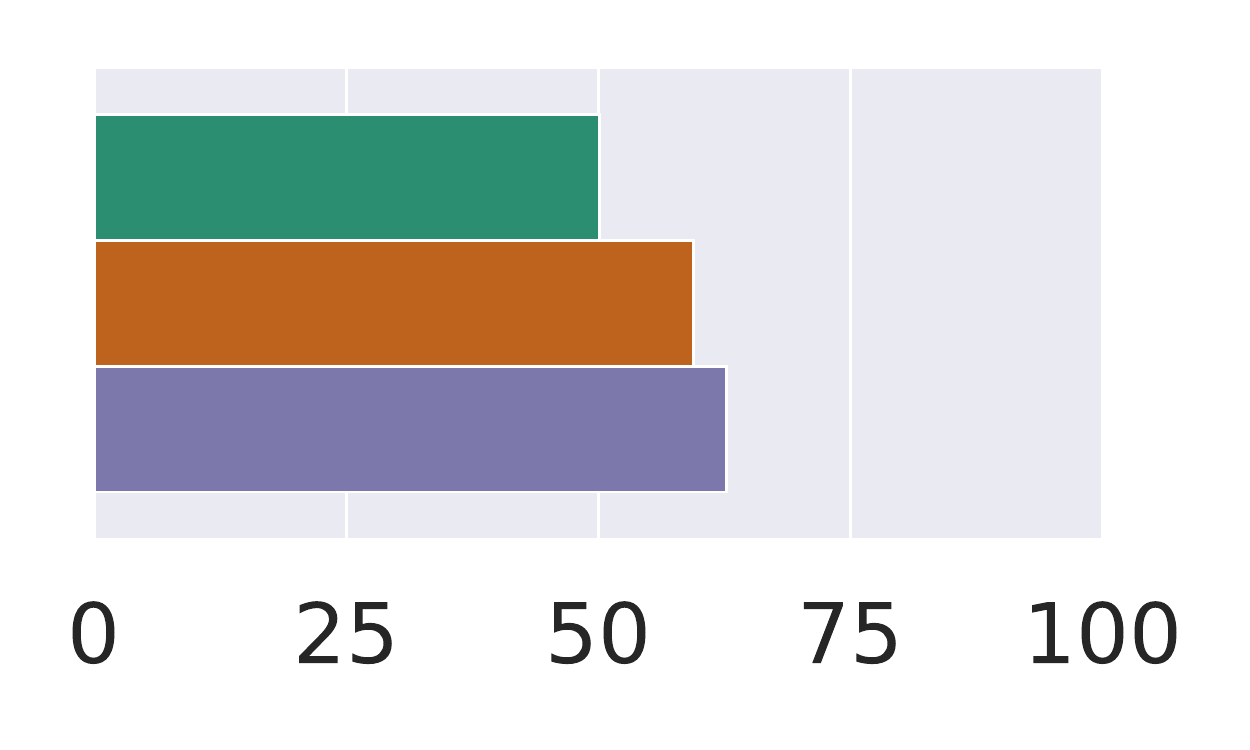}\end{minipage}
\\
\UseMacro{TH-metric-rouge_l_f}
 & \begin{minipage}{.12\textwidth}\includegraphics[width=\textwidth]{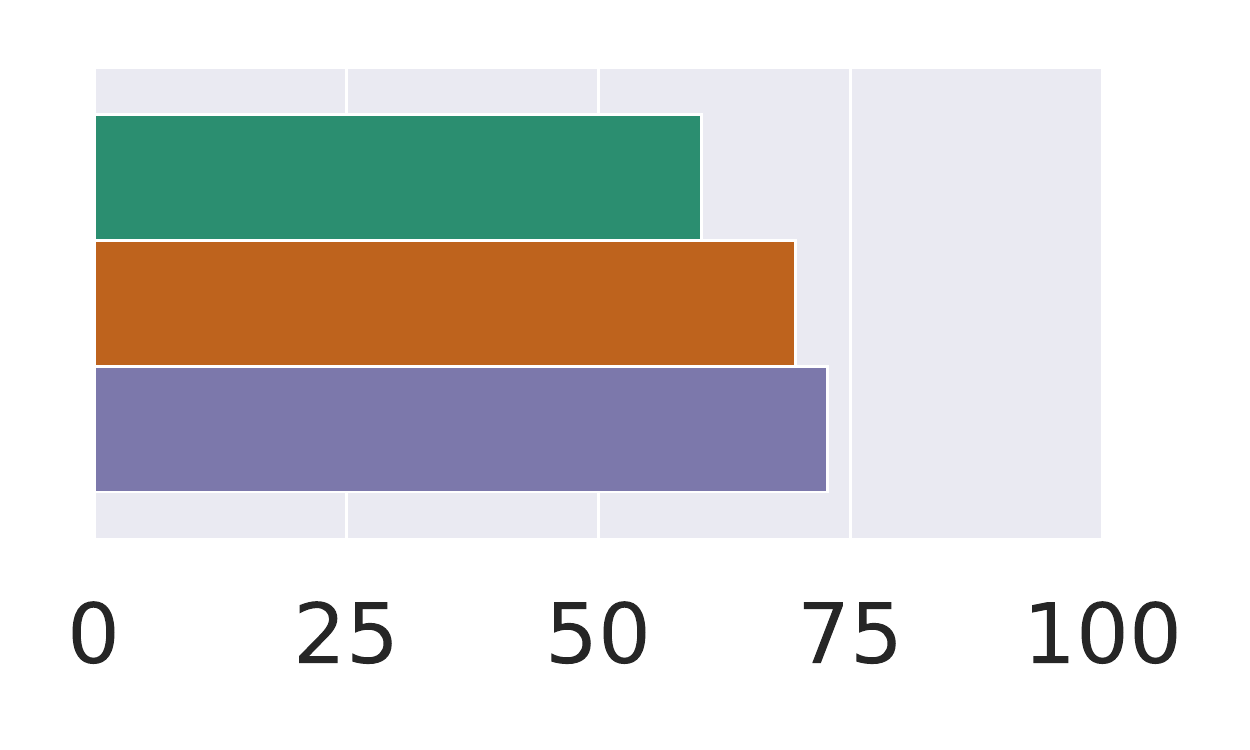}\end{minipage}
 & \begin{minipage}{.12\textwidth}\includegraphics[width=\textwidth]{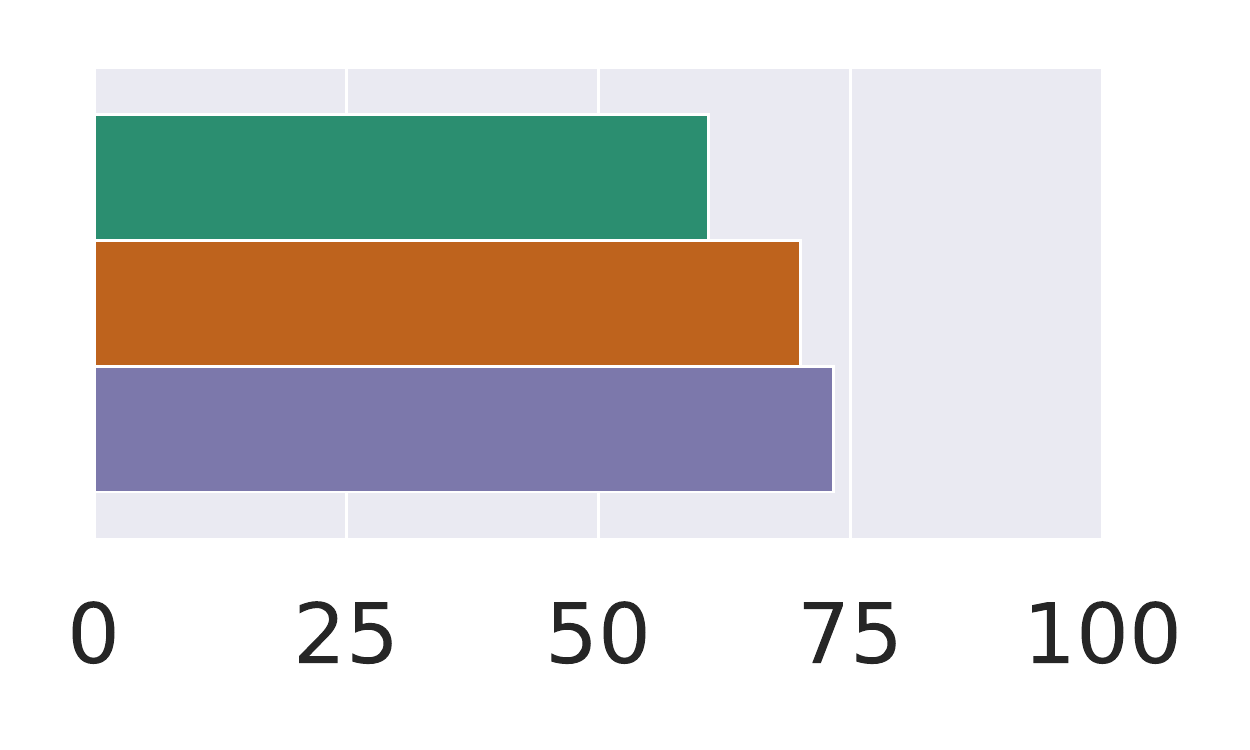}\end{minipage}
 & \begin{minipage}{.12\textwidth}\includegraphics[width=\textwidth]{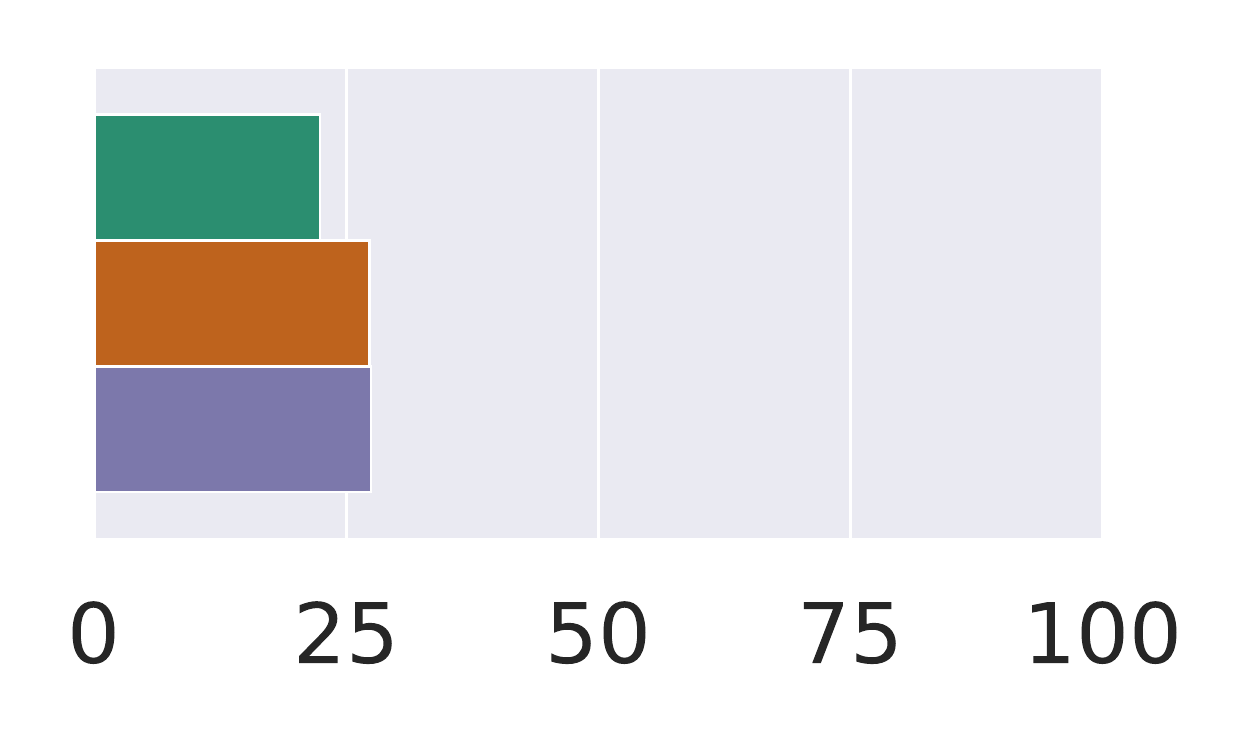}\end{minipage}
 & \begin{minipage}{.12\textwidth}\includegraphics[width=\textwidth]{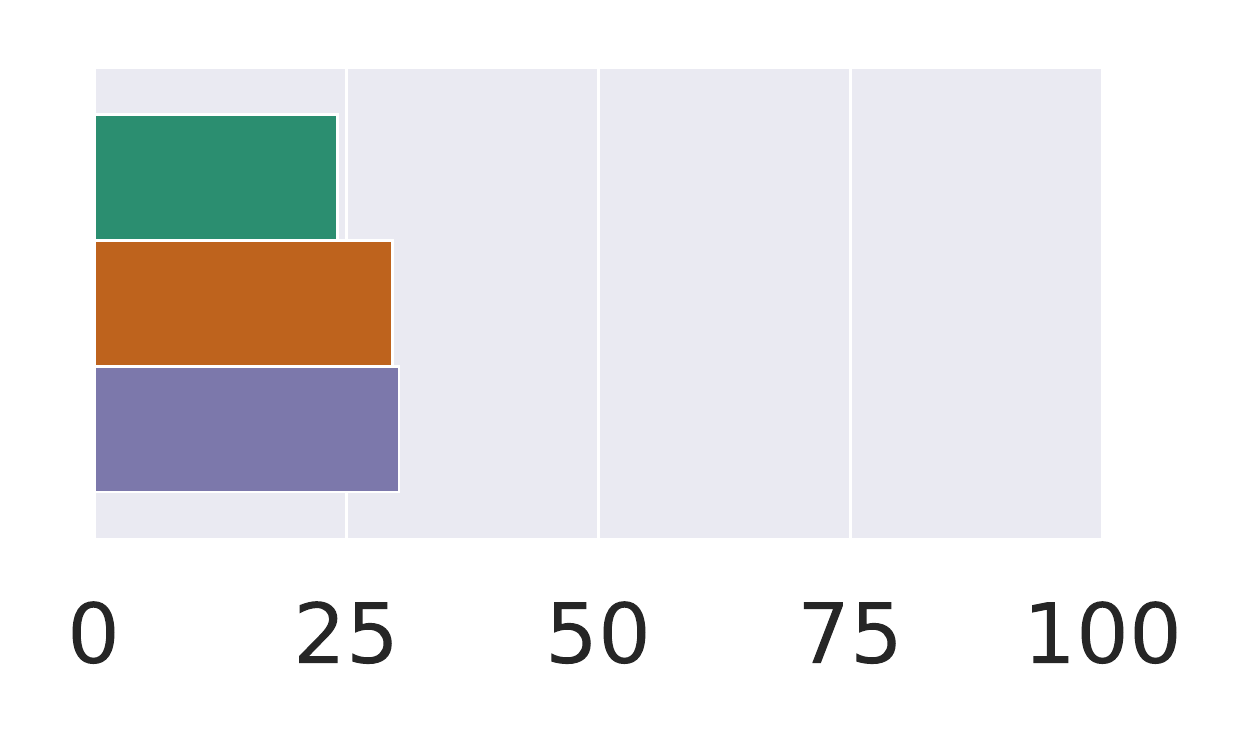}\end{minipage}
 & \begin{minipage}{.12\textwidth}\includegraphics[width=\textwidth]{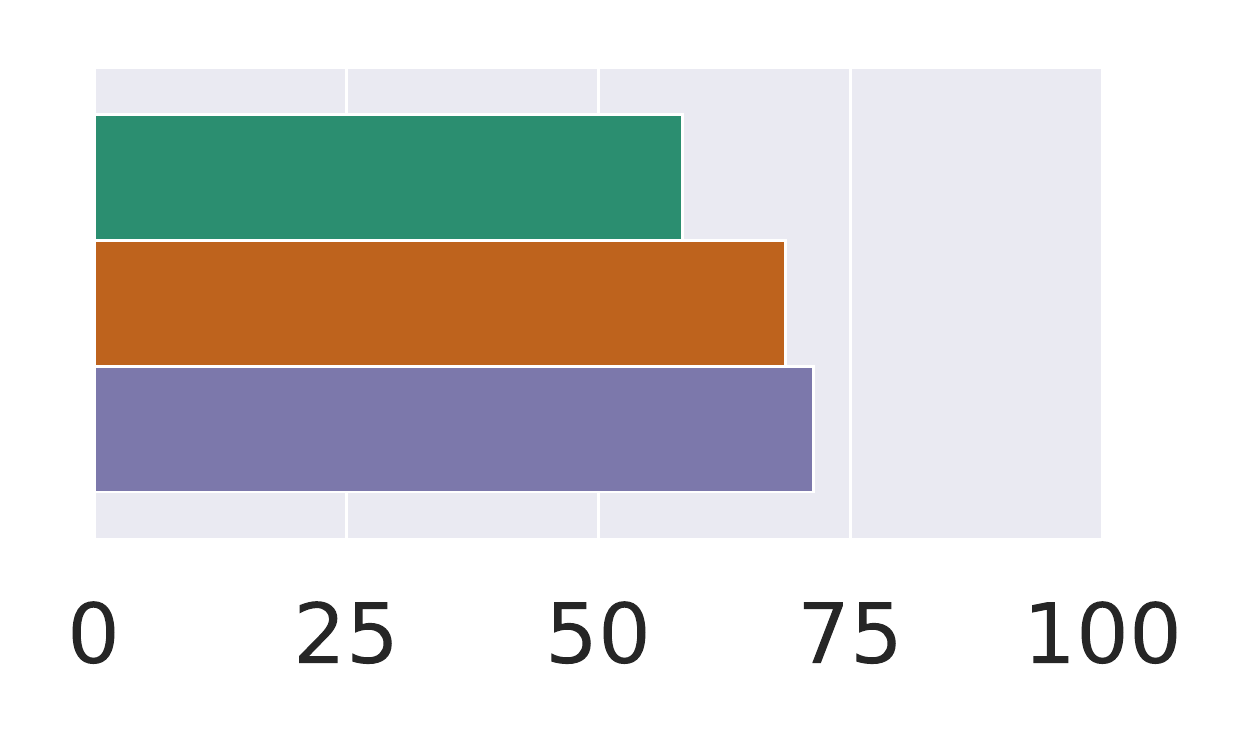}\end{minipage}
 & \begin{minipage}{.12\textwidth}\includegraphics[width=\textwidth]{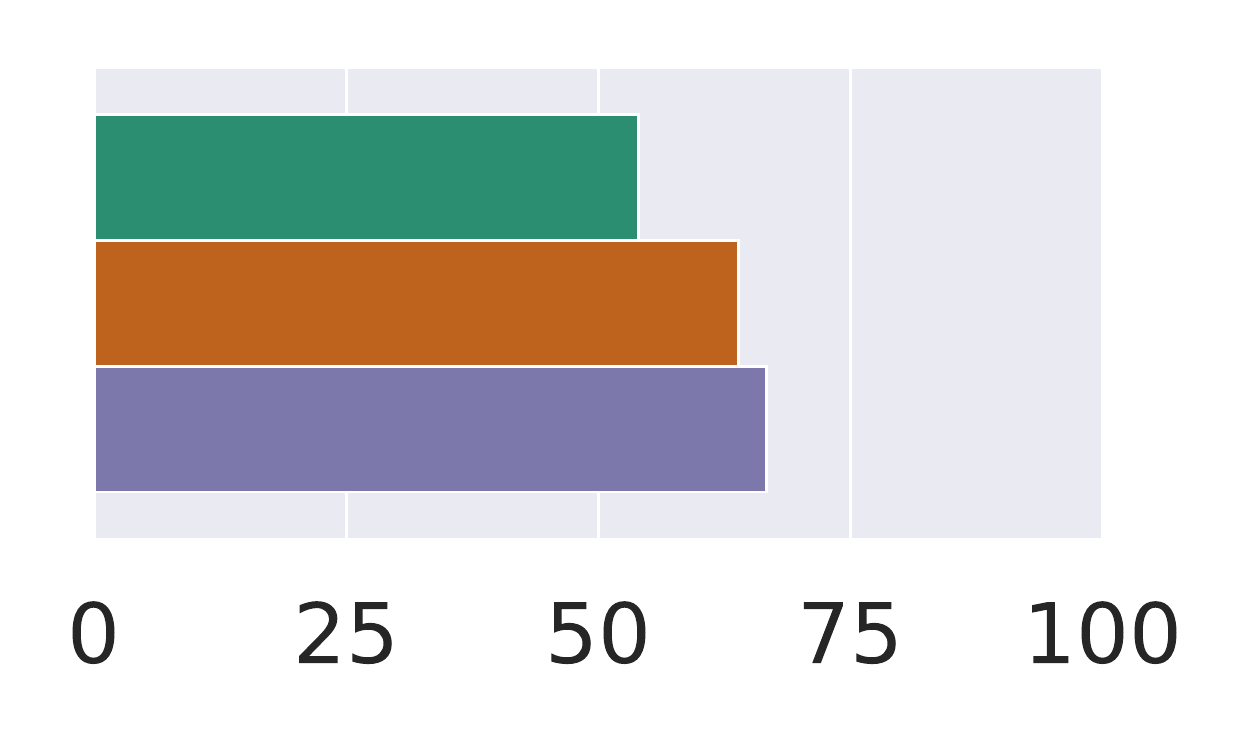}\end{minipage}
\\
\UseMacro{TH-metric-exact_match}
 & \begin{minipage}{.12\textwidth}\includegraphics[width=\textwidth]{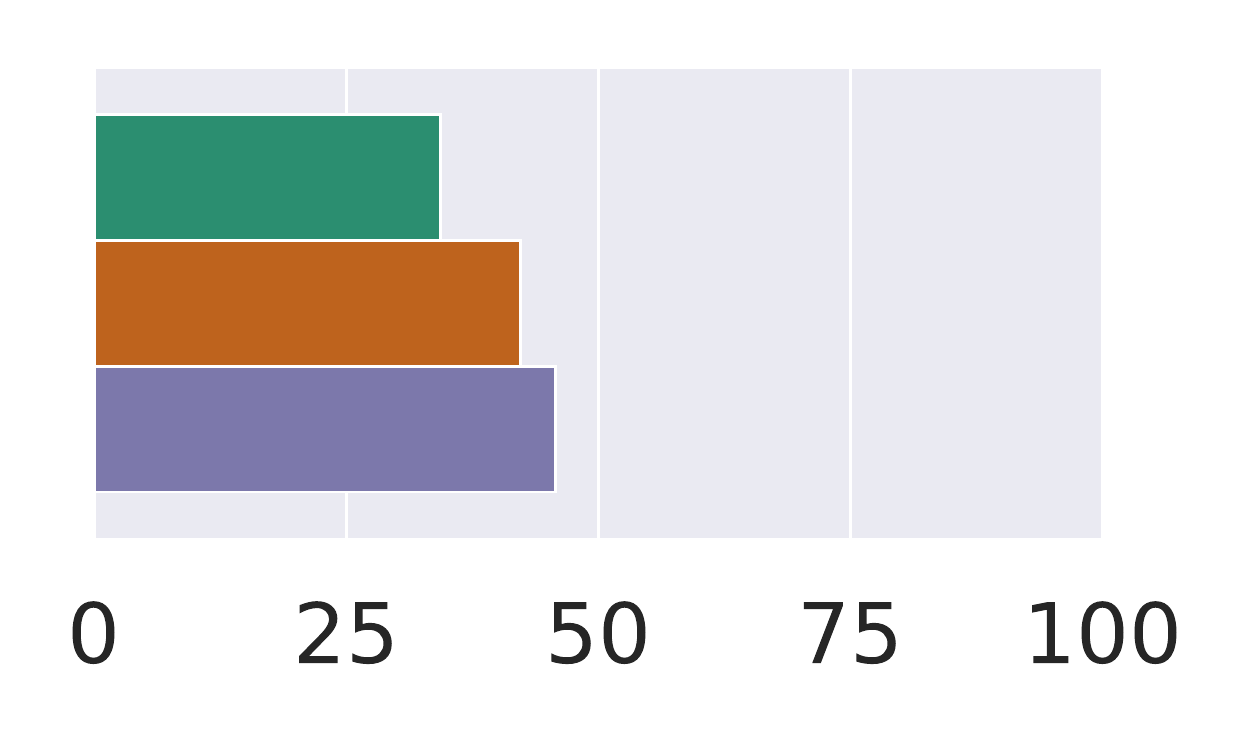}\end{minipage}
 & \begin{minipage}{.12\textwidth}\includegraphics[width=\textwidth]{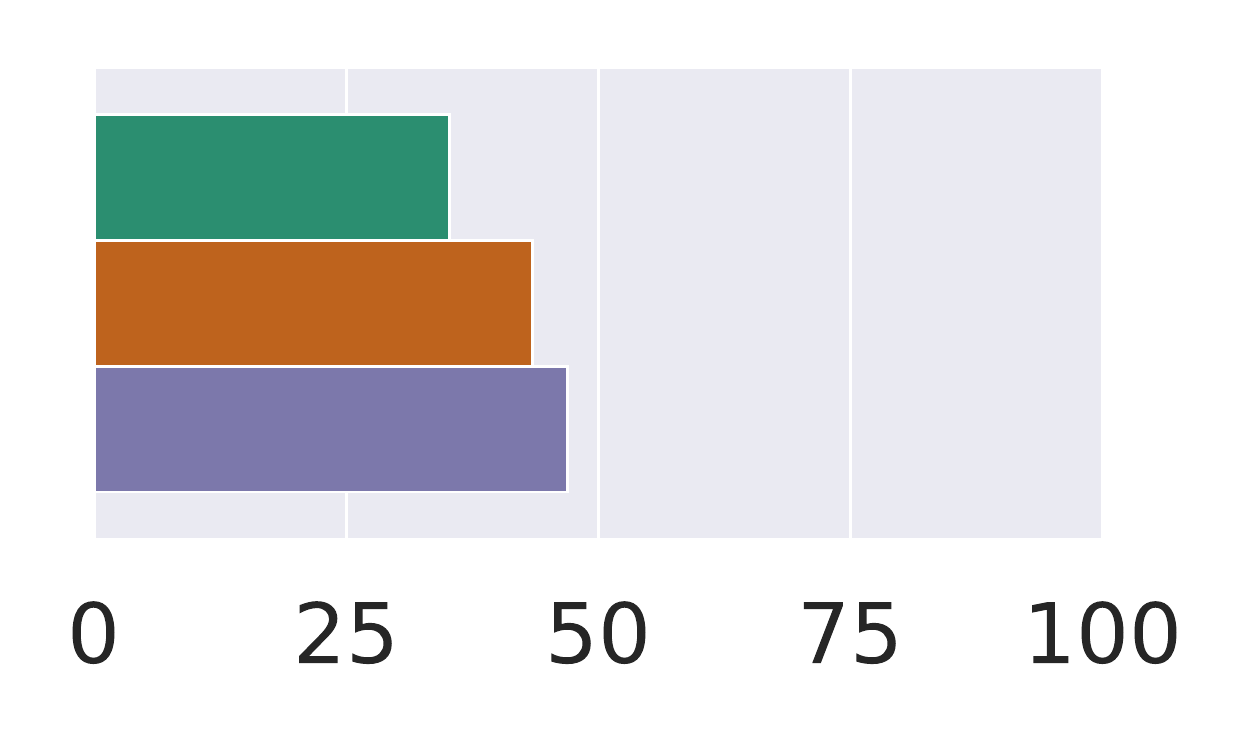}\end{minipage}
 & \begin{minipage}{.12\textwidth}\includegraphics[width=\textwidth]{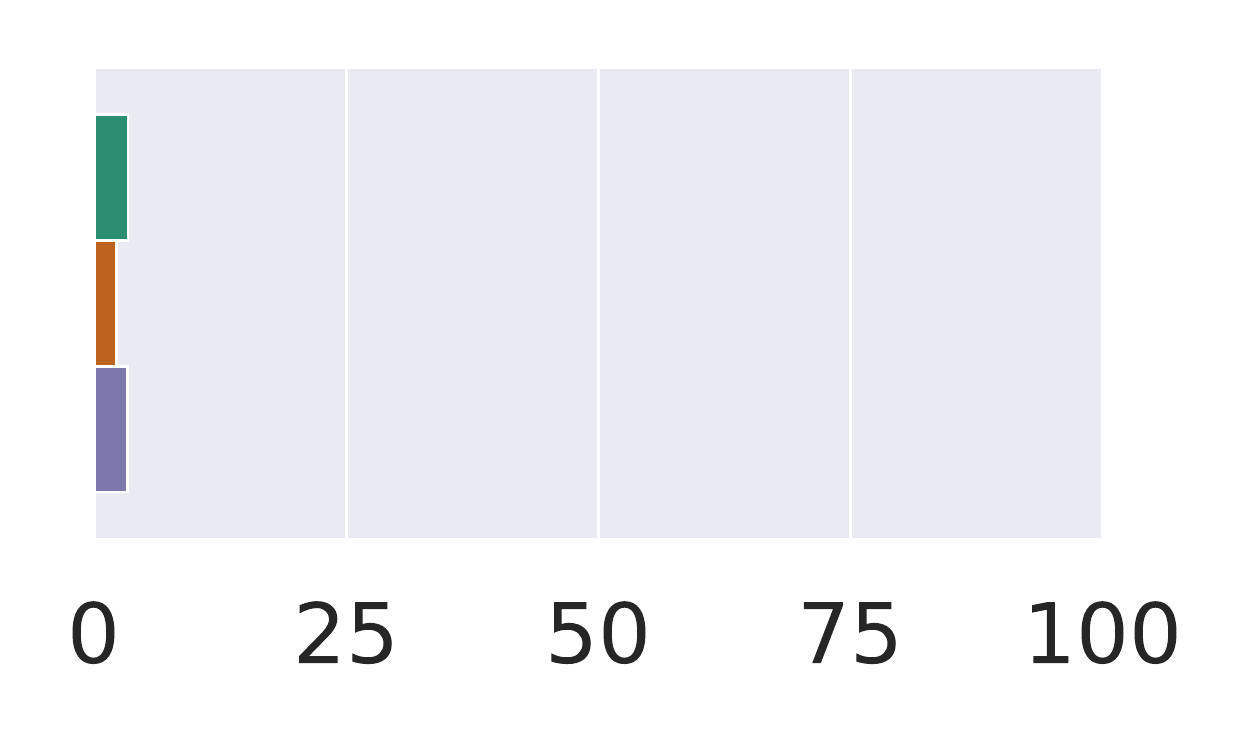}\end{minipage}
 & \begin{minipage}{.12\textwidth}\includegraphics[width=\textwidth]{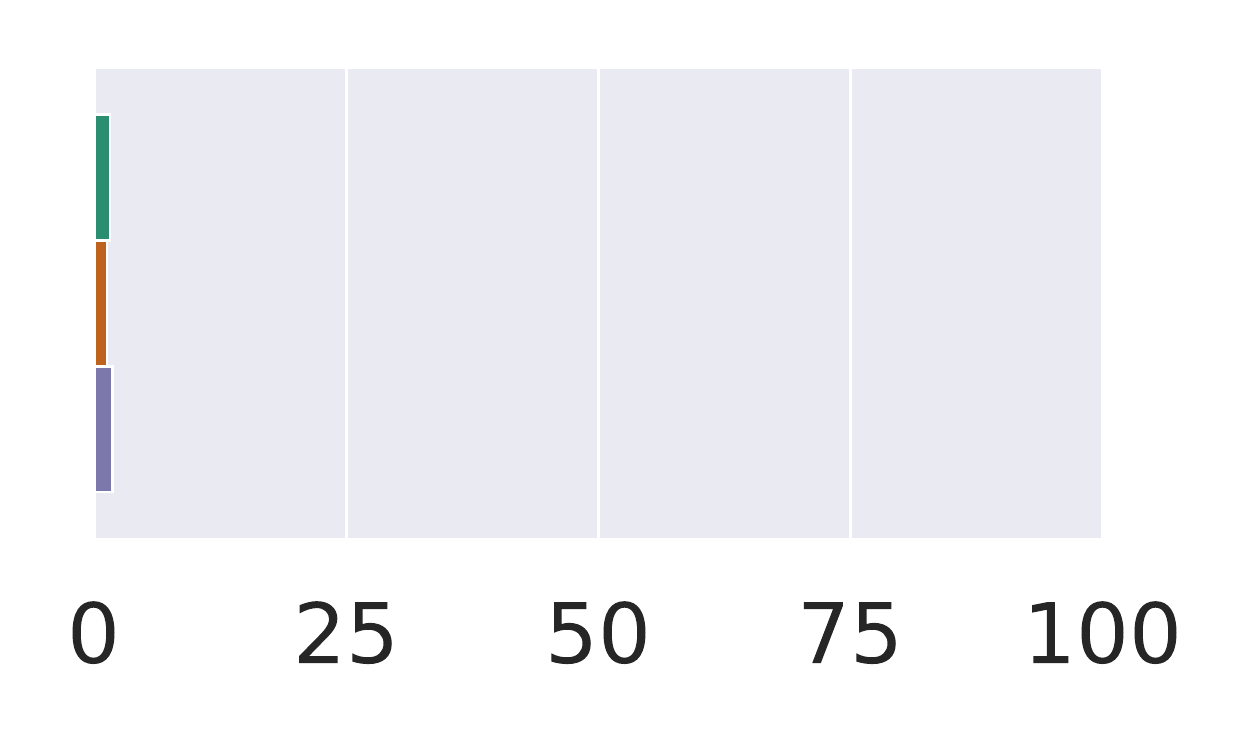}\end{minipage}
 & \begin{minipage}{.12\textwidth}\includegraphics[width=\textwidth]{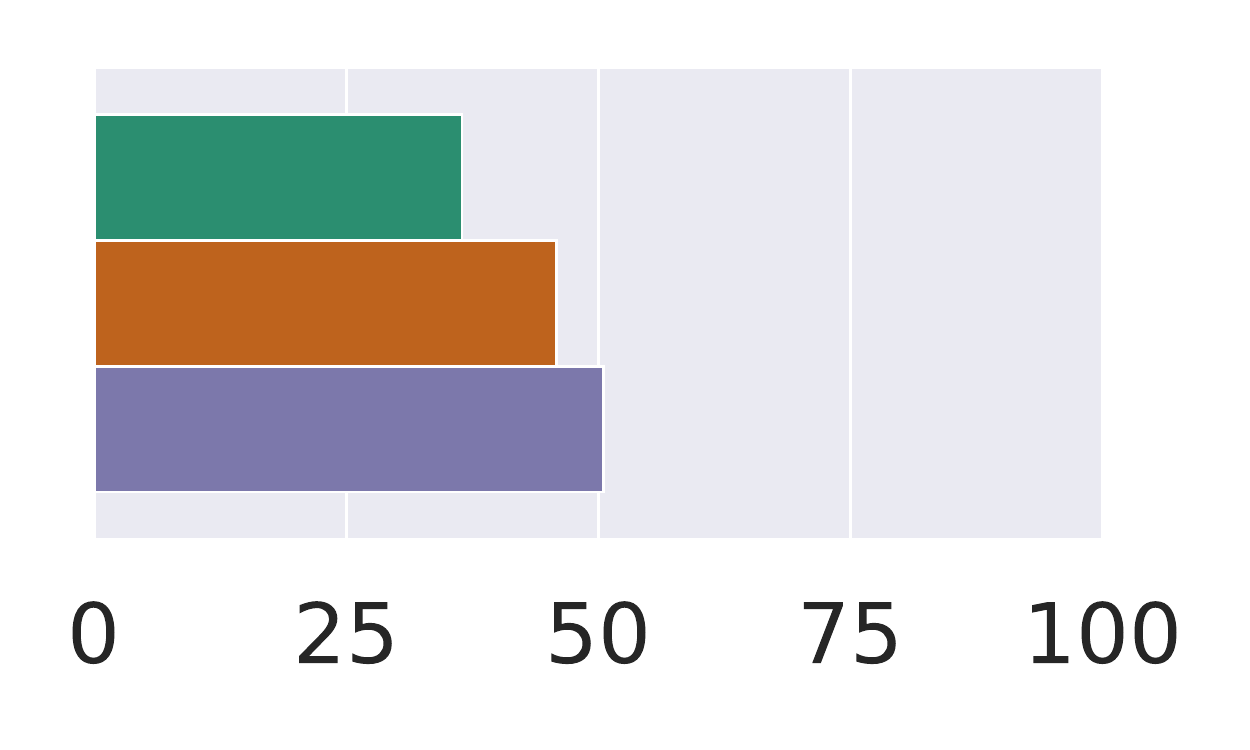}\end{minipage}
 & \begin{minipage}{.12\textwidth}\includegraphics[width=\textwidth]{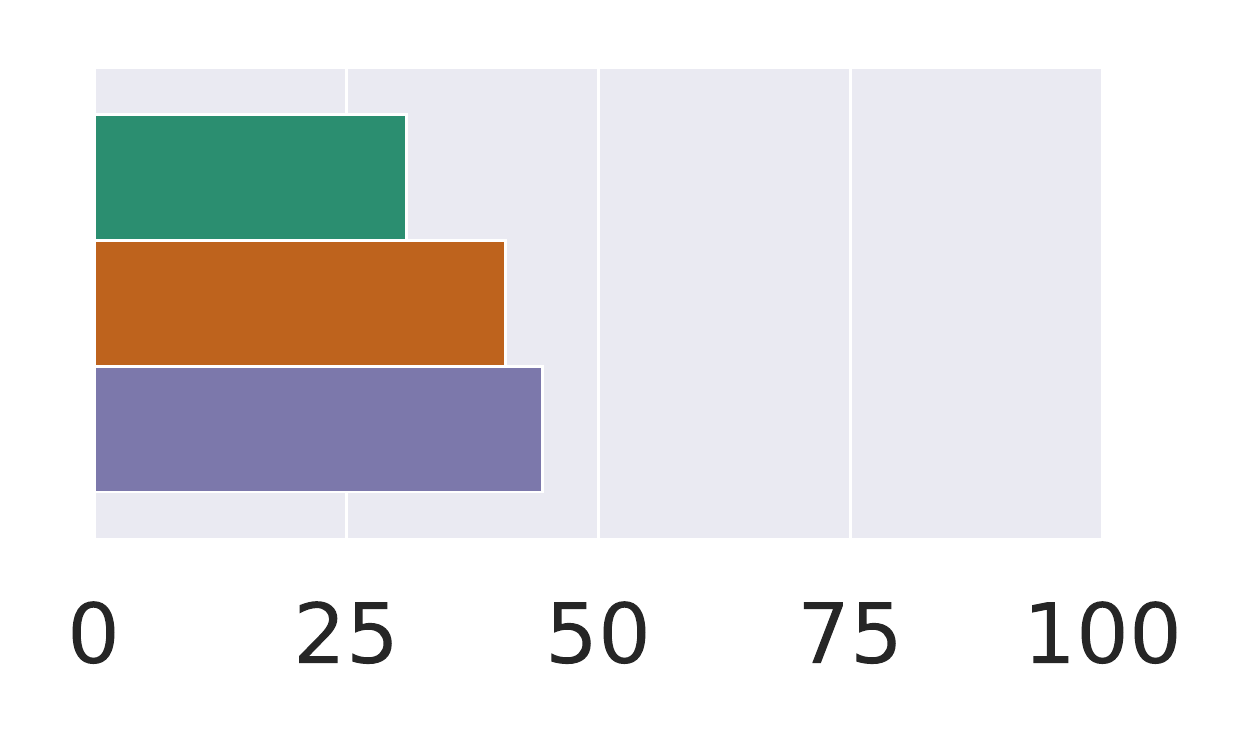}\end{minipage}
\\
\hline
\multicolumn{7}{c}{\includegraphics[scale=0.3]{figs/results-CG/legend}} \\
\end{tabular}
\end{footnotesize}
\end{center}

\vspace{-15pt}
\caption{Results of \comgen models on \val and \tests sets.\label{fig:results-aux-cg}}
\end{figure*}

\begin{figure*}

\begin{center}
\begin{footnotesize}
\begin{tabular}{|r|@{}c@{}c@{}|@{}c@{}c@{}|@{}c@{}c@{}|}
\hline
\makecell[c]{\UseMacro{TH-train-on}}
 & \makecell[c]{\UseMacro{TH-MP}}
 & \makecell[c]{\UseMacro{TH-CP}}
 & \makecell[c]{\UseMacro{TH-MP}}
 & \makecell[c]{\UseMacro{TH-T}}
 & \makecell[c]{\UseMacro{TH-CP}}
 & \makecell[c]{\UseMacro{TH-T}}
\\
\hline
\makecell[c]{\UseMacro{TH-test-on}}
 & \multicolumn{2}{c|}{\UseMacro{TH-MP-CP}}
 & \multicolumn{2}{c|}{\UseMacro{TH-MP-T}}
 & \multicolumn{2}{c|}{\UseMacro{TH-CP-T}}
\\
\hline
\UseMacro{TH-metric-set_match_p}
 & \begin{minipage}{.12\textwidth}\includegraphics[width=\textwidth]{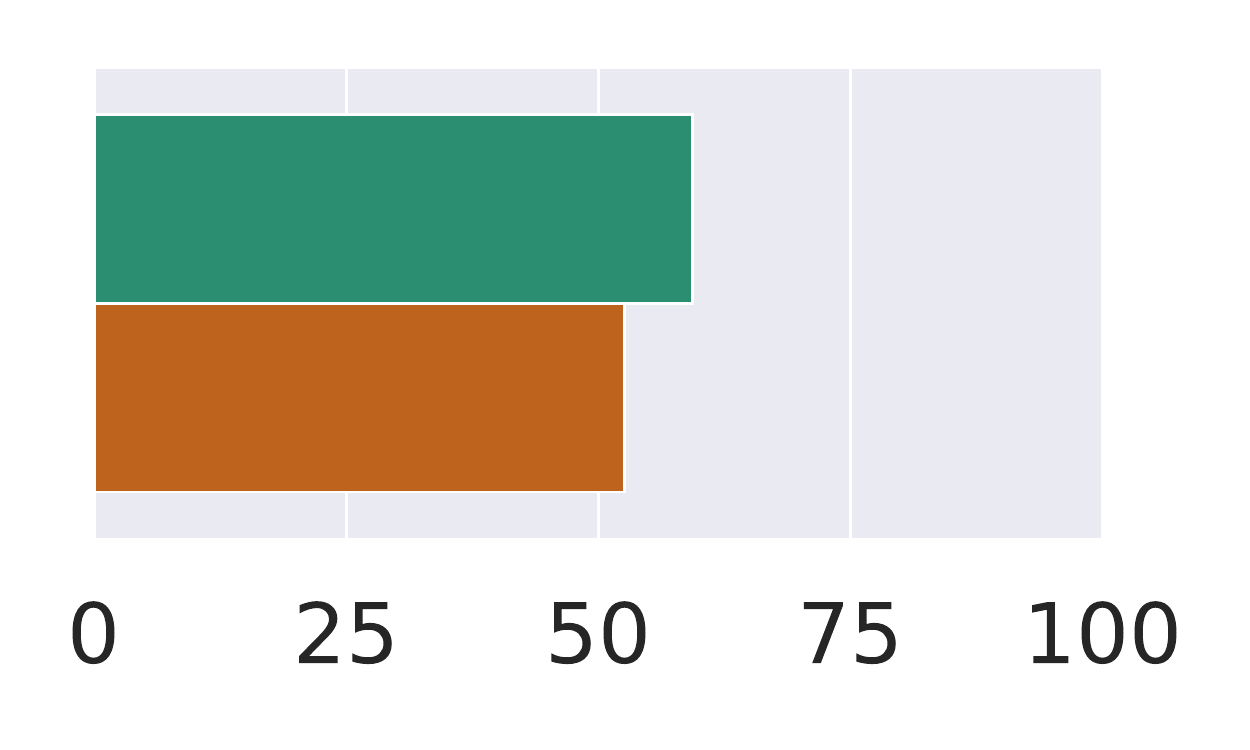}\end{minipage}
 & \begin{minipage}{.12\textwidth}\includegraphics[width=\textwidth]{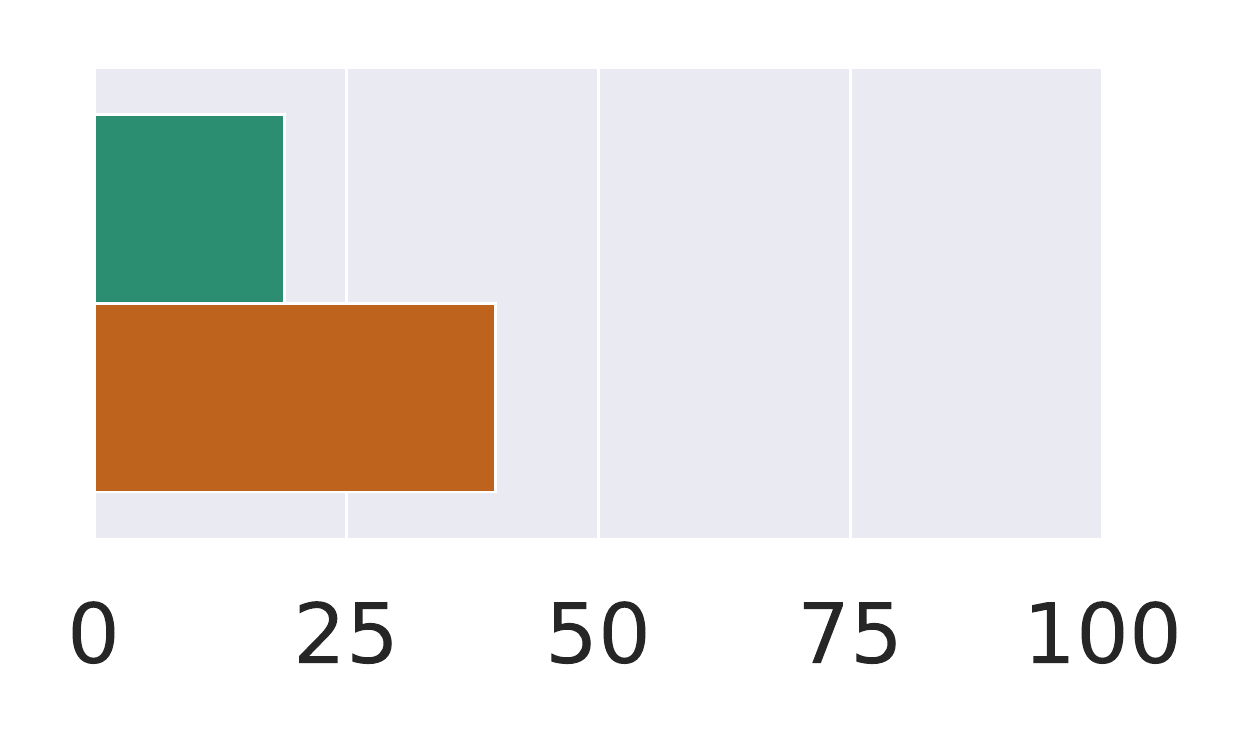}\end{minipage}
 & \begin{minipage}{.12\textwidth}\includegraphics[width=\textwidth]{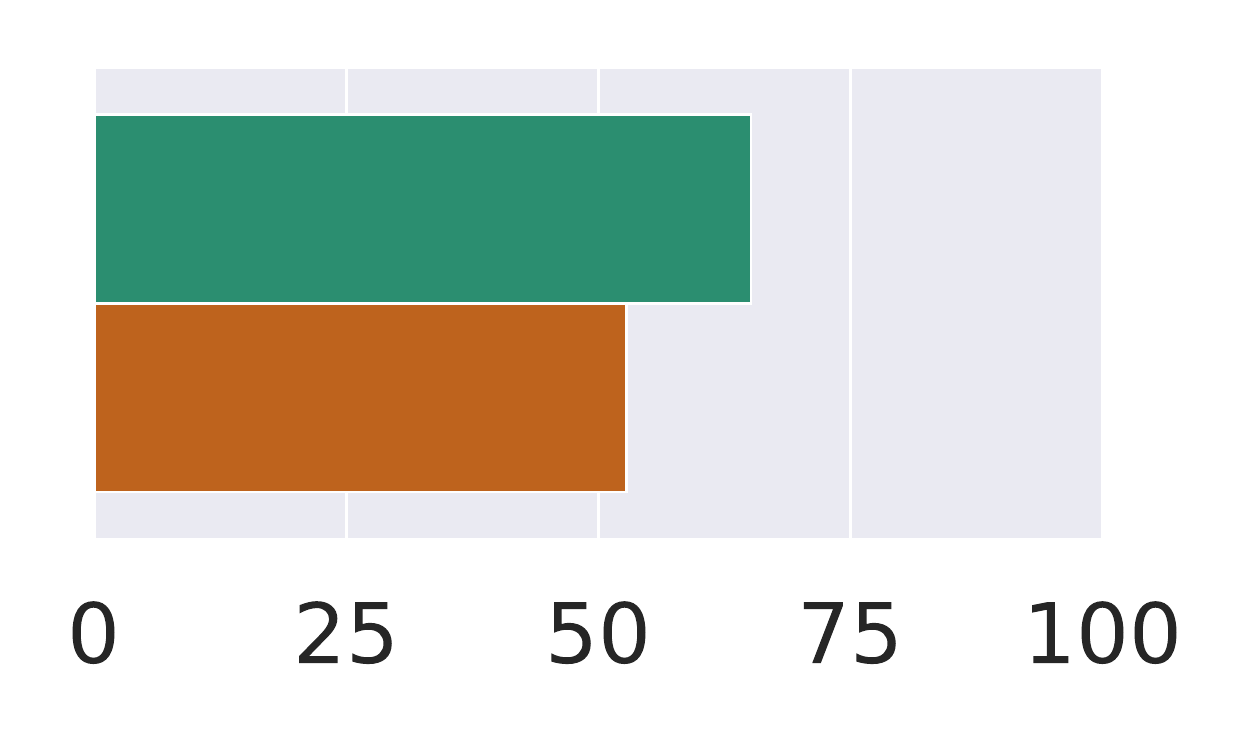}\end{minipage}
 & \begin{minipage}{.12\textwidth}\includegraphics[width=\textwidth]{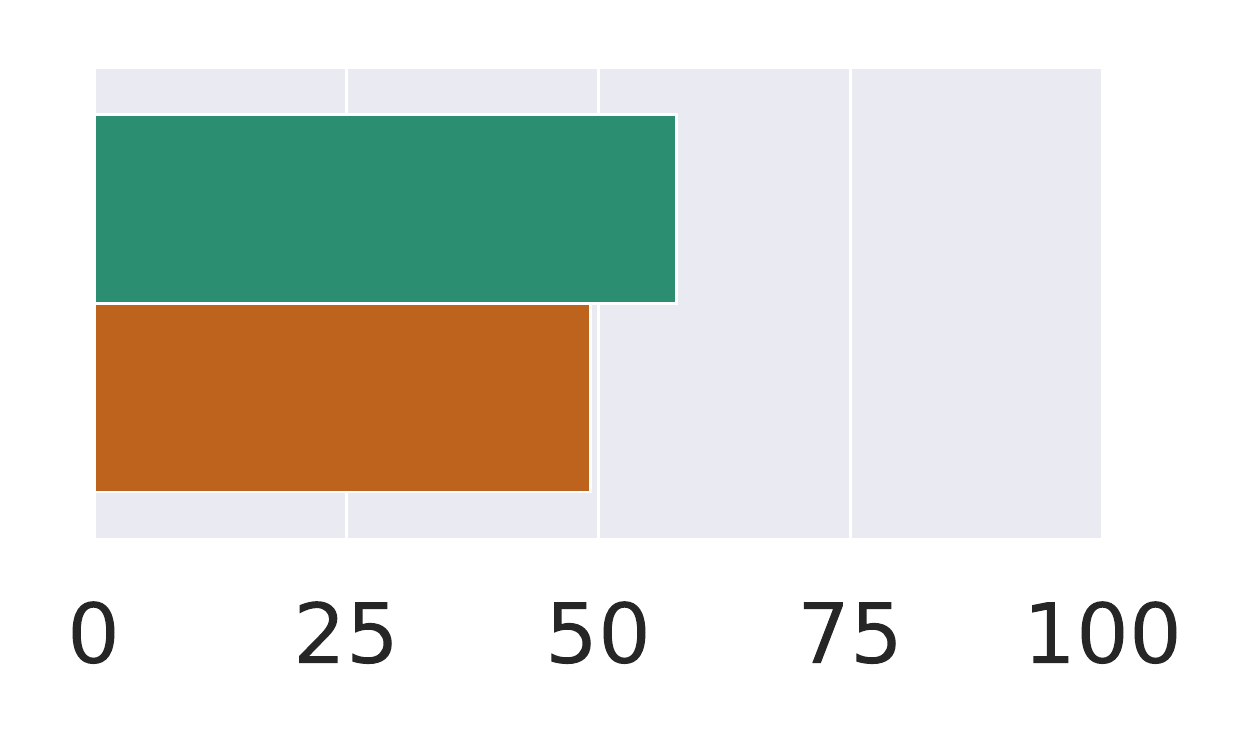}\end{minipage}
 & \begin{minipage}{.12\textwidth}\includegraphics[width=\textwidth]{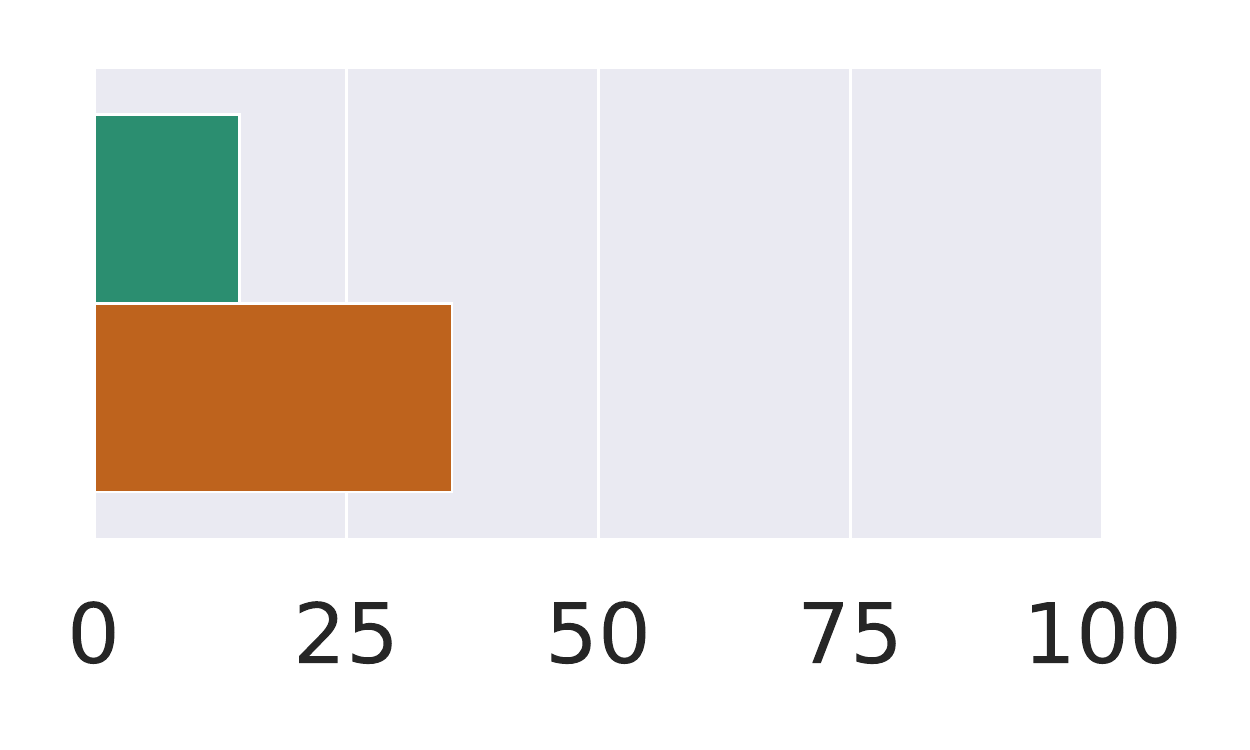}\end{minipage}
 & \begin{minipage}{.12\textwidth}\includegraphics[width=\textwidth]{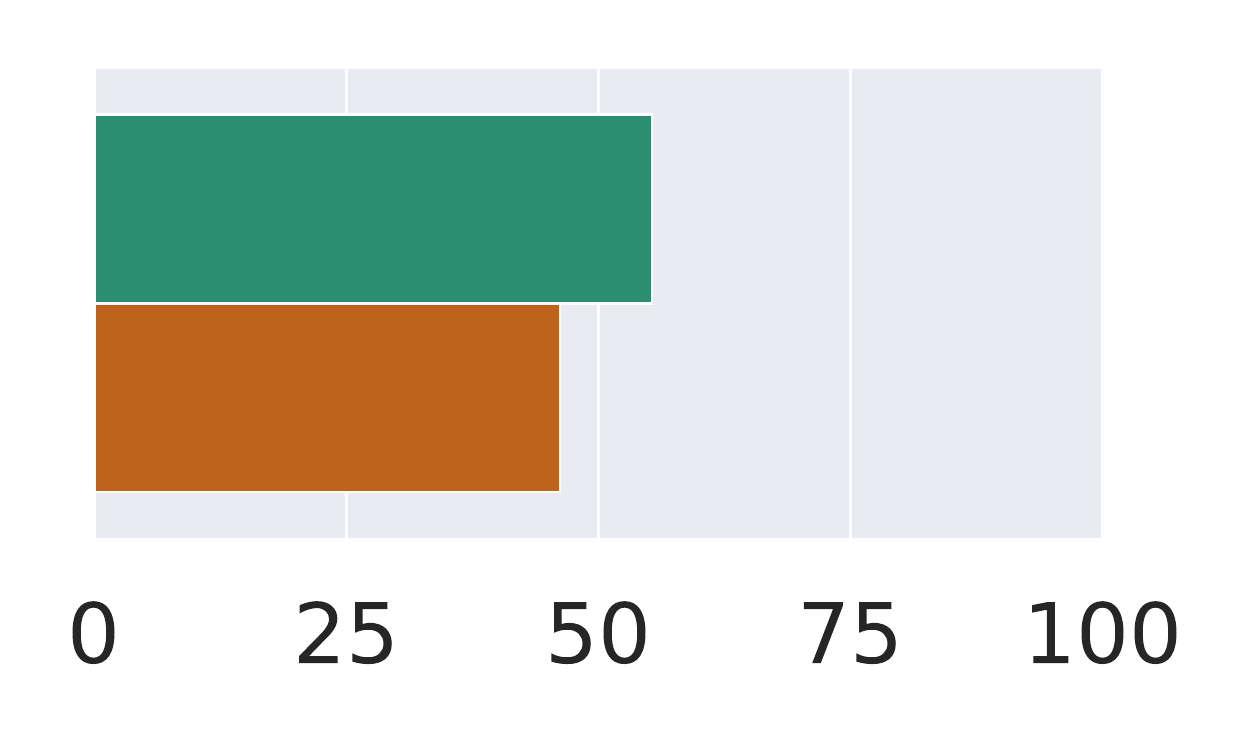}\end{minipage}
\\
\UseMacro{TH-metric-set_match_r}
 & \begin{minipage}{.12\textwidth}\includegraphics[width=\textwidth]{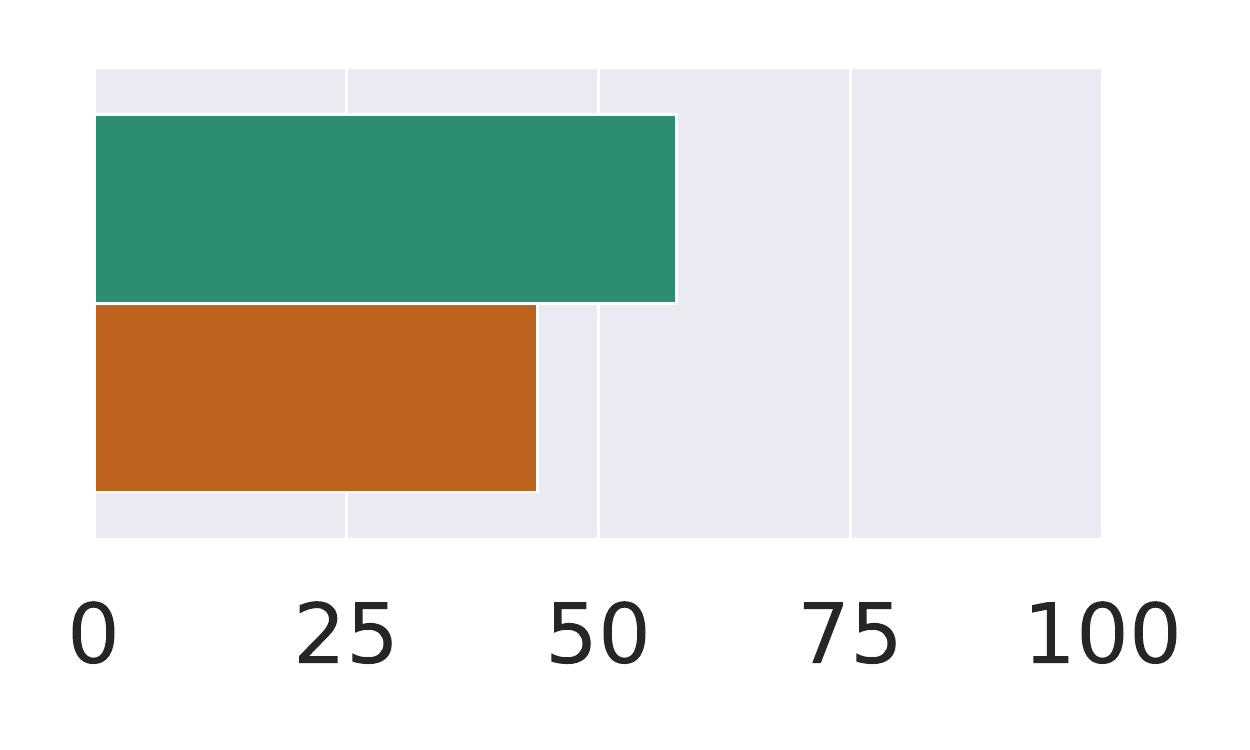}\end{minipage}
 & \begin{minipage}{.12\textwidth}\includegraphics[width=\textwidth]{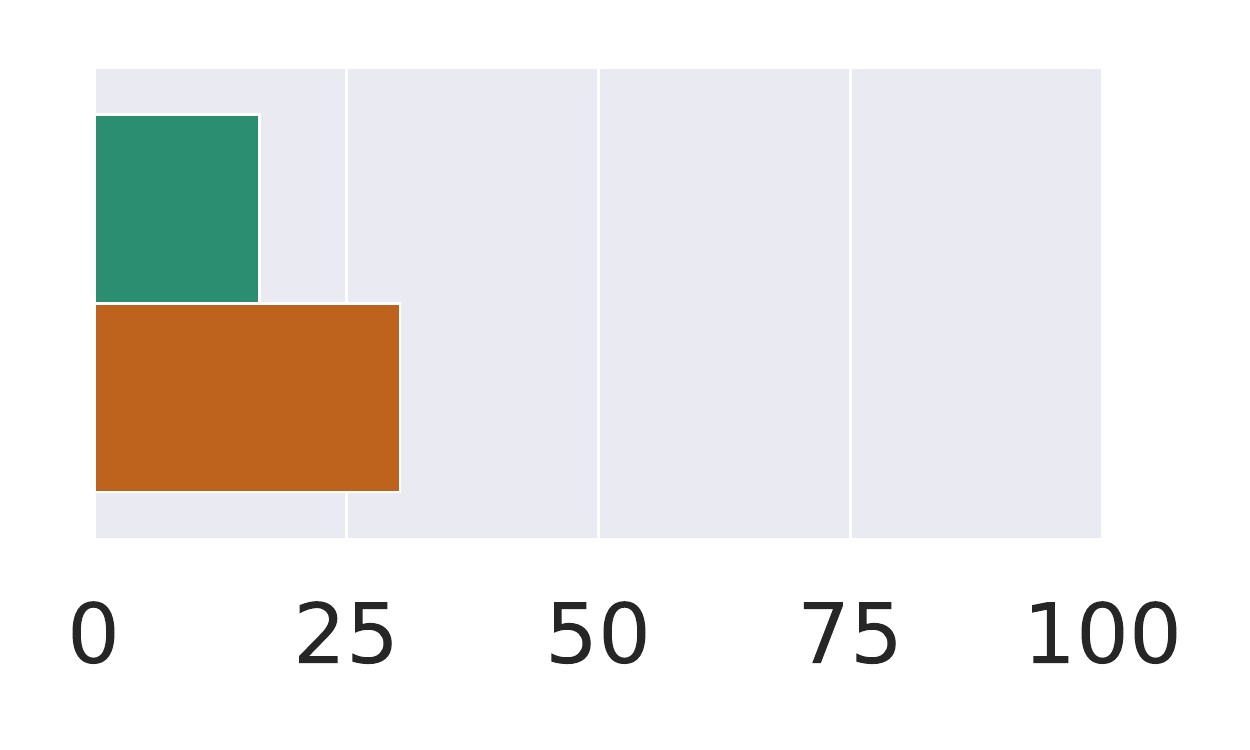}\end{minipage}
 & \begin{minipage}{.12\textwidth}\includegraphics[width=\textwidth]{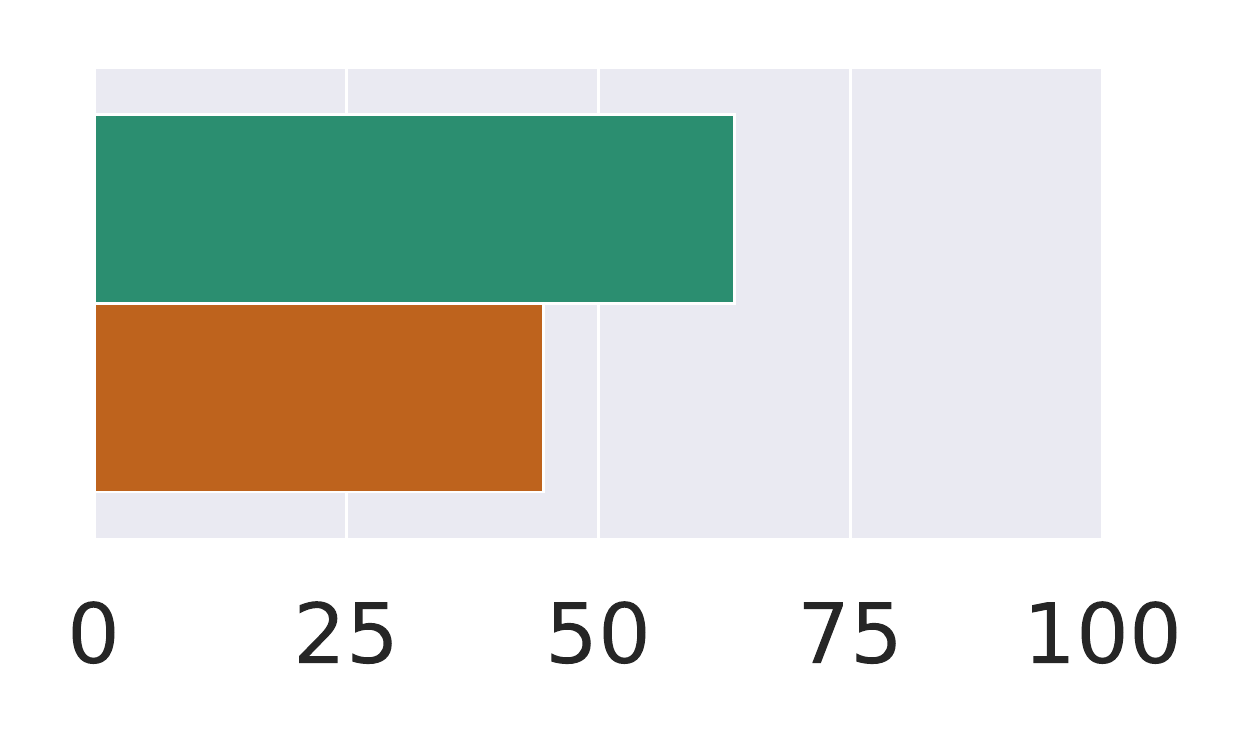}\end{minipage}
 & \begin{minipage}{.12\textwidth}\includegraphics[width=\textwidth]{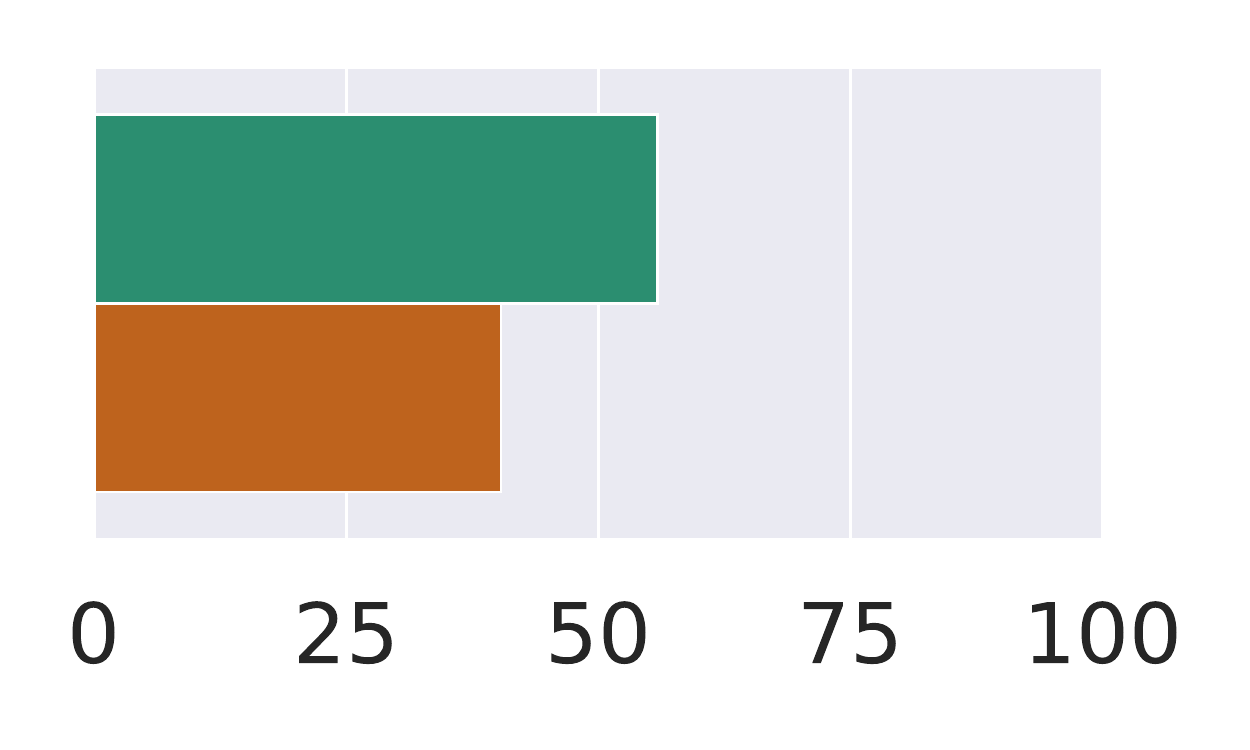}\end{minipage}
 & \begin{minipage}{.12\textwidth}\includegraphics[width=\textwidth]{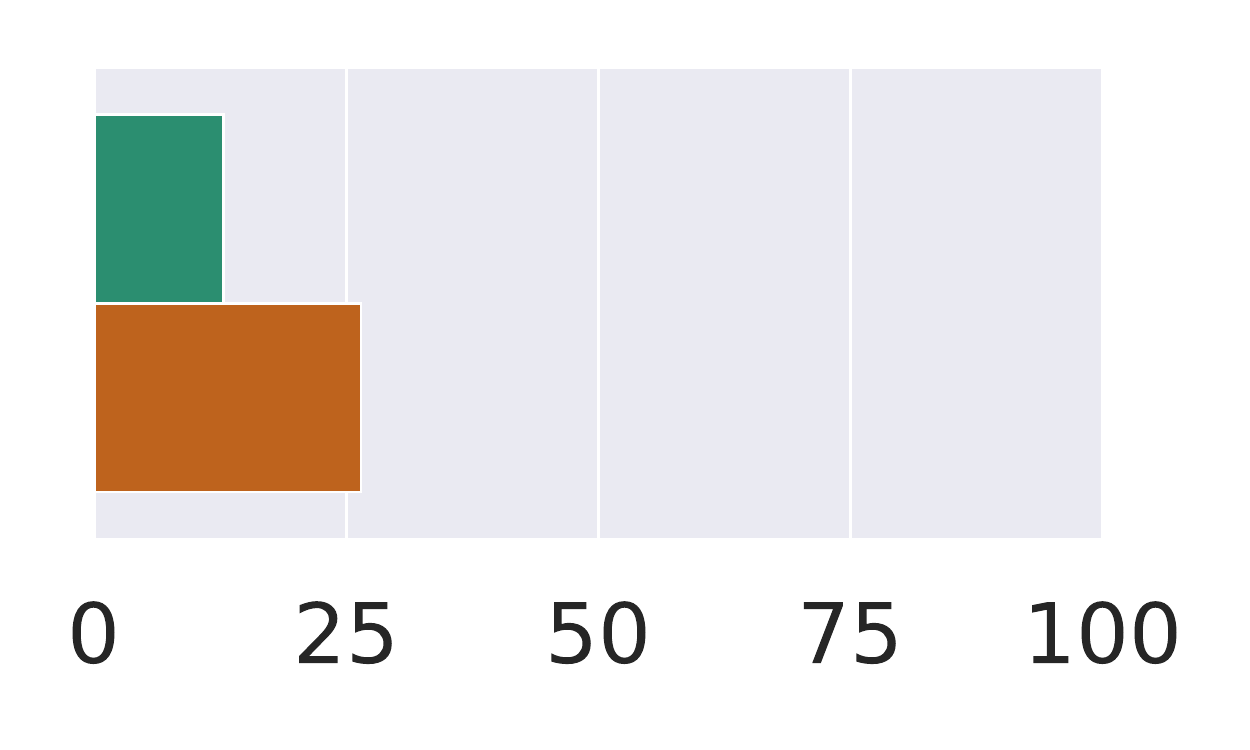}\end{minipage}
 & \begin{minipage}{.12\textwidth}\includegraphics[width=\textwidth]{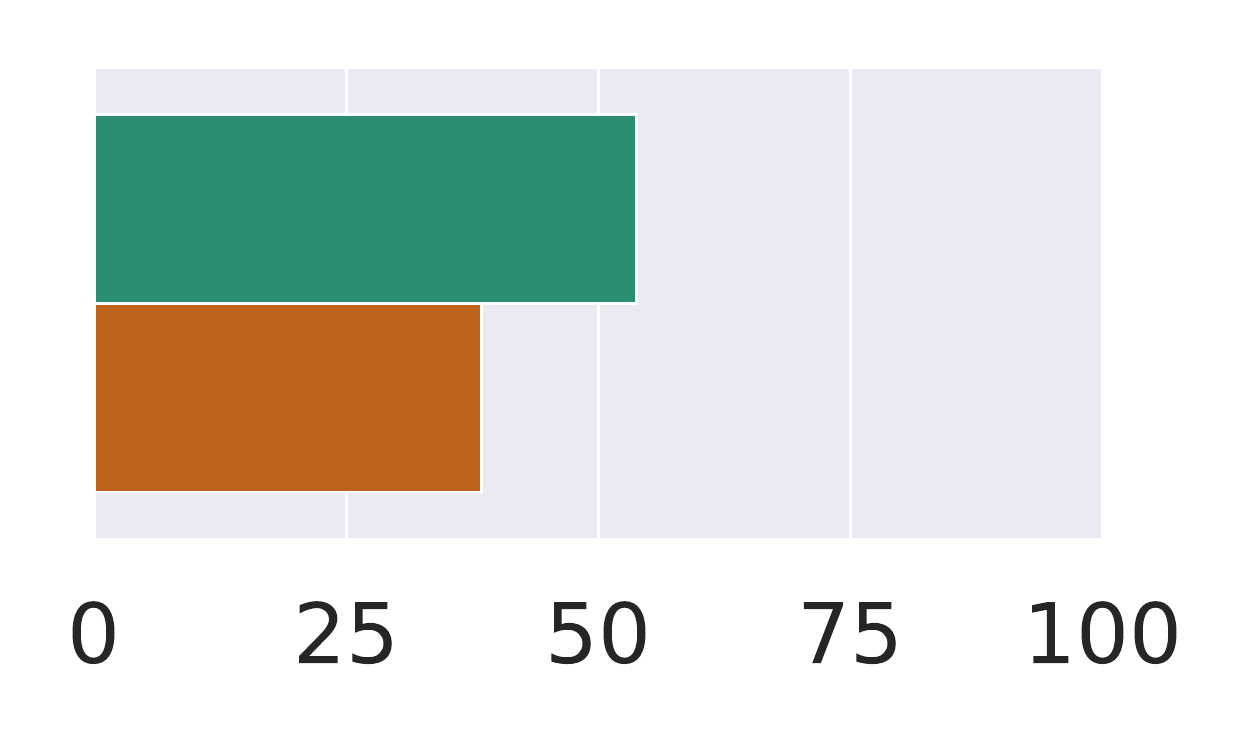}\end{minipage}
\\
\UseMacro{TH-metric-set_match_f}
 & \begin{minipage}{.12\textwidth}\includegraphics[width=\textwidth]{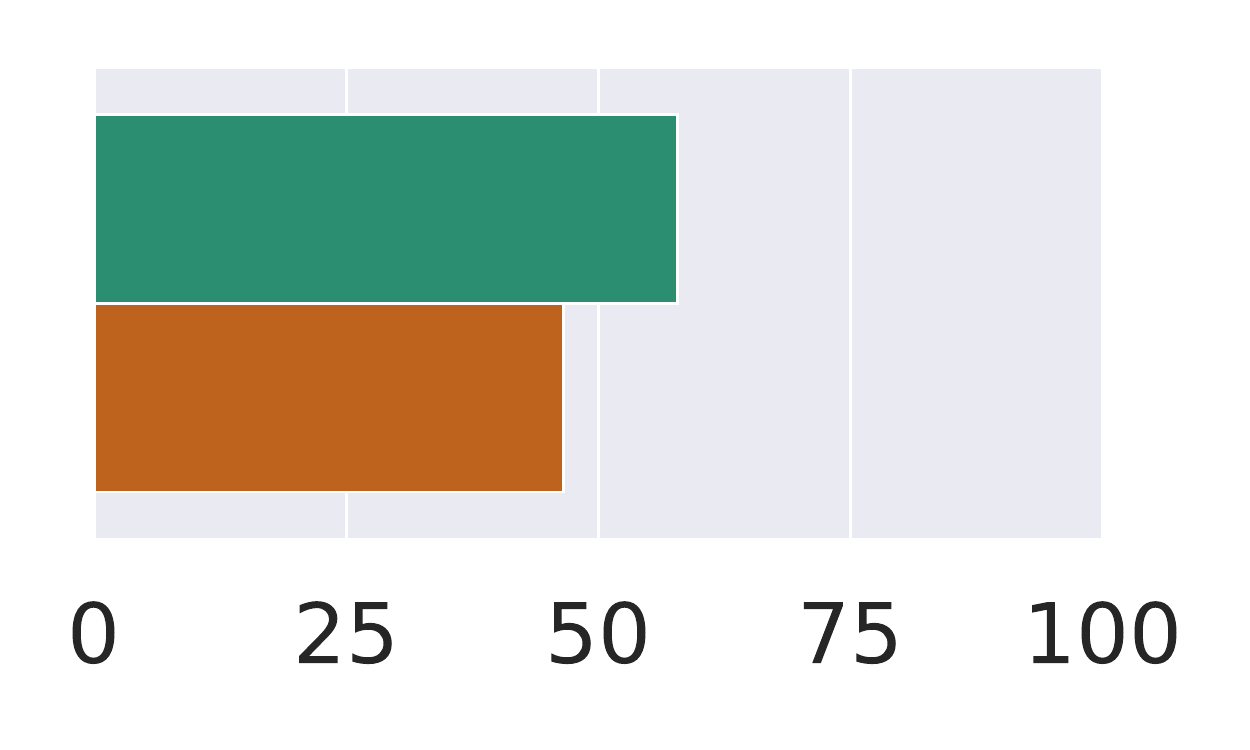}\end{minipage}
 & \begin{minipage}{.12\textwidth}\includegraphics[width=\textwidth]{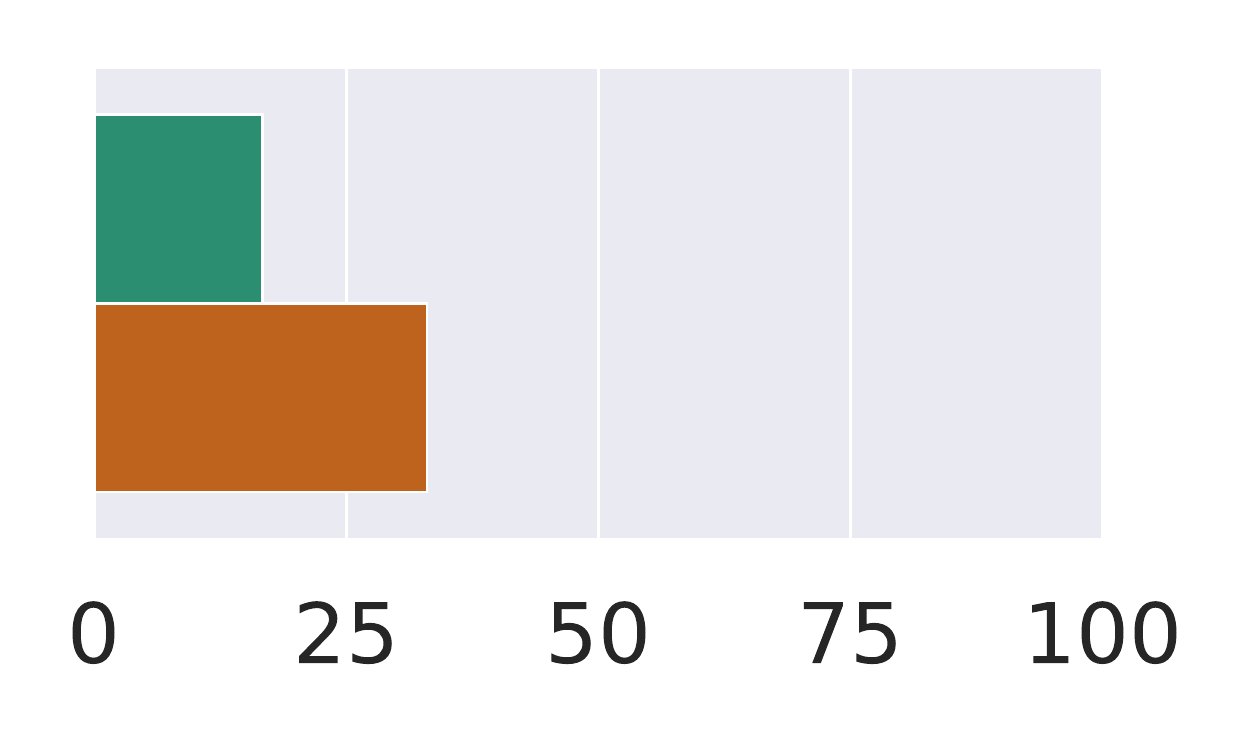}\end{minipage}
 & \begin{minipage}{.12\textwidth}\includegraphics[width=\textwidth]{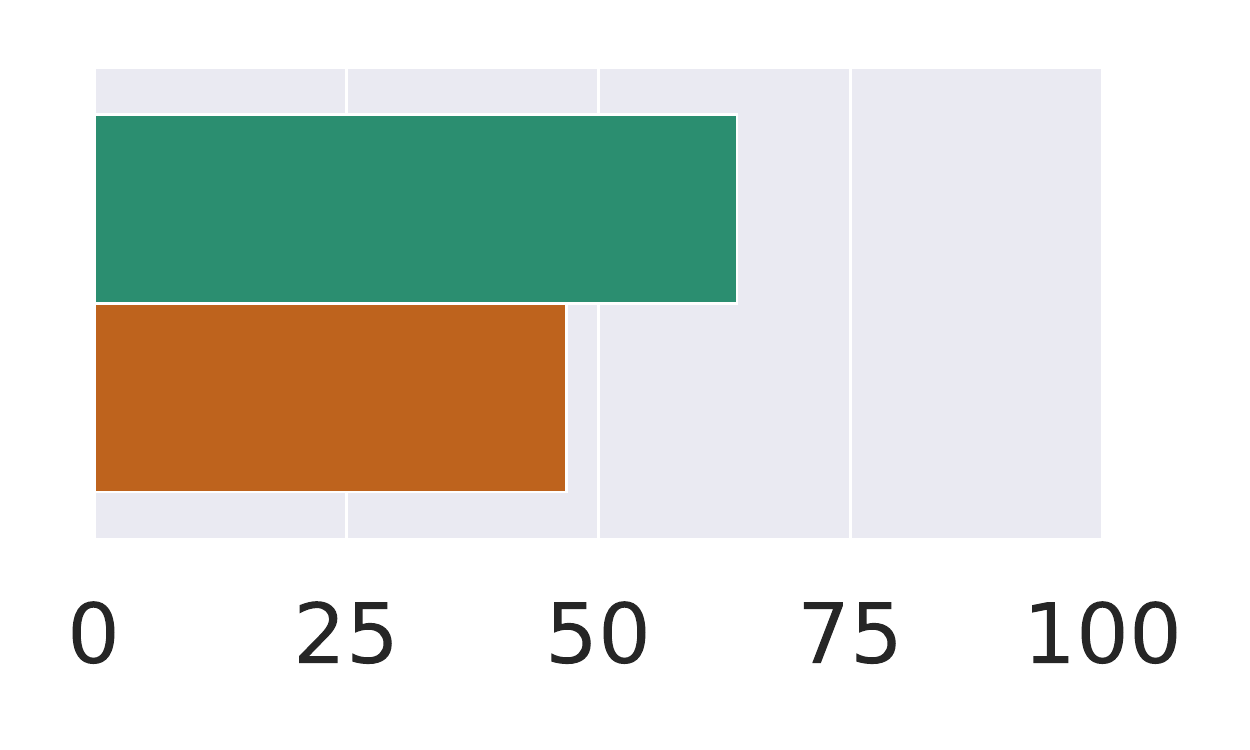}\end{minipage}
 & \begin{minipage}{.12\textwidth}\includegraphics[width=\textwidth]{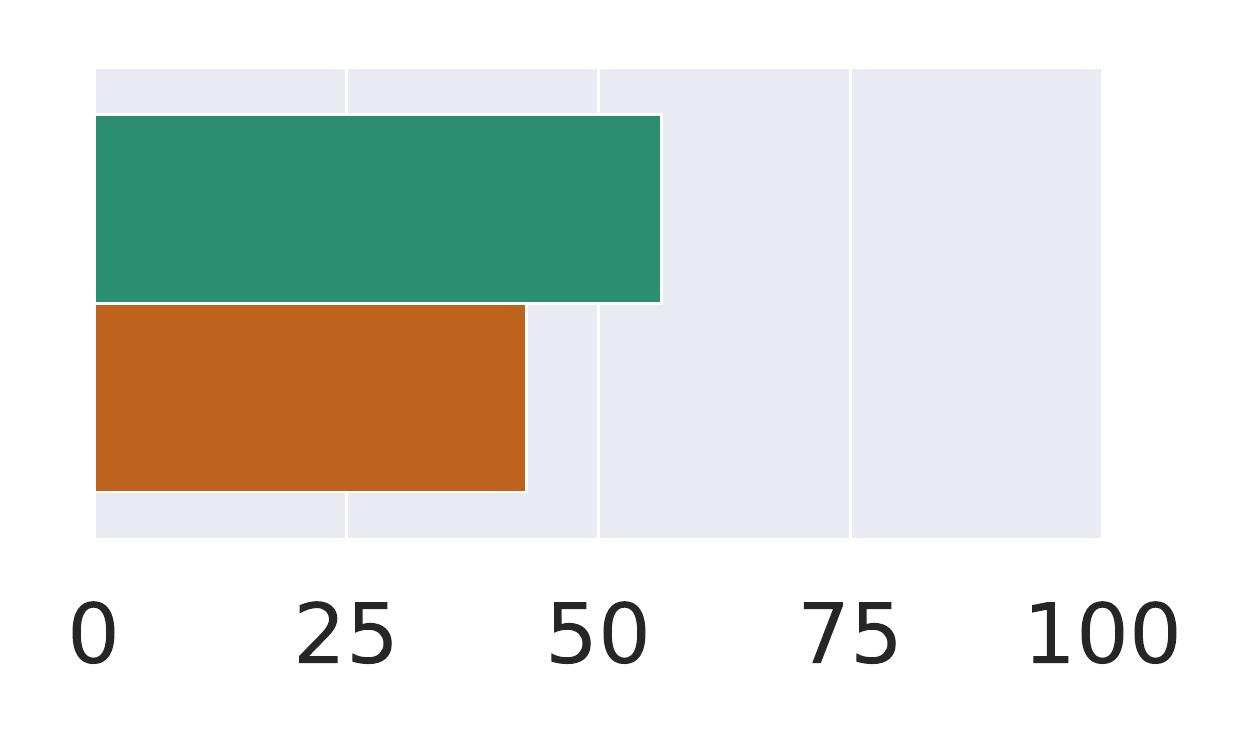}\end{minipage}
 & \begin{minipage}{.12\textwidth}\includegraphics[width=\textwidth]{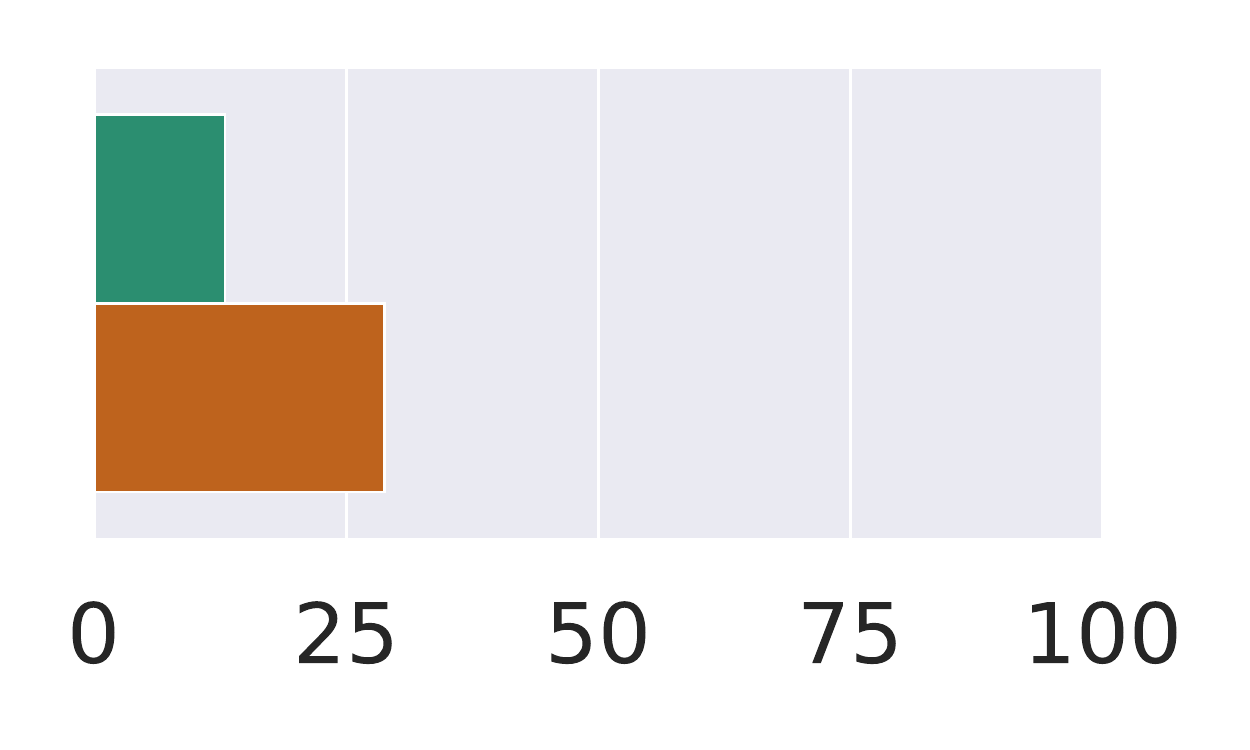}\end{minipage}
 & \begin{minipage}{.12\textwidth}\includegraphics[width=\textwidth]{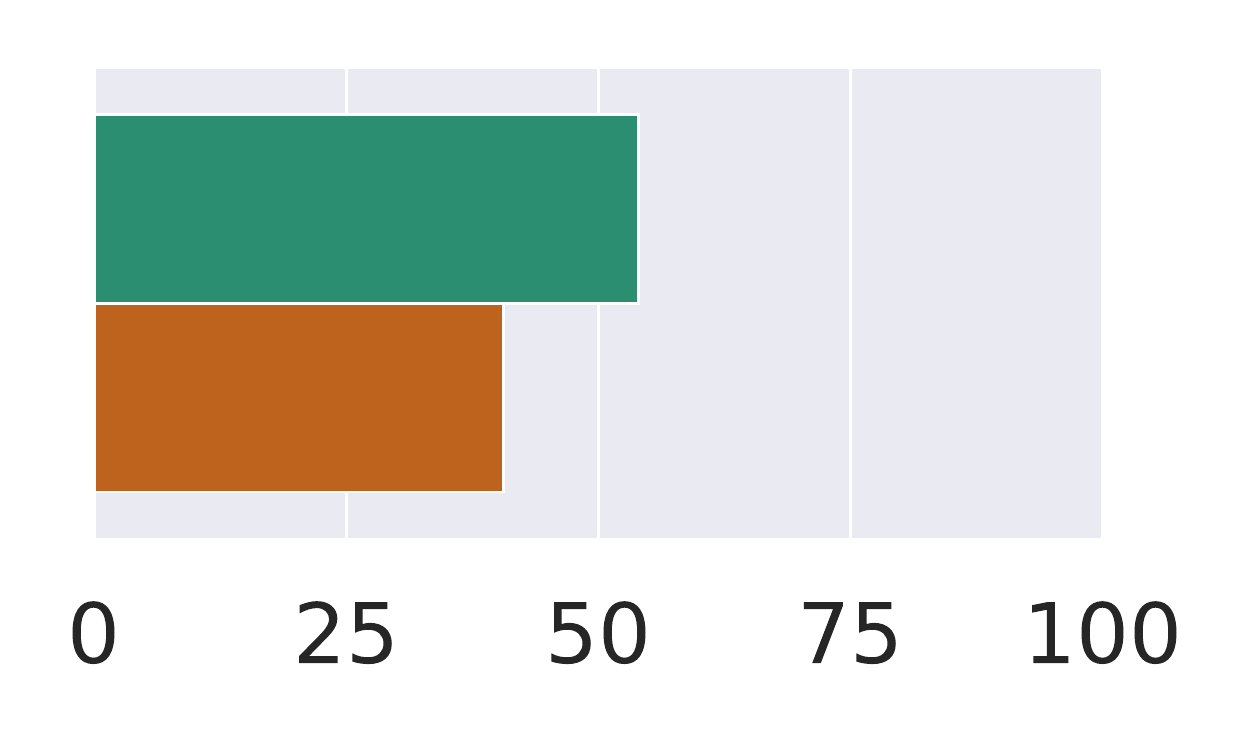}\end{minipage}
\\
\UseMacro{TH-metric-exact_match}
 & \begin{minipage}{.12\textwidth}\includegraphics[width=\textwidth]{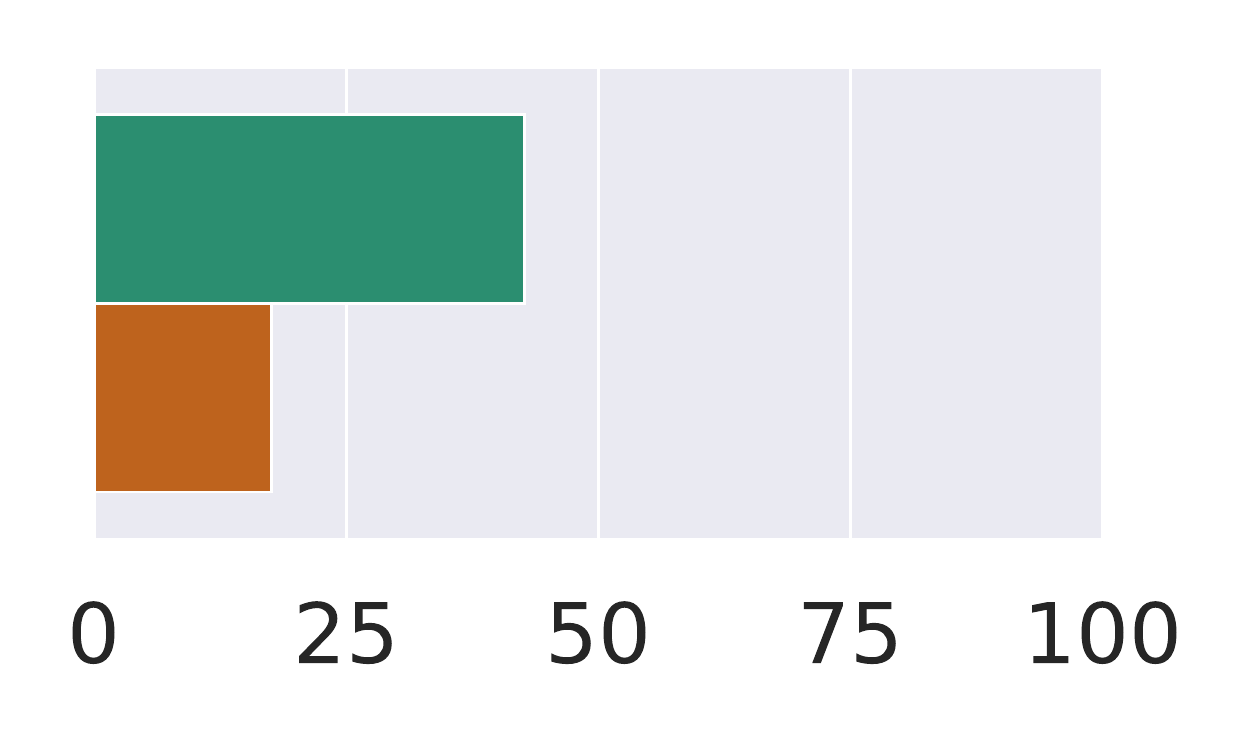}\end{minipage}
 & \begin{minipage}{.12\textwidth}\includegraphics[width=\textwidth]{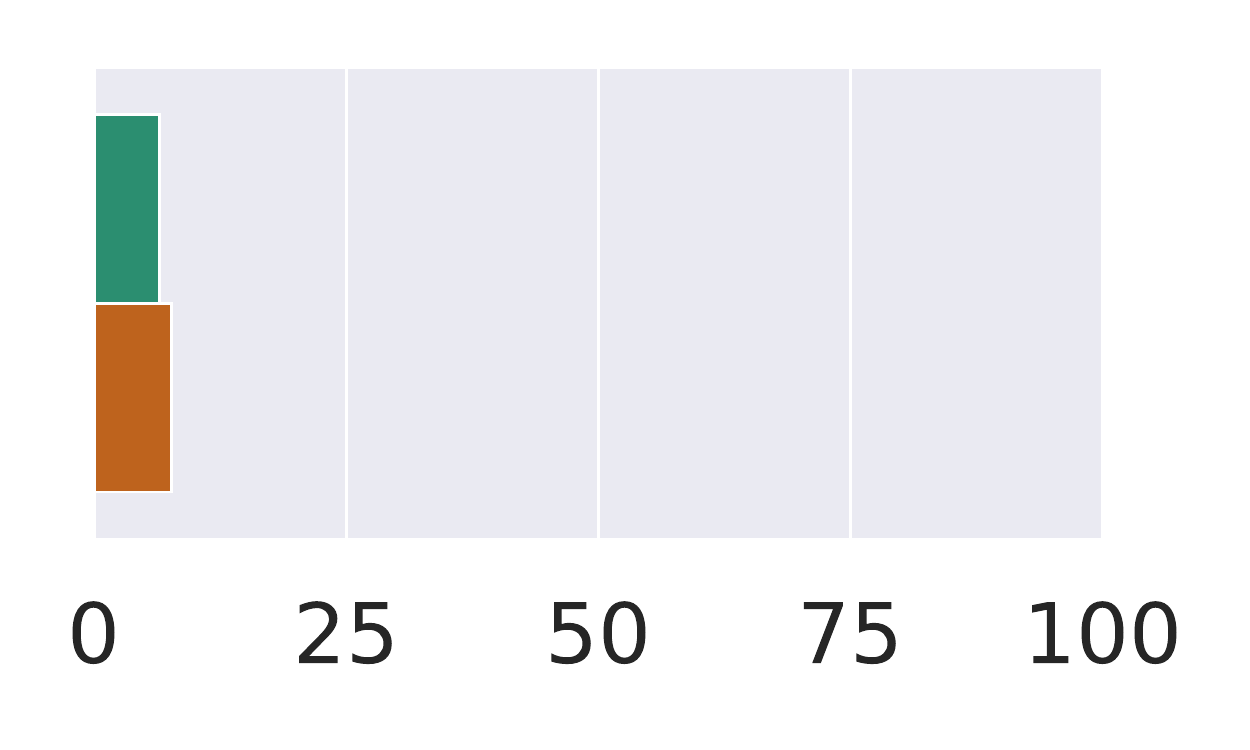}\end{minipage}
 & \begin{minipage}{.12\textwidth}\includegraphics[width=\textwidth]{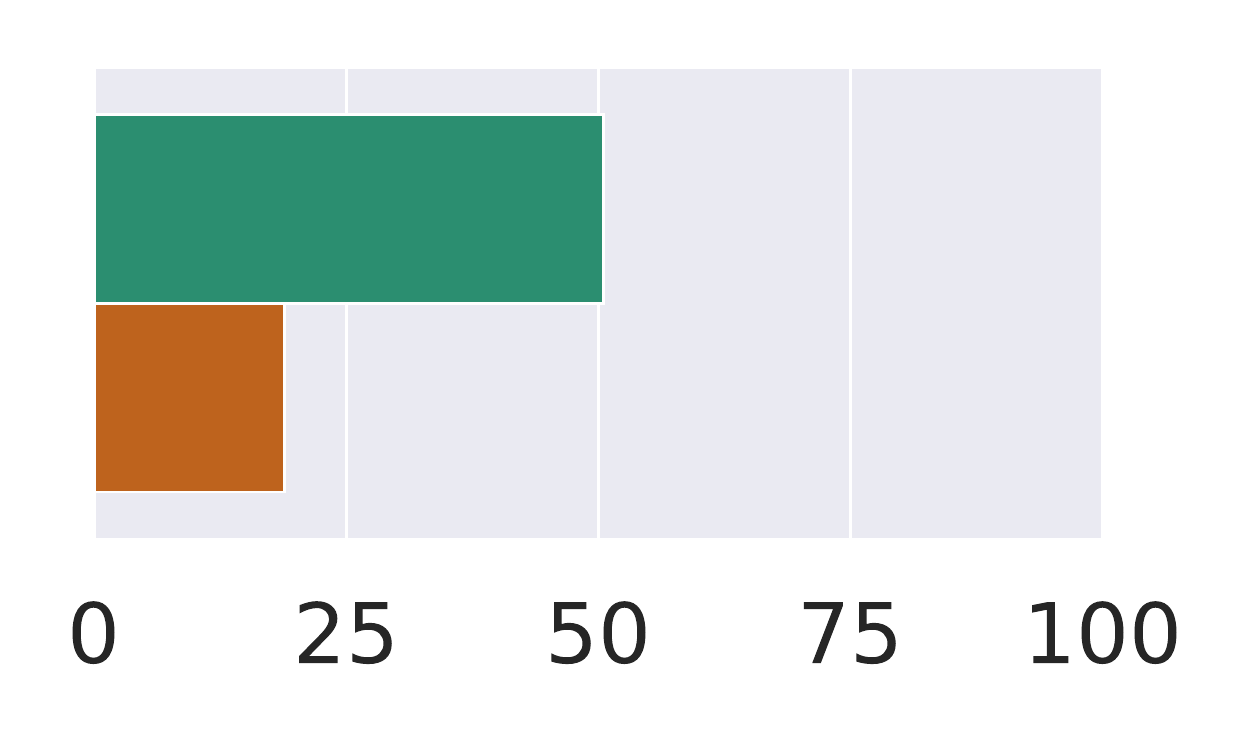}\end{minipage}
 & \begin{minipage}{.12\textwidth}\includegraphics[width=\textwidth]{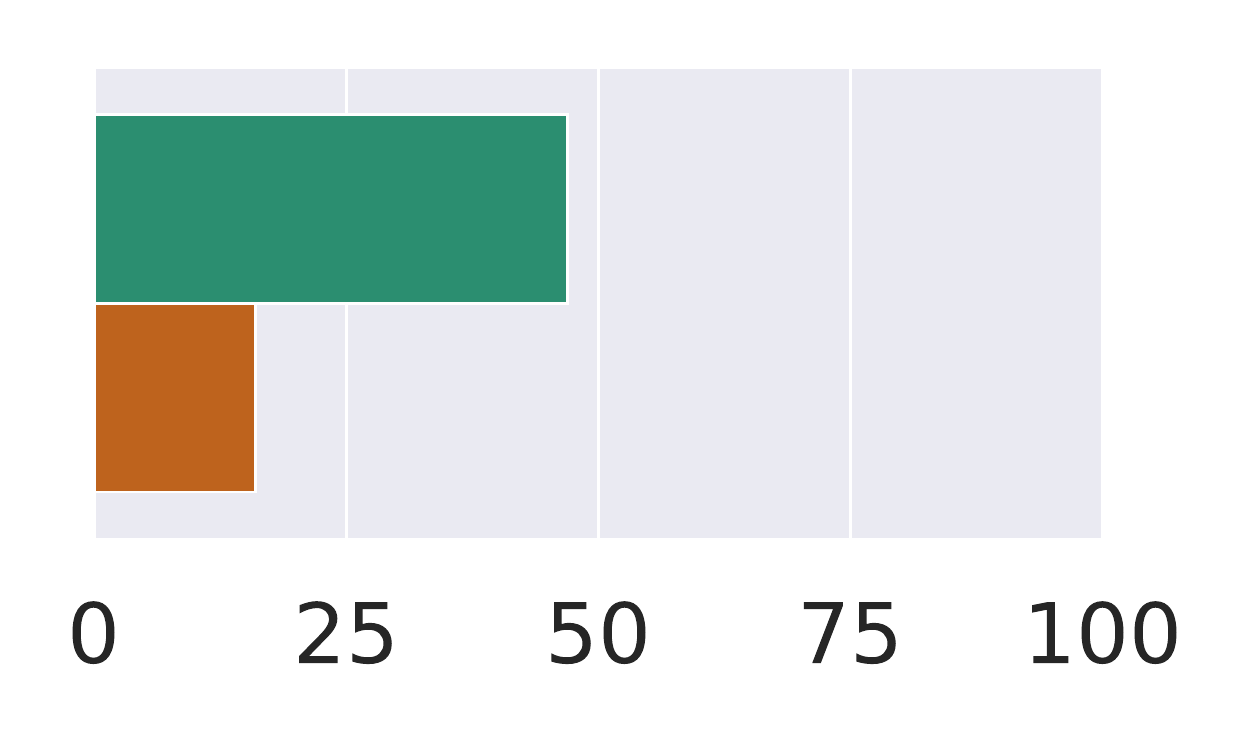}\end{minipage}
 & \begin{minipage}{.12\textwidth}\includegraphics[width=\textwidth]{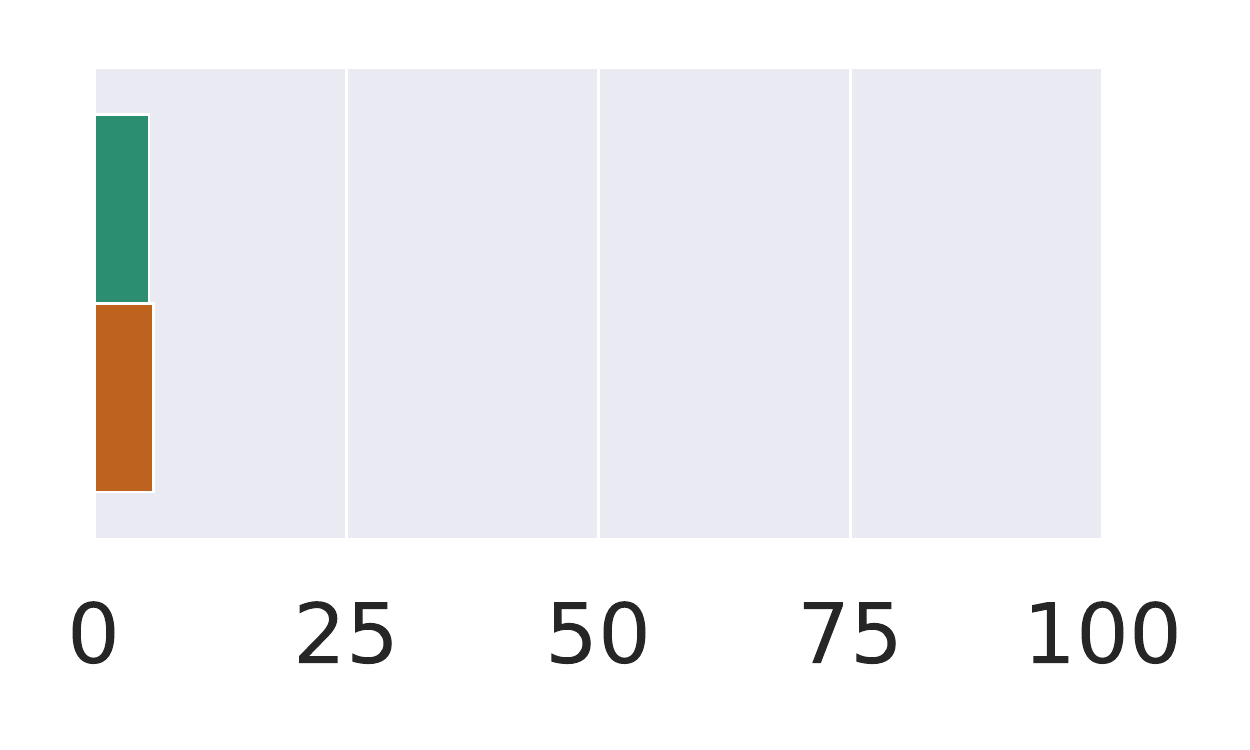}\end{minipage}
 & \begin{minipage}{.12\textwidth}\includegraphics[width=\textwidth]{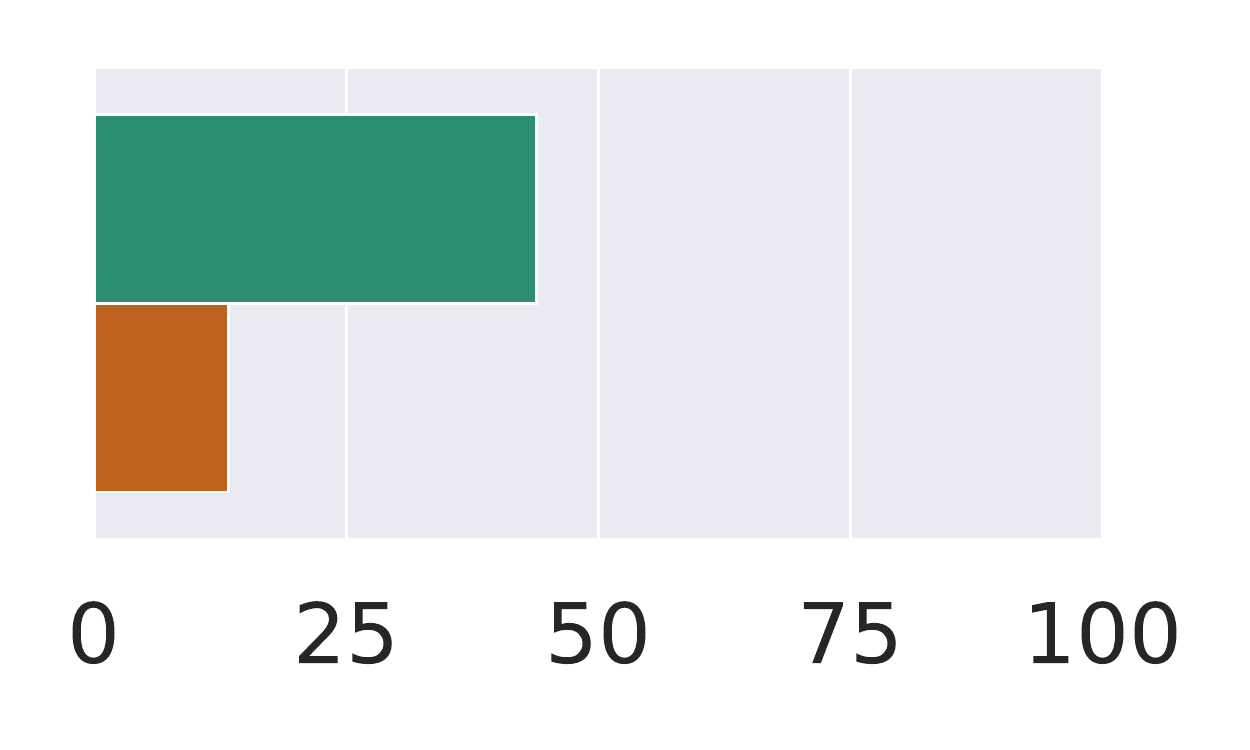}\end{minipage}
\\
\hline
\multicolumn{7}{c}{\includegraphics[scale=0.3]{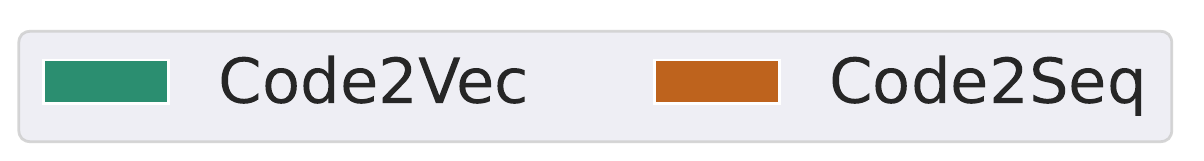}} \\
\end{tabular}
\end{footnotesize}
\end{center}

\vspace{-15pt}
\caption{Results of \metnam models on \testc sets.\label{fig:results-mn}}
\end{figure*}

\begin{figure*}

\begin{center}
\begin{footnotesize}
\begin{tabular}{|r|c|c|c|c|c|c|}
\hline
\makecell[c]{\UseMacro{TH-train-on}}
 & \multicolumn{2}{c|}{\UseMacro{TH-MP}}
 & \multicolumn{2}{c|}{\UseMacro{TH-CP}}
 & \multicolumn{2}{c|}{\UseMacro{TH-T}}
\\
\hline
\makecell[c]{\UseMacro{TH-test-on}}
 & \UseMacro{TH-val}
 & \UseMacro{TH-test_standard}
 & \UseMacro{TH-val}
 & \UseMacro{TH-test_standard}
 & \UseMacro{TH-val}
 & \UseMacro{TH-test_standard}
\\
\hline
\UseMacro{TH-metric-set_match_p}
 & \begin{minipage}{.12\textwidth}\includegraphics[width=\textwidth]{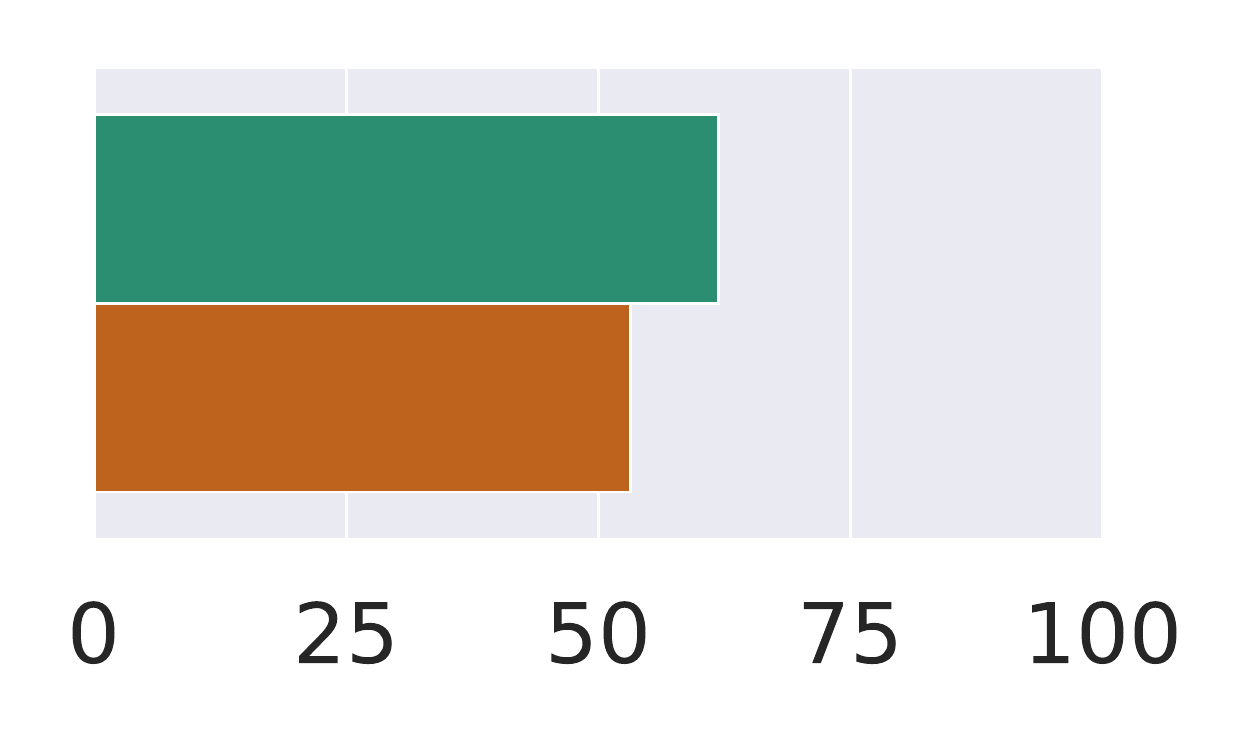}\end{minipage}
 & \begin{minipage}{.12\textwidth}\includegraphics[width=\textwidth]{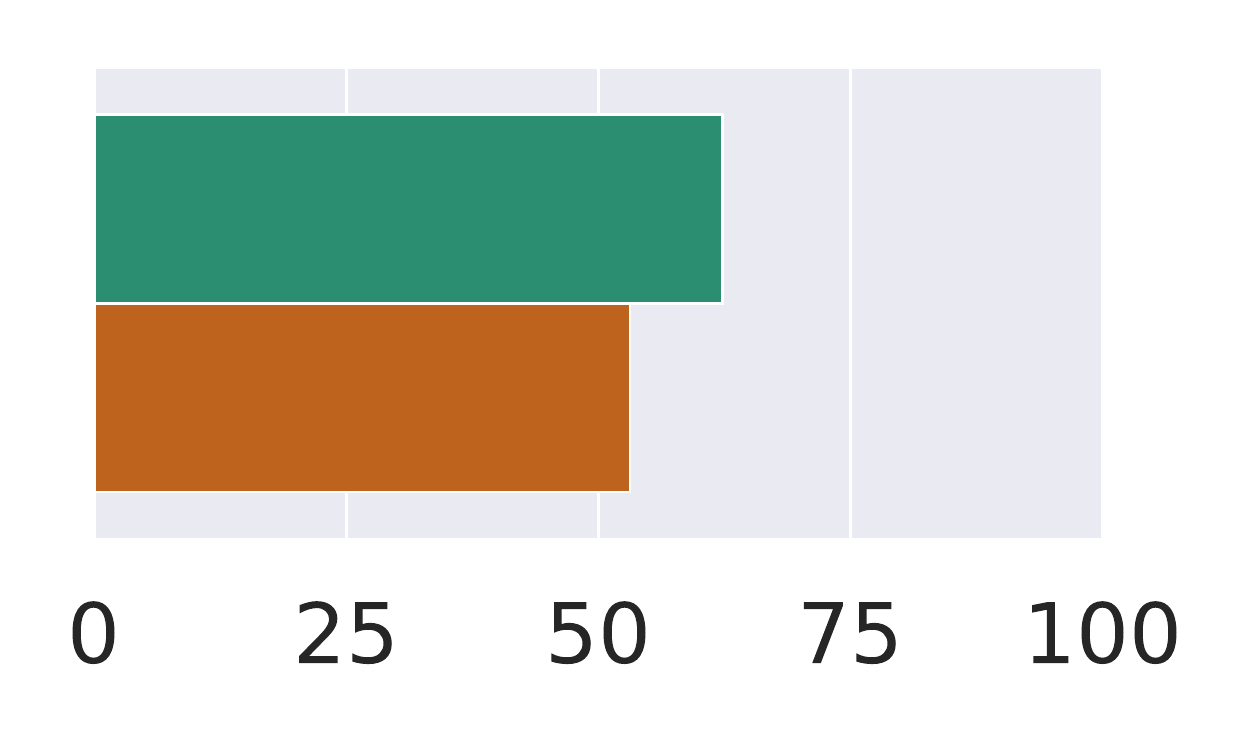}\end{minipage}
 & \begin{minipage}{.12\textwidth}\includegraphics[width=\textwidth]{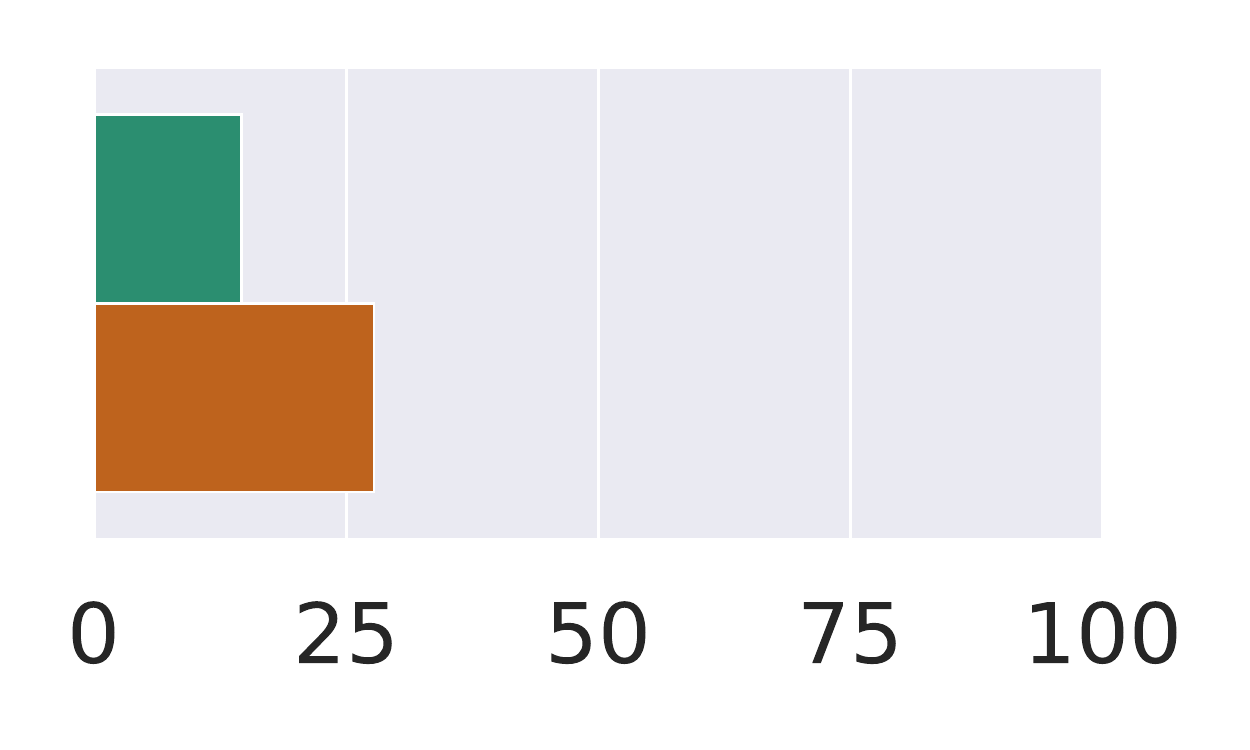}\end{minipage}
 & \begin{minipage}{.12\textwidth}\includegraphics[width=\textwidth]{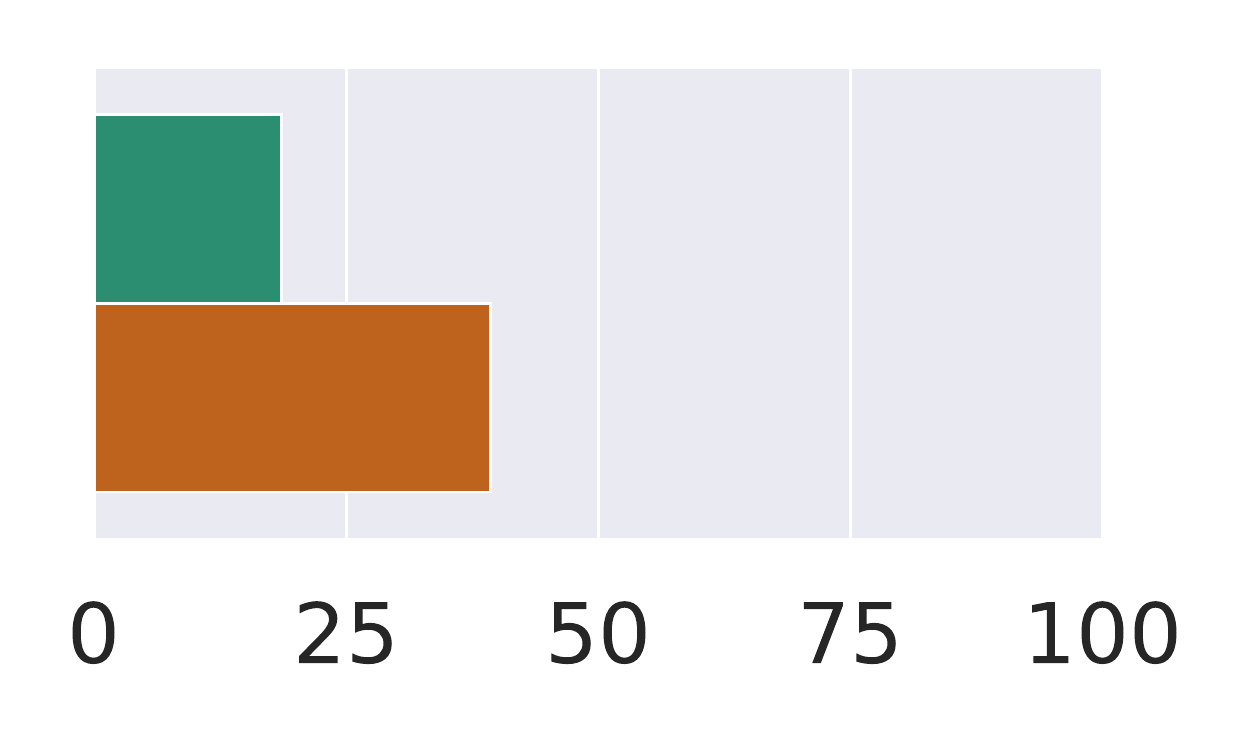}\end{minipage}
 & \begin{minipage}{.12\textwidth}\includegraphics[width=\textwidth]{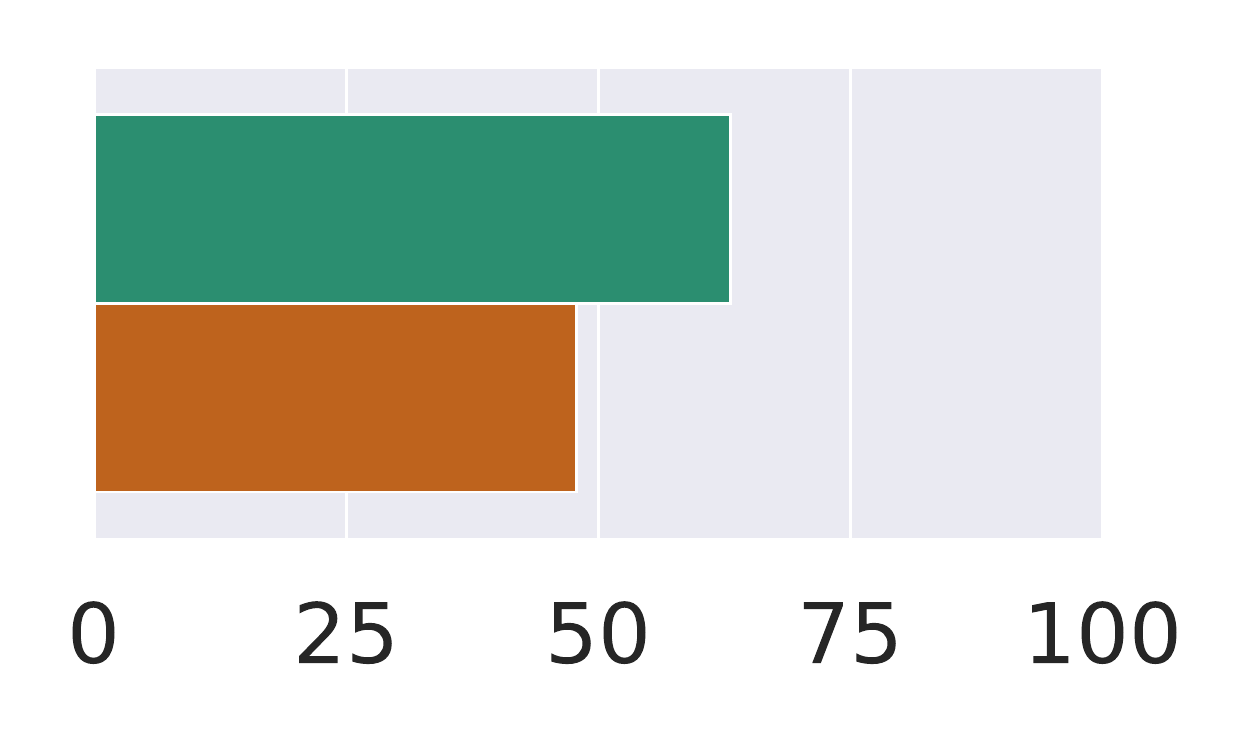}\end{minipage}
 & \begin{minipage}{.12\textwidth}\includegraphics[width=\textwidth]{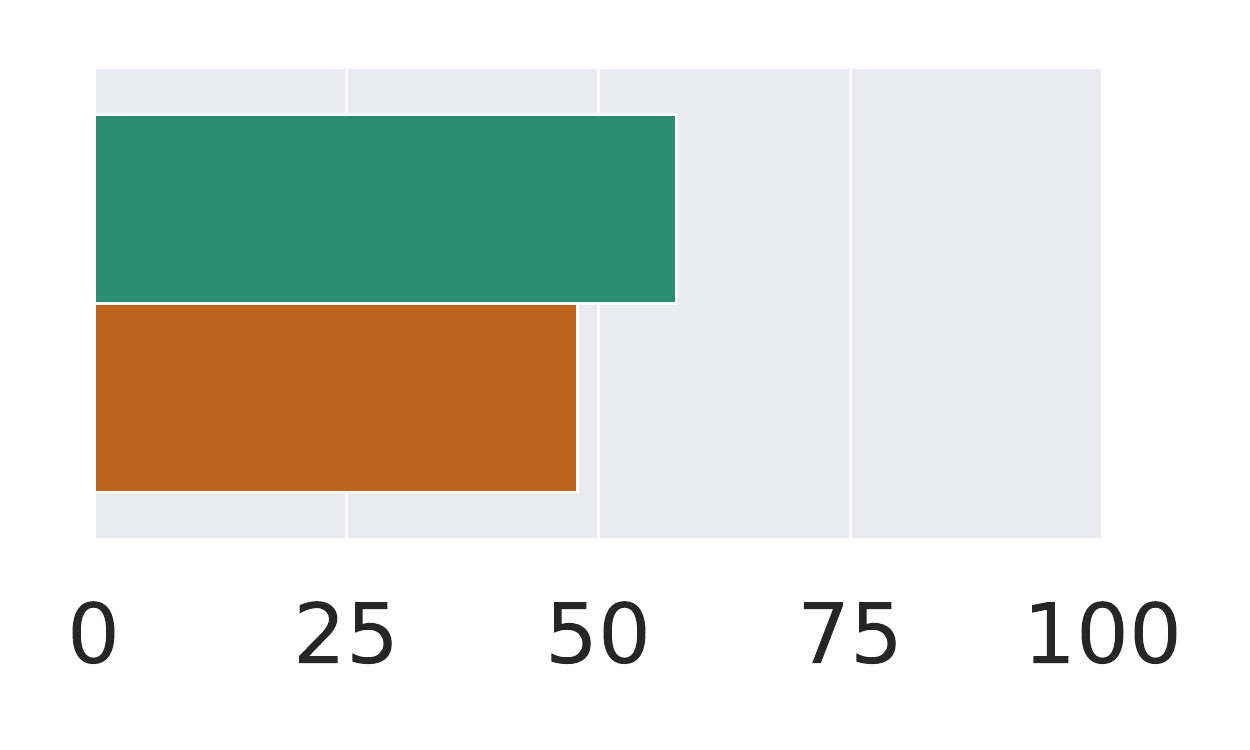}\end{minipage}
\\
\UseMacro{TH-metric-set_match_r}
 & \begin{minipage}{.12\textwidth}\includegraphics[width=\textwidth]{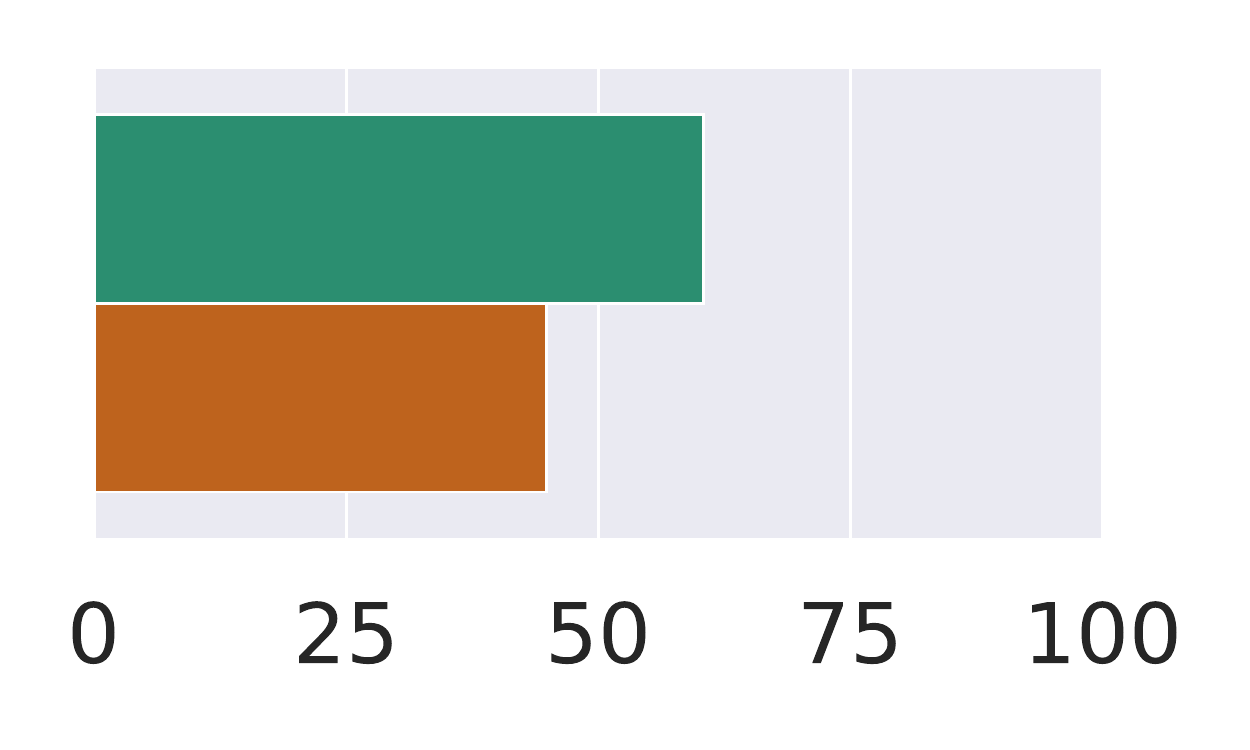}\end{minipage}
 & \begin{minipage}{.12\textwidth}\includegraphics[width=\textwidth]{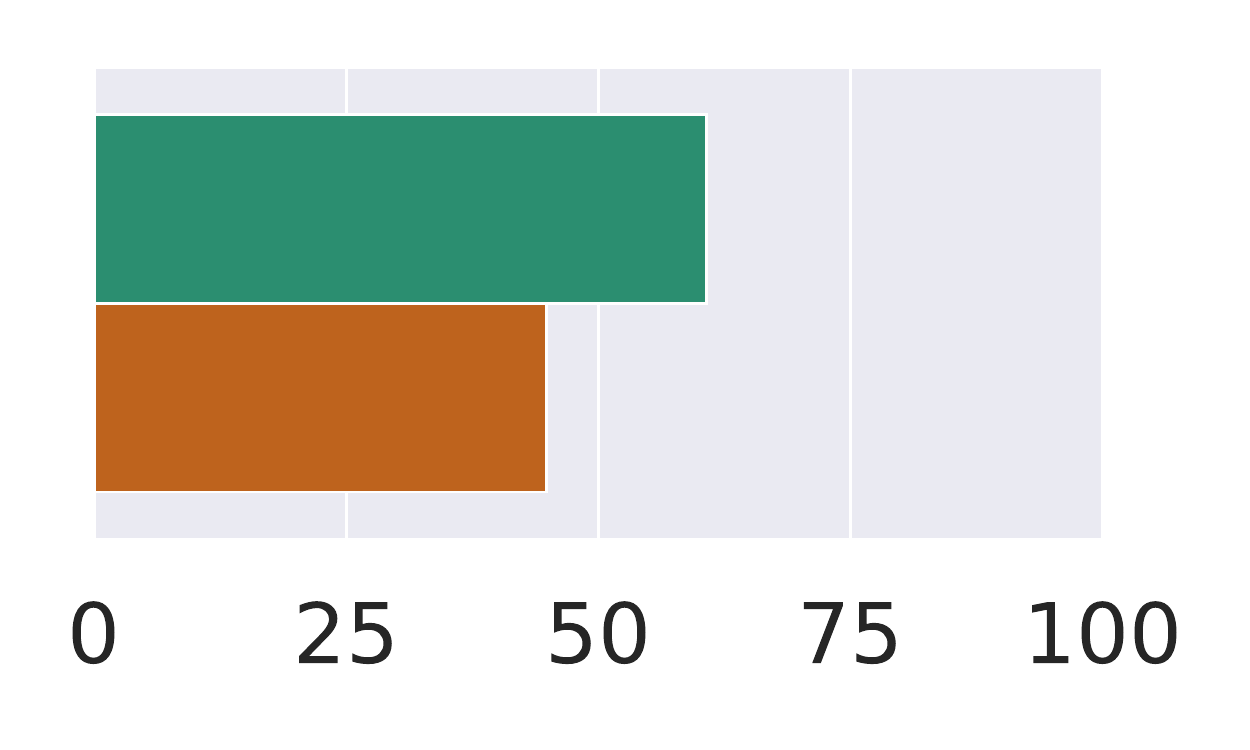}\end{minipage}
 & \begin{minipage}{.12\textwidth}\includegraphics[width=\textwidth]{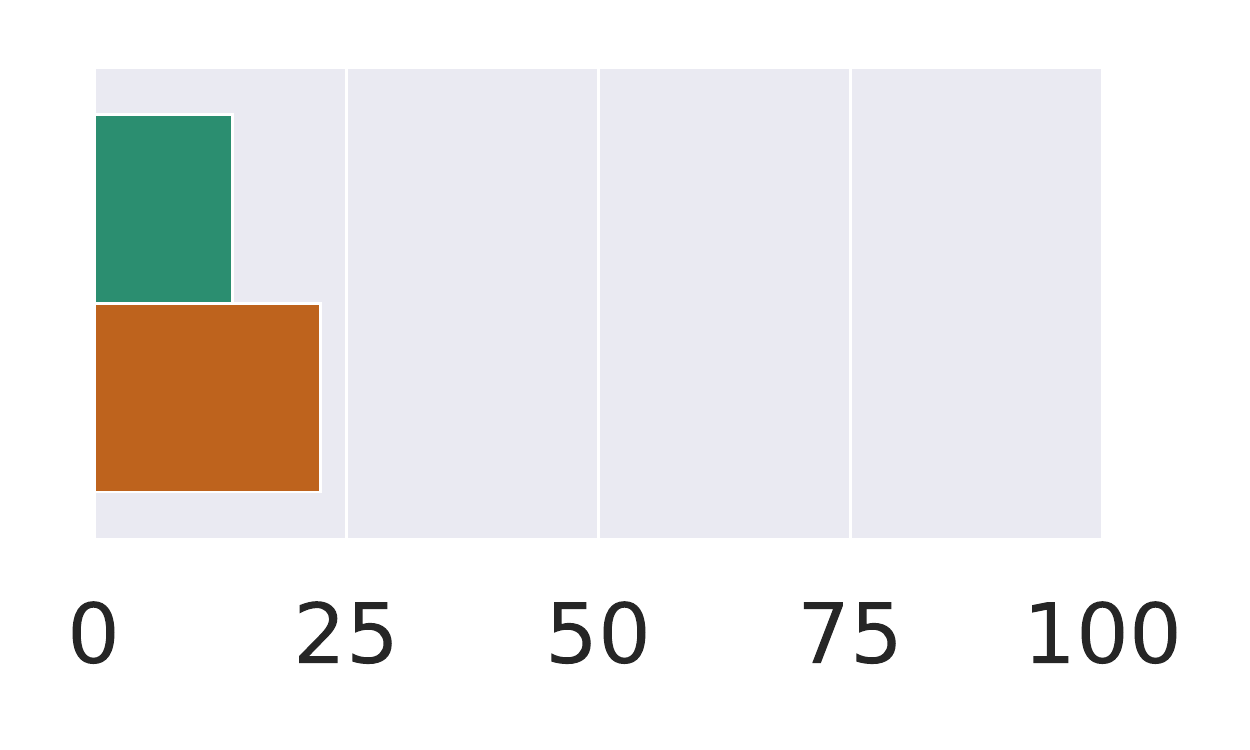}\end{minipage}
 & \begin{minipage}{.12\textwidth}\includegraphics[width=\textwidth]{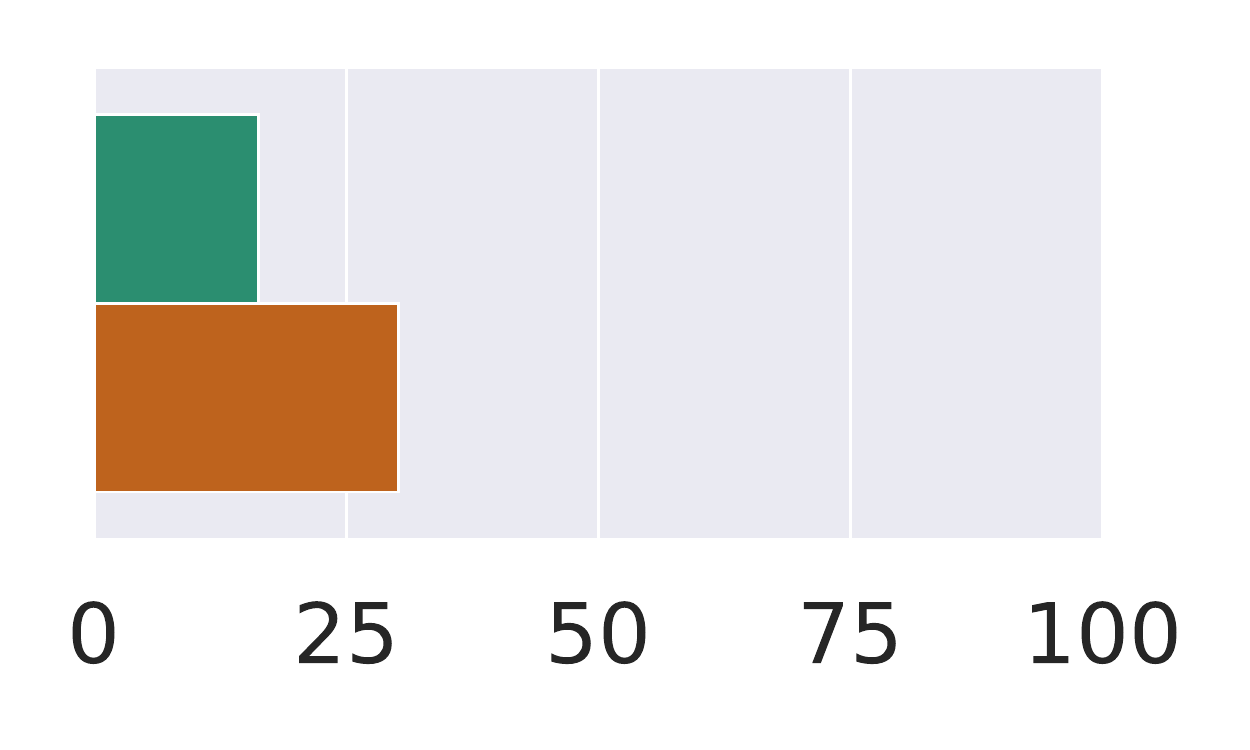}\end{minipage}
 & \begin{minipage}{.12\textwidth}\includegraphics[width=\textwidth]{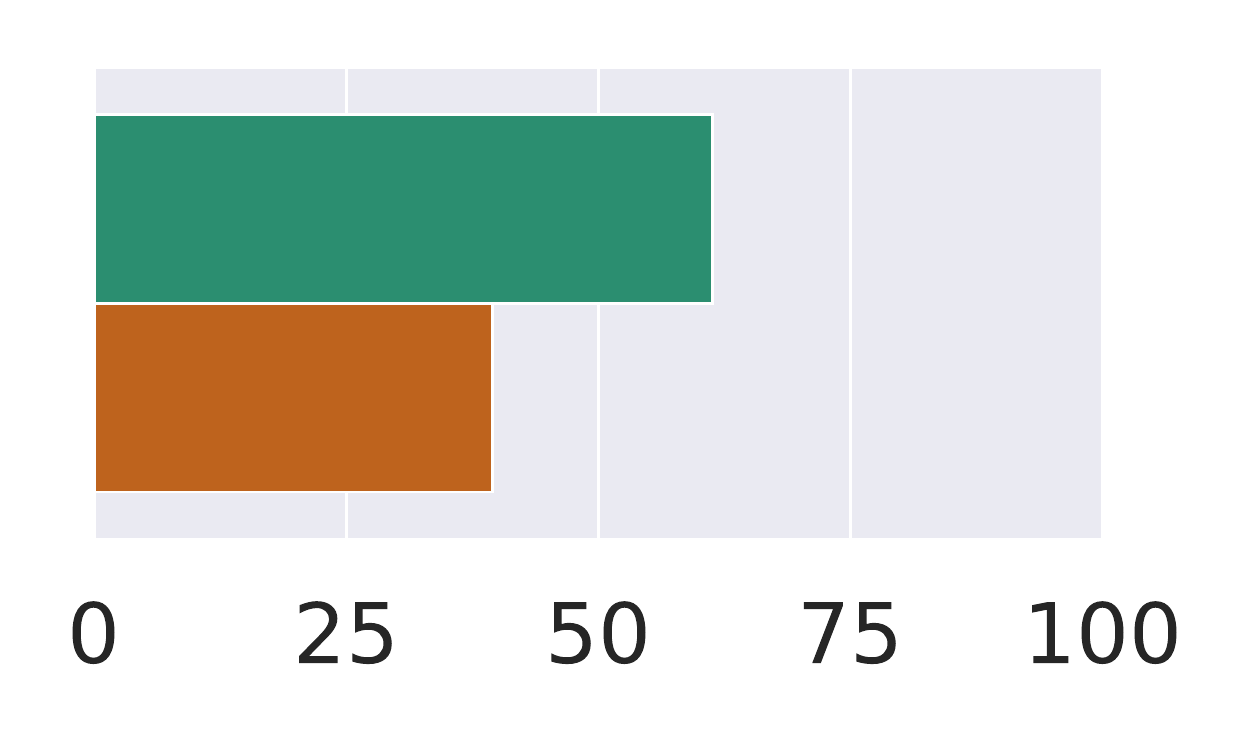}\end{minipage}
 & \begin{minipage}{.12\textwidth}\includegraphics[width=\textwidth]{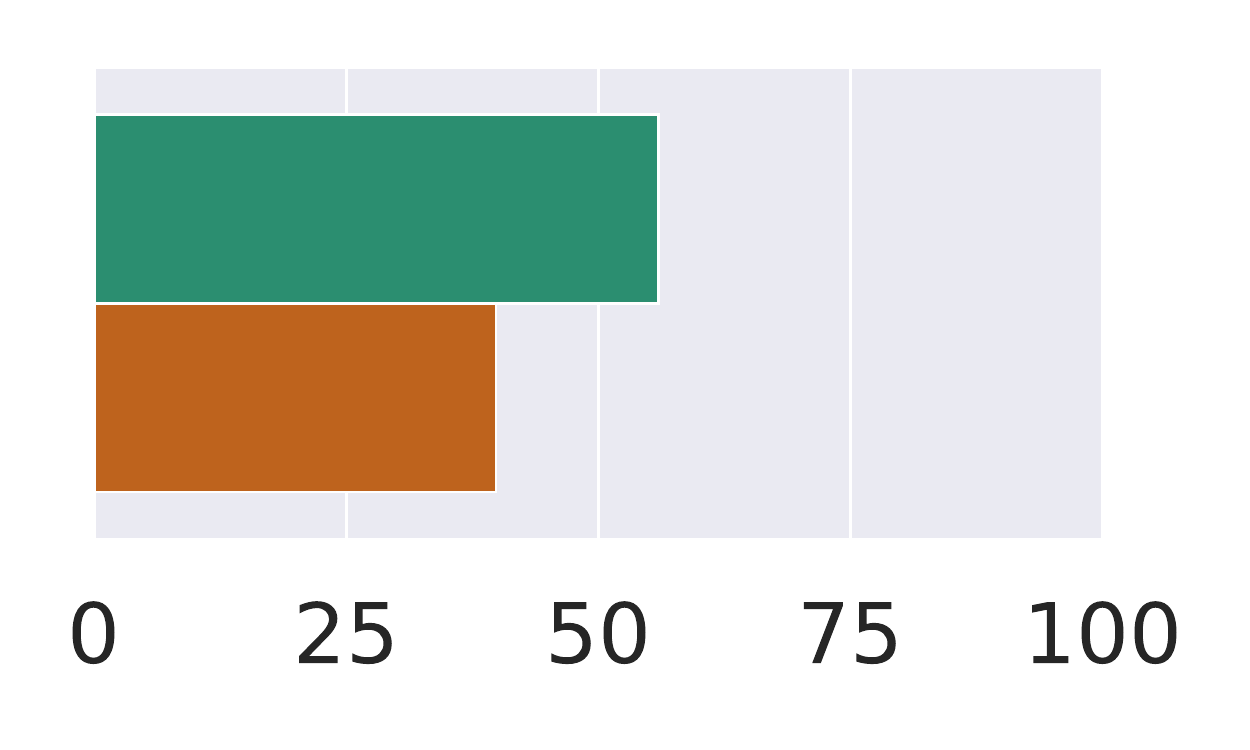}\end{minipage}
\\
\UseMacro{TH-metric-set_match_f}
 & \begin{minipage}{.12\textwidth}\includegraphics[width=\textwidth]{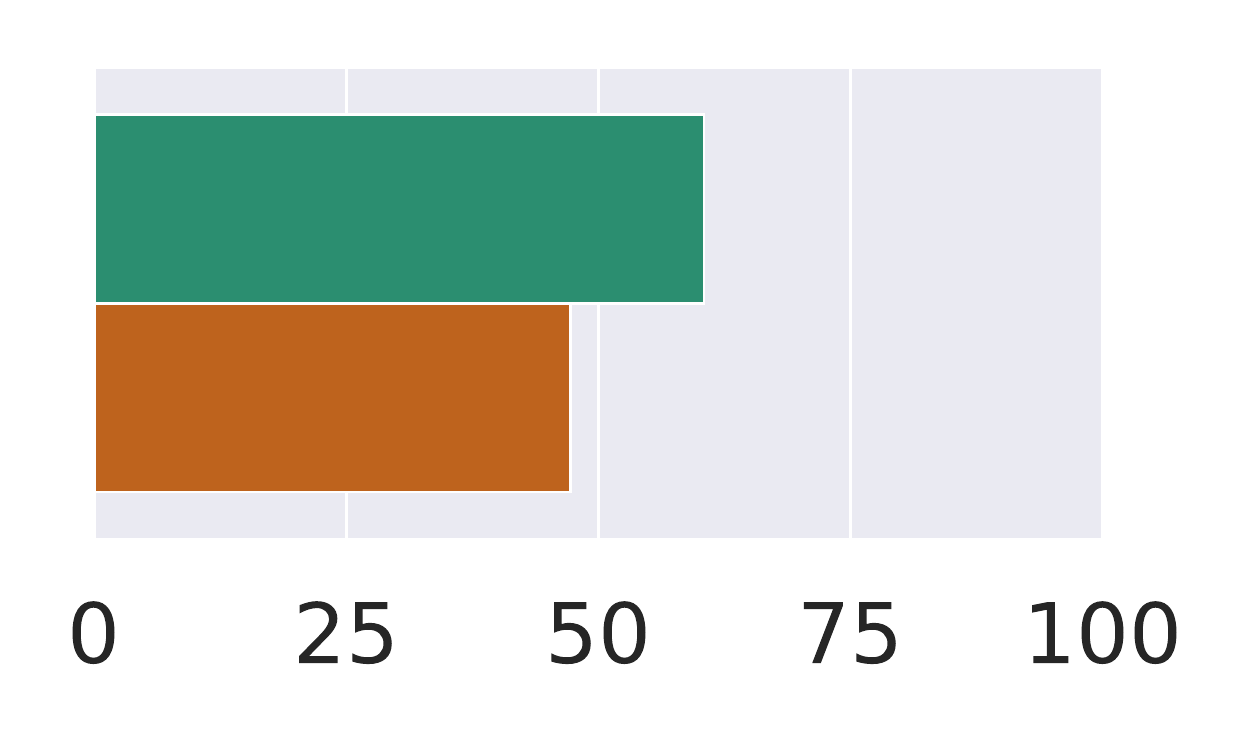}\end{minipage}
 & \begin{minipage}{.12\textwidth}\includegraphics[width=\textwidth]{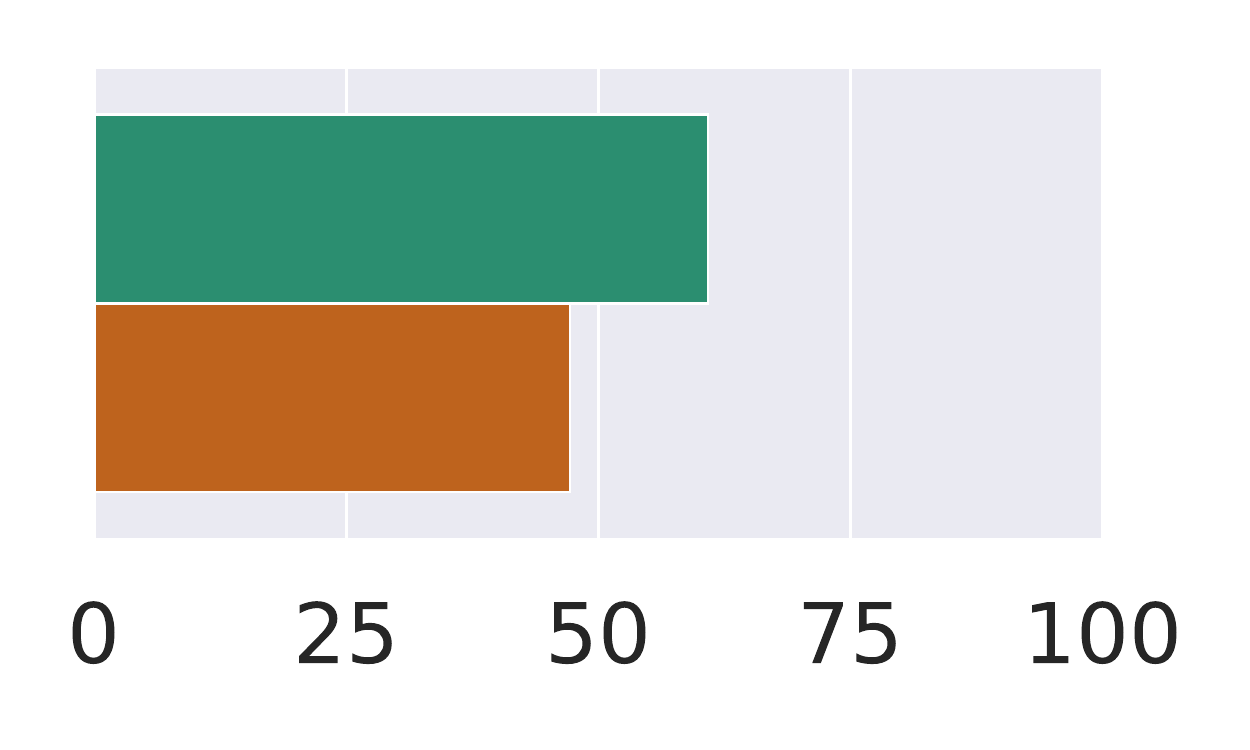}\end{minipage}
 & \begin{minipage}{.12\textwidth}\includegraphics[width=\textwidth]{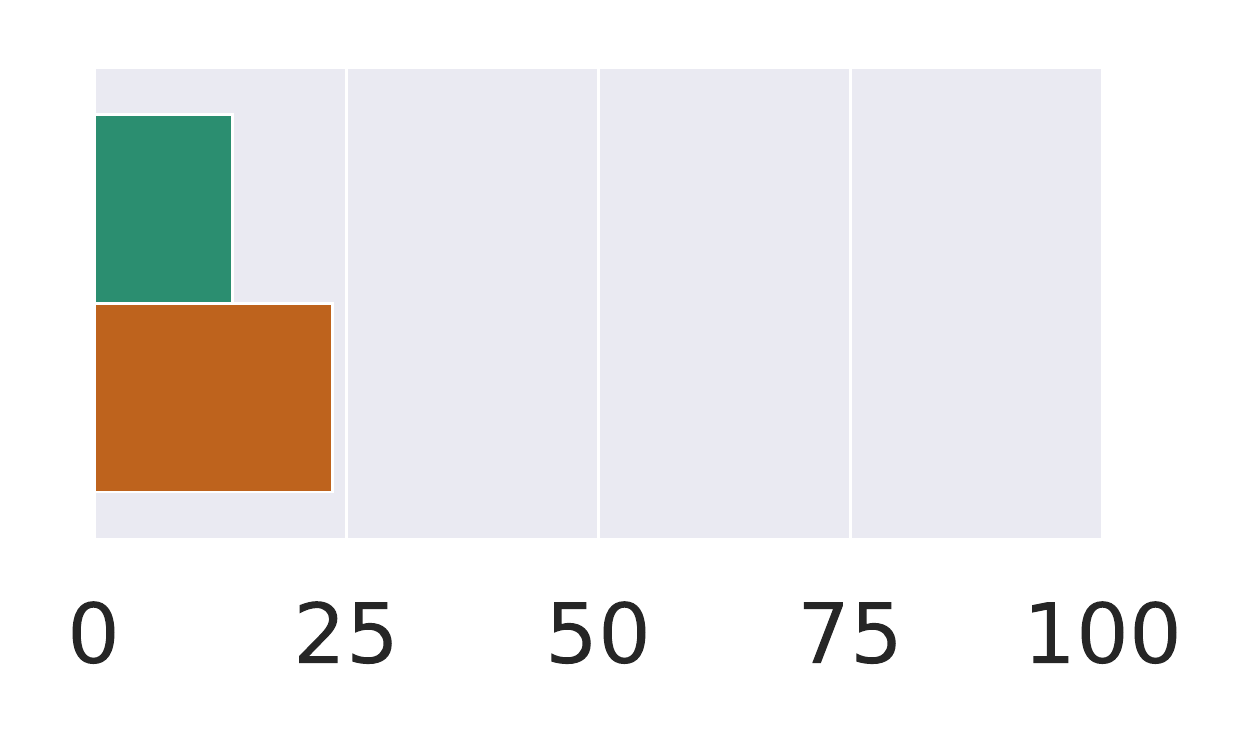}\end{minipage}
 & \begin{minipage}{.12\textwidth}\includegraphics[width=\textwidth]{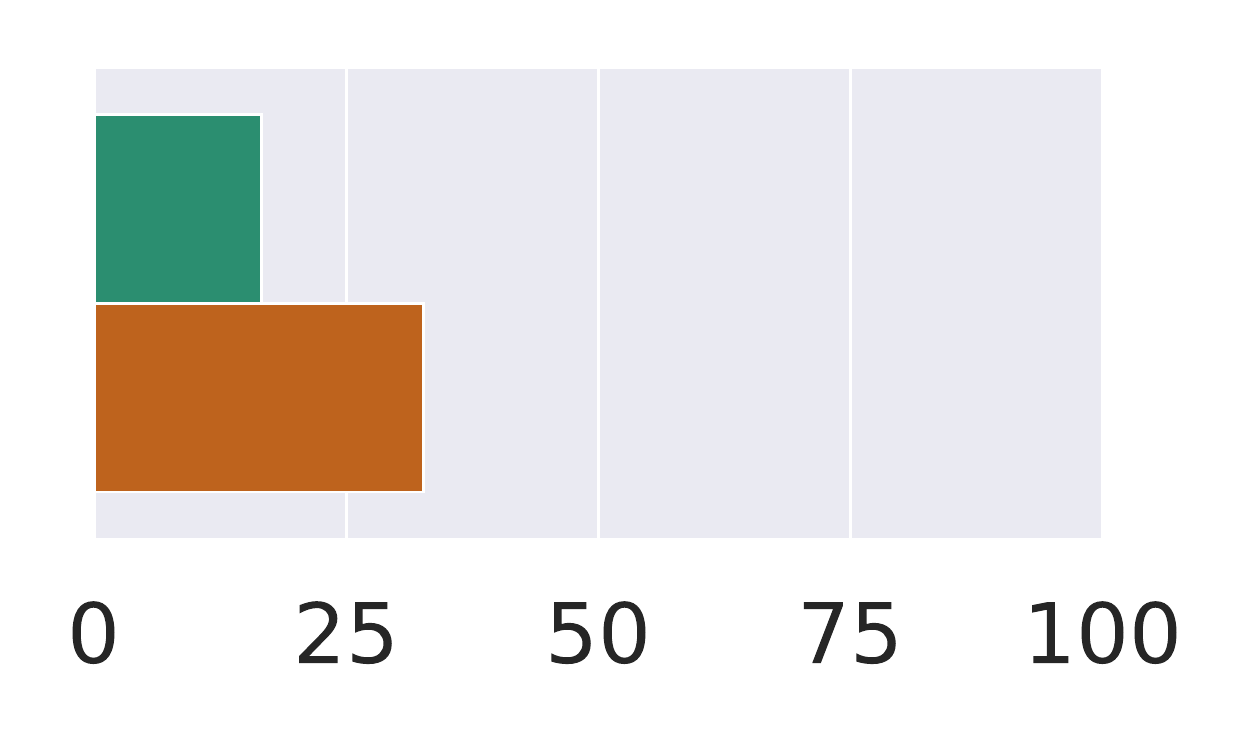}\end{minipage}
 & \begin{minipage}{.12\textwidth}\includegraphics[width=\textwidth]{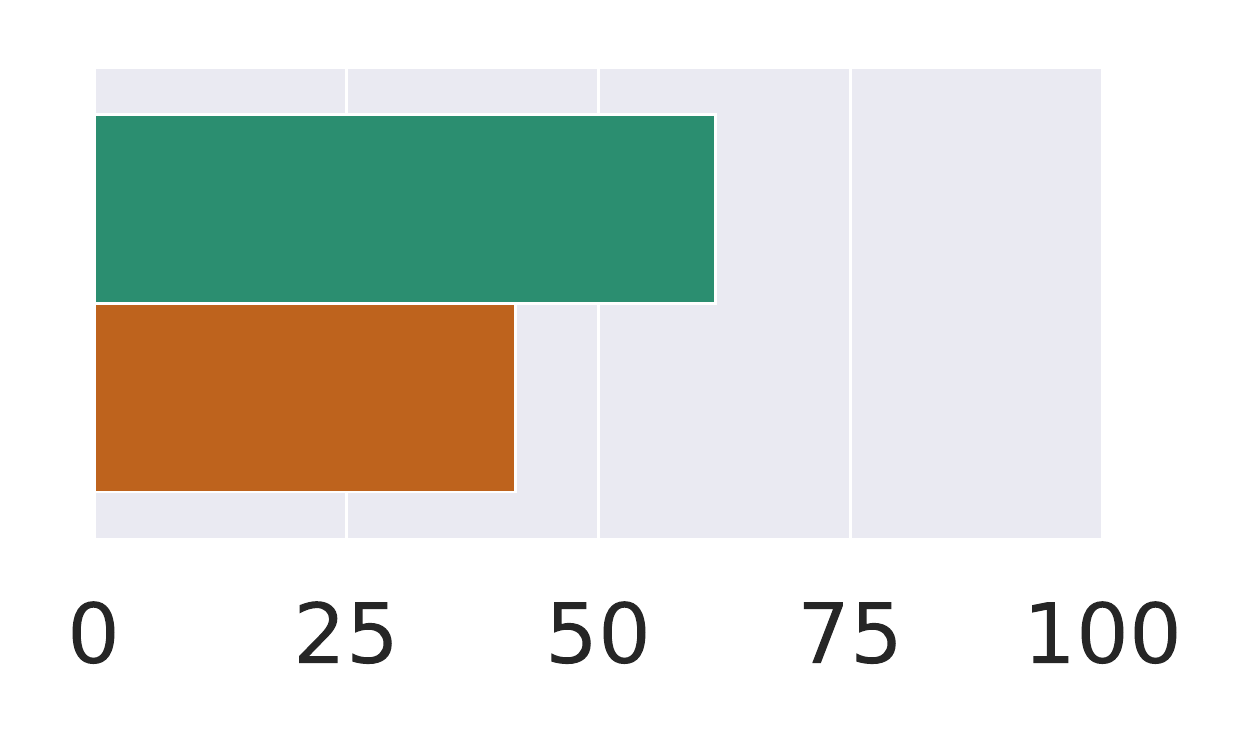}\end{minipage}
 & \begin{minipage}{.12\textwidth}\includegraphics[width=\textwidth]{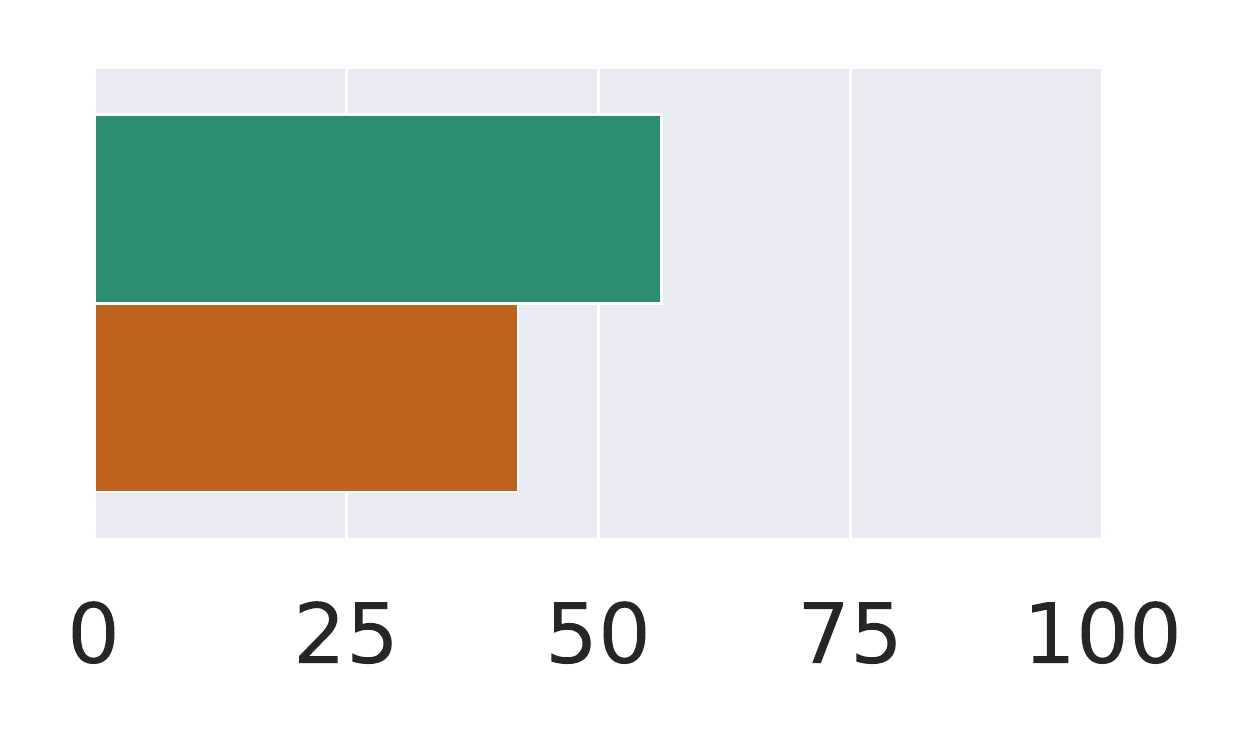}\end{minipage}
\\
\UseMacro{TH-metric-exact_match}
 & \begin{minipage}{.12\textwidth}\includegraphics[width=\textwidth]{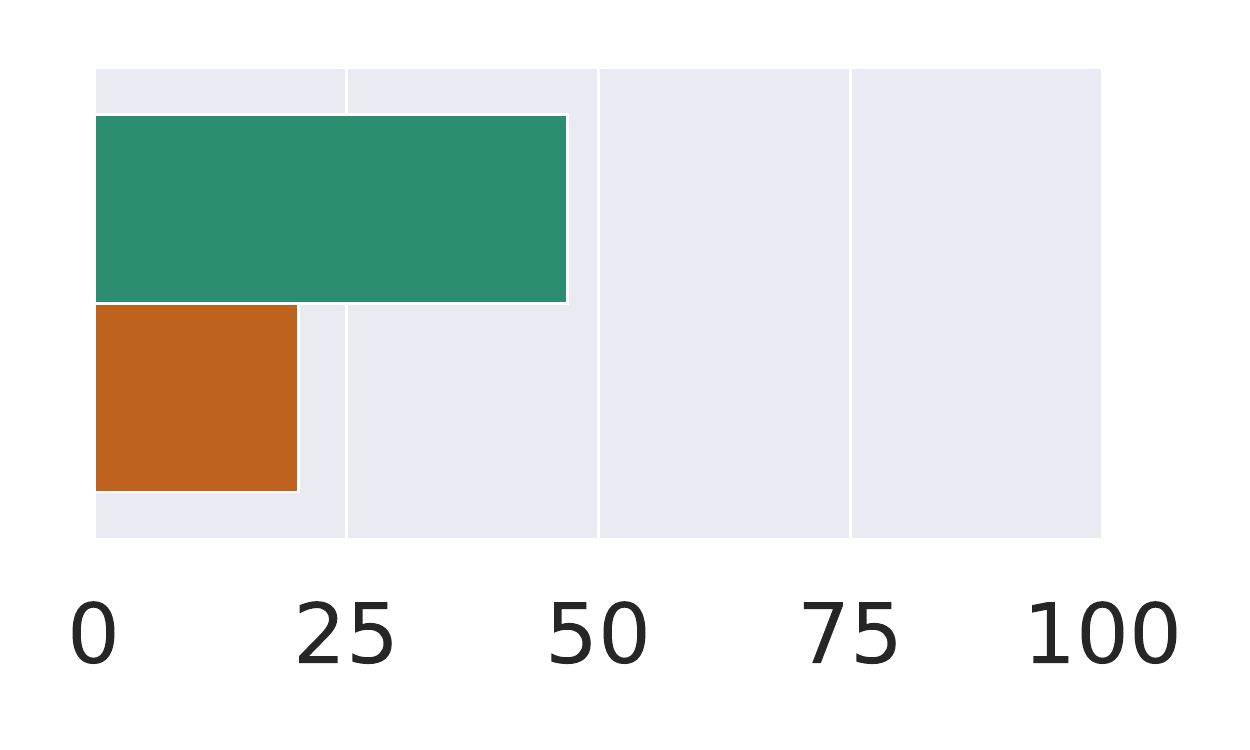}\end{minipage}
 & \begin{minipage}{.12\textwidth}\includegraphics[width=\textwidth]{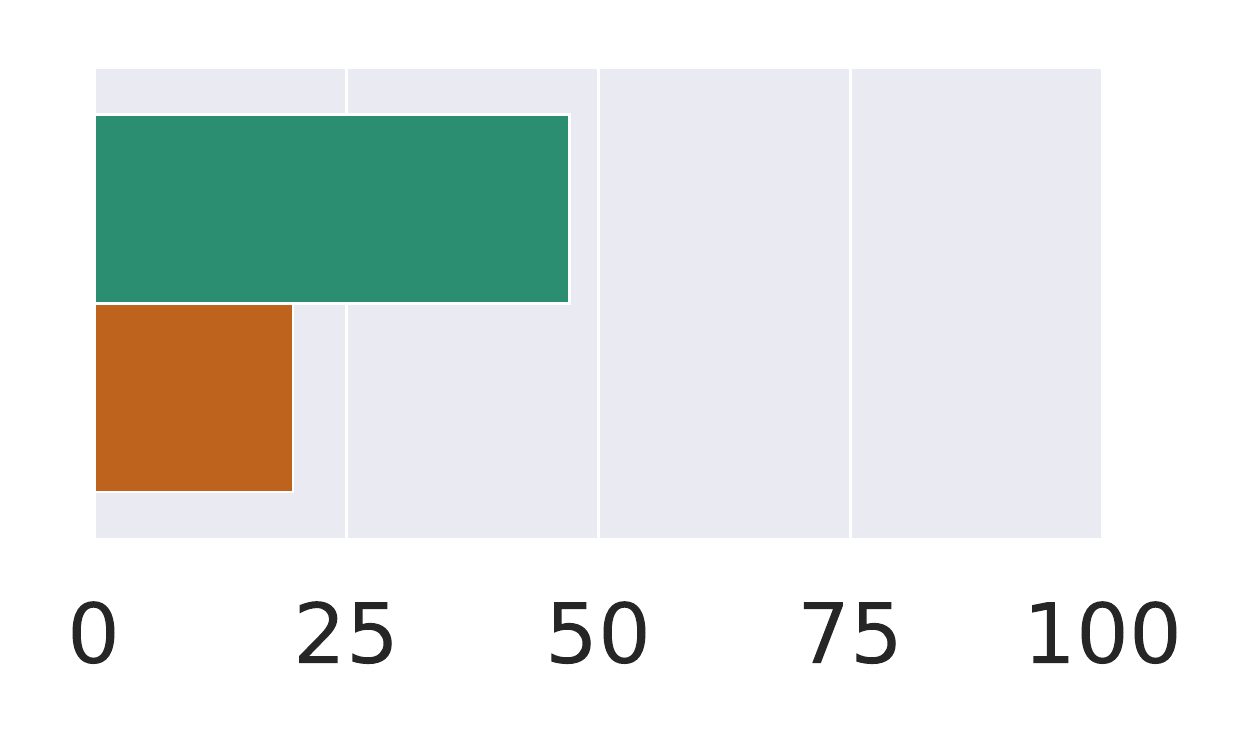}\end{minipage}
 & \begin{minipage}{.12\textwidth}\includegraphics[width=\textwidth]{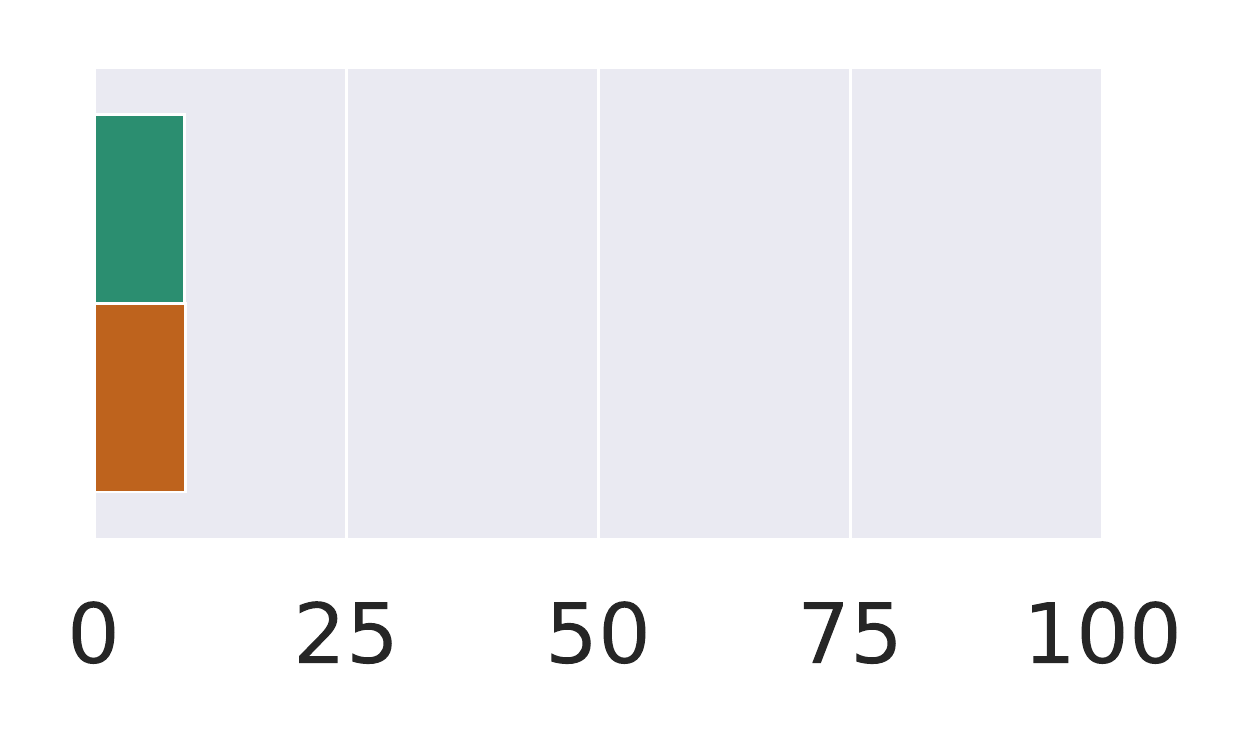}\end{minipage}
 & \begin{minipage}{.12\textwidth}\includegraphics[width=\textwidth]{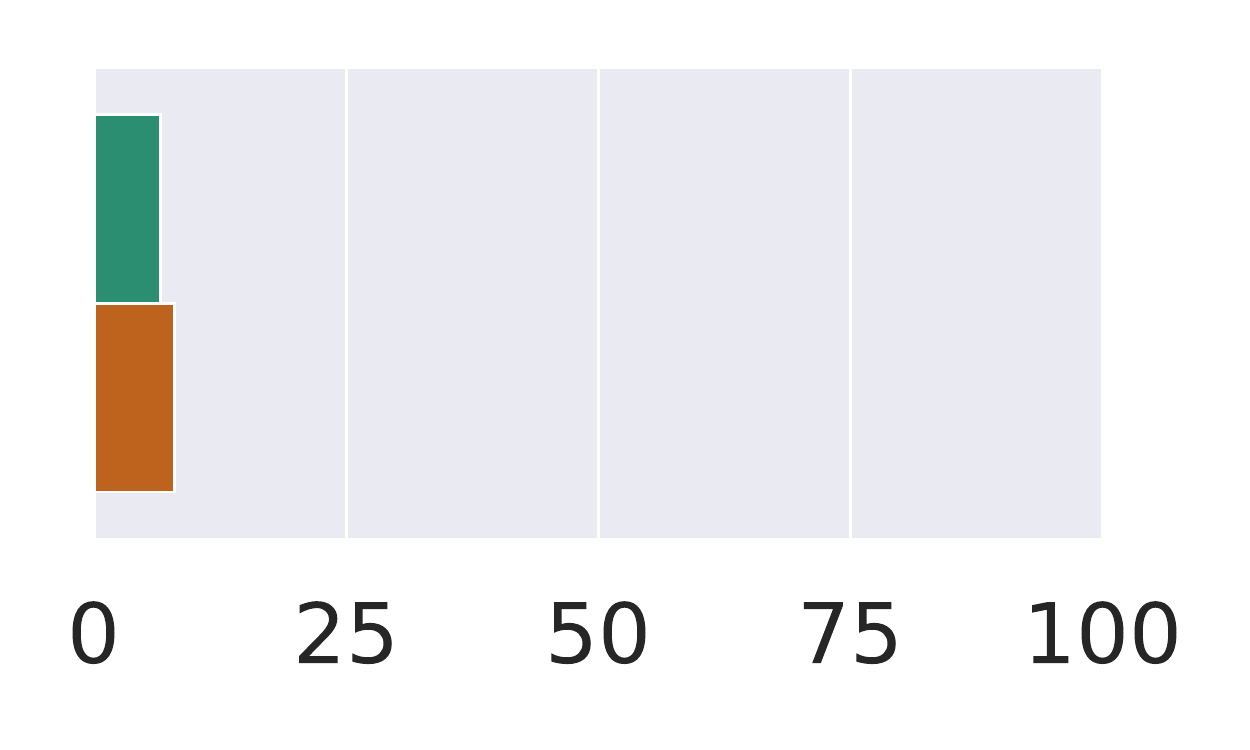}\end{minipage}
 & \begin{minipage}{.12\textwidth}\includegraphics[width=\textwidth]{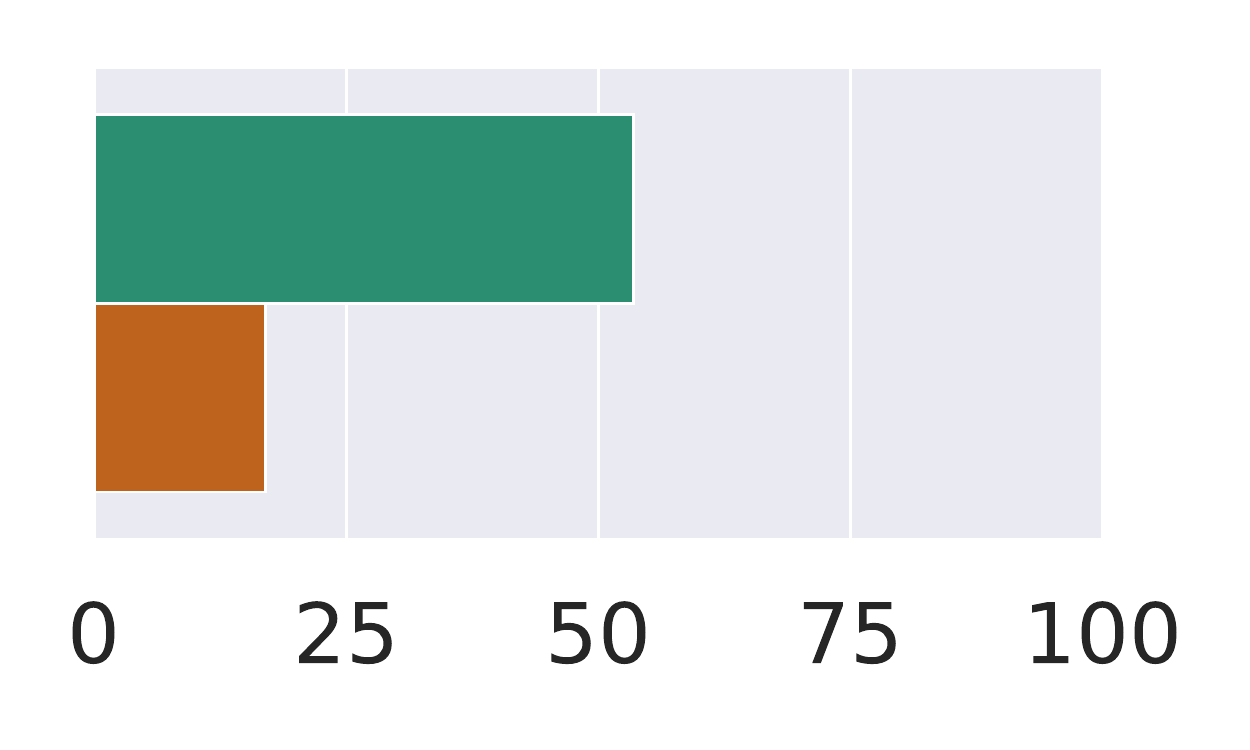}\end{minipage}
 & \begin{minipage}{.12\textwidth}\includegraphics[width=\textwidth]{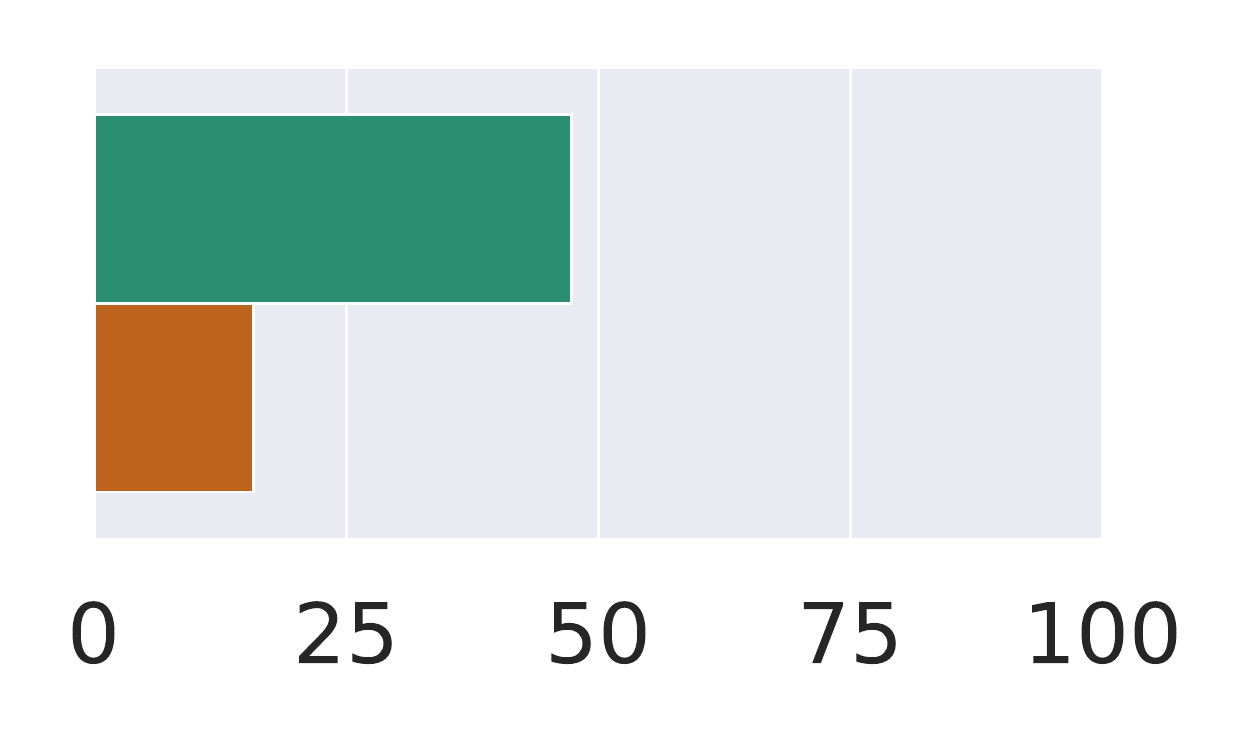}\end{minipage}
\\
\hline
\multicolumn{7}{c}{\includegraphics[scale=0.3]{figs/results-MN/legend}} \\
\end{tabular}
\end{footnotesize}
\end{center}

\vspace{-15pt}
\caption{Results of \metnam models on \val and \tests sets.\label{fig:results-aux-mn}}
\end{figure*}

We present the following additional tables and figures to help
characterize our experiments results and support our findings:
\begin{itemize}[topsep=3pt,itemsep=3pt,partopsep=0ex,parsep=0ex,leftmargin=*]
\item Evaluation results on the \AVal and \ATestS sets.
  \begin{itemize}
  \item \comgen: Table~\ref{table:results-aux-cg}.
  \item \metnam: Table~\ref{table:results-aux-mn}.
  \end{itemize}
\item Bar plots of the automatic metrics per \example.
  \begin{itemize}
  \item \comgen:
    \begin{itemize}
    \item on the \ATestC sets: Figure~\ref{fig:results-cg}.
    \item on the \AVal and \ATestS sets: Figure~\ref{fig:results-aux-cg}.
    \end{itemize}
  \item \metnam:
    \begin{itemize}
    \item on the \ATestC sets: Figure~\ref{fig:results-mn}.
    \item on the \AVal and \ATestS sets: Figure~\ref{fig:results-aux-mn}.
    \end{itemize}
  \end{itemize}
\end{itemize}

\clearpage
\newpage

\renewcommand{\textfraction}{0}
\renewcommand{\floatpagefraction}{0.1}

\onecolumn

\foreach \ndconfig in {same_code, same_nl, sim_90}{
\foreach \task in {CG, MN}{

\DefMacro{TC-results-\task_\ndconfig}{\UseMacro{TaskC_\task} models'
  results with filtering near-duplicates using the
  \UseMacro{nd_\ndconfig} configuration, on \ATestC sets.\label{table:nd-results:\task-\ndconfig}}
\DefMacro{TV-results-\task_\ndconfig}{-5pt}

\input{tables/\task_nd_\ndconfig/numbers-results-\task}
\input{tables/\task_nd_\ndconfig/table-results-\task_\ndconfig}

\begin{figure}[H]


\begin{center}
\begin{footnotesize}
\begin{tabular}{|l|c|c|c || l|c|}
\hline
& \textbf{ATrain} & \textbf{\AVal} & \textbf{\ATestS} & & \textbf{\ATestC} \\
\hline
\UseMacro{TH-ds-MP}
 & \begin{minipage}{.18\textwidth}\includegraphics[width=\textwidth]{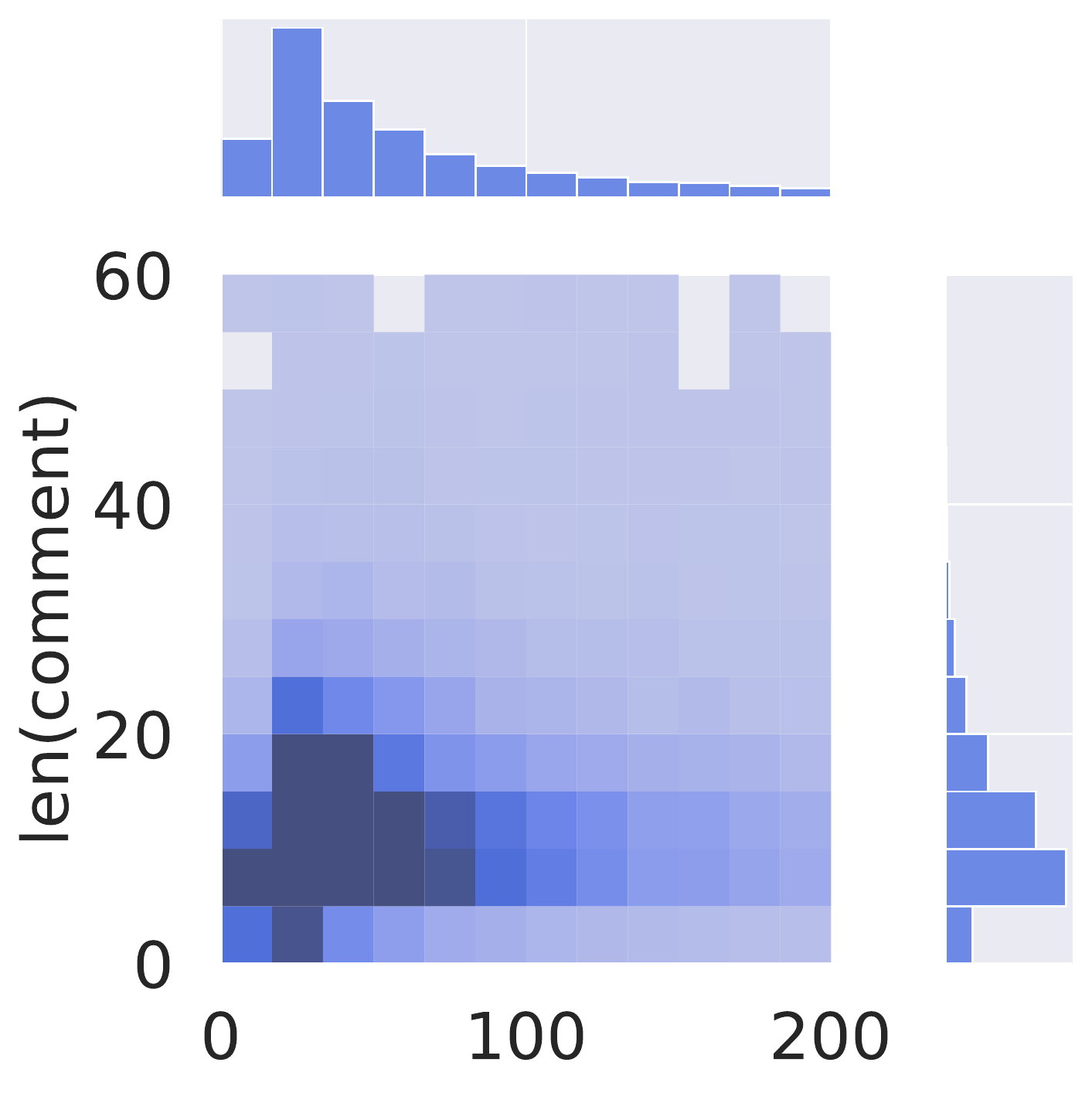}\end{minipage}
 & \begin{minipage}{.18\textwidth}\includegraphics[width=\textwidth]{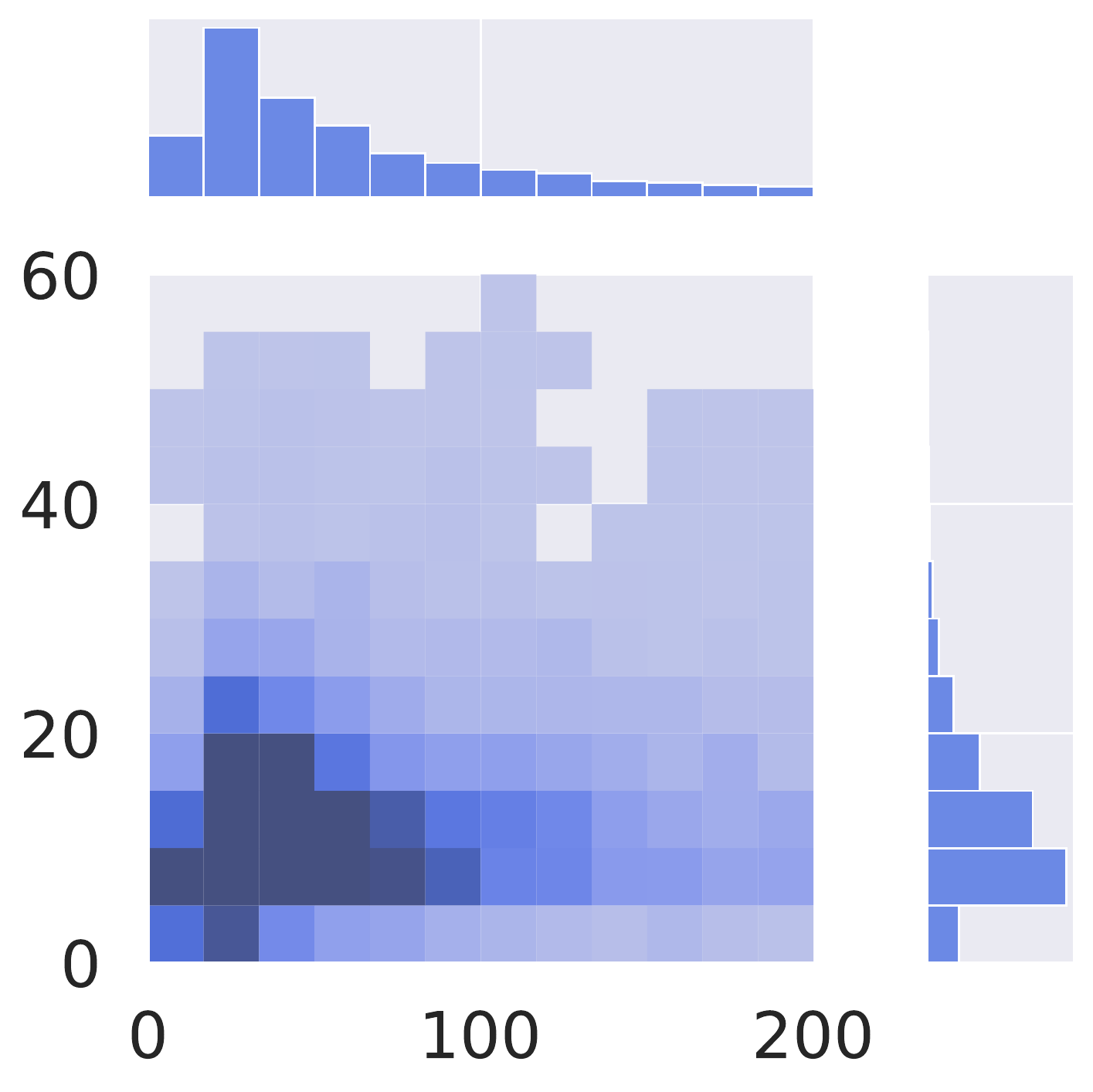}\end{minipage}
 & \begin{minipage}{.18\textwidth}\includegraphics[width=\textwidth]{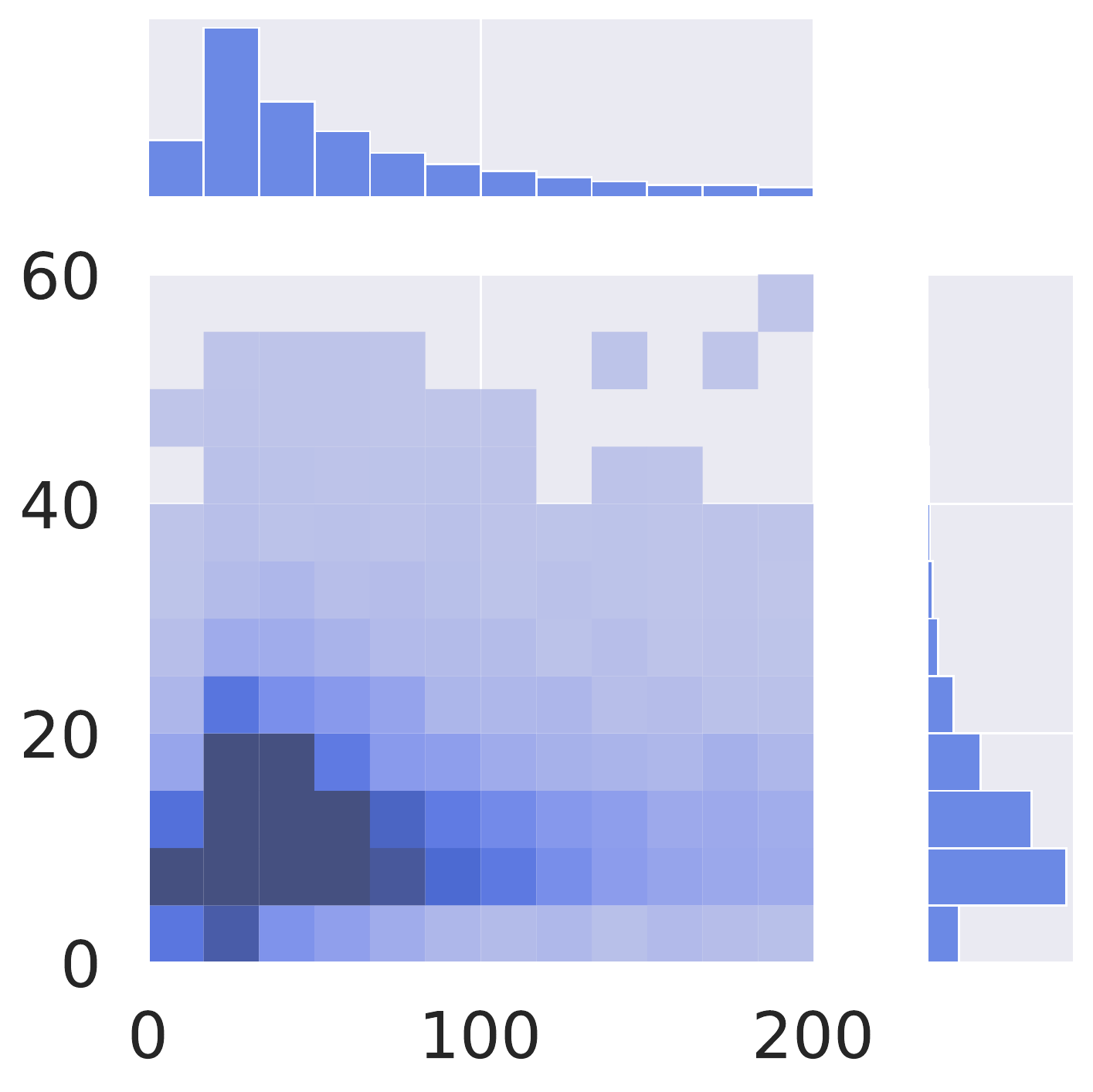}\end{minipage}
 & \UseMacro{TH-ds-MP-CP}
 & \begin{minipage}{.18\textwidth}\includegraphics[width=\textwidth]{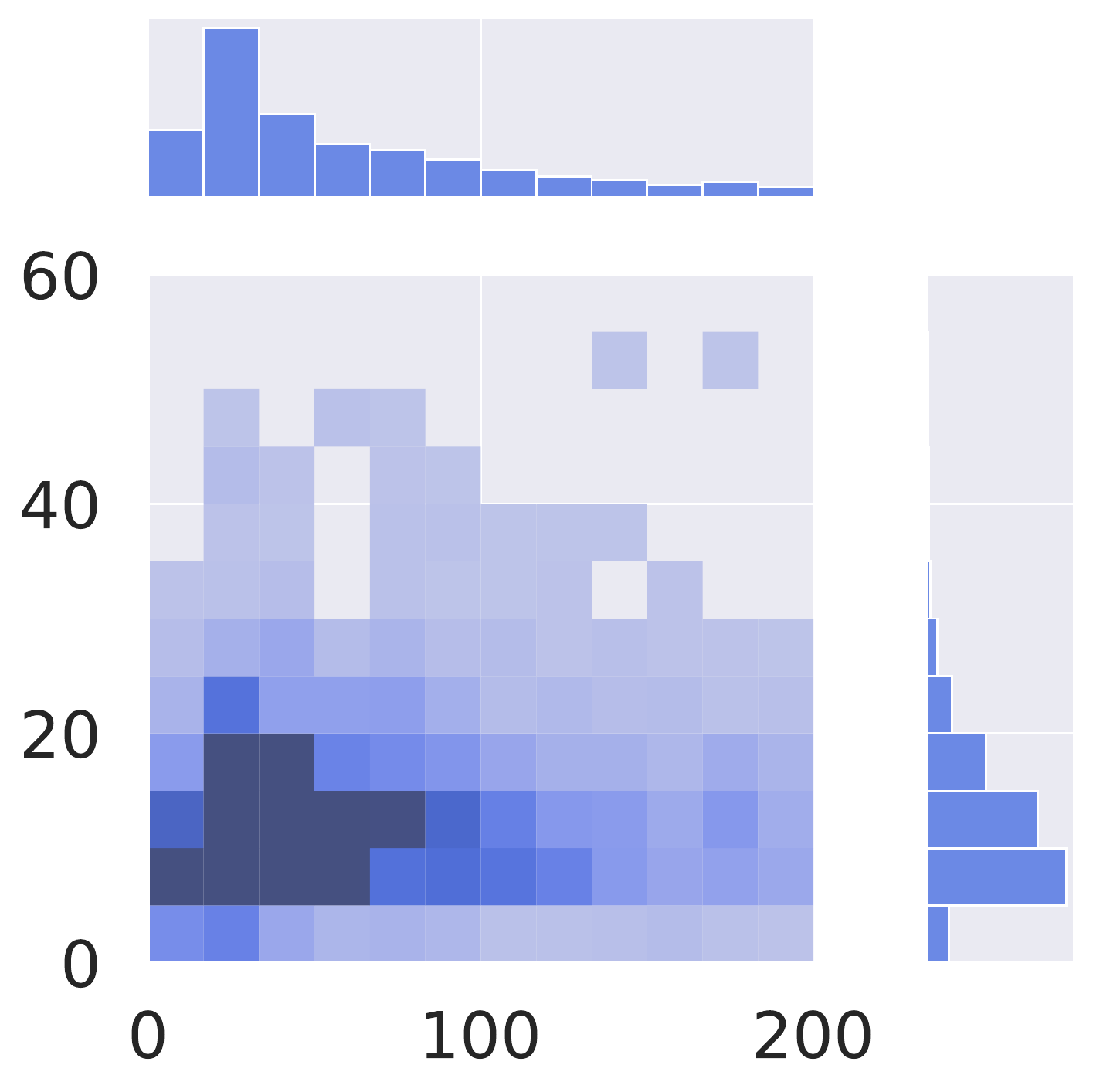}\end{minipage}
\\
\hline
\UseMacro{TH-ds-CP}
 & \begin{minipage}{.18\textwidth}\includegraphics[width=\textwidth]{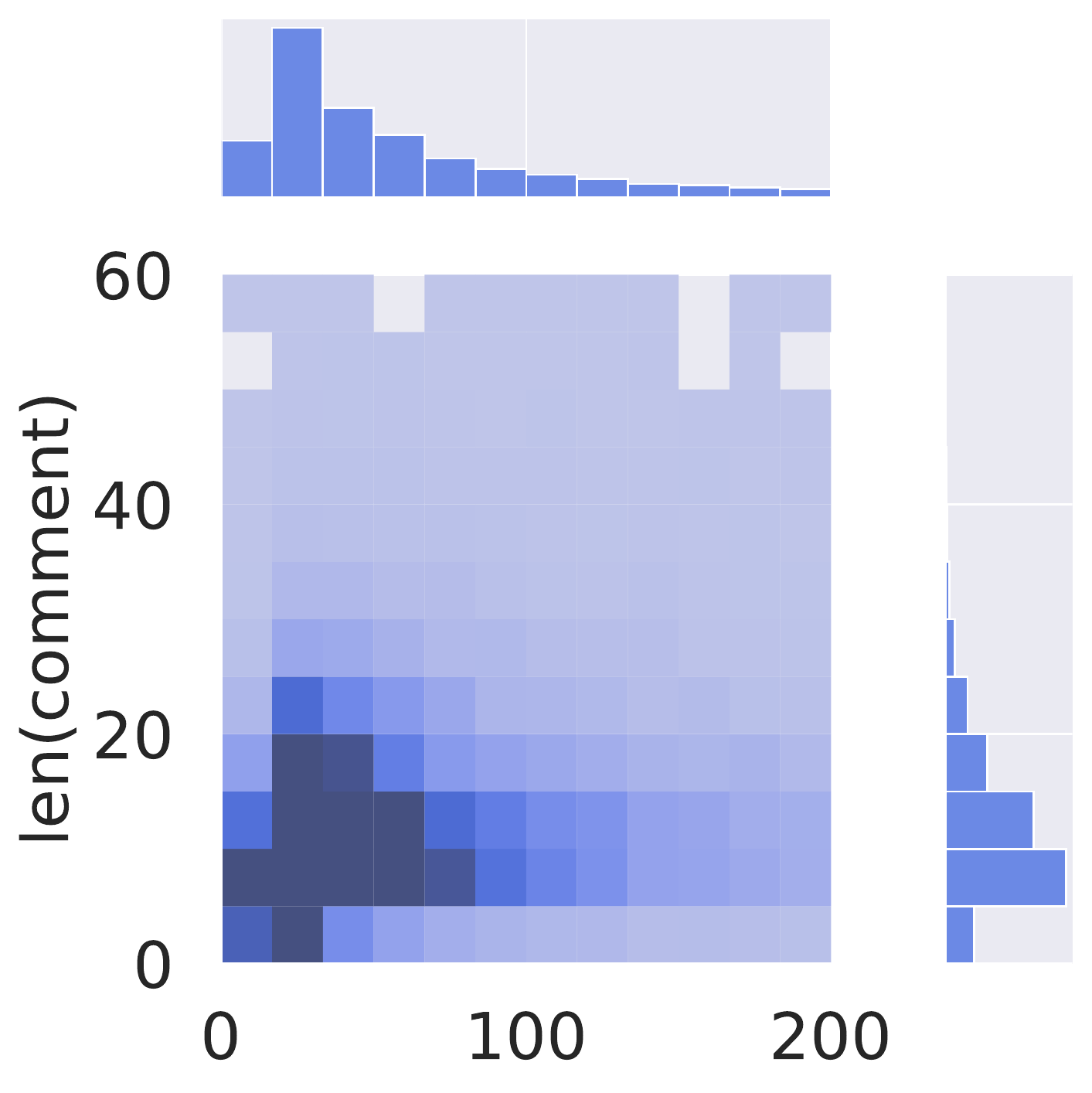}\end{minipage}
 & \begin{minipage}{.18\textwidth}\includegraphics[width=\textwidth]{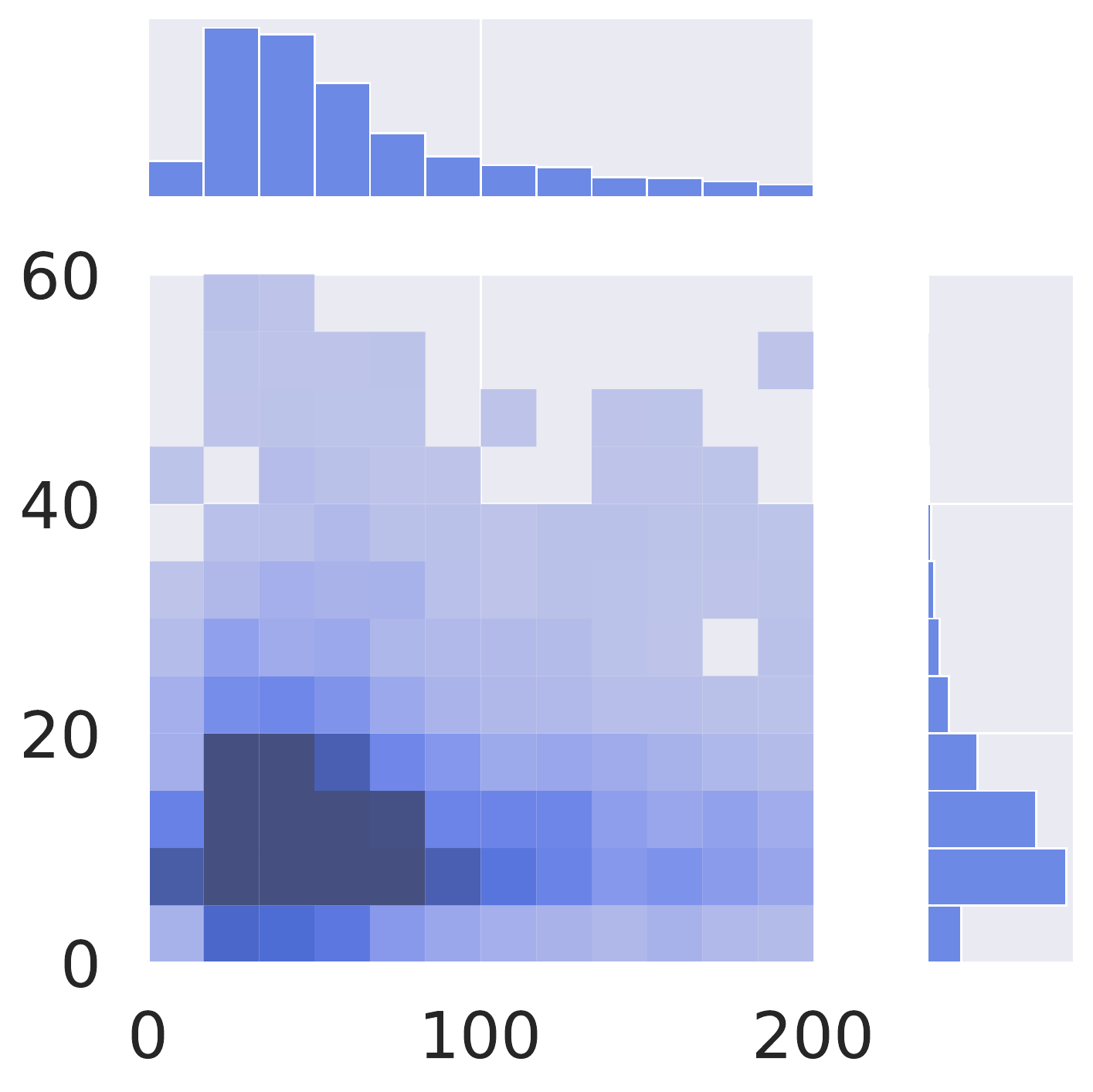}\end{minipage}
 & \begin{minipage}{.18\textwidth}\includegraphics[width=\textwidth]{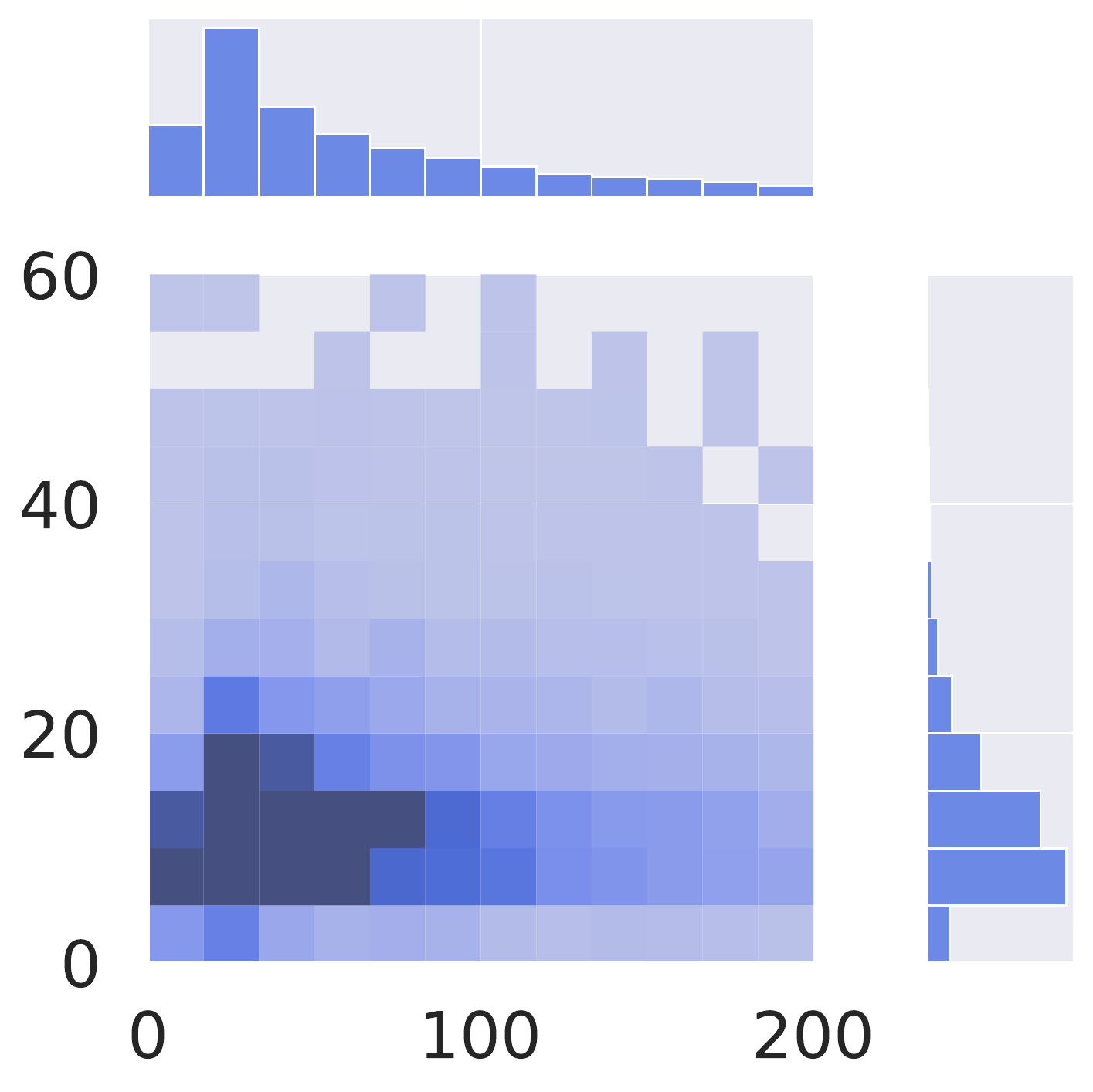}\end{minipage}
 & \UseMacro{TH-ds-MP-T}
 & \begin{minipage}{.18\textwidth}\includegraphics[width=\textwidth]{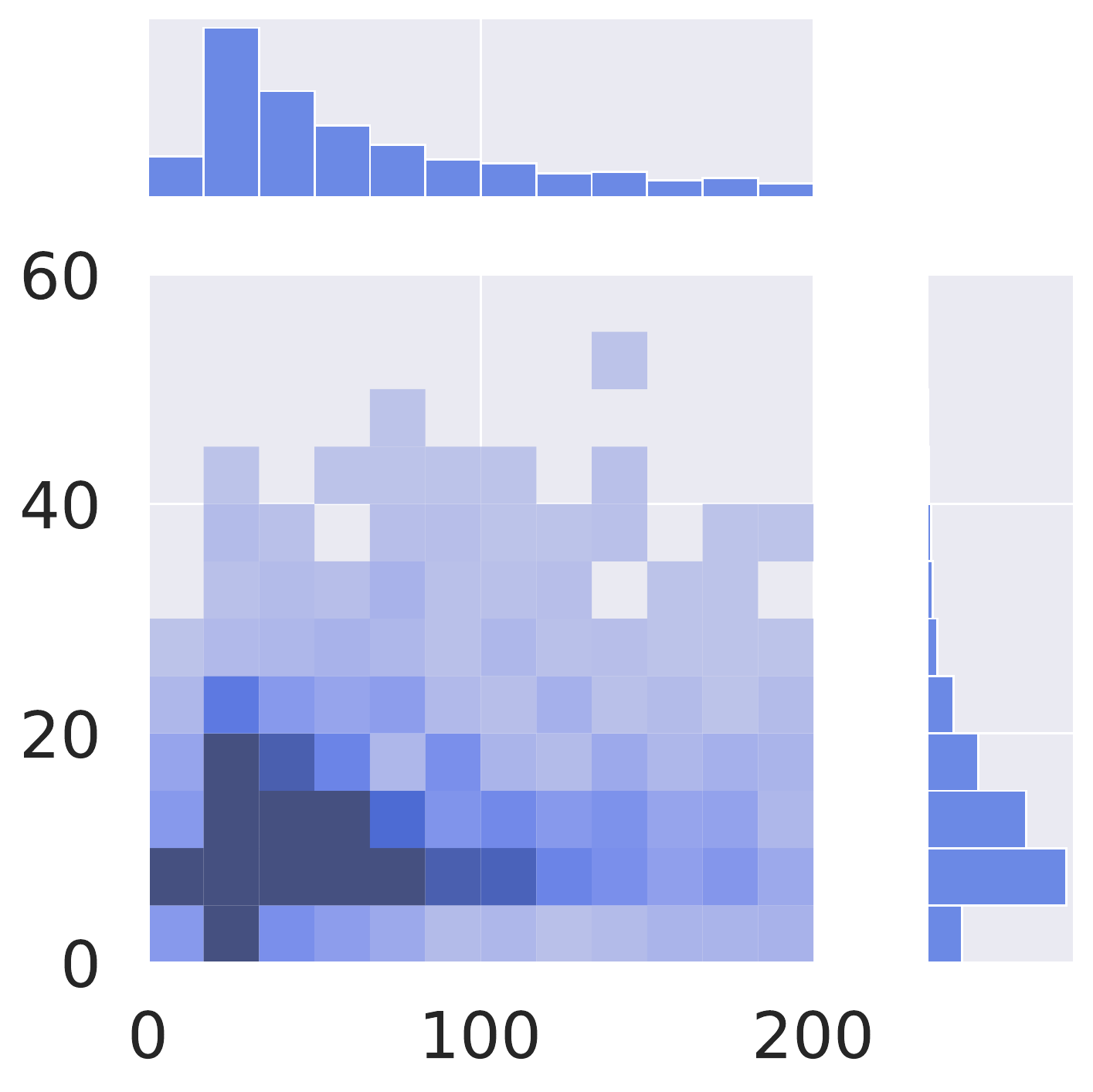}\end{minipage}
\\
\hline
\UseMacro{TH-ds-T}
 & \begin{minipage}{.18\textwidth}\includegraphics[width=\textwidth]{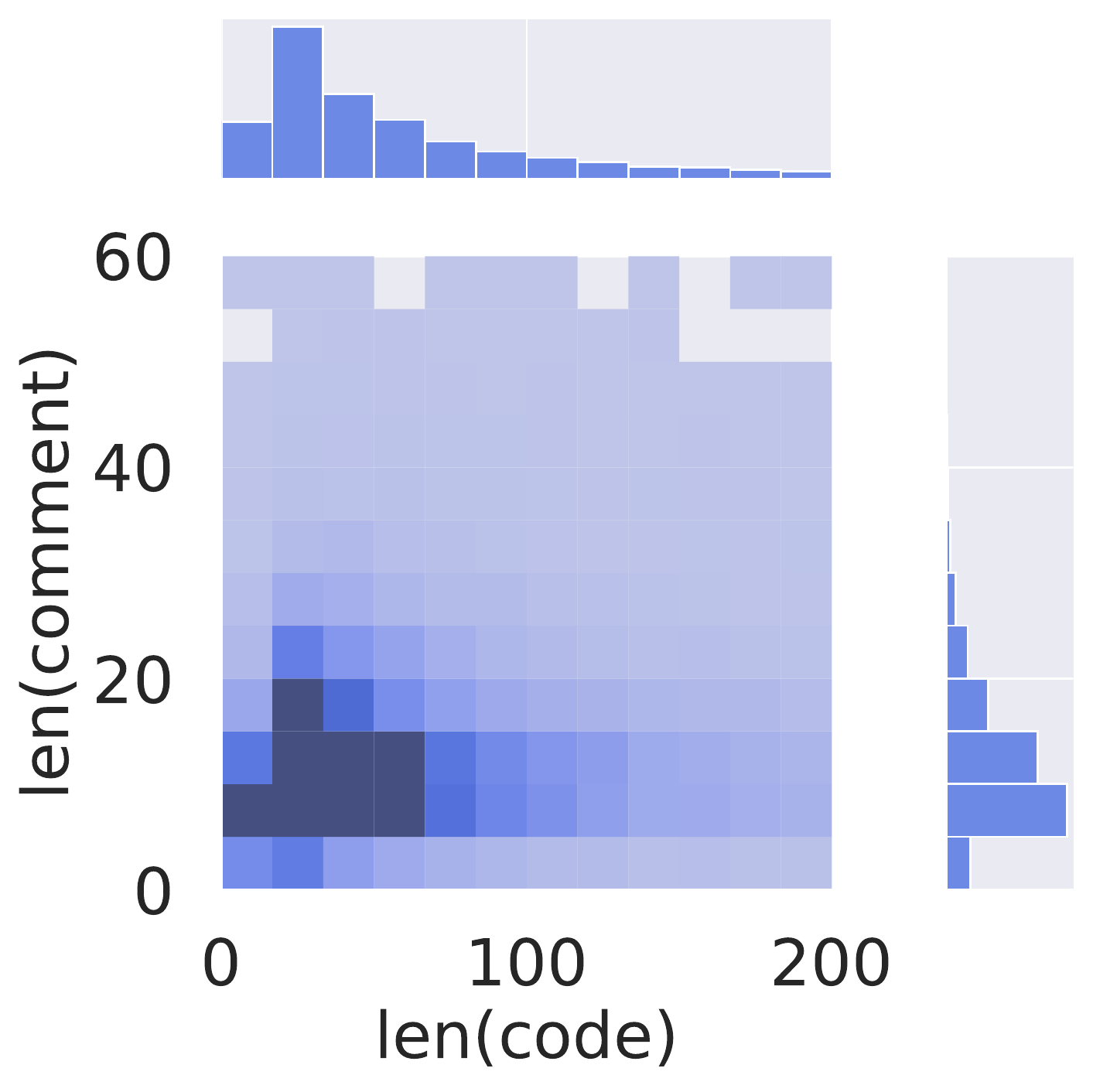}\end{minipage}
 & \begin{minipage}{.18\textwidth}\includegraphics[width=\textwidth]{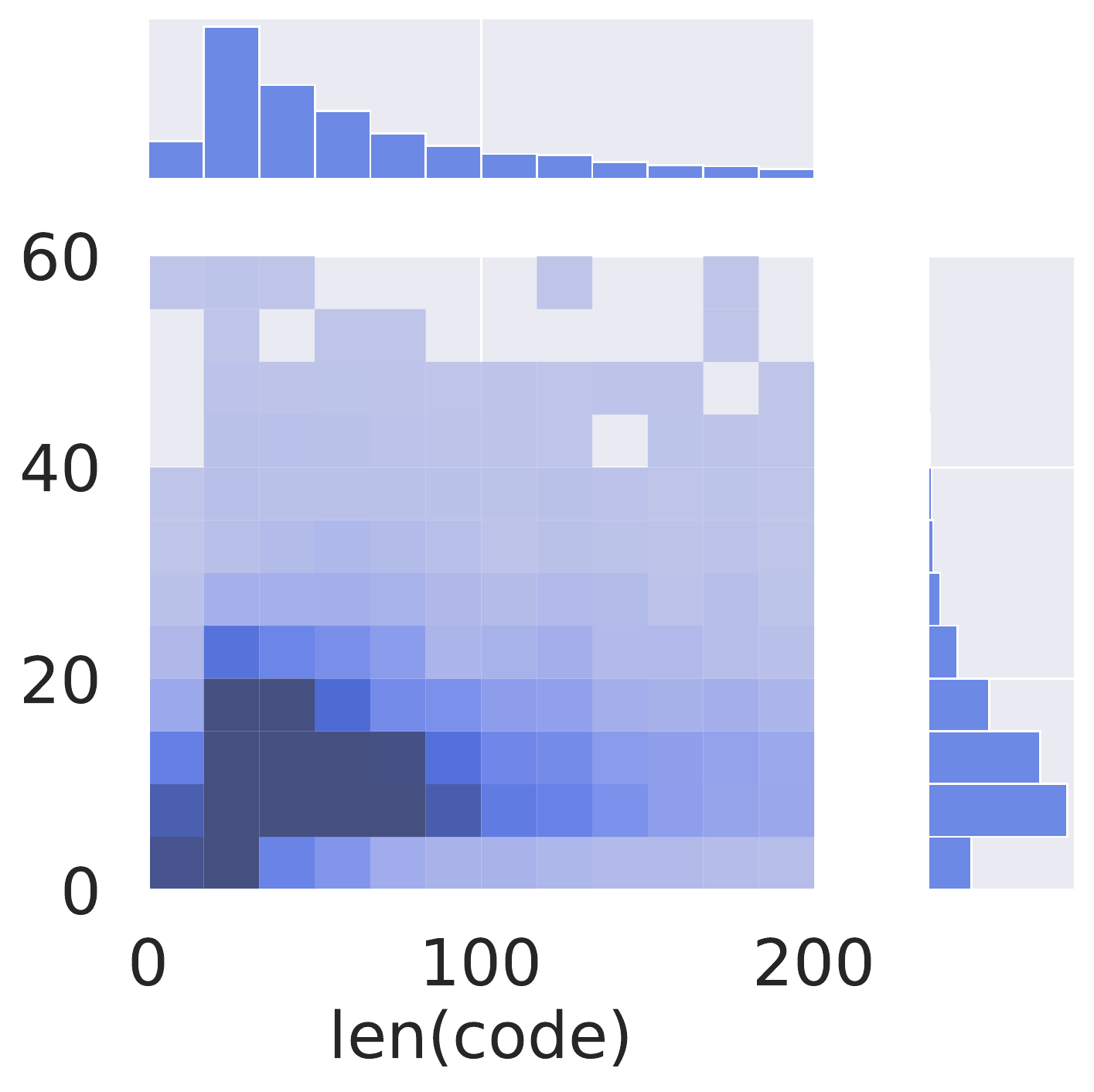}\end{minipage}
 & \begin{minipage}{.18\textwidth}\includegraphics[width=\textwidth]{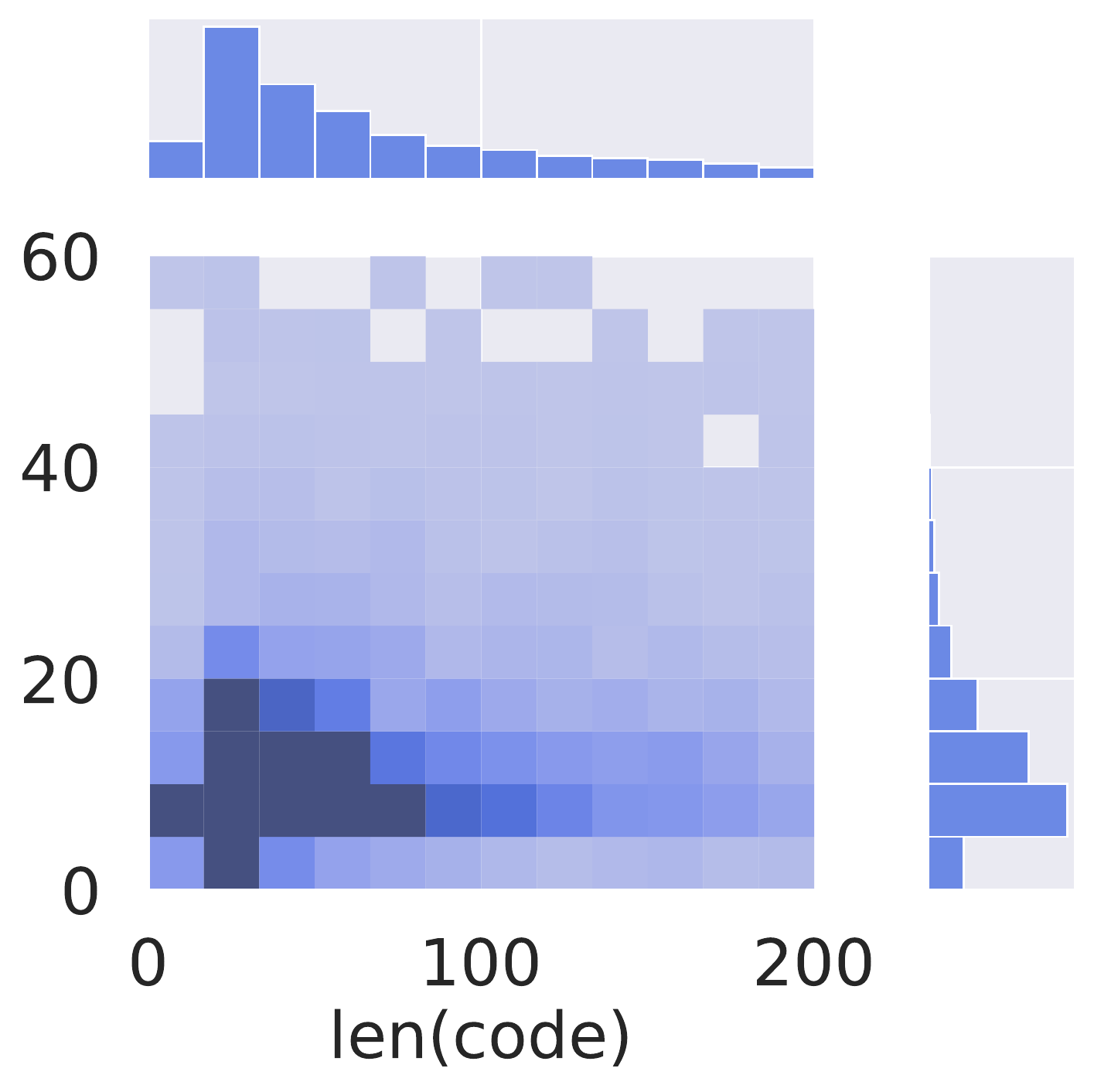}\end{minipage}
 & \UseMacro{TH-ds-CP-T}
 & \begin{minipage}{.18\textwidth}\includegraphics[width=\textwidth]{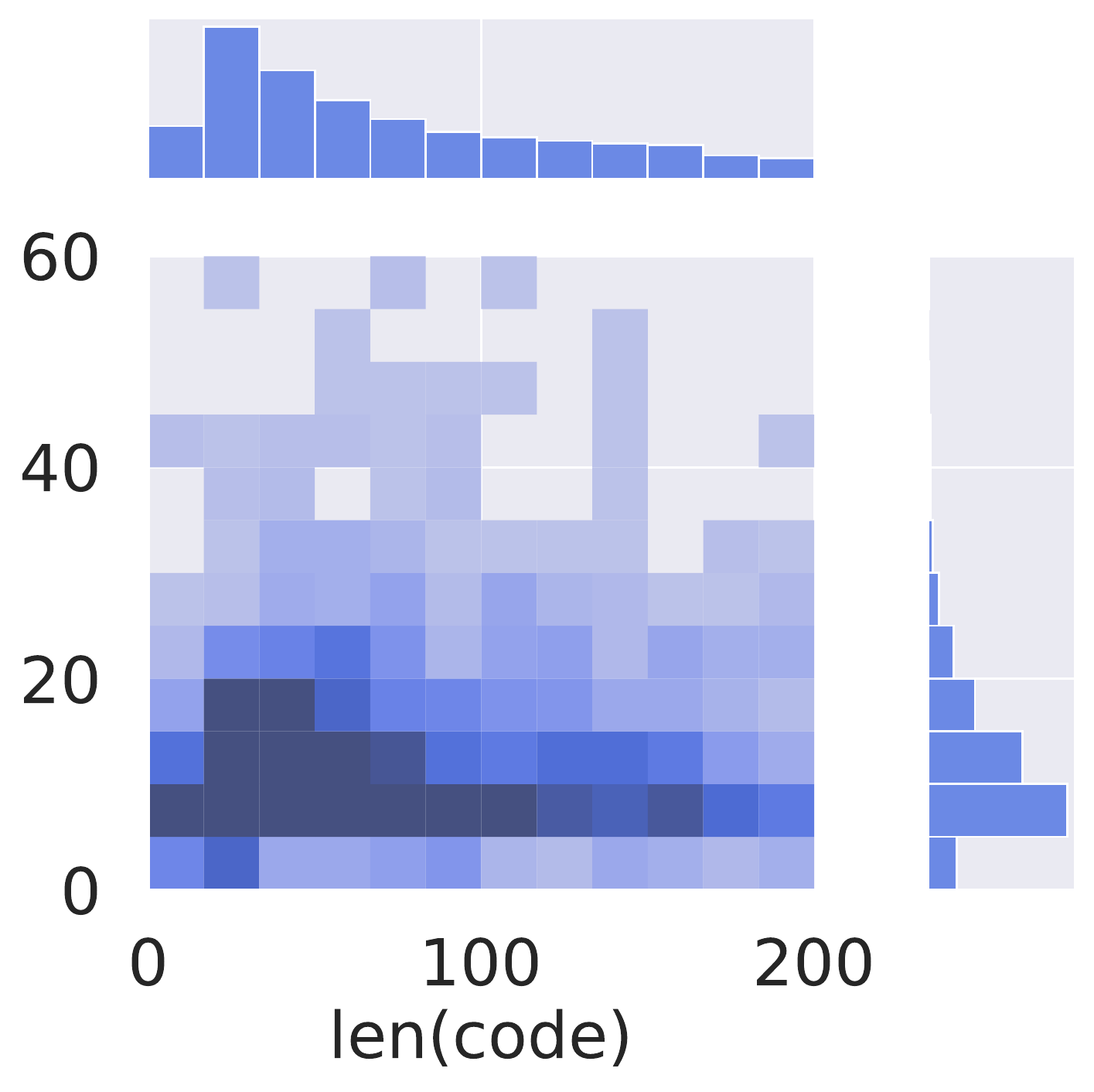}\end{minipage}
\\
\hline
\end{tabular}
\end{footnotesize}
\end{center}

\vspace{-15pt}
\caption{Results of \UseMacro{task_\task} models with filtering
  near-duplicates using the \UseMacro{nd_\ndconfig} configuration, on
  \ATestC sets.\label{fig:nd-results:\task-\ndconfig}}
\end{figure}

\clearpage

}}

\begin{table}[t]
\begin{footnotesize}
\begin{center}
\begin{tabular}{@{\hspace{0pt}} l @{\hspace{0pt}} | @{\hspace{2pt}}r@{\hspace{2pt}} | @{\hspace{2pt}}r@{\hspace{2pt}} | r@{\hspace{2pt}} | @{\hspace{2pt}}r@{\hspace{2pt}} | r@{\hspace{2pt}} | @{\hspace{2pt}}r @{\hspace{0pt}}}
\toprule
\makecell[c]{\UseMacro{TH-train-on}}
 & \multicolumn{2}{c|}{\UseMacro{TH-MP}}
 & \multicolumn{2}{c|}{\UseMacro{TH-CP}}
 & \multicolumn{2}{c}{\UseMacro{TH-T}}
\\ \cline{2-3} \cline{4-5} \cline{6-7}
\makecell[c]{\UseMacro{TH-test-on}}
 & \UseMacro{TH-val}
 & \UseMacro{TH-test_standard}
 & \UseMacro{TH-val}
 & \UseMacro{TH-test_standard}
 & \UseMacro{TH-val}
 & \UseMacro{TH-test_standard}
\\
\midrule
\midrule
\multicolumn{7}{c}{\UseMacro{TH-metric-table-bleu}} \\
\midrule
\UseMacro{TH-model-DeepComHybridESE19}
 & \UseMacro{result-CG_MP_DeepComHybridESE19_val_bleu}
 & \UseMacro{result-CG_MP_DeepComHybridESE19_test_standard_bleu}
 & \UseMacro{result-CG_CP_DeepComHybridESE19_val_bleu}
 & \UseMacro{result-CG_CP_DeepComHybridESE19_test_standard_bleu}
 & \UseMacro{result-CG_T_DeepComHybridESE19_val_bleu}
 & \UseMacro{result-CG_T_DeepComHybridESE19_test_standard_bleu}
\\
\UseMacro{TH-model-RNNBaseline}
 & \UseMacro{result-CG_MP_RNNBaseline_val_bleu}
 & \UseMacro{result-CG_MP_RNNBaseline_test_standard_bleu}
 & $^{\alpha}$\UseMacro{result-CG_CP_RNNBaseline_val_bleu}
 & $^{\beta}$\UseMacro{result-CG_CP_RNNBaseline_test_standard_bleu}
 & \UseMacro{result-CG_T_RNNBaseline_val_bleu}
 & \UseMacro{result-CG_T_RNNBaseline_test_standard_bleu}
\\
\UseMacro{TH-model-TransformerACL20}
 & \textbf{\UseMacro{result-CG_MP_TransformerACL20_val_bleu}}
 & \textbf{\UseMacro{result-CG_MP_TransformerACL20_test_standard_bleu}}
 & $^{\alpha}$\textbf{\UseMacro{result-CG_CP_TransformerACL20_val_bleu}}
 & $^{\beta}$\textbf{\UseMacro{result-CG_CP_TransformerACL20_test_standard_bleu}}
 & \textbf{\UseMacro{result-CG_T_TransformerACL20_val_bleu}}
 & \textbf{\UseMacro{result-CG_T_TransformerACL20_test_standard_bleu}}
\\
\midrule
\midrule
\multicolumn{7}{c}{\UseMacro{TH-metric-table-meteor}} \\
\midrule
\UseMacro{TH-model-DeepComHybridESE19}
 & \UseMacro{result-CG_MP_DeepComHybridESE19_val_meteor}
 & \UseMacro{result-CG_MP_DeepComHybridESE19_test_standard_meteor}
 & \UseMacro{result-CG_CP_DeepComHybridESE19_val_meteor}
 & \UseMacro{result-CG_CP_DeepComHybridESE19_test_standard_meteor}
 & \UseMacro{result-CG_T_DeepComHybridESE19_val_meteor}
 & \UseMacro{result-CG_T_DeepComHybridESE19_test_standard_meteor}
\\
\UseMacro{TH-model-RNNBaseline}
 & \UseMacro{result-CG_MP_RNNBaseline_val_meteor}
 & \UseMacro{result-CG_MP_RNNBaseline_test_standard_meteor}
 & $^{\gamma}$\textbf{\UseMacro{result-CG_CP_RNNBaseline_val_meteor}}
 & $^{\delta}$\textbf{\UseMacro{result-CG_CP_RNNBaseline_test_standard_meteor}}
 & \UseMacro{result-CG_T_RNNBaseline_val_meteor}
 & \UseMacro{result-CG_T_RNNBaseline_test_standard_meteor}
\\
\UseMacro{TH-model-TransformerACL20}
 & \textbf{\UseMacro{result-CG_MP_TransformerACL20_val_meteor}}
 & \textbf{\UseMacro{result-CG_MP_TransformerACL20_test_standard_meteor}}
 & $^{\gamma}$\UseMacro{result-CG_CP_TransformerACL20_val_meteor}
 & $^{\delta}$\UseMacro{result-CG_CP_TransformerACL20_test_standard_meteor}
 & \textbf{\UseMacro{result-CG_T_TransformerACL20_val_meteor}}
 & \textbf{\UseMacro{result-CG_T_TransformerACL20_test_standard_meteor}}
\\
\midrule
\midrule
\multicolumn{7}{c}{\UseMacro{TH-metric-table-rouge_l_f}} \\
\midrule
\UseMacro{TH-model-DeepComHybridESE19}
 & \UseMacro{result-CG_MP_DeepComHybridESE19_val_rouge_l_f}
 & \UseMacro{result-CG_MP_DeepComHybridESE19_test_standard_rouge_l_f}
 & \UseMacro{result-CG_CP_DeepComHybridESE19_val_rouge_l_f}
 & \UseMacro{result-CG_CP_DeepComHybridESE19_test_standard_rouge_l_f}
 & \UseMacro{result-CG_T_DeepComHybridESE19_val_rouge_l_f}
 & \UseMacro{result-CG_T_DeepComHybridESE19_test_standard_rouge_l_f}
\\
\UseMacro{TH-model-RNNBaseline}
 & \UseMacro{result-CG_MP_RNNBaseline_val_rouge_l_f}
 & \UseMacro{result-CG_MP_RNNBaseline_test_standard_rouge_l_f}
 & $^{\epsilon}$\UseMacro{result-CG_CP_RNNBaseline_val_rouge_l_f}
 & $^{\zeta}$\UseMacro{result-CG_CP_RNNBaseline_test_standard_rouge_l_f}
 & \UseMacro{result-CG_T_RNNBaseline_val_rouge_l_f}
 & \UseMacro{result-CG_T_RNNBaseline_test_standard_rouge_l_f}
\\
\UseMacro{TH-model-TransformerACL20}
 & \textbf{\UseMacro{result-CG_MP_TransformerACL20_val_rouge_l_f}}
 & \textbf{\UseMacro{result-CG_MP_TransformerACL20_test_standard_rouge_l_f}}
 & $^{\epsilon}$\textbf{\UseMacro{result-CG_CP_TransformerACL20_val_rouge_l_f}}
 & $^{\zeta}$\textbf{\UseMacro{result-CG_CP_TransformerACL20_test_standard_rouge_l_f}}
 & \textbf{\UseMacro{result-CG_T_TransformerACL20_val_rouge_l_f}}
 & \textbf{\UseMacro{result-CG_T_TransformerACL20_test_standard_rouge_l_f}}
\\
\midrule
\midrule
\multicolumn{7}{c}{\UseMacro{TH-metric-table-exact_match}} \\
\midrule
\UseMacro{TH-model-DeepComHybridESE19}
 & \UseMacro{result-CG_MP_DeepComHybridESE19_val_exact_match}
 & \UseMacro{result-CG_MP_DeepComHybridESE19_test_standard_exact_match}
 & $^{\eta}$\textbf{\UseMacro{result-CG_CP_DeepComHybridESE19_val_exact_match}}
 & $^{\theta}$$^{\iota}$\UseMacro{result-CG_CP_DeepComHybridESE19_test_standard_exact_match}
 & \UseMacro{result-CG_T_DeepComHybridESE19_val_exact_match}
 & \UseMacro{result-CG_T_DeepComHybridESE19_test_standard_exact_match}
\\
\UseMacro{TH-model-RNNBaseline}
 & \UseMacro{result-CG_MP_RNNBaseline_val_exact_match}
 & \UseMacro{result-CG_MP_RNNBaseline_test_standard_exact_match}
 & \UseMacro{result-CG_CP_RNNBaseline_val_exact_match}
 & $^{\theta}$$^{\kappa}$\UseMacro{result-CG_CP_RNNBaseline_test_standard_exact_match}
 & \UseMacro{result-CG_T_RNNBaseline_val_exact_match}
 & \UseMacro{result-CG_T_RNNBaseline_test_standard_exact_match}
\\
\UseMacro{TH-model-TransformerACL20}
 & \textbf{\UseMacro{result-CG_MP_TransformerACL20_val_exact_match}}
 & \textbf{\UseMacro{result-CG_MP_TransformerACL20_test_standard_exact_match}}
 & $^{\eta}$\UseMacro{result-CG_CP_TransformerACL20_val_exact_match}
 & $^{\iota}$$^{\kappa}$\textbf{\UseMacro{result-CG_CP_TransformerACL20_test_standard_exact_match}}
 & \textbf{\UseMacro{result-CG_T_TransformerACL20_val_exact_match}}
 & \textbf{\UseMacro{result-CG_T_TransformerACL20_test_standard_exact_match}}
\\
\bottomrule
\end{tabular}
\end{center}
\end{footnotesize}
\vspace{\UseMacro{TV-results-aux-CG}}
\caption{\UseMacro{TC-results-aux-CG}}
\end{table}

\begin{table}[t]
\begin{footnotesize}
\begin{center}
\begin{tabular}{@{\hspace{0pt}} l @{\hspace{0pt}} | @{\hspace{2pt}}r@{\hspace{2pt}} | @{\hspace{2pt}}r@{\hspace{2pt}} | r@{\hspace{2pt}} | @{\hspace{2pt}}r@{\hspace{2pt}} | r@{\hspace{2pt}} | @{\hspace{2pt}}r @{\hspace{0pt}}}
\toprule
\makecell[c]{\UseMacro{TH-train-on}}
 & \multicolumn{2}{c|}{\UseMacro{TH-MP}}
 & \multicolumn{2}{c|}{\UseMacro{TH-CP}}
 & \multicolumn{2}{c}{\UseMacro{TH-T}}
\\ \cline{2-3} \cline{4-5} \cline{6-7}
\makecell[c]{\UseMacro{TH-test-on}}
 & \UseMacro{TH-val}
 & \UseMacro{TH-test_standard}
 & \UseMacro{TH-val}
 & \UseMacro{TH-test_standard}
 & \UseMacro{TH-val}
 & \UseMacro{TH-test_standard}
\\
\midrule
\midrule
\multicolumn{7}{c}{\UseMacro{TH-metric-table-set_match_p}} \\
\midrule
\UseMacro{TH-model-Code2VecPOPL19}
 & \textbf{\UseMacro{result-MN_MP_Code2VecPOPL19_val_set_match_p}}
 & \textbf{\UseMacro{result-MN_MP_Code2VecPOPL19_test_standard_set_match_p}}
 & \UseMacro{result-MN_CP_Code2VecPOPL19_val_set_match_p}
 & \UseMacro{result-MN_CP_Code2VecPOPL19_test_standard_set_match_p}
 & \textbf{\UseMacro{result-MN_T_Code2VecPOPL19_val_set_match_p}}
 & \textbf{\UseMacro{result-MN_T_Code2VecPOPL19_test_standard_set_match_p}}
\\
\UseMacro{TH-model-Code2SeqICLR19}
 & \UseMacro{result-MN_MP_Code2SeqICLR19_val_set_match_p}
 & \UseMacro{result-MN_MP_Code2SeqICLR19_test_standard_set_match_p}
 & \textbf{\UseMacro{result-MN_CP_Code2SeqICLR19_val_set_match_p}}
 & \textbf{\UseMacro{result-MN_CP_Code2SeqICLR19_test_standard_set_match_p}}
 & \UseMacro{result-MN_T_Code2SeqICLR19_val_set_match_p}
 & \UseMacro{result-MN_T_Code2SeqICLR19_test_standard_set_match_p}
\\
\midrule
\midrule
\multicolumn{7}{c}{\UseMacro{TH-metric-table-set_match_r}} \\
\midrule
\UseMacro{TH-model-Code2VecPOPL19}
 & \textbf{\UseMacro{result-MN_MP_Code2VecPOPL19_val_set_match_r}}
 & \textbf{\UseMacro{result-MN_MP_Code2VecPOPL19_test_standard_set_match_r}}
 & \UseMacro{result-MN_CP_Code2VecPOPL19_val_set_match_r}
 & \UseMacro{result-MN_CP_Code2VecPOPL19_test_standard_set_match_r}
 & \textbf{\UseMacro{result-MN_T_Code2VecPOPL19_val_set_match_r}}
 & \textbf{\UseMacro{result-MN_T_Code2VecPOPL19_test_standard_set_match_r}}
\\
\UseMacro{TH-model-Code2SeqICLR19}
 & \UseMacro{result-MN_MP_Code2SeqICLR19_val_set_match_r}
 & \UseMacro{result-MN_MP_Code2SeqICLR19_test_standard_set_match_r}
 & \textbf{\UseMacro{result-MN_CP_Code2SeqICLR19_val_set_match_r}}
 & \textbf{\UseMacro{result-MN_CP_Code2SeqICLR19_test_standard_set_match_r}}
 & \UseMacro{result-MN_T_Code2SeqICLR19_val_set_match_r}
 & \UseMacro{result-MN_T_Code2SeqICLR19_test_standard_set_match_r}
\\
\midrule
\midrule
\multicolumn{7}{c}{\UseMacro{TH-metric-table-set_match_f}} \\
\midrule
\UseMacro{TH-model-Code2VecPOPL19}
 & \textbf{\UseMacro{result-MN_MP_Code2VecPOPL19_val_set_match_f}}
 & \textbf{\UseMacro{result-MN_MP_Code2VecPOPL19_test_standard_set_match_f}}
 & \UseMacro{result-MN_CP_Code2VecPOPL19_val_set_match_f}
 & \UseMacro{result-MN_CP_Code2VecPOPL19_test_standard_set_match_f}
 & \textbf{\UseMacro{result-MN_T_Code2VecPOPL19_val_set_match_f}}
 & \textbf{\UseMacro{result-MN_T_Code2VecPOPL19_test_standard_set_match_f}}
\\
\UseMacro{TH-model-Code2SeqICLR19}
 & \UseMacro{result-MN_MP_Code2SeqICLR19_val_set_match_f}
 & \UseMacro{result-MN_MP_Code2SeqICLR19_test_standard_set_match_f}
 & \textbf{\UseMacro{result-MN_CP_Code2SeqICLR19_val_set_match_f}}
 & \textbf{\UseMacro{result-MN_CP_Code2SeqICLR19_test_standard_set_match_f}}
 & \UseMacro{result-MN_T_Code2SeqICLR19_val_set_match_f}
 & \UseMacro{result-MN_T_Code2SeqICLR19_test_standard_set_match_f}
\\
\midrule
\midrule
\multicolumn{7}{c}{\UseMacro{TH-metric-table-exact_match}} \\
\midrule
\UseMacro{TH-model-Code2VecPOPL19}
 & \textbf{\UseMacro{result-MN_MP_Code2VecPOPL19_val_exact_match}}
 & \textbf{\UseMacro{result-MN_MP_Code2VecPOPL19_test_standard_exact_match}}
 & $^{\alpha}$\UseMacro{result-MN_CP_Code2VecPOPL19_val_exact_match}
 & $^{\beta}$\UseMacro{result-MN_CP_Code2VecPOPL19_test_standard_exact_match}
 & \textbf{\UseMacro{result-MN_T_Code2VecPOPL19_val_exact_match}}
 & \textbf{\UseMacro{result-MN_T_Code2VecPOPL19_test_standard_exact_match}}
\\
\UseMacro{TH-model-Code2SeqICLR19}
 & \UseMacro{result-MN_MP_Code2SeqICLR19_val_exact_match}
 & \UseMacro{result-MN_MP_Code2SeqICLR19_test_standard_exact_match}
 & $^{\alpha}$\textbf{\UseMacro{result-MN_CP_Code2SeqICLR19_val_exact_match}}
 & $^{\beta}$\textbf{\UseMacro{result-MN_CP_Code2SeqICLR19_test_standard_exact_match}}
 & \UseMacro{result-MN_T_Code2SeqICLR19_val_exact_match}
 & \UseMacro{result-MN_T_Code2SeqICLR19_test_standard_exact_match}
\\
\bottomrule
\end{tabular}
\end{center}
\end{footnotesize}
\vspace{\UseMacro{TV-results-aux-MN}}
\caption{\UseMacro{TC-results-aux-MN}}
\end{table}

\twocolumn

\end{document}